\documentclass[ 11pt, a4paper]{article}
               
\usepackage[T1]{fontenc}
\usepackage{mathptmx}
\usepackage{amsmath,amsfonts,amssymb}
\usepackage{graphicx}\graphicspath{{Graphics/}}
\usepackage[onehalfspacing]{setspace}
\usepackage{ragged2e} %\justifying
\usepackage{textcomp}
\usepackage{caption}
\usepackage[square,comma,sort&compress]{natbib}
\usepackage[bottom]{footmisc}
 
\usepackage[toc,page]{appendix}
\usepackage{bm}
\usepackage{float}
\usepackage{booktabs,multirow}
\usepackage[table]{xcolor}
\usepackage{fancyhdr}
\usepackage{pifont}
\usepackage{marvosym}
\usepackage{hyperref}
 \hypersetup{
     bookmarks=true,         % show bookmarks     
     colorlinks=true,        % false: boxed links; true: colored links     
     citecolor=blue,         % color of links to bibliography
     linkcolor=black,
    } 
\usepackage{cleveref}
\usepackage{fancybox}
\usepackage{epsfig}
\usepackage{epstopdf}
\usepackage[top=1in, bottom=1in, left=1.25in, right=1in]{geometry} 
\usepackage{subcaption}
\usepackage{color}
\definecolor{forestgreen}{rgb}{0.13, 0.55, 0.13}
\definecolor{brown}{rgb}{0.24, 0.17, 0.12}
\definecolor{orange}{rgb}{0.93, 0.53, 0.18}
\newcommand{\mi}{\boldsymbol{-} \mathrel{\mkern -16mu} \boldsymbol{-}}
\title{Large deformation electrohydrodynamics of an elastic capsule in DC electric field}
\author{Sudip Das and Rochish M. Thaokar\\
Department of Chemical Engineering, Indian Institute of Technology Bombay,
Mumbai,\\ 400076, India
}
\def\l[{{[\![}}
\def\r]{{]\!]}}

\def\be{\begin{equation}}
\def\ee{\end{equation}}
\def\.{\cdot}

\date{}

\begin{document}
\maketitle

%===================================
% \pagenumbering{roman}
% \pagestyle{plain}

% \include{Frontmatter/title}
% \include{Frontmatter/dec}

%===================================
% % %
% \clearpage\mbox{\thispagestyle{empty}}\clearpage
% \pagenumbering{arabic}
% 
% 
% \setcounter{page}{1}

% \pagestyle{fancy}
% \include{Frontmatter/abs}
\begin{abstract}
The dynamics of a spherical elastic capsule, containing a Newtonian fluid bounded by an elastic membrane and immersed in another Newtonian fluid, in a uniform DC electric field is investigated. Discontinuity of electrical properties such as conductivities of the internal and external fluid media as well as capacitance and conductance of the membrane lead to a net interfacial Maxwell stress which can cause the deformation of such an elastic capsule. We investigate this problem considering well established membrane laws for a thin elastic membrane, with fully resolved hydrodynamics in the Stokes flow limit and describe the electrostatics using the capacitor model. In the limit of small deformation, the analytical theory predicts the dynamics fairly satisfactorily. Large deformations at high capillary number though necessitate a numerical approach (Boundary element method in the present case) to solve this highly non-linear problem. Akin to vesicles, at intermediate times, highly nonlinear biconcave shapes along with squaring and hexagon like shapes are observed when the outer medium is more conducting. The study identifies the essentiality of parameters such as high membrane capacitance, low membrane conductance, low hydrodynamic time scales and high capillary number for observation of these shape transitions. The transition is due to large compressive  Maxwell stress at the poles at intermediate times. Thus such shape transition can be seen in spherical globules admitting electrical capacitance, possibly, irrespective of the nature of the interfacial restoring force. 
\end{abstract}
\section{Introduction}
An elastic capsule can be considered to be a drop of a liquid separated from another liquid by a very thin elastic membrane~\citep{barthesbiesel81}. The role of such an elastic interface is not only to prevent permeation of one fluid into another, thereby preventing mixing of contents, but also maintain the integrity of the capsule. The membrane can enclose a complex functional fluid, for example, biological cells which are capsules with membranes enclosing cell organelles embedded in a thick cytoplasm. These membranes are made up of amphiphilic molecules such as lipids which form bilayers, block copolymers which can form monolayers or crosslinked surface active moieties~\citep{karyappa14,kataoka01}. They are popular as encapsulating agents and are used as carriers for drugs, genes and other biotechnologically relevant materials~\citep{kataoka01}. 

Biologically important capsules such as Red-blood cells or capsules used in technological applications, are subject to large external forces, such as shear or electric forces, during their life-span. These forces can cause significant deformation in the capsules resulting in membrane stresses and shape deformation. From a mechanics point of view, a membrane of an elastic capsule can be considered as a two dimensional interface, characterized by an elastic modulus, that undergoes in-plane deformation (strain) to generate in-plane elastic stresses~\citep{barthesbiesel81} etc. These elastic stresses resist the applied forces such that a force balance and thereby a steady shape is attained and the capsule retains its integrity. Different kinds of constitutive relationships relating in-plane finite deformation, strain measures to in-plane stresses have been developed for a variety of membrane materials having technological or biological significance. Previously used elastic solid~\citep{lighthill68} and liquid droplet models~\citep{hyman72} have now been replaced by thin viscoelastic membrane models with properties, such as elastic and viscous moduli of the membrane, taking account of the microscopic response corresponding to the microstructure of the capsule. Among several membrane constitutive laws, the Mooney-Rivlin law~\citep{green60} for a strain softening membrane and Skalak law~\citep{skalak73} for a strain hardening membrane are popularly used especially in the computational analyses of elastic capsules~\citep{gao01,fery07,keller2006,dubreuil2003,chang93}.

The deformation of elastic capsules subjected to weak forces (shear or electric) have been studied using asymptotic theories that assume an expansion parameter which is typically the capillary number (a ratio of destabilizing force typically shear or electric to the stabilizing elastic force) and the deformation is assumed to be proportional to the capillary number~\citep{barthesbiesel80,barthesbiesel81,BARTHESBIESEL81a,barthesbiesel91,jong00,rt16,kessler09,seifert11}. In most examples of practical relevance though, the capillary number, and thereby the deformation, can be quite high. The nonlinear elastic response becomes relevant and an accurate description is possible only by a suitable numerical method.

In the last three decades, several computational studies addressing a variety of issues in large deformation problems in elastic capsules have been reported. \citet{zhou95} studied the deformation of a capsule with an area incompressible membrane in simple shear flow using the boundary integral method. Such a membrane develops tension because of the imposed area incompressibility under applied shear flow. Later several other investigations reported the study on the deformation of capsules and red blood cells in shear flow and looked at aspects such as deformation under small and moderate capillary numbers (proportional to the rate of shear)~\citet{seifert11}, effects of viscosity ratio on deformation as well as tumbling and tanktreading~\citet{rao94,pozrikidis95,ramanujan98,pozrikidis01,seifert11,duc10,kessler08,sui08}, effects of initial shape of the capsule as well as membrane constitutive laws on deformation and instability beyond a critical capillary number~\citet{sui07,sui10,skotheim07,navot98}. Significant sensitivity to membrane constitutive laws has also been investigated, for example the computational studies on the deformation of a capsule with strain hardening and strain softening membrane under linear shear flow~\citep{barthesbiesel91,lac04,lac05,sui08a} indicates that membrane resistance in a strain hardening capsule (obeying the Skalak law~\citep{skalak73}) may prevent its burst. 

In the large deformation limit, the numerical analysis of elastic interfaces interacting with fluids lead to a coupled solid-fluid interaction problem, which is known to present formidable numerical challenges. For instance, elastic instabilities~\citep{seifert06, dodson09} are commonly encountered in their numerical solution, thereby severely limiting analysis of large deformation problems. A membrane that offers resistance to bending can suppress these numerical elastic instabilities~\citep{kwak01,sui07,dupont15,pozrikidis01}. Several computational studies on the dynamics of elastic capsules ignore the bending resistance~\citep{zhou95,pozrikidis95,lac04}, predominantly because the bending force depends upon the fourth derivative of the position vector, thereby introducing numerical errors. Thus, although highly relevant, a good numerical model for addressing large deformation problems remains a challenge. 

The present work addresses the large deformation of capsules induced by strong uniform DC electric fields. The response of capsules to electric fields is important in biomedical applications such as electroporation~\citep{crowley73,tieleman04}, electrofusion~\citep{neumann89}, cell separation and manipulation~\citep{ronald96,kang08}. Moreover, the response of spherical globules such as drops, vesicles and capsules to electric field can be used to understand the interfacial properties as well as novel aspects of electrohydrodynamics~\citep{dimova10,rtsd10,shivrajrt12,deshmukh_thaokar_2013,karyappa14,karyappa_deshmukh14,skd15,mcconnell15sm,vlahovska15,rt16}. Uniform field also implies axisymmetry in most cases, thereby allowing easy-to-analyze experiments as well as elegant analytical theory and simplified numerical models, such as using the boundary element method. 

Relatively few studies have been reported on the deformation of elastic capsules under DC~\citep{jong00} and AC~\citep{rt16} electric fields, both restricted to small deformation limit. Experiments, small deformation theory and boundary integral calculations of a capsule, with a conducting membrane, enclosing a conducting fluid and suspended in a dielectric external fluid medium, showed that its deformation can be quite different from that of drops, and that it can be used to estimate linear and nonlinear elastic properties of the capsules~\citep{karyappa14}. 

Experimental investigations~\citep{karin06,dimova07,dimova09,dimova10} on the deformation of a vesicle in a strong electric field showed squaring of the vesicle with high values of tension at the corners. This was also reaffirmed and explained by a few 2-D~\citep{mcconnell15sm,mcconnell15,mcconnell13} and 3-D~\citep{Veerapaneni16,ebrahim15,ebrahim15a} computational analyses. Vesicles and capsules are inherently different at least on a few counts: vesicles are area preserving, which means a spherical vesicle cannot undergo deformation. Thus a vesicle can deform only if it has an excess area, and it resists deformation via the bending forces as well as uniform and non-uniform tensions generated to conserve the area. Capsules on the other hand can resist deformation by elastic forces which arise as deviation from a reference, unstressed shape which can also be a sphere. 

In this work we consider a capsule with a membrane that exhibits Skalak or neo-Hookean (a special case of Mooney-Rivlin model) membrane constitutive law along with the capacitor model for the electric field to study the dynamics of capsule deformation under strong electric field and ask the following questions: what is the response of a capsule characterized by a capacitance and conductance to externally applied uniform DC fields, using analytical theory and Boundary integral calculations. What is the effect of these electrical properties of membrane as well as competition of the various timescales in the problem such as membrane charging time, Maxwell-Wagner relaxation time, membrane deformation time, hydrodynamic relaxation times on the deformation? Is there a possibility of squaring and other large deformation modes in elastic capsules, akin to that reported in vesicles and the necessary conditions thereof? and what are the possible breakup modes and their dependence on the type of membrane models?
 
\section{Problem description}
\begin{figure}[H]
\centering
  \includegraphics[width=0.3\textwidth]{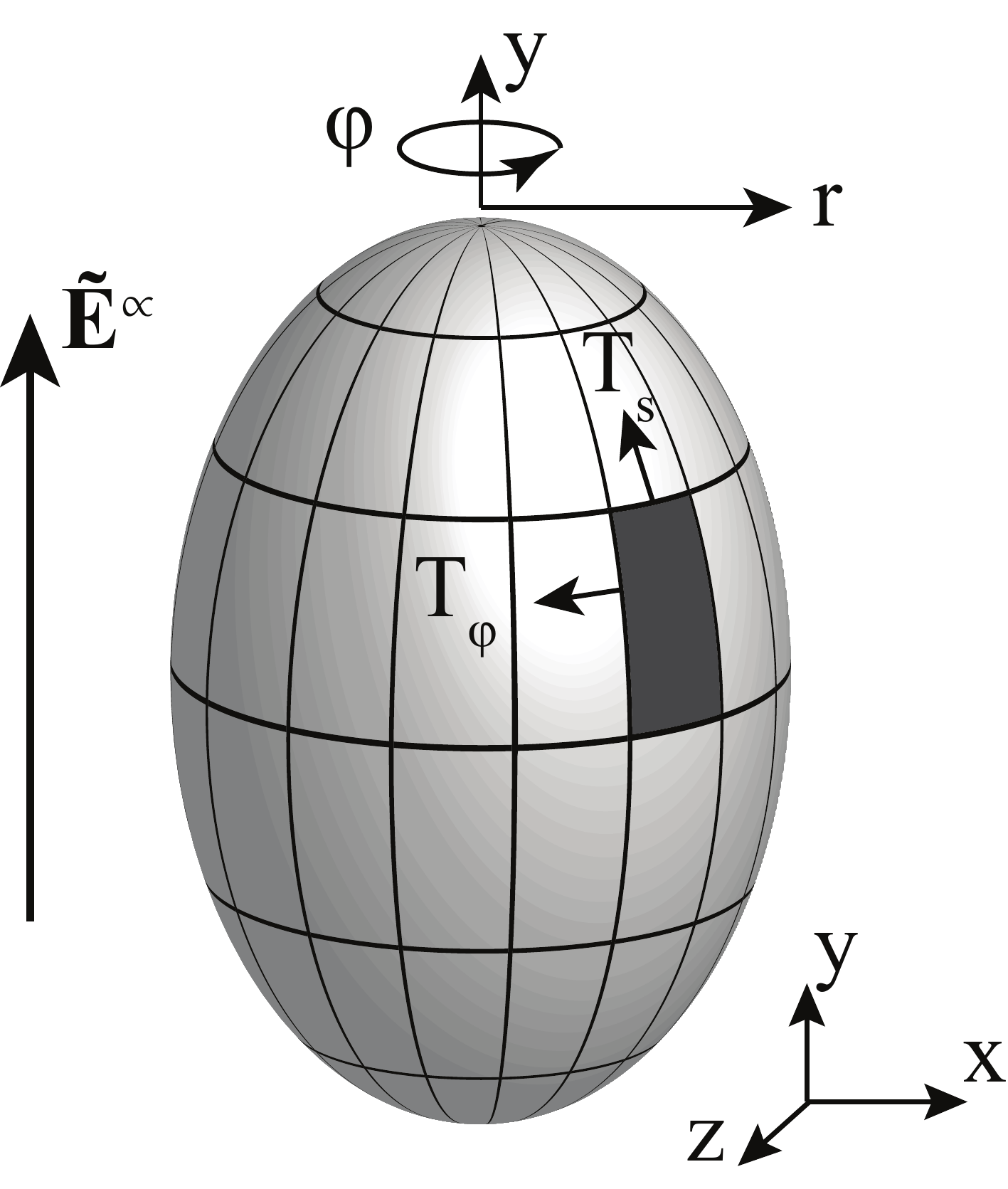}
  \caption{Schematic representation of a capsule in a uniform electric field.}
  \label{fig1}
\end{figure}

The schematic representation of a deformed elastic capsule in uniform electric field is shown in \cref{fig1}. A uniform DC electric field is applied parallel to the $y$ axis which is the axis of symmetry of the capsule. This field is defined by $\tilde{{{\bf E}}}^{\infty}=\tilde{E}_0 {\bf e}_y$, where $\tilde{E}_0$ is the strength of the electric field and thus acts in the $+$ve $y$ direction. All the notations, in this article, with a tilde (\ $\tilde{}$\ ) represent dimensional quantities, whereas without (\ $\tilde{}$\ ) represent their nondimensional counterpart. 

The position vector, as measured from the origin is given by  $\tilde{{\bf x}}=\tilde x {\bf e_x}+\tilde y {\bf e_y}+ \tilde z {\bf e_z}$, where ${\bf {e_x}},\ {\bf {e_y}},$ and  ${\bf {e_z}}$ are the unit vectors in the $x,\ y,$ and  $z$ directions, respectively. The axis of symmetry of the capsule is assumed to coincide with the $y$ axis, such that its centroid is at the origin. In the axisymmetric configuration (cylindrical coordinates), $y$ is the axis of the cylindrical coordinate system, where as $\phi$ is the azimuthal angle. The radial coordinate $r=\sqrt{x^2+z^2}$. When local spherical coordinate system is used in the analytical theory, presented in the appendix, $r=\sqrt{x^2+y^2+z^2}$ and the azimuthal and polar angles are $\phi$ and $\theta$ respectively, where $\theta$ is measured from the $y$ axis in the clockwise direction. ${\bf n}$ and ${\bf t}$ are considered to be the outward unit normal and clockwise unit tangent to the surface of the capsule, respectively.  The dielectric constant and electrical conductivity of internal fluid medium are $\epsilon_i$ and  $\sigma_i$, respectively, whereas the external medium has a dielectric constant and an electrical conductivity given by $\epsilon_e$ and $\sigma_e$, respectively. Subscripts $i$ and $e$ stand for internal and the external fluid media of the capsule, respectively. The ratios of electrical properties of the two media are defined as: $\sigma_r=\sigma_i/\sigma_e$ and $\epsilon_r=\epsilon_i/\epsilon_e$. Both the external and the internal fluids are Newtonian, and the viscosities of the external and internal fluids are denoted by $\mu$ and  $\lambda\mu$ respectively, where $\lambda$ is the viscosity ratio. The mechanical properties of the membrane are elasticity, $E_s$ and the bending modulus, $\kappa_b$ and the electrical properties are capacitance, $ C_m$ and conductance, $ G_m$. The membrane is considered to be purely elastic and viscous resistance is neglected. The time is scaled by the charge relaxation time of the outer fluid, $\tilde {t}_e={\epsilon_e\epsilon_0}/{\sigma_e}$, where $\epsilon_0$ is the permittivity of the free space and the velocity is scaled by $E_s/\mu_e$. The electric field $({\bf \tilde E})$, potential $(\tilde \phi)$ and the stress are scaled by $E_0$, $E_0 a$ and $E_s/a$. Other associated timescales in this problem are hydrodynamic response time $\tilde{t}_H=\mu_ea/E_s$, Maxwell-Wagner charge relaxation time $\tilde{t}_{MW}=\epsilon_0(\epsilon_i+2\epsilon_e)/(\sigma_i+2\sigma_e)$ and membrane charging time $\tilde{t}_{cap}=aC_m(\sigma_i+\sigma_e/2)$. Nondimensional counterparts of all these timescales are  $t_e=1$, $t_H=\tilde{t_H}/\tilde{t}_e$, $t_{MW}=\tilde{t}_{MW}/\tilde{t}_e=(2+\epsilon_r)/(2 +\sigma_r)$ and $t_{cap}=\tilde{t}_{cap}/\tilde{t}_e=\hat{C}_m(1/2+1/\sigma_r)$ where $\hat C_m={a {C_m}}/{\epsilon_e\epsilon_0}$ is the nondimensional membrane capacitance~\citep{grosse92}.

A balance between the Maxwell stresses caused by the electric field, the elastic forces and hydrodynamic forces result in the evolution of a capsule into a new shape. It is of interest to establish a quantitative relationship between the deformation, elasticity and the strength of the electric field. Capsule deformation is analyzed by considering neo-Hookean and Skalak type of membranes representing strain softening and hardening membranes, respectively. For modelling electric response of the capsule to the applied electric field a capacitive membrane model is used which incorporates the capacitance and conductance of the membrane as well as the contrast in conductivity and dielectric constant between the internal and external fluid media. For simplicity of computations the interface is considered to be impermeable, thereby preventing osmotic flows across the membrane. Since the elastic forces at large deformations are known to result in the formation of sharp edges and high curvatures, bending forces are included along with the elastic forces. Moreover, inclusion of resistance to  bending also allows removal of spurious  elastic instabilities known to occur in elastic interfaces. 

\section{Problem formulation}
The deformation of an elastic capsule in uniform electric field is assumed to remain  axisymmetric with the axis parallel to the electric field lines. Therefore, to study the electrohydrodynamics of capsule deformation axisymmetric boundary integral method is considered. The electric, elastic and hydrodynamic forces as well as the bending resistance of the membrane are in equilibrium at the interface for an inertialess membrane and these are discussed next.

\subsection{Electrostatics}\label{sec:electrostatics}
The electric potentials inside and outside of the capsule satisfy the Laplace equation $\nabla^2 \phi_{i,e}=0$. Therefore the electric potential can be obtained as an integral equation by using Green's theorem for the solution of the Laplace equation, to yield,~\citep{mcconnell15,mcconnell15sm,mcconnell13}: 
\begin{eqnarray}\label{eq:pot}
 \frac{1}{2}\phi_i({\bf x}_0) &=& \int_s \left[G^E({\bf x},{\bf x}_0){\bf \nabla}\phi_i({\bf x})\cdot {\bf n}({\bf x})-\phi_i({\bf x}){\bf n}({\bf x})\cdot {\bf \nabla}G^E({\bf x},{\bf x}_0)\right] ds({\bf x}),\\ 
 \frac{1}{2}\phi_e({\bf x}_0) &=& \phi^\infty ({\bf x}_0)-\int_s \left[G^E({\bf x},{\bf x}_0){\bf \nabla}\phi_e({\bf x})\cdot {\bf n}({\bf x})-\phi_e({\bf x}){\bf n}({\bf x})\cdot {\bf \nabla}G^E({\bf x},{\bf x}_0)\right] ds({\bf x}),
\end{eqnarray}
 where $G^E({\bf x}_0,{\bf x})=\frac{1}{4\pi |\hat{{\bf x}}|}$ is the Green function for the Laplace equation, $\phi^{\infty}=-y$ is the applied potential ($\tilde \phi^{\infty} =-E_o \tilde y$). 

 The transmembrane potential, defined as the discontinuity in electrical potential across the interface, is given by 
\begin{equation}\label{eq:phim}
\phi_m=\phi_i-\phi_e 
\end{equation}
 and it can be calculated from the electrical current continuity (a sum of displacement and Ohmic currents) in the direction normal to the plane of the membrane~\citep{debruin99}
 \begin{equation}\label{eq:currentcont}
 \sigma_r E_{n,i}+\epsilon_r\frac{dE_{n,i}}{dt}=E_{n,e}+\frac{dE_{n,e}}{dt}=\hat C_m \frac{d \phi_m}{dt}+ \hat G_m \phi_m,
\end{equation}
where $E_{n,i}$ and $E_{n,e}$ are normal electric fields at inside and outside of be membrane and $ \hat G_m={a {G}_m}/{\sigma_e}$ is the nondimensional membrane conductance. Solving \cref{eq:pot} along with \cref{eq:phim} and \cref{eq:currentcont}, the tangential (derivative of potential with respect to the arc-length) and normal electric fields can be calculated (termed as boundary integral method with capacitor model, BEM-CM). 

 The discontinuity of the electric traction across the interface is given by ${\bf f}^E= (\tilde {\bf \tau}^E_e-\tilde {\bf \tau}^E_i)\cdot{\bf n}=\tilde\tau^{E}_n{\bf n}+\tilde\tau^{E}_t{\bf t}$, where the dimensional Maxwell electric stress  $\tilde{\bf \tau}^E=\epsilon \epsilon_0 \left(\tilde {\bf E}\tilde{\bf  E}-\frac{1}{2} \tilde E^2{\bf I} \right)$, considering ${\bf I}$ as the identity tensor and $\epsilon$ as the dielectric constant of the fluid. The normal and tangential components of the electric stress tensor can be expressed in nondimensional form as
 \begin{eqnarray}
 \tau_n^E &=& \frac{1}{2}Ca\left[(E_{n,e}^2-E_{t,e}^2)-\epsilon_r(E_{n,i}^2-E_{t,i}^2)\right],\\
 \tau_t^E &=& Ca\left(E_{n,e}E_{t,e}-\epsilon_rE_{n,i}E_{t,i}\right),
 \end{eqnarray}
where $Ca=a \epsilon_e\epsilon_0E_0^2/E_s$ is the ratio of the electric force to the elastic force.
 
\subsection{Membrane laws and elastic stress}\label{sec:elasticity}
For an isotropic and area dilating membrane, elastic tensions can be calculated using the Mooney Rivlin model~\citep{green60}. According to the Mooney-Rivlin law, 
the membrane tension in the principal direction $i$ can be expressed as 
\begin{equation}\label{eq:mlnd}
{\tilde T}_{i}^{MR} =\frac{\tilde G_{MR}}{\lambda_i\lambda_j}\left(\lambda_i^2-\frac{1}{\lambda_i^2\lambda_j^2}\right)\left[\alpha+\lambda_j^2(1-\alpha)\right],
\end{equation}
where $G^{MR}$ and $\alpha$ are the surface elastic modulus and measure of the nonlinearity of the Mooney-Rivlin law, respectively and $\lambda_{i,j}$ are stretch ratios in principal directions $i$ and $j$. Membrane nonlinearity parameter $\alpha$ varies in a range of $[0,1]$, a special case corresponding to $\alpha=1$ is termed as neo-Hookean law which is used in this work~\citep{barthesbiesel02}.

The relation between the non-dimensional membrane elastic tensions and the stretch ratios for a strain hardening membrane in the principal direction $i$ is given by~\citet{skalak73} 
\begin{equation}\label{eq:skalaknd}
{\tilde T}_{i}^{SK} = \frac{\tilde G^{SK}}{\lambda_i\lambda_j}\left[\lambda_i^2(\lambda_i^2-1)+C(\lambda_i\lambda_j)^2\{(\lambda_i\lambda_j)^2-1\}\right],
\end{equation}
% $\frac{ 1}{2}\frac{1+C}{1+2C}$
where the first term at the right hand side accounts for the shear deformation associated with the modulus $G^{SK}$ and the second term accounts for the area dilatation with the modulus $C G^{SK}$. The area dilatation parameter $C$ controls the extent of change in area, higher the value of $C$ lower is the membrane dilatation, a small value of $C=1$ is chosen to allow the membrane to change its area during deformation. Membrane tension in the principal direction $j$ can be expressed by interchanging indices in the corresponding constitutive relations (\cref{eq:mlnd} and \cref{eq:skalaknd}). In our consideration $i$ is the meridional and $j$ is the azimuthal principal directions and the corresponding tensions are considered as $T_{s}$ and $T_{\phi}$. In the small deformation limit, $G^{MR}$ and $G^{SK}$ are related to the surface Young modulus, $E_s$,~\citep{barthesbiesel02} as $E_s=3G^{MR}=2G^{SK}(1+2C)/(1+C)$. 

 A combined contribution of the elastic tensions (force/length) obtained from the constitutive laws govern the membrane elastic traction (force/area), given by ${{\bf f}^{el}}=\tau_n^{el} {\bf n}+\tau_t^{el} {\bf t}$, exerted at the interface. The components of the elastic stresses can be obtained as
\begin{eqnarray}
 \tau_n^{el} &=& -(K_sT_{s}+K_{\phi}T_{\phi})\\
 \tau_t^{el} &=& \frac{d T_{s}}{ds}+\frac{1}{r}\frac{dr}{ds}(T_{s}-T_{\phi})
 \end{eqnarray}
where $K_s=\left|\frac{d{\bf t}}{ds}\right|$ and $K_\phi=\frac{n_r}{r}$ are principal curvatures of the meridional surface. $t_y=n_r=\frac{dy}{ds}$ and $t_r=-n_y=\frac{dr}{ds}$ are the $y$ and $r$ components of the unit normal and tangent vectors at the interface, respectively. 

\subsection{Resistance due to bending}
Even though thin elastic membranes have very low resistance to  bending deformations, it can have significant contribution to the restoring forces especially at the regions of high curvature. The overall dimensionless bending force (force/area) is given by
\begin{equation}
 {\bf f}^b =\hat \kappa_b \left[2{ \Delta_s} H+4 H(H^2-K_G)\right] {\bf n},
\end{equation}
where $H$ and $K_G$ are the mean curvature ($H=\frac{1}{2} (K_s+ K_{\phi}))$ and Gaussian curvature ($K_G=K_s K_{\phi}$), respectively and the nondimensional bending rigidity, $\hat \kappa_b=\kappa_b/a^2E_s $ is the ratio of the bending force to the elastic force. $\Delta_sH$ is the Laplace Beltrami of the mean curvature in cylindrical coordinate system, defined as 
\begin{equation}
 \Delta_sH=\nabla_s\cdot(\nabla_sH)=\frac{1}{r\mid{\bf X}_s\mid}\frac{\partial}{\partial s}\left(\frac{r}{\mid{\bf X}_s\mid}\frac{\partial H}{\partial s}\right),
\end{equation}
where $\mid{\bf X}_s\mid=\sqrt{\frac{\partial r}{\partial s}^2+\frac{\partial y}{\partial s}^2}$~\citep{hu14}. Higher order derivatives of mean curvature with respect to arc length are calculated using spectral method for  better accuracy in estimating bending force.
 
\subsection{Hydrodynamics}
Typically capsule sizes range from $0.1$ to few hundreds of microns which means that the flow inside and outside the interface can be considered to obey low Reynolds number hydrodynamics and therefore the Stokes equations suffice to describe the problem. Thus
\begin{equation}\label{eq:stokes}
{\bf \nabla} .{\bf u}=0,\ {\bf \nabla} .{\bf \tau^H}=0,
\end{equation}
where viscous stress for the internal fluid domain $V$ and external fluid domain $V^*$ are given by
\begin{eqnarray}
  {\bf \tau^H} &=& -P{\bf I}+\lambda({\bf \nabla} {\bf u}+{\bf \nabla} {\bf u}^T)\quad \mbox{at\ }\quad V,\\
   {\bf \tau^H} &=& -P{\bf I}+({\bf \nabla} {\bf u}+{\bf \nabla} {\bf u}^T)\quad \mbox{at\ }\quad V^*,
\end{eqnarray}
where ${\bf \tau^H}$ is the hydrodynamic stress tensor, ${\bf u}$, and $P$ are the velocity and pressure fields, respectively. 

The solution to \cref{eq:stokes} results in the interfacial velocity to be described in an integral form as
\begin{equation}\label{eq:veleqn}
{\bf u}({\bf x}_0)=-\frac{1}{1+\lambda}\frac{1}{4\pi}\int_s\Delta{{\bf f}}({\bf x})\cdot{\bf G}({\bf x}_0,{\bf x})  ds({\bf x})-\frac{1}{4\pi}\frac{1-\lambda}{1+\lambda}\int_s {\bf u}({\bf x})\cdot{\bf Q}({\bf x}_0,{\bf x}) \cdot {\bf n}({\bf x})ds({\bf x}),  
\end{equation}
where ${\bf G}({\bf x}_0,{\bf x})$ and ${\bf Q}({\bf x}_0,{\bf x})$ are Green's functions for velocity (stokeslet) and stress (stresslet) respectively and $\Delta{\bf f}$ is the unbalanced non-hydrodynamic traction acting on the interface~\citep{rallison78}. For an unbounded three dimensional flow, the stokeslet and the stresslet are defined as,

\begin{equation}
 {\bf G}({\bf x}_0,{\bf x})=\frac{{\bf I}}{| \hat{{\bf x}}|}-\frac{\hat{\bf x}\hat{\bf x}}{{| \hat{{\bf x}}|}^3}, \quad Q_{ijk}({\bf x}_0,{\bf x}) = -6\frac{\hat{\bf x}\hat{\bf x}\hat{\bf x}}{{|\hat{{\bf x}}|}^5}\quad\mbox{where\ }\quad\hat{{\bf x}}={\bf x}-{\bf x}_0,
 \end{equation}
where ${\bf x}_0$ and ${\bf x}$ are the source point and the observation point, respectively. The viscosity of the two fluids is assumed to be the same, i.e.  $\lambda=1$, whereby the double layer potential, second term in the right hand side of \cref{eq:veleqn},  drops out.  The boundary integral equation (\cref{eq:veleqn}) is solved to obtain the interfacial velocity of an axisymmetric deformable body, capsule, in applied uniform electric field~\citep{shivraj12,shivraj13,karyappa14, poz90}.  The non-hydrodynamic tractions responsible for the deformation of capsule, in this case, include the elastic traction, electric traction and traction due to the resistance to bending. The net non-hydrodynamic traction acting at the interface is given by
\begin{equation}
 \Delta {\bf f}={\bf f}^{el}+{\bf f}^E+{\bf f}^b.
\end{equation} 
Eventually, the deformed shape of the capsule, ${\bf x}(t)$, can be calculated from the kinematic condition given by
 \begin{equation}
  {\bf x}(t)={\bf x}(t-1)+\frac{1}{t_H}{\bf u}\Delta t,
 \end{equation}
 where ${\bf x}(t-1)$ is the shape of the capsule at previous time, $(t-1)$,  $\Delta t$ is the time step used in boundary integral calculation.
\section{Results}
Results are presented for the dynamics of axisymmetric deformation of an elastic capsule, placed in a uniform DC electric field, by considering membrane conductance $\hat G_m=0$, capacitance $\hat C_m=50$, ratio of dielectric constant $\epsilon_r=1$ and the hydrodynamic response time $t_H=1$, unless explicitly mentioned. Assuming very small contribution of the resistance due to bending, a small value of dimensionless rigidity modulus $\kappa_b=0.001$ is considered and the criticality of incorporating bending is clearly demonstrated in~\cref{sec:bending}. Axisymmetric boundary integral method is used to simulate in which the azimuthal elliptic integrals are solved using analytical theory and integrations in meridional plane are carried out numerically using $16-$point Gaussian quadrature considering $60$-collocation points over the arc length between poles (north - south).  
\subsection{Validation of the numerical code}
\begin{figure}[H]
\centering
  \includegraphics[width=0.5\textwidth]{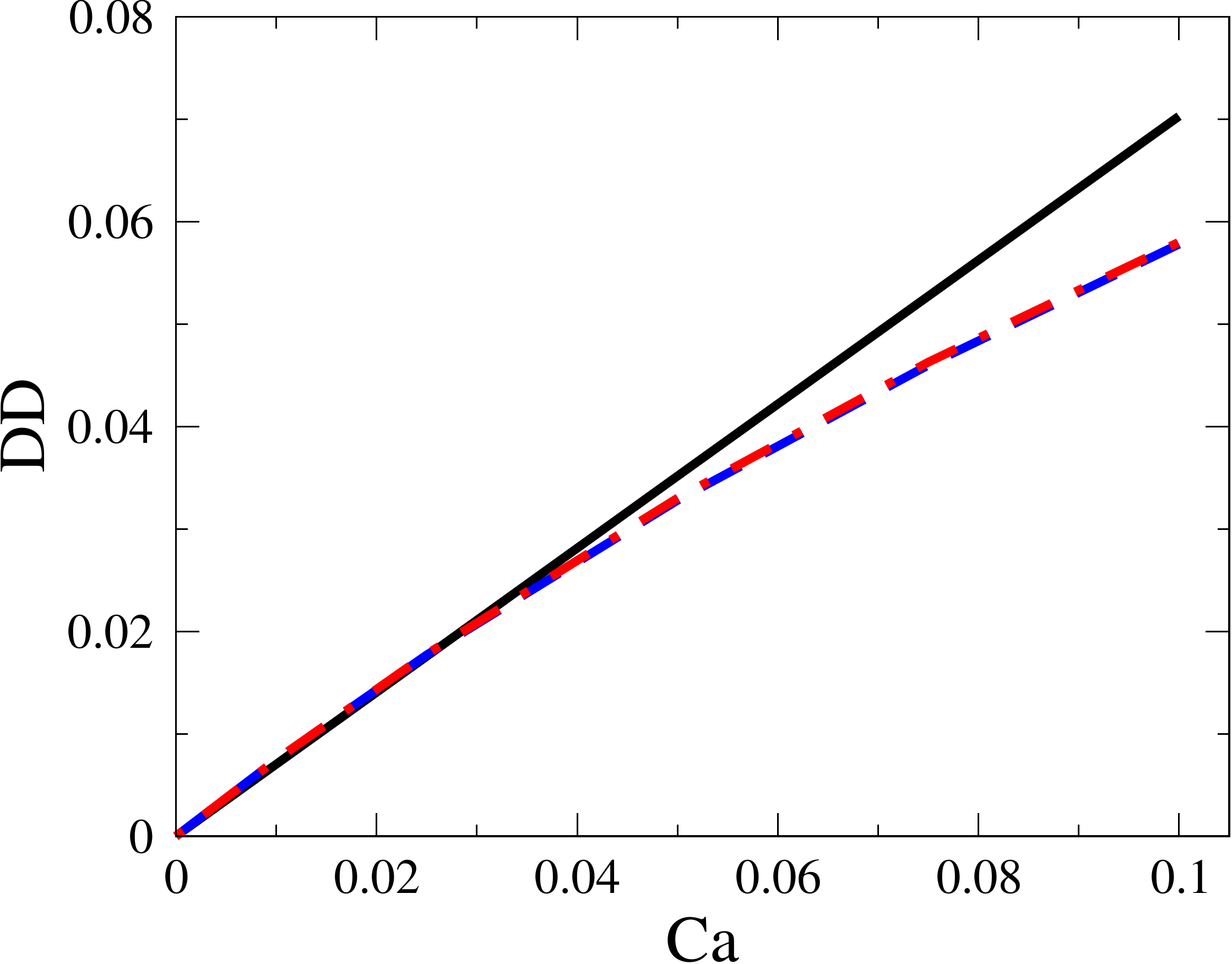}
  \caption{Comparison of degree of deformation as a function of $Ca$ obtained from analytical theory (${\pmb{\mi}}$), boundary integral simulation considering simplified electrostatic model with Skalak law    (\textcolor{red}{$\pmb{-\cdot -}$}) and neo-Hookean law (\textcolor{blue}{$\pmb{--}$}) considering $\hat C_m=50$, $\hat G_m=0$, $\epsilon_r=1$ and $\sigma_r=1$.}
  \label{fig2}
\end{figure}

The deformation of an elastic capsule in electric field can be analytically solved in the small capillary number limit. Steady state deformation obtained from boundary integral method for a neo-Hookean membrane and a Skalak membrane are compared (\cref{fig2}) with the results obtained from simplified electrostatic theory (SEM) (\cref{aapAsm}) considering $\hat C_m=50$, $\hat G_m=0$, $\epsilon_r=1$ and $\sigma_r=1$, (although it does not depend upon $\sigma_r$ which is discussed in~\cref{sec:casel0p1}). It can be observed that the deformations considering neo-Hookean and Skalak membrane law are in good agreement with the analytical theory in the  small deformation limit and both kind of membranes exhibit same degree of deformation which in turn is identical to that given by a Hookean membrane. A deviation is observed at higher capillary numbers because of corrections to both electrostatics as well as elastic deformation. 
 
 Numerical boundary integral method to solve the electrostatics is further validated with the analytical electrostatics theory (AET), discussed in \cref{aapAcm}, which essentially  solves electrostatics equations over an undeformed sphere (\cref{fgr:vmcheck,fgr:taunl1,fgr:tautl1,fgr:vmcheckl0p1,fgr:taunl0p1,fgr:tautl0p1,fgr:vmcheckl10,fgr:taunl10,fgr:tautl10} in  \cref{sec:fieldvalidation}). Analytical theory for  deformation though is not provided in this limit owing to complicated expressions. It should be mentioned here that in the analytical theory as well as in the numerical model, the choice of initial condition on the inner and outer electric field is critical, and is a characteristic of the capacitor model. In this work, we assume that the non-dimensional inner and outer normal electric field at $t=0$ is given by $\cos{\theta}$. This pre-supposes that the capacitor is uncharged at $t=0$, and offers no impedance, thereby the two fluids which have the same dielectric constants (as assumed in this work) have the same initial electric fields.

A simplified electrostatics model (SEM) is also suggested (\cref{aapAsm}) by assuming that the Maxwell-Wagner charge relaxation times is much smaller than the capacitor charging time i.e $t_{MW}<t_{cap}$. In this limit though, analytical solution for the deformation of a capsule is obtained and compared with numerical computations (\cref{fgr:thnumddfl0p1,fgr:thnumddfl1,fgr:thnumddfl10} in \cref{sec:fieldvalidation}). The variation of transmembrane potential obeys $\phi_m=\frac{3}{2}(1-e^{-t/t_{cap}})\cos{\theta},$ which suggests that the charging time is proportional to capacitance and equal to $3 \hat C_m/2$ for $\sigma_r=1$~\citep{grosse92,Schwalbe11}. The membrane becomes fully charged at long times with equal and opposite non-dimensional charge densities on either sides of the membrane (equal to $3 \hat C_m/2$) such that the net charge is zero. 

\subsection{Dynamics of capsule deformation at small and intermediate capillary numbers}
\begin{figure}[H]
\centering
  \includegraphics[width=0.5\textwidth]{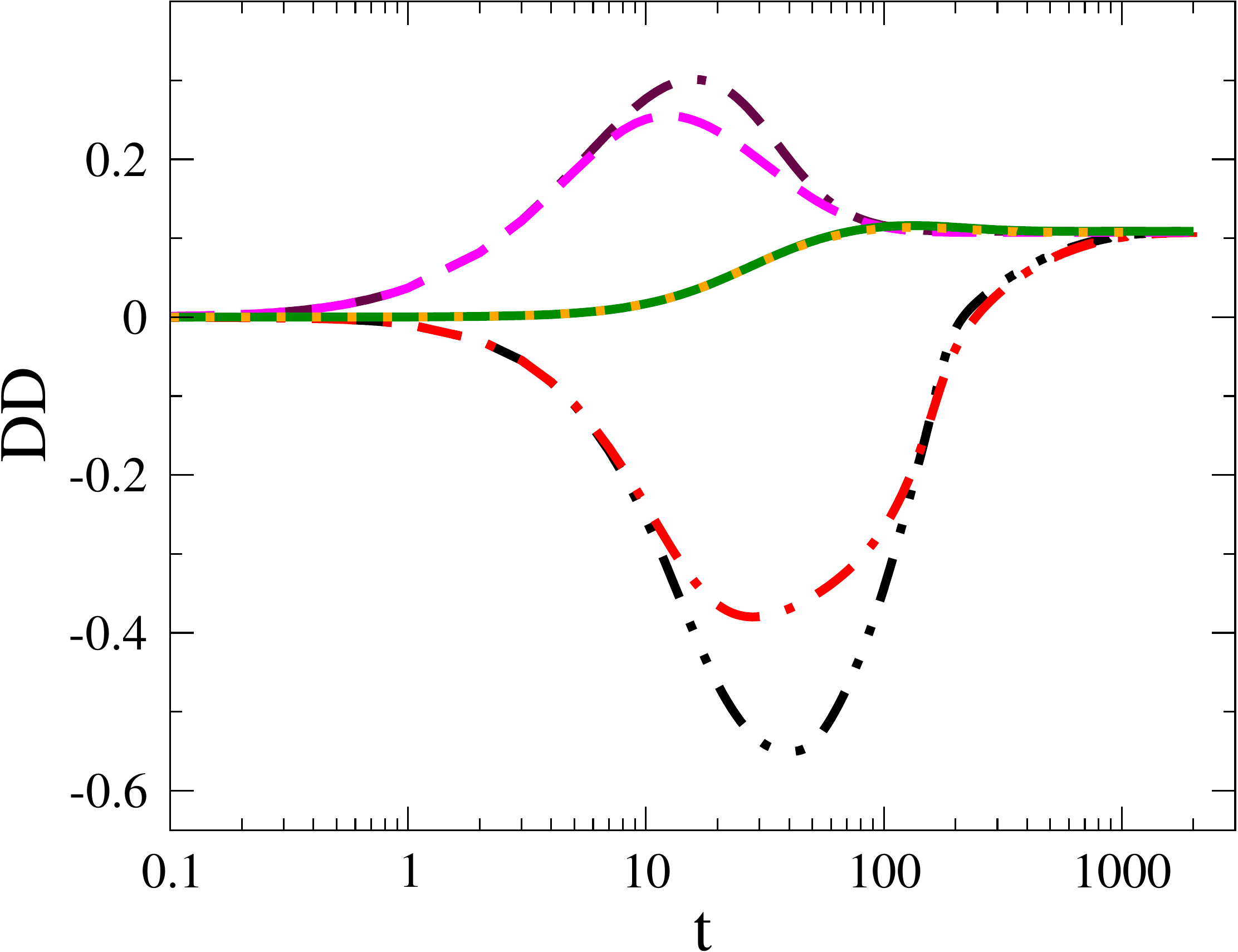}
  \caption{Dynamics for the deformation of a capsule with Skalak membrane at $\sigma_r=10$, $\sigma_r=1$ and $\sigma=0.1$ are shown by curve with line-style (\textcolor{magenta}{$\pmb{--}$}), (\textcolor{orange}{$\pmb{\cdots}$}) and (\textcolor{red}{$\pmb{-\cdot-}$}), respectively where as the deformation for a capsule with neo-Hookean membrane at $\sigma_r=10$, $\sigma_r=1$ and $\sigma=0.1$ are shown by curve with line-style (\textcolor{brown}{$\pmb{\cdot--}$}), (\textcolor{forestgreen}{$\pmb{\mi}$}) and (\textcolor{black}{$\pmb{-\cdot\cdot}$}), respectively at $Ca=0.25$.}
  \label{fgr:dyl0p1}
\end{figure}

The dynamics of deformation of spherical capsules with Skalak or a neo-Hookean membrane at $\sigma_r=1,\ 10$ and $0.1$ is  shown in \cref{fgr:dyl0p1}. These results are obtained using boundary integral calculations considering $\epsilon_r=1$, $\hat C_m=50$ and $\hat G_m=0$ at a small capillary number, $Ca=0.25$. The distinctive behavior of the dynamics at different conductivity ratios and for two different membrane types are explained in the following subsections.   
 
\subsubsection{Case of $\sigma_r=1$}
For $\sigma_r=1$, the  deformation  of a  spherical capsule with either a  Skalak or a neo-Hookean membrane attains an intermediate maximum prolate deformation at $t=132$ and slowly reaches a steady state deformation at $Ca=0.25$ considering $\epsilon_r=1$, $\hat C_m=50$ and $\hat G_m=0$ (\cref{fgr:dyl0p1}). Both Skalak and neo-Hookean capsule show nearly identical  dynamics of deformation. Although there is no dielectric constant or conductivity contrast between the internal and external fluids, a steady state prolate deformation is observed. The normal electric fields are identical in the two fluids at all times, the tangential field though is discontinuous, such that the external fluid has higher magnitude than the inner fluid owing to the transmembrane potential. At t=0, the normal (outward) and tangential fields (anticlockwise) are equal for both internal and external fluids on account of same conductivities and dielectric constants and hence no net Maxwell stress is generated and the deformation is absent. As time progresses the normal electric field falls to zero, over the membrane charging time, since at long times a fully charged non conducting membrane leads to zero current. The tangential field on the other hand continues to be negative (anti-clockwise), with the inner field always lower than the outer and the difference between the two being proportional to the transmembrane potential. At long times, the inner tangential field is zero and the outer is $-3/2\sin\theta$, corresponding to the transmembrane potential (\cref{fgr:vmcheck} in \cref{sec:fieldvalidation}). Thus, the normal Maxwell's stresses arise purely due to the tangential field (normal stress is proportional to $E_n^2-E_t^2$, and $E_n$ is same in the inner and outer surface). Therefore the normal stresses are always compressive, since the tangential electric field on the outer surface is always higher (\cref{fgr:taunl1} in \cref{sec:fieldvalidation}).   The net normal compressive stress can be written as an isotropic compressive part and a tensile part corresponding to the $2^{nd}$ Legendre polynomial of $\cos\theta$. This tensile part leads to prolate deformation.  The tangential stresses are zero at $t=0$, and $t=\infty$, because of absence of electric contrast, and  normal electric fields being zero, respectively. The tangential stresses are from equator to pole at  intermediate times.

\subsubsection{Case of $\sigma_r>1$}
\Cref{fgr:dyl0p1} shows that for $\sigma_r=10$ the capsule initially  undergoes a high prolate deformation and then it slowly relaxes back to a steady state prolate shape. At very short times $t<t_{MW}$, the two fluids act as dielectrics, and since there is no dielectric contrast, the deformation is zero. The two fluids act as leaky dielectrics at time $t\sim O(t_{MW})$ and their conductivity ratio determines their deformation, and increases as $\sigma_r$ increases. The membrane impedance is small since $t_{MW} <  t_{cap}$, and the membrane is not charged. Therefore, the electric field distribution as well as the Maxwell's stresses are similar to that observed in a liquid drop suspended in an ambient fluid and subjected to an electric field. The current continuity condition means that when the inner fluid is more conducting ($\sigma_r=10$), the polarization vector inside the drop is aligned in the direction of electric field, whereby there are net positive charges at the north pole and net negative charges at the south pole. Thus, unlike the $\sigma_r=1$ case, even the normal field contributes to normal stresses and the normal stresses are predominantly tensile (\cref{fgr:taunl10,fgr:vmcheckl10,fgr:taunl10,fgr:tautl10} in \cref{sec:fieldvalidation}), hence a strong prolate deformation is observed. At $t>t_{cap}$, normal electric stress at poles  and tangential stress over the interface disappear as the membrane becomes fully charged. Therefore the prolate steady state deformation is mainly due to the compressive electric stress which is maximum at the equator ( \cref{fgr:vmcheckl10,fgr:taunl10,fgr:tautl10} in \cref{sec:fieldvalidation}). Note that prolate deformation is observed even in the absence of conductivity contrast as discussed earlier for $\sigma_r=1$ and later elaborated for the case of $\sigma_r<1$.

\begin{figure}[H]
\begin{center}
\begin{subfigure}{.22\textwidth}
  \centering
  \includegraphics[width=1\textwidth]{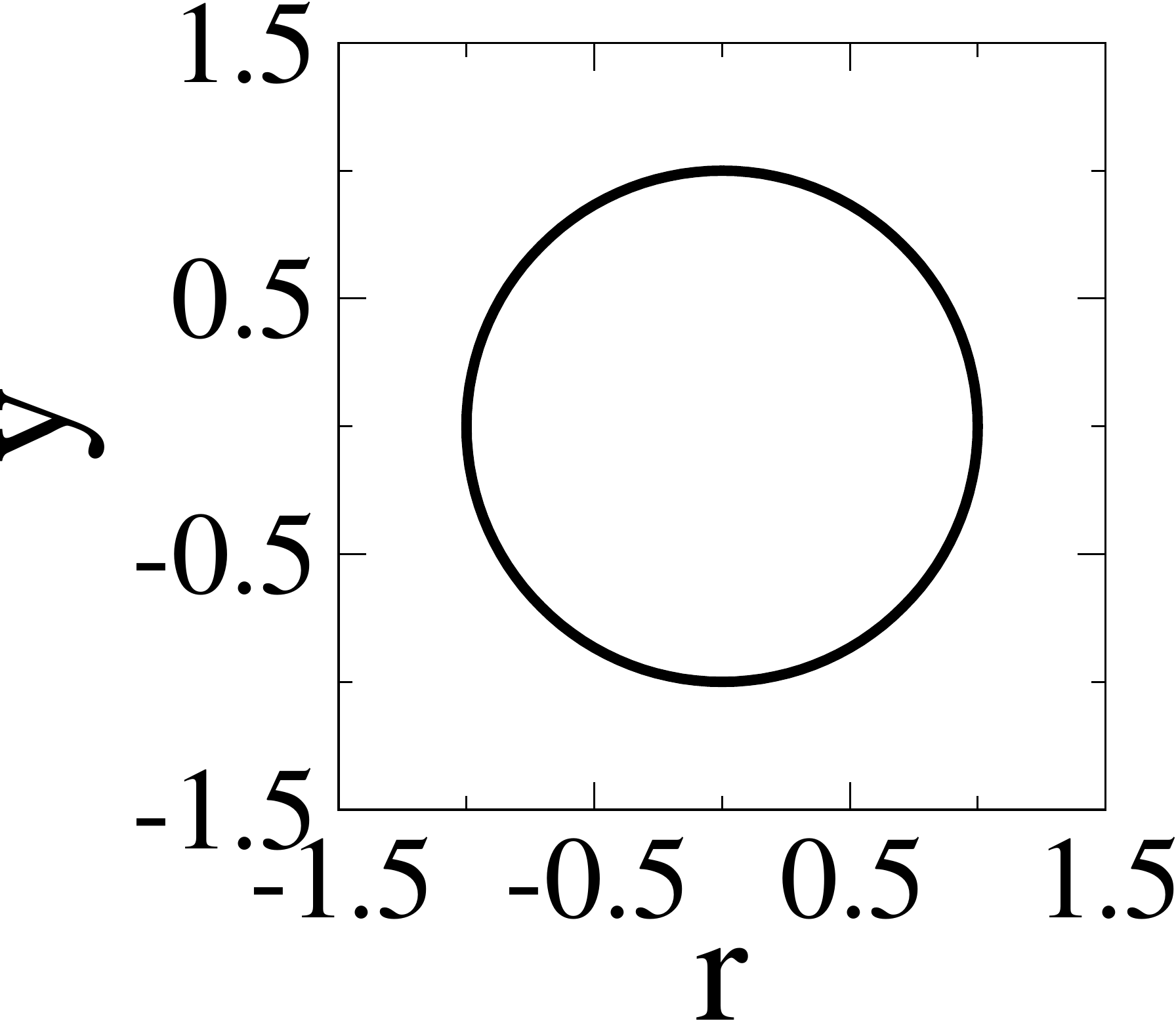}
  \caption{$t=0$}
  \label{fgr:lcal10a}
\end{subfigure}
\begin{subfigure}{.22\textwidth}
  \centering
  \includegraphics[width=1\textwidth]{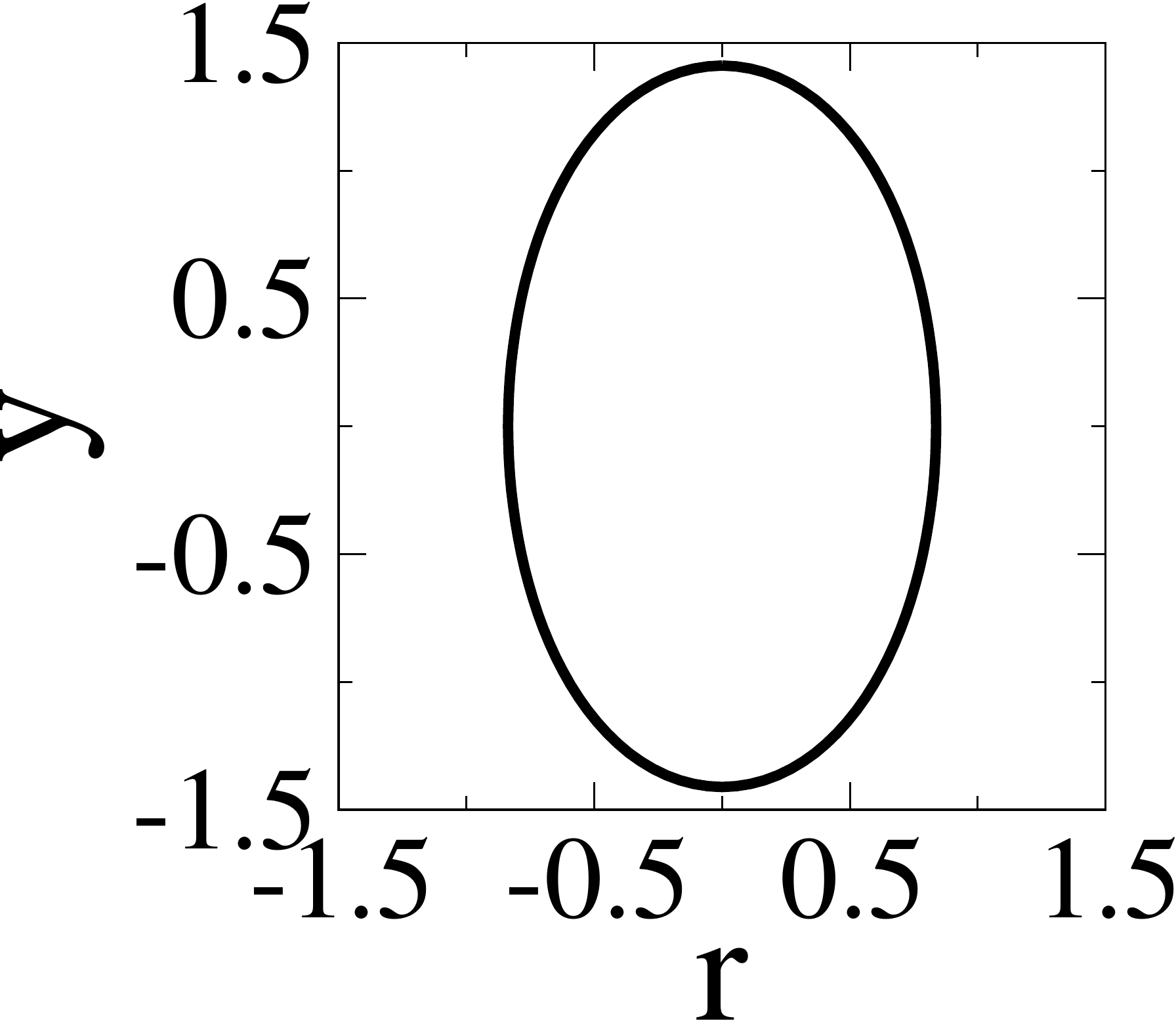}
  \caption{$t=12$}
  \label{fgr:lcal10b}
\end{subfigure}
\begin{subfigure}{.22\textwidth}
  \centering
  \includegraphics[width=1\textwidth]{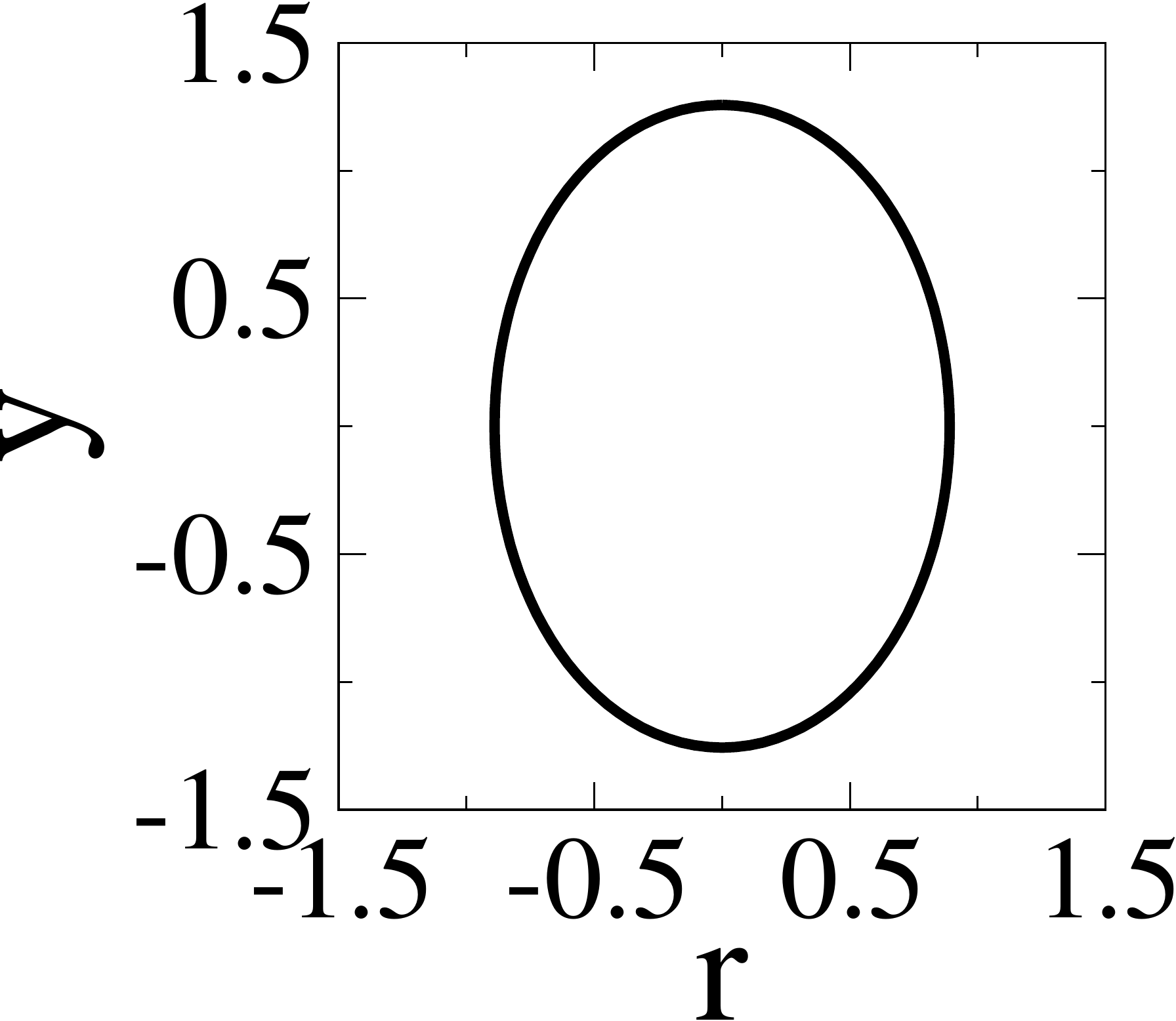}
  \caption{$t=40$}
  \label{fgr:lcal10c}
\end{subfigure}
\begin{subfigure}{.22\textwidth}
  \centering
  \includegraphics[width=1\textwidth]{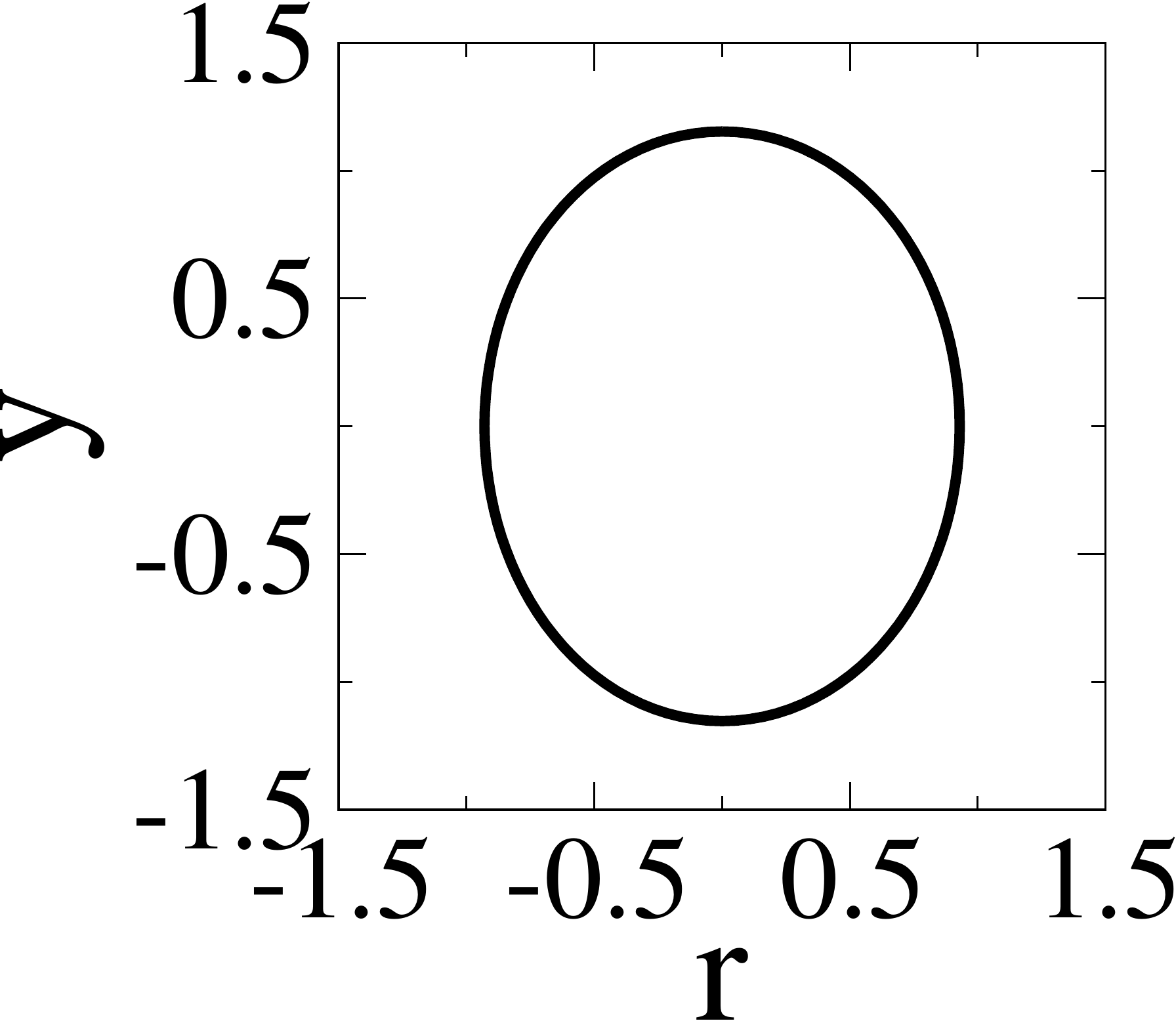}
  \caption{$t=\infty$}
  \label{fgr:lcal10d}
\end{subfigure}
\caption{Shape evolution of a capsule with Skalak membrane at $\sigma_r=10$ for $Ca=0.25$.}
\label{fgr:ca0p25shape10}
\end{center}
\end{figure} 

Intermediate shapes in the dynamics of the deformation of an elastic capsule from an initial spherical shape to the final steady state shape at $Ca=0.25$ are shown in \cref{fgr:ca0p25shape10} for $\sigma_r=10$. A spherical capsule undergoes an intermediate maximum proloate deformation (\cref{fgr:lcal10b}) before reaching to the steady state deformation (\cref{fgr:lcal10d}).

\subsubsection{Case of $\sigma_r<1$}\label{sec:casel0p1}
 When the external fluid is more conducting ($\sigma_r=0.1$), at short times, the polarization vector is from north pole to south pole, opposite in direction to the applied field, that is the north pole has a net negative charge and the south pole has a net positive charge.  Thus for $t\sim O(t_{MW})$, the net Maxwell stress at the membrane interface is  compressive and leads to an oblate shape.  Note that the normal electric field leads to a compressive normal stress, and the isotropic electric pressure is also compressive (see \cref{fgr:vmcheckl0p1,fgr:taunl0p1,fgr:tautl0p1} in \cref{sec:fieldvalidation} for variation of transmembrane potential, normal and tangential electric stresses with time on a static sphere).  \\

Interestingly, irrespective of the conductivity ratio, the final steady state deformations are the same, and are prolate (\cref{fgr:dyl0p1}). This independence of the final steady state shape and deformation with respect to $\sigma_r$ can be easily understood: since at long times the nonconducting membrane acts like a dielectric, thereby rendering the outer normal electric field zero as well as the corresponding tangential stress also zero (\cref{fgr:tautl1,fgr:tautl0p1,fgr:tautl10} in \cref{sec:fieldvalidation}). The Maxwell stress is, therefore, only normal. Moreover, since the normal electric field is zero, the normal electric stress is compressive (\cref{fgr:taunl1,fgr:taunl0p1,fgr:taunl10} in \cref{sec:fieldvalidation}) and is maximum at the poles, the deformation is always prolate . 

\begin{figure}[H]
\begin{center}
\begin{subfigure}{.4\textwidth}
  \centering
  \includegraphics[width=1\textwidth]{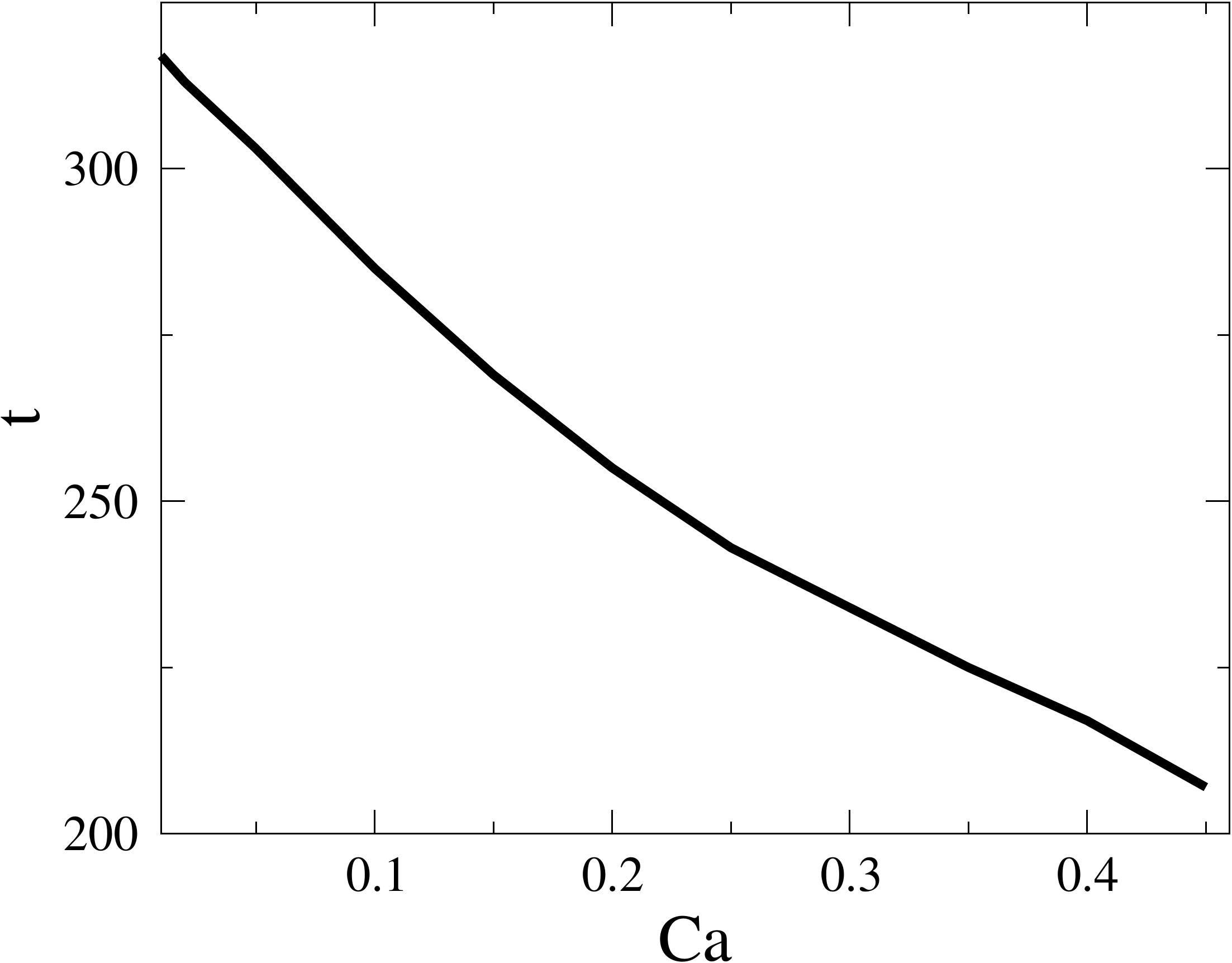}
  \caption{}
  \label{fgr:maxdda}
\end{subfigure}
\begin{subfigure}{.4\textwidth}
  \centering
  \includegraphics[width=1\textwidth]{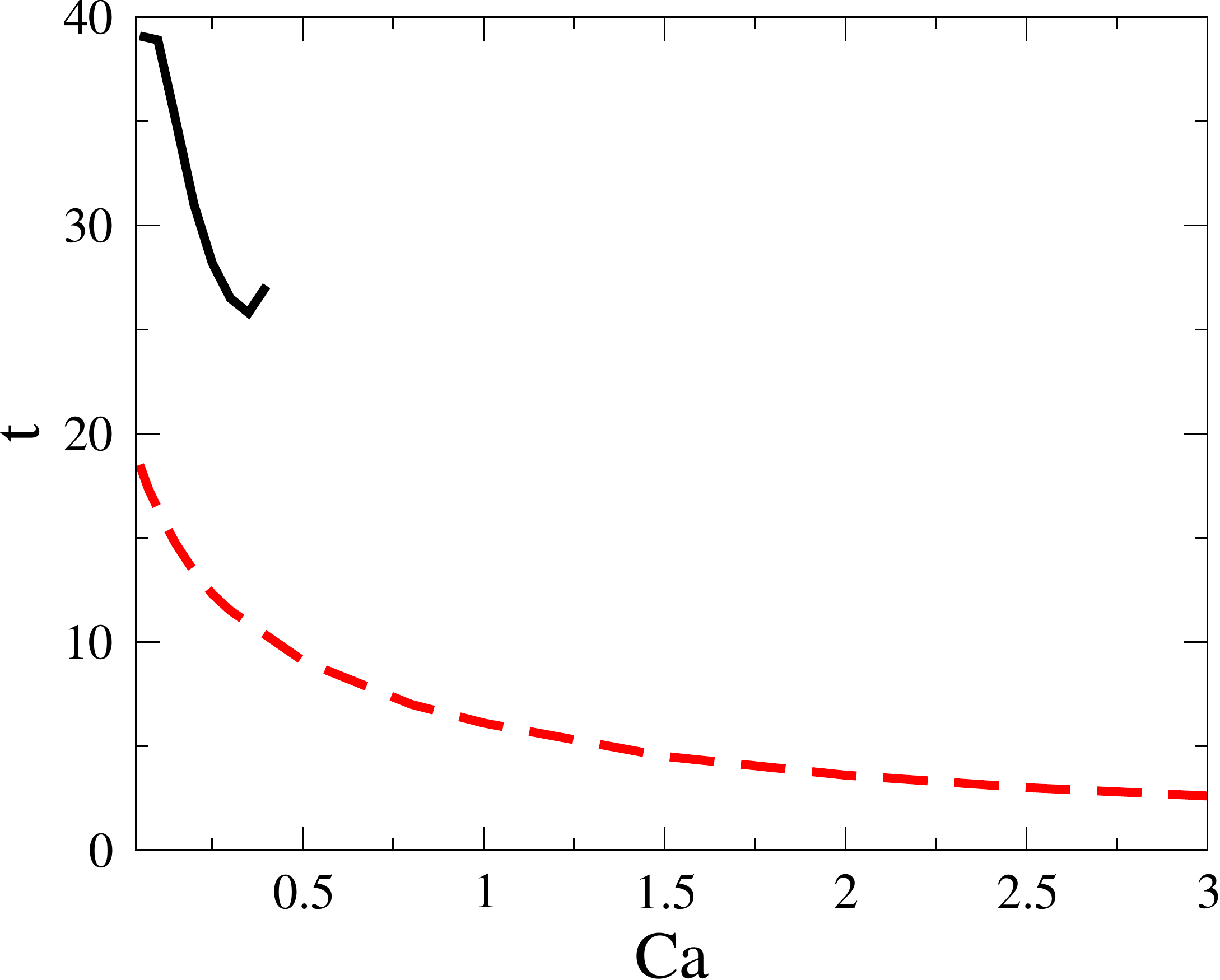}
  \caption{}
  \label{fgr:maxddb}
\end{subfigure}
\caption{(a) Time required for oblate to prolate shape transition as a function of capillary number at $\sigma_r=0.1$. (b) Time required to attain intermediate maximum deformation at $\sigma_r=10$ (\textcolor{red}{$\pmb{--}$}) and $\sigma_r=0.1$ (\textcolor{black}{$\pmb{\mi}$}). }
\label{fgr:maxdd}
\end{center}
\end{figure}

Thus a oblate-prolate transition is observed at long times for $\sigma_r<1$. The exact expression although complicated, the physics seems to be governed by the normal Maxwell's stress going to zero, which in the case of SEM, leads to  
 \begin{equation} 
 t \varpropto t_{cap} \log{\frac{\sigma_r+2\sqrt{2-\sigma_r^2}}{(2+ \sigma_r)}}. 
 \end{equation}
 This leads to a value of around $t=173$ for $\sigma_r$=0.1 which is  of the same order as $t=243$ seen in figure \cref{fgr:dyl0p1,fgr:maxdda}.  Thus the transition time can be expected to be of this order, although not exactly since it also depends on the capillary number as well as on the ratio of hydrodynamic and electric timescales. 

The Skalak and the neo-Hookean models do not show much difference for $\sigma_r>1$, but show significant differences in the maximum deformation for $\sigma_r<1$ and expectedly the strain softening neo-Hookean model shows higher deformation than the Skalak model. The time required to reach the maximum deformation is a complex interplay of the membrane charging and the fluids charging time scales, that is $t_{MW}$ and $t_{cap}$ and is found to obey $t_{MW}<t<t_{cap}$. Thus it is higher for $\sigma_r=0.1$ as compared to that for $\sigma_r=10$. The time  required to reach the maximum deformation decreases with an  increase in the capillary number (\cref{fgr:maxddb}) since typically the deforming forces are larger than the restoring forces.  For $\sigma_r=0.1$, at high capillary number an increase in the time required to reach the maximum deformation is observed which is essentially due to the formation of biconcave intermediate shapes (discussed later).
 
\begin{figure}[H]
\begin{center}
\begin{subfigure}{.22\textwidth}
  \centering
  \includegraphics[width=1\textwidth]{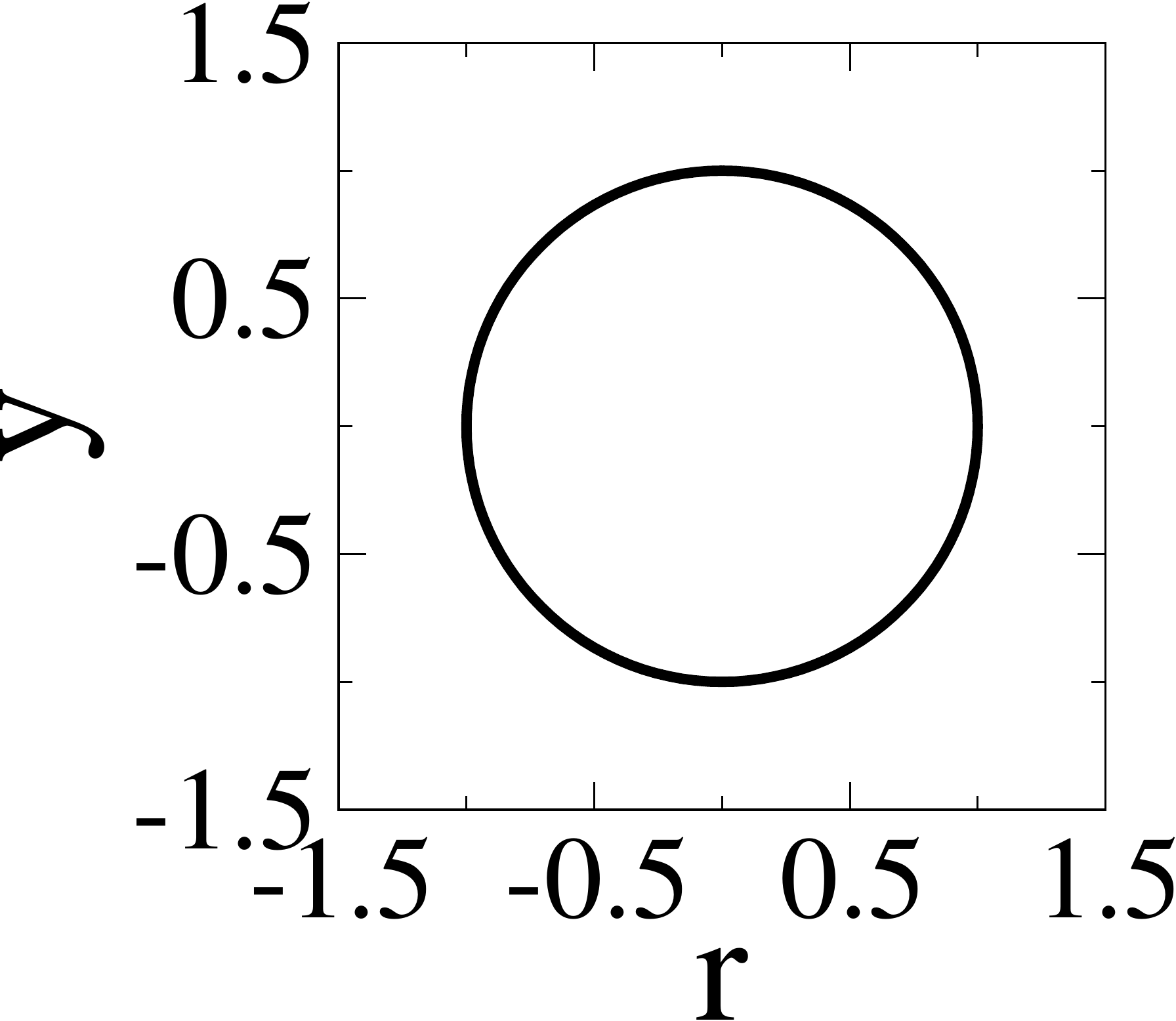}
  \caption{$t=0$}
  \label{fgr:lcal0p1a}
\end{subfigure}
\begin{subfigure}{.22\textwidth}
  \centering
  \includegraphics[width=1\textwidth]{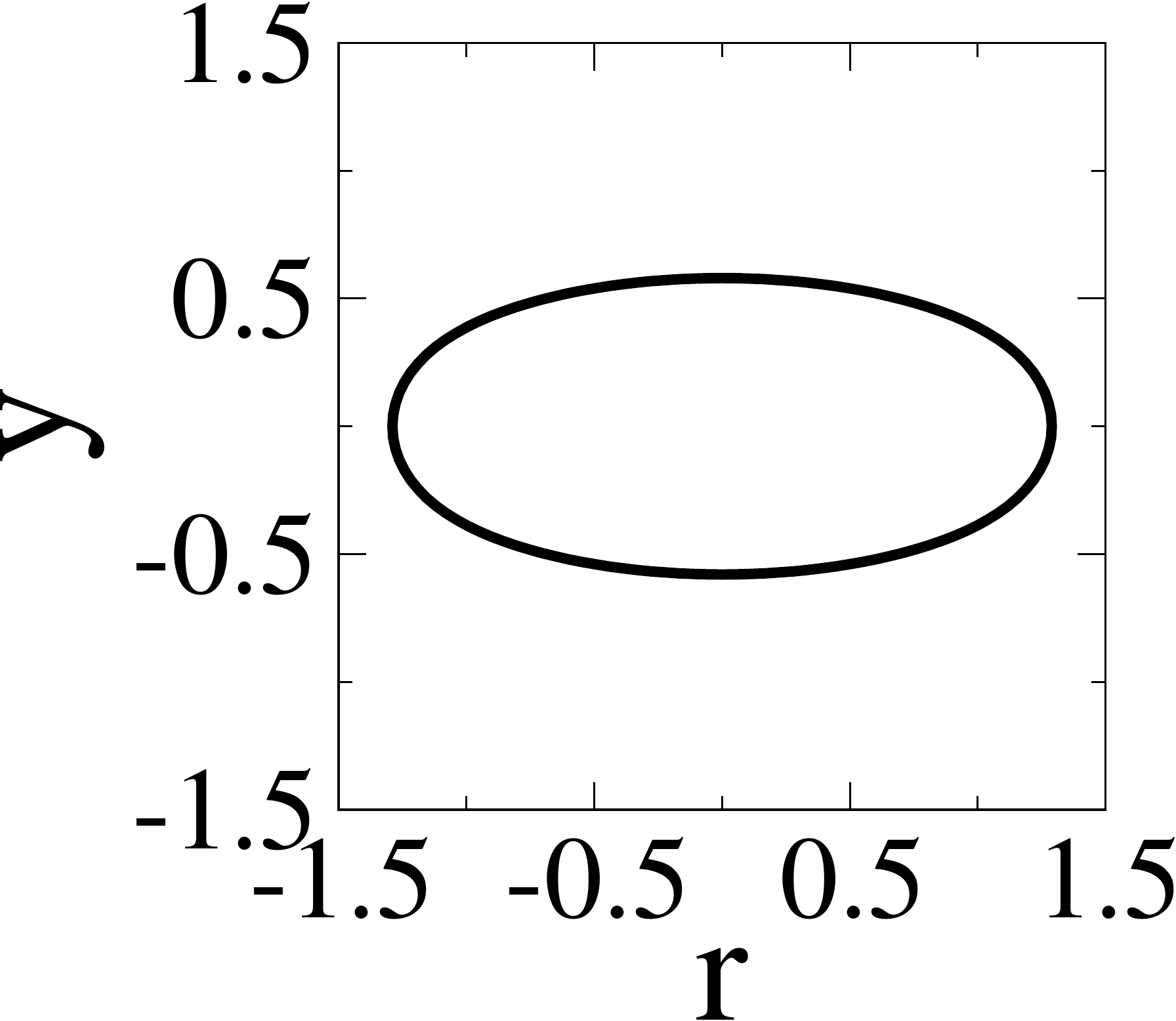}
  \caption{$t=28$}
  \label{fgr:lcal0p1b}
\end{subfigure}
\begin{subfigure}{.22\textwidth}
  \centering
  \includegraphics[width=1\textwidth]{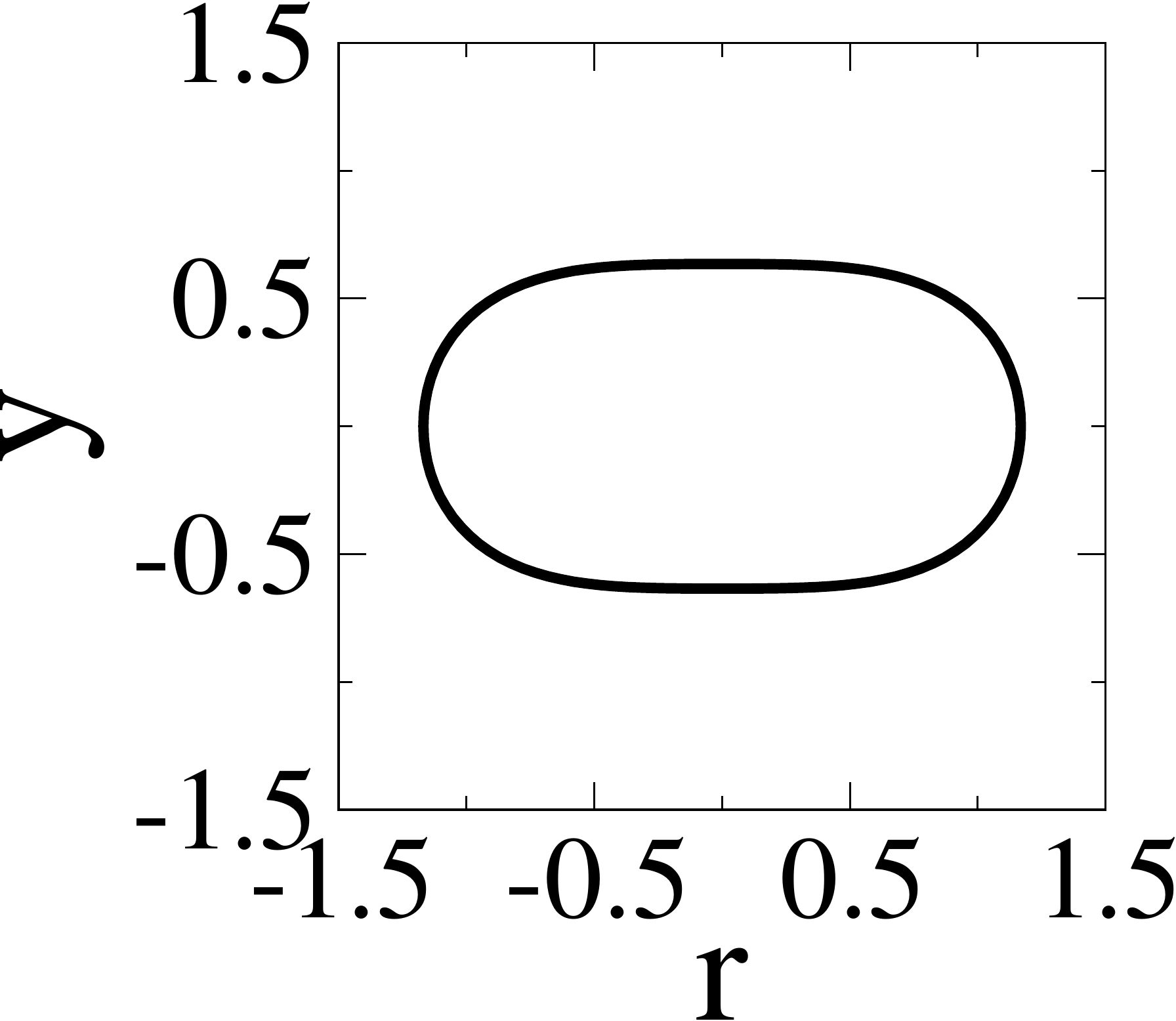}
  \caption{$t=85$}
  \label{fgr:lcal0p1c}
\end{subfigure}
\begin{subfigure}{.22\textwidth}
  \centering
  \includegraphics[width=1\textwidth]{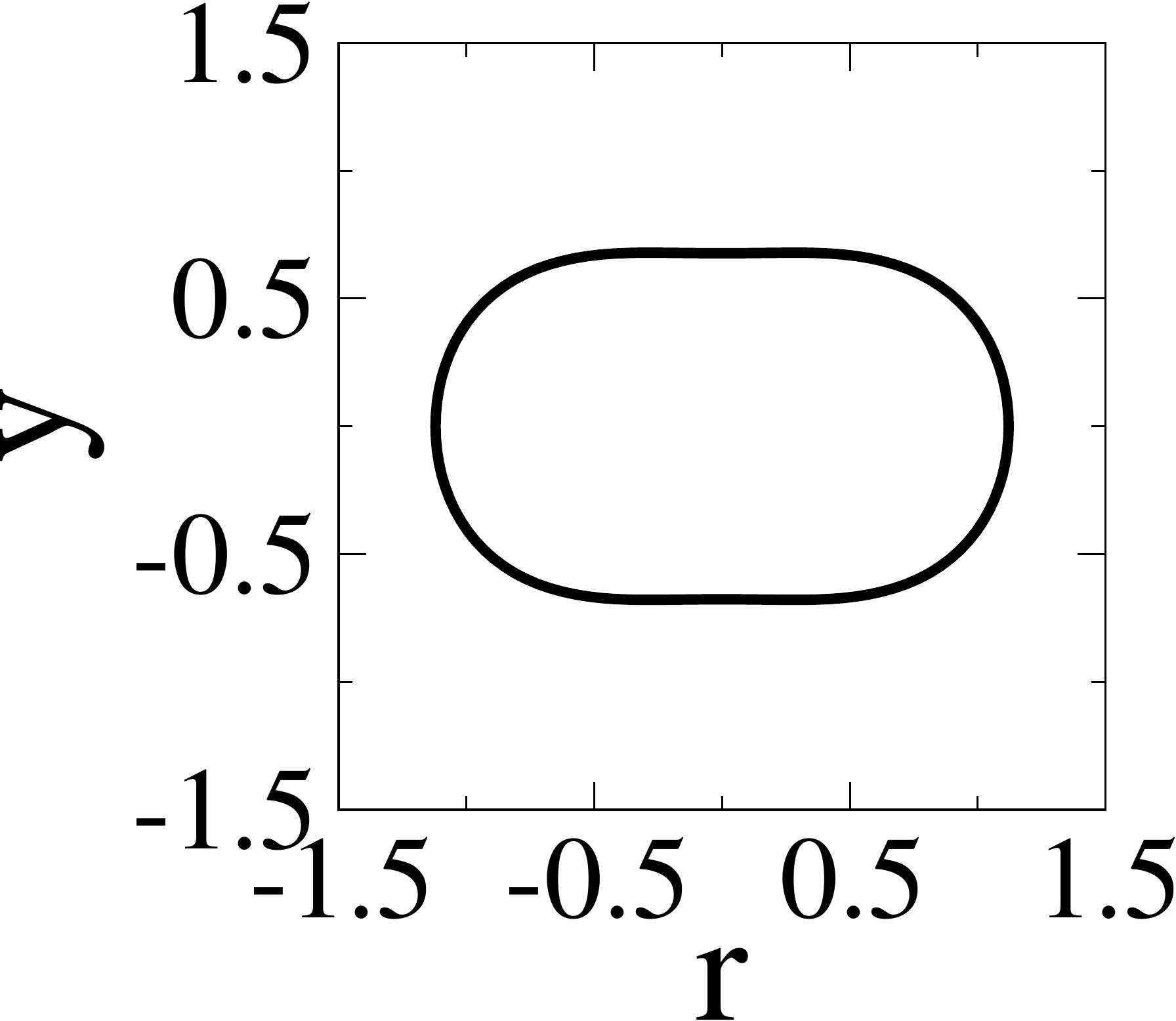}
  \caption{$t=110$}
  \label{fgr:lcal0p1d}
\end{subfigure}
\begin{subfigure}{.22\textwidth}
  \centering
  \includegraphics[width=1\textwidth]{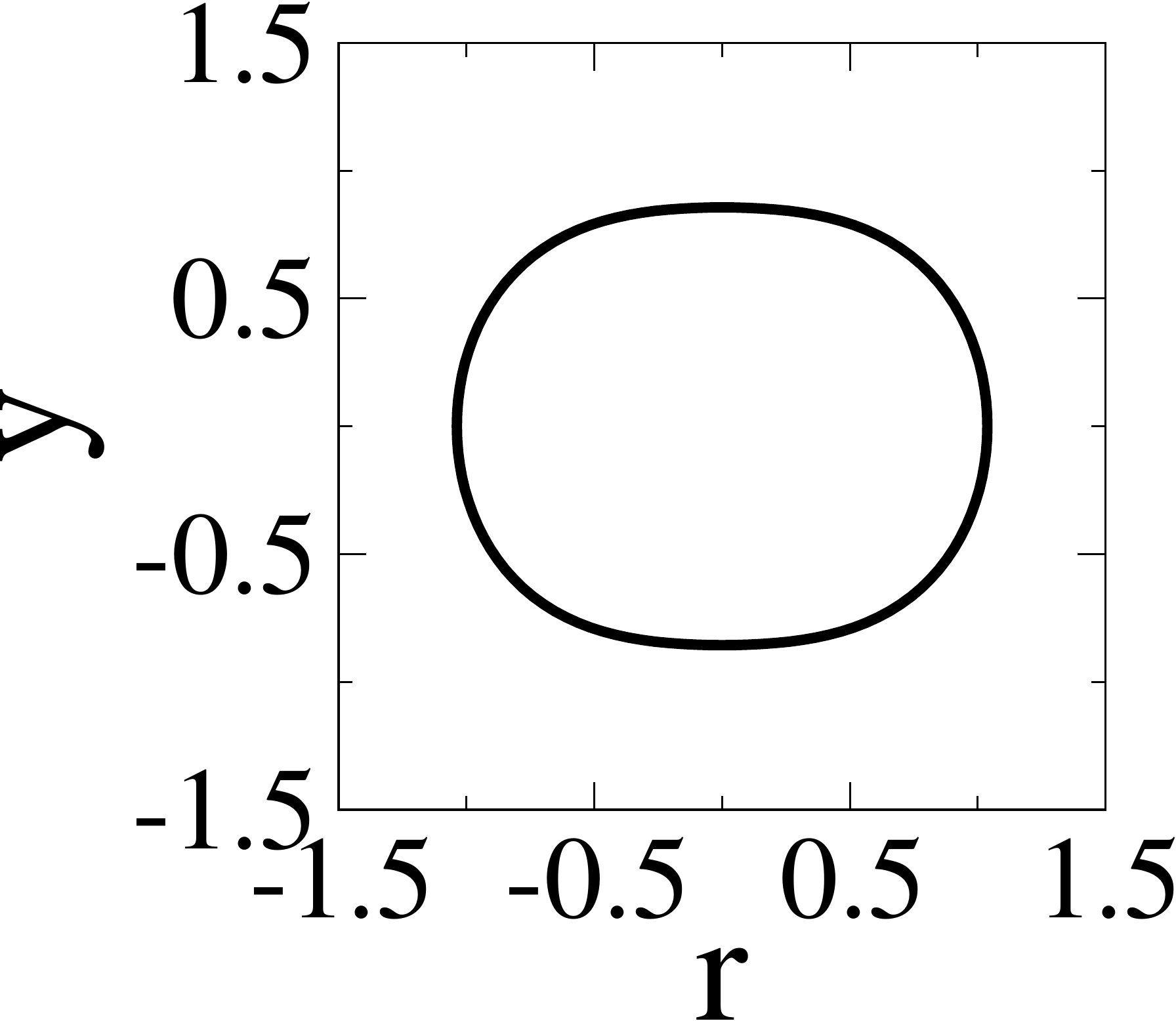}
  \caption{$t=170$}
  \label{fgr:lcal0p1e}
\end{subfigure}
\begin{subfigure}{.22\textwidth}
  \centering
  \includegraphics[width=1\textwidth]{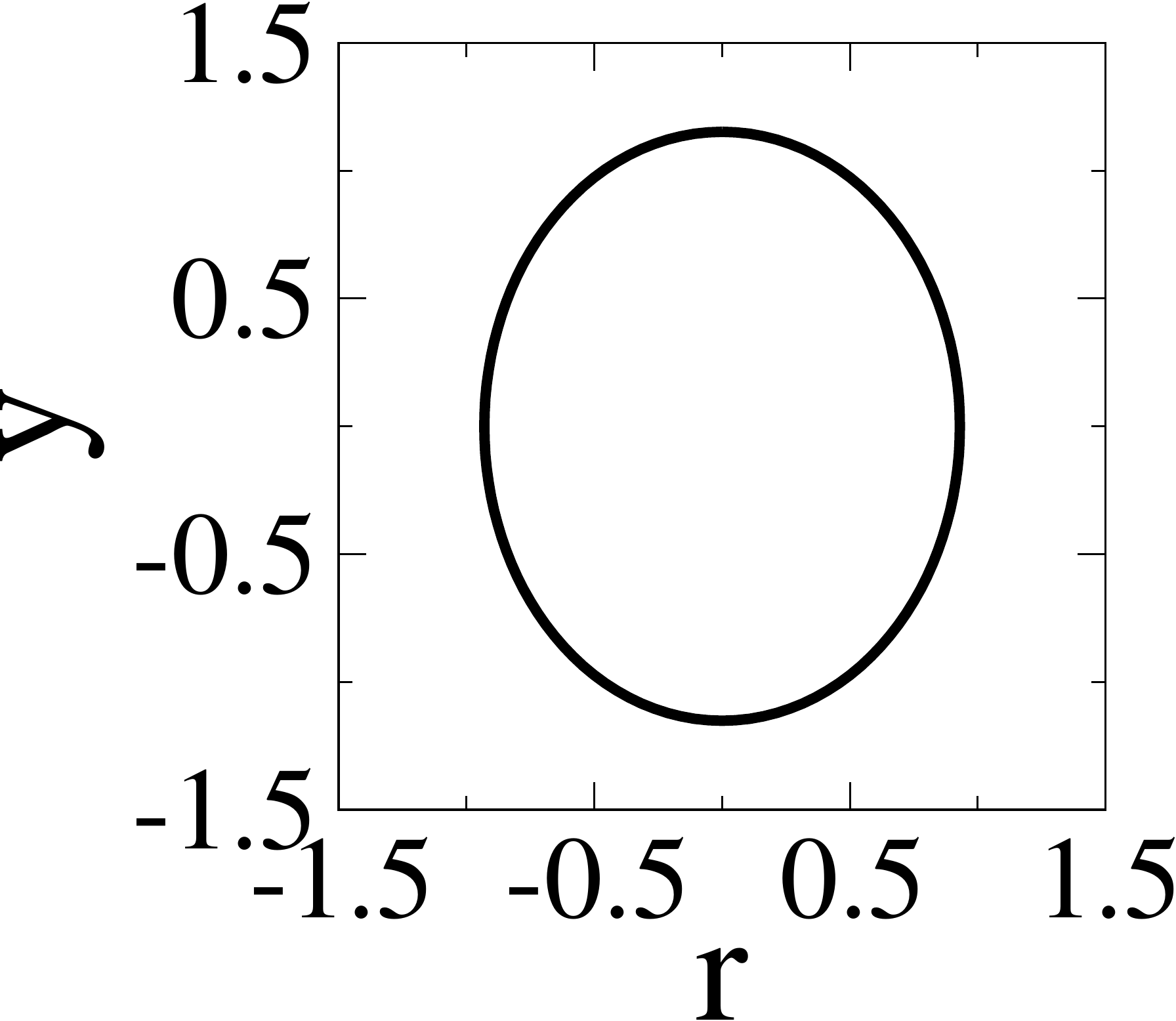}
  \caption{$t=\infty$}
  \label{fgr:lcal0p1f}
\end{subfigure}
\caption{Shape evolution of a capsule with Skalak membrane at $\sigma_r=0.1$ for $Ca=0.25$.}
\label{fgr:lcal0p1}
\end{center}
\end{figure}  

Intermediate shapes for the dynamics of the deformation of an elastic capsule from a spherical shape to the steady state shape at $Ca=0.25$ are shown in \cref{fgr:lcal0p1} for $\sigma_r=0.1$. For $\sigma_r>1$ prolate shapes are observed whereas oblate shapes are seen (\cref{fgr:lcal0p1b,fgr:lcal0p1c,fgr:lcal0p1d}) for $\sigma_r<1$ until the membrane is fully charged. To understand this better, deformation studies at a still higher capillary number (Ca=0.45) are conducted with an emphasis on $\sigma_r<1$. 

\subsection{Dynamics of capsule deformation at large capillary number}
\begin{figure}[H]
\begin{center}
\begin{subfigure}{.22\textwidth}
  \centering
  \includegraphics[width=1\textwidth]{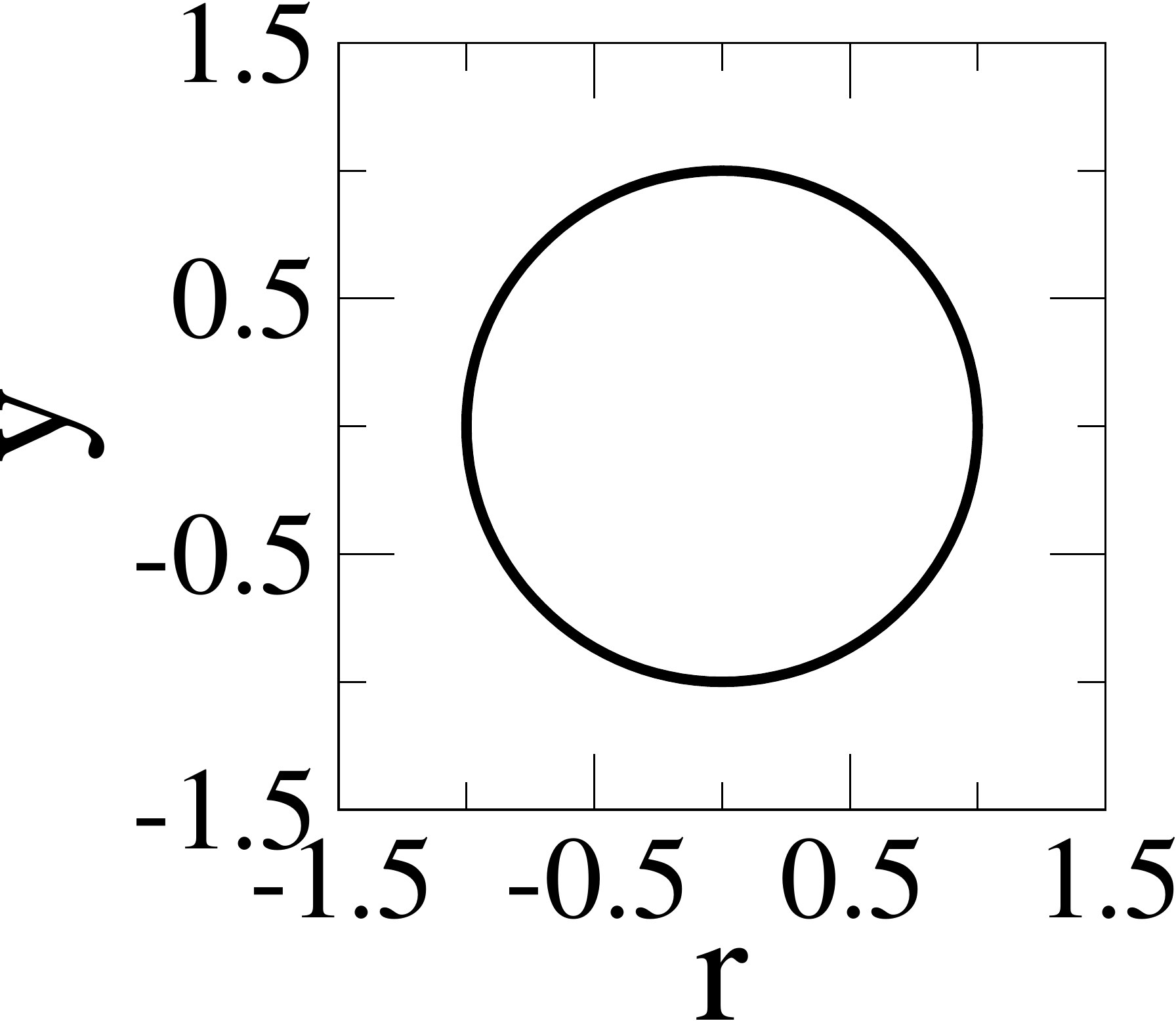}
  \caption{$t=0$}
  \label{fgr:shapellhcaa}
\end{subfigure}
\begin{subfigure}{.22\textwidth}
  \centering
  \includegraphics[width=1\textwidth]{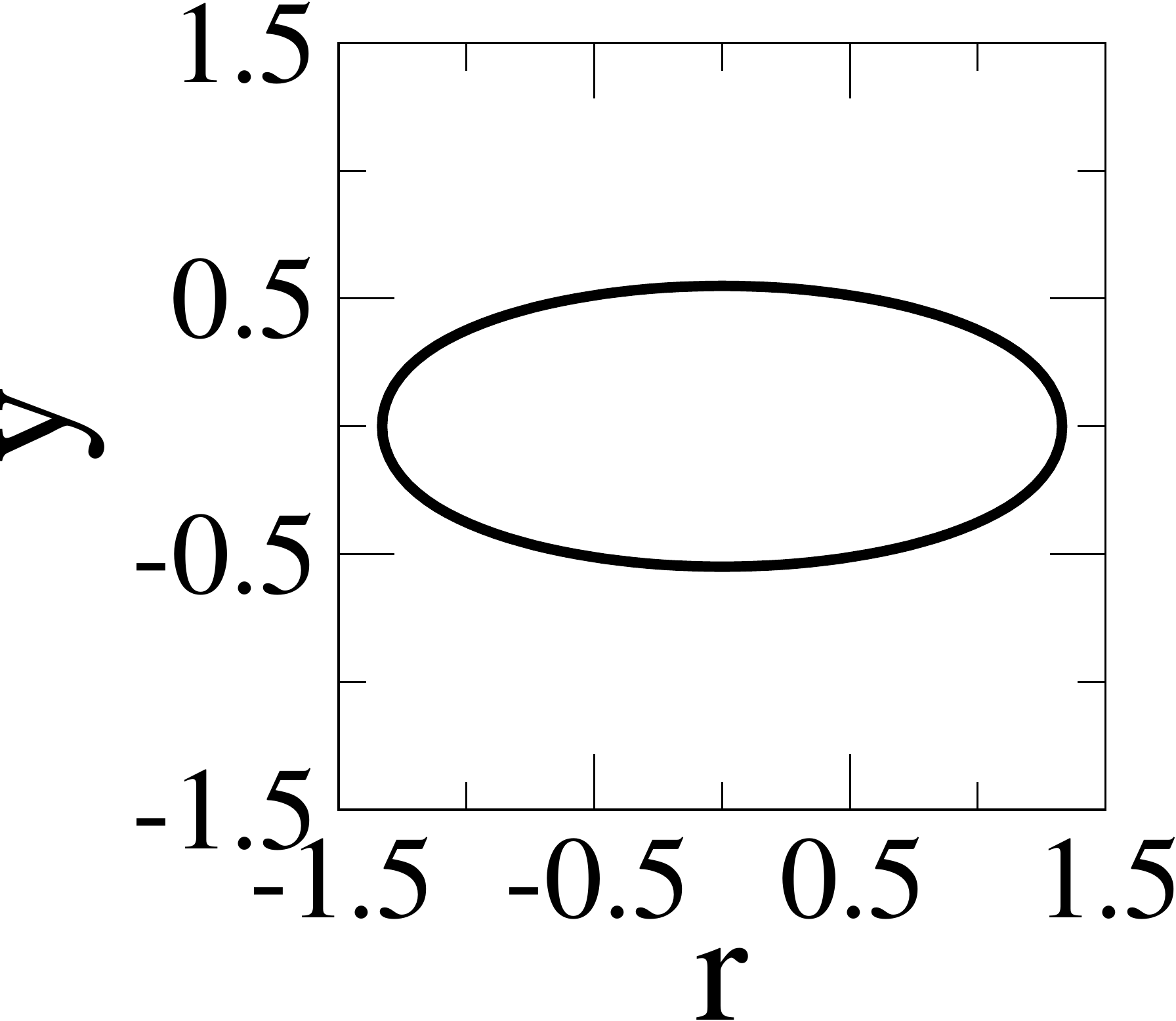}
  \caption{$t=10$}
  \label{fgr:shapellhcab}
\end{subfigure}
\begin{subfigure}{.22\textwidth}
  \centering
  \includegraphics[width=1\textwidth]{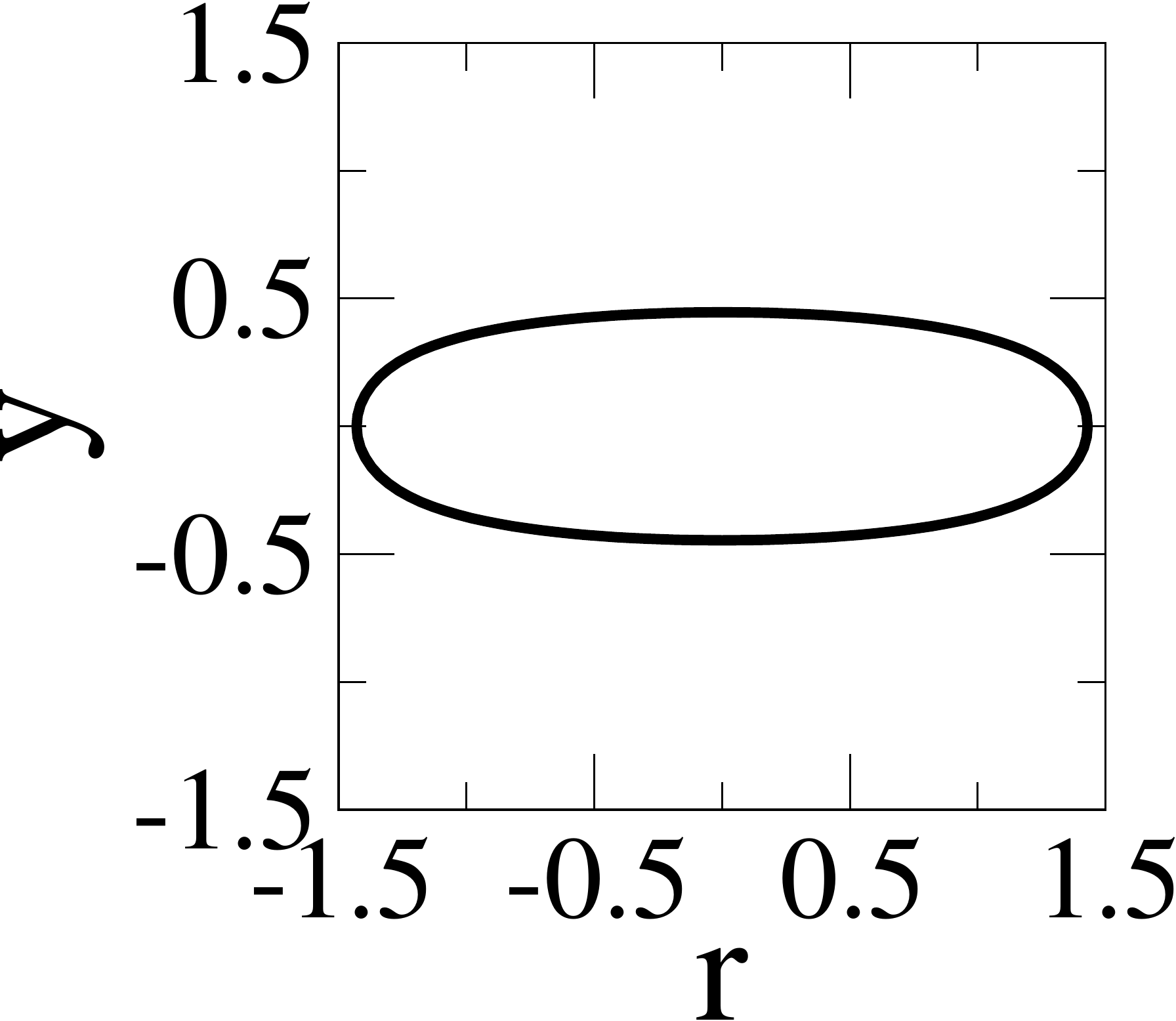}
  \caption{$t=23$}
  \label{fgr:shapellhcac}
\end{subfigure}
\begin{subfigure}{.22\textwidth}
  \centering
  \includegraphics[width=1\textwidth]{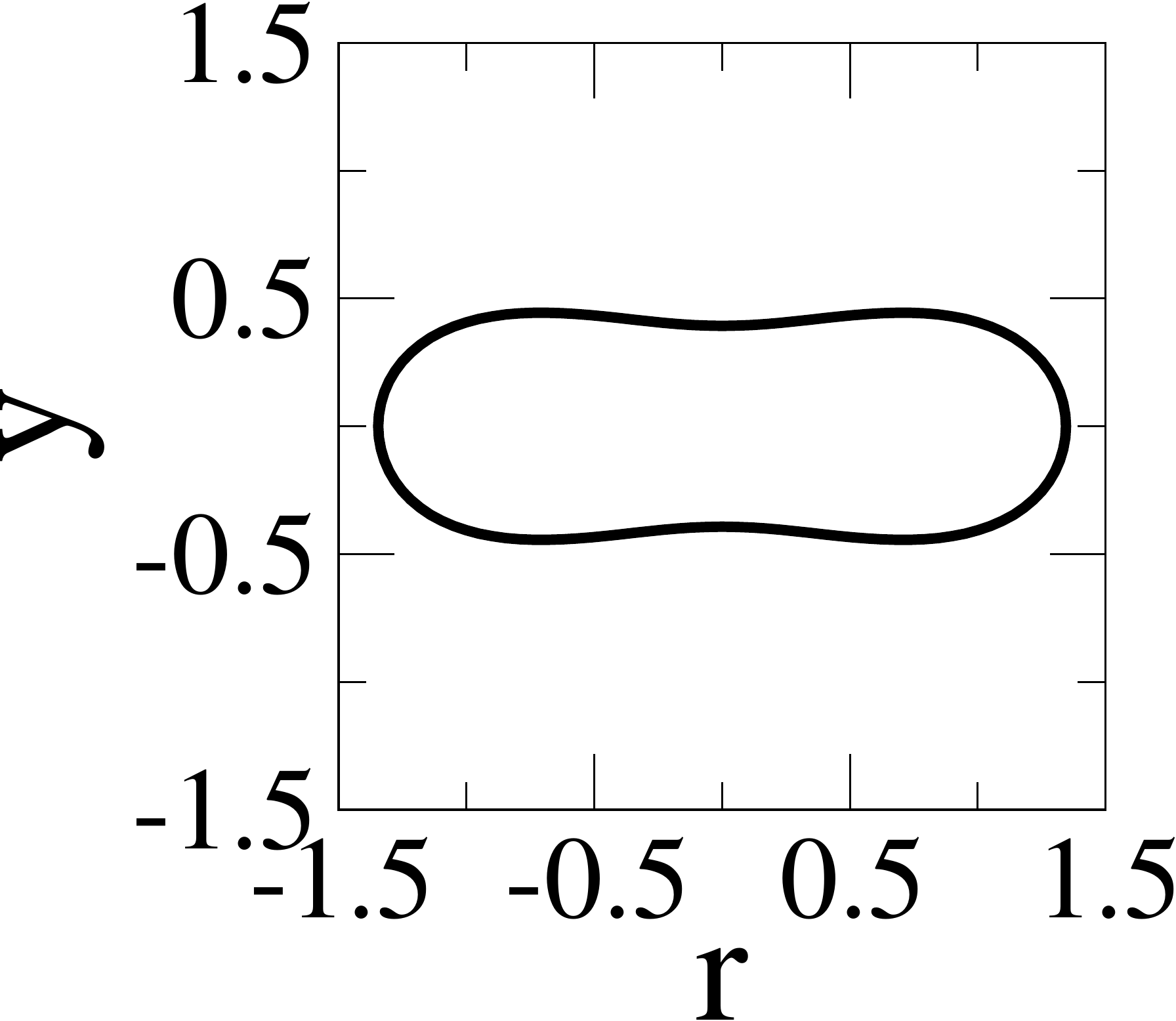}
  \caption{$t=50$}
  \label{fgr:shapellhcad}
\end{subfigure}
\begin{subfigure}{.22\textwidth}
  \centering
  \includegraphics[width=1\textwidth]{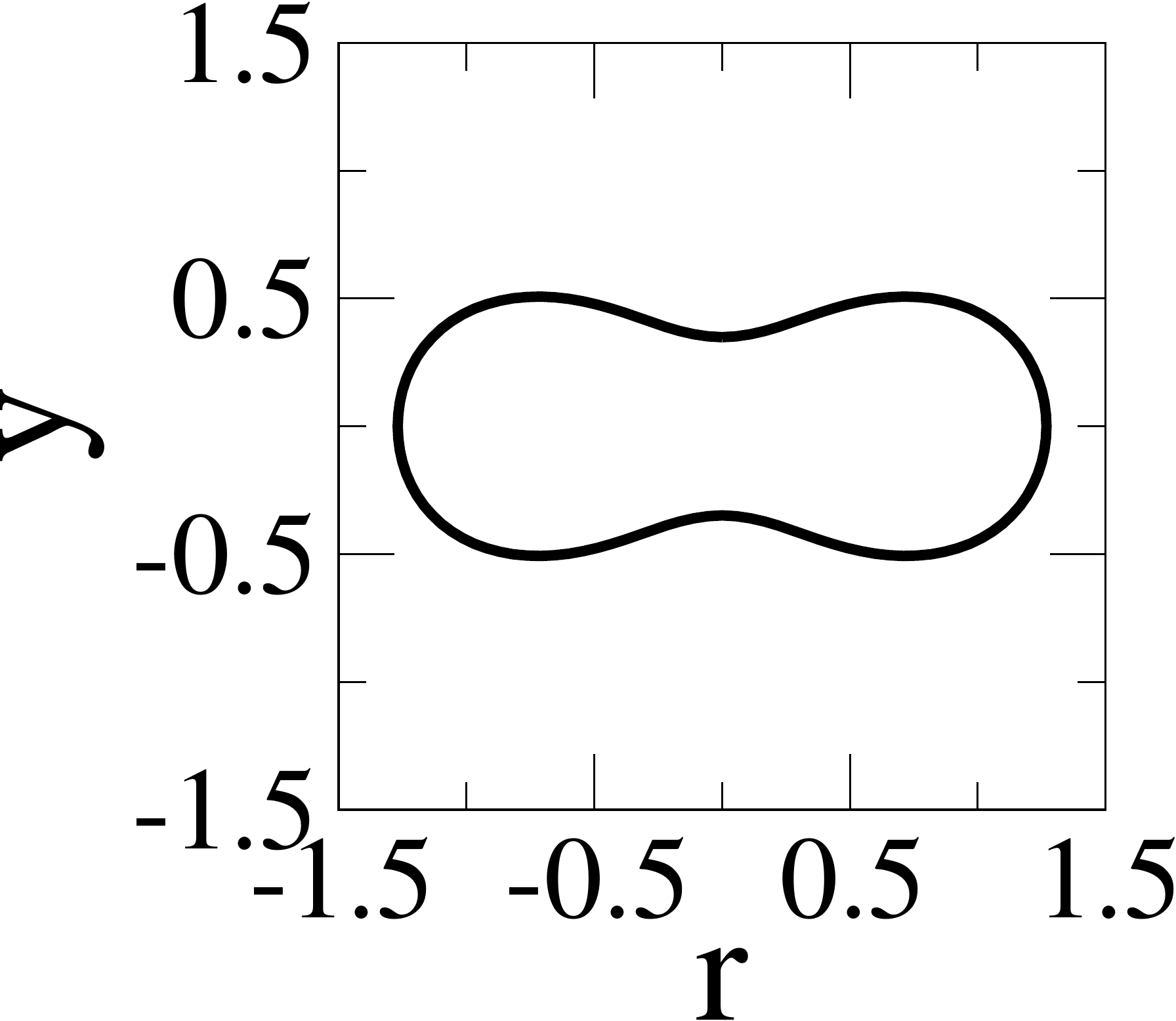}
  \caption{$t=75$}
  \label{fgr:shapellhcae}
\end{subfigure}
\begin{subfigure}{.22\textwidth}
  \centering
  \includegraphics[width=1\textwidth]{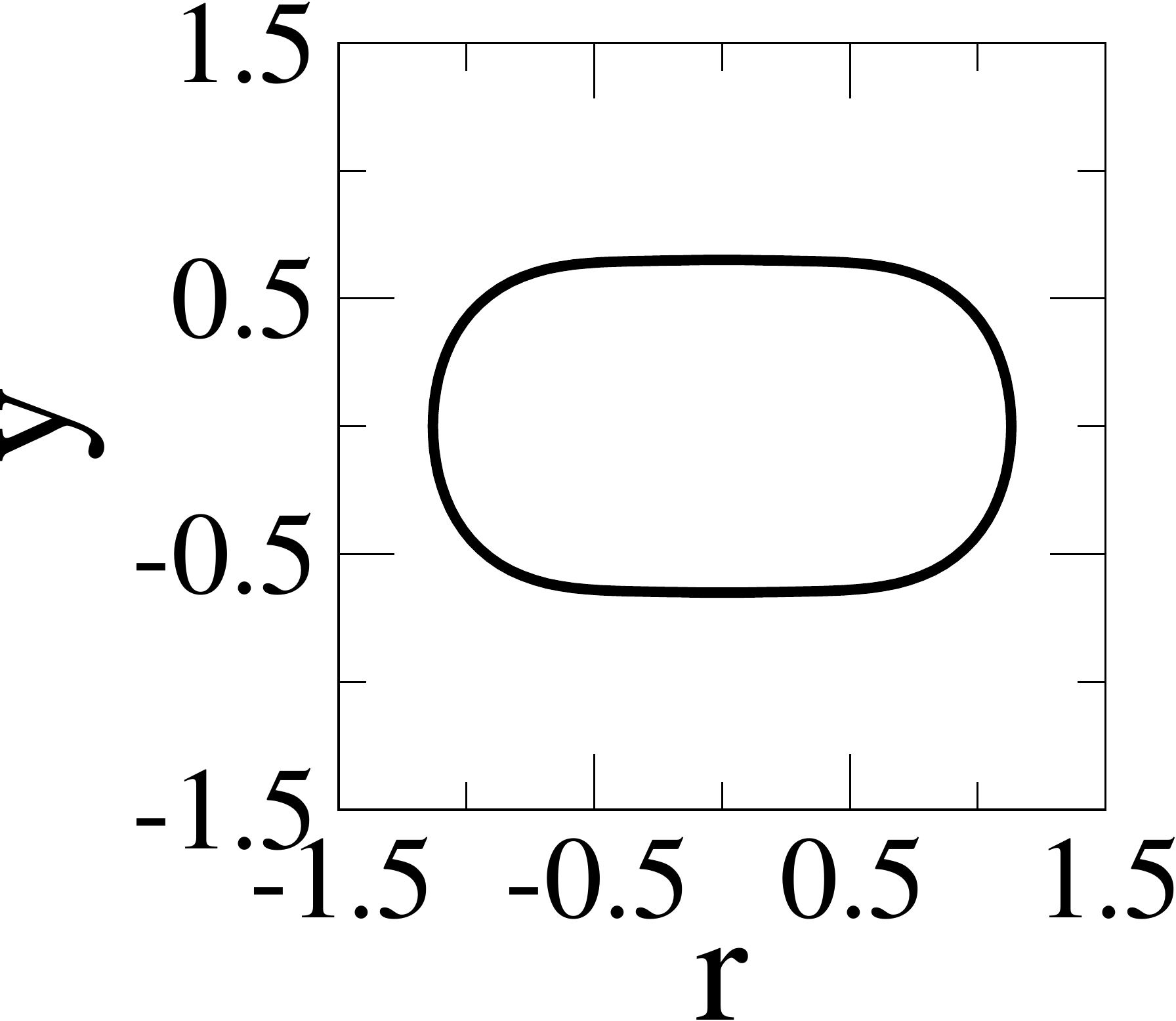}
  \caption{$t=120$}
  \label{fgr:shapellhcaf}
\end{subfigure}
\begin{subfigure}{.22\textwidth}
  \centering
  \includegraphics[width=1\textwidth]{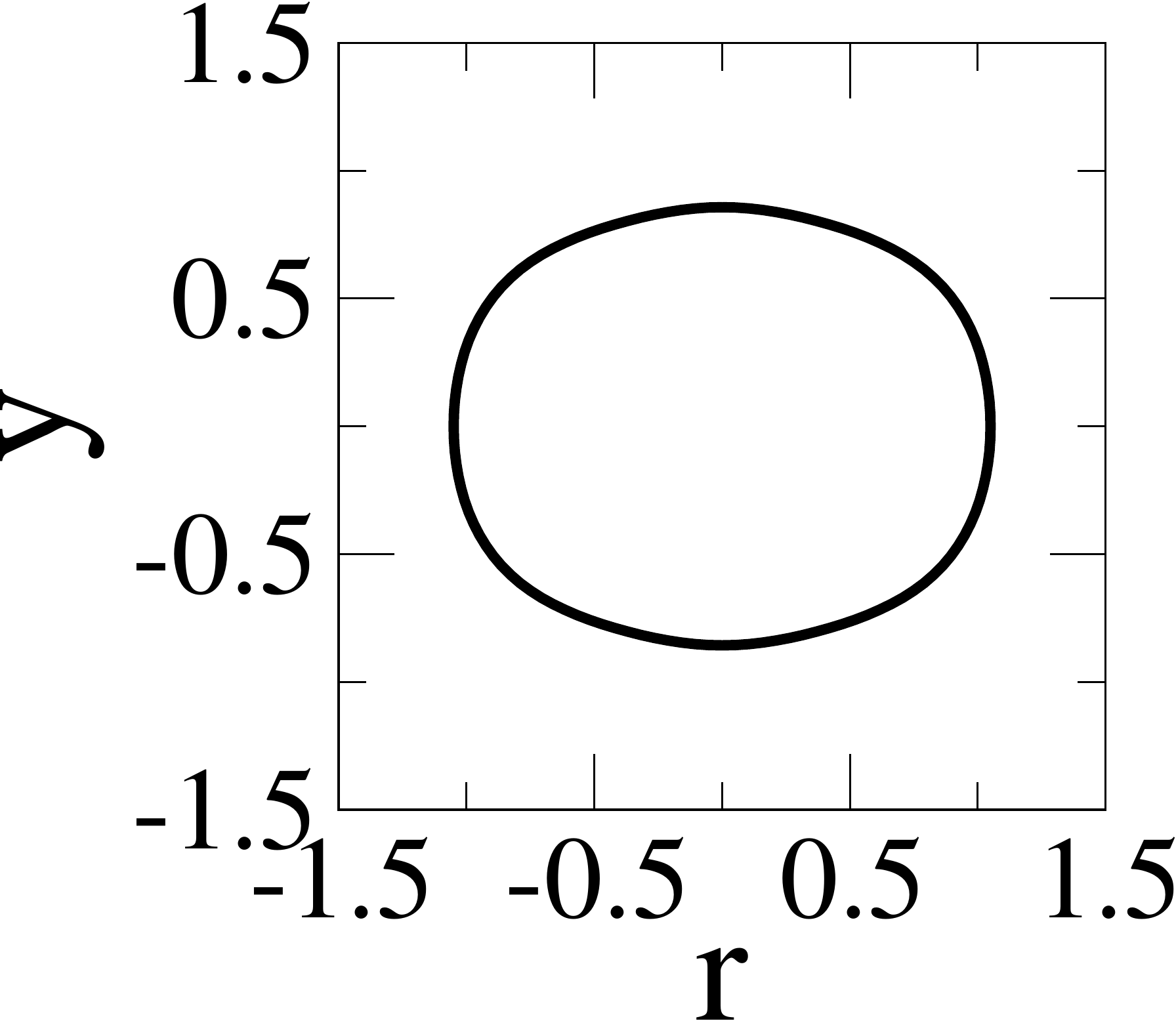}
  \caption{$t=150$}
  \label{fgr:shapellhcag}
\end{subfigure}
\begin{subfigure}{.22\textwidth}
  \centering
  \includegraphics[width=1\textwidth]{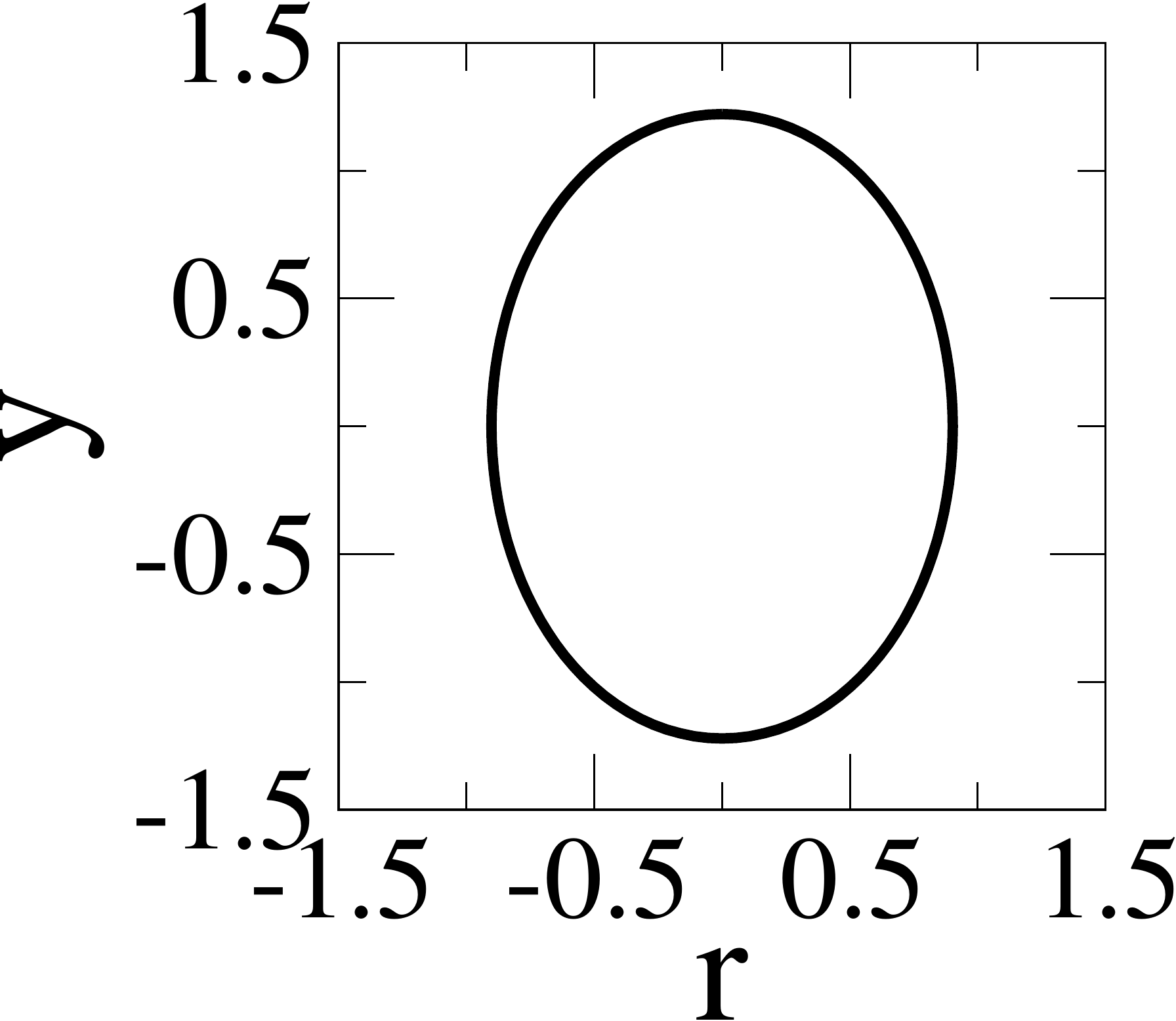}
  \caption{$t=\infty$}
  \label{fgr:shapellhcah}
\end{subfigure}
\caption{Shape evolution of a capsule with Skalak membrane at $\sigma_r=0.1$ for $Ca=0.45$.}
\label{fgr:shapellhca}
\end{center}
\end{figure}

\begin{figure}[H]
\begin{center}
\begin{subfigure}{.32\textwidth}
  \centering
  \includegraphics[width=1\textwidth]{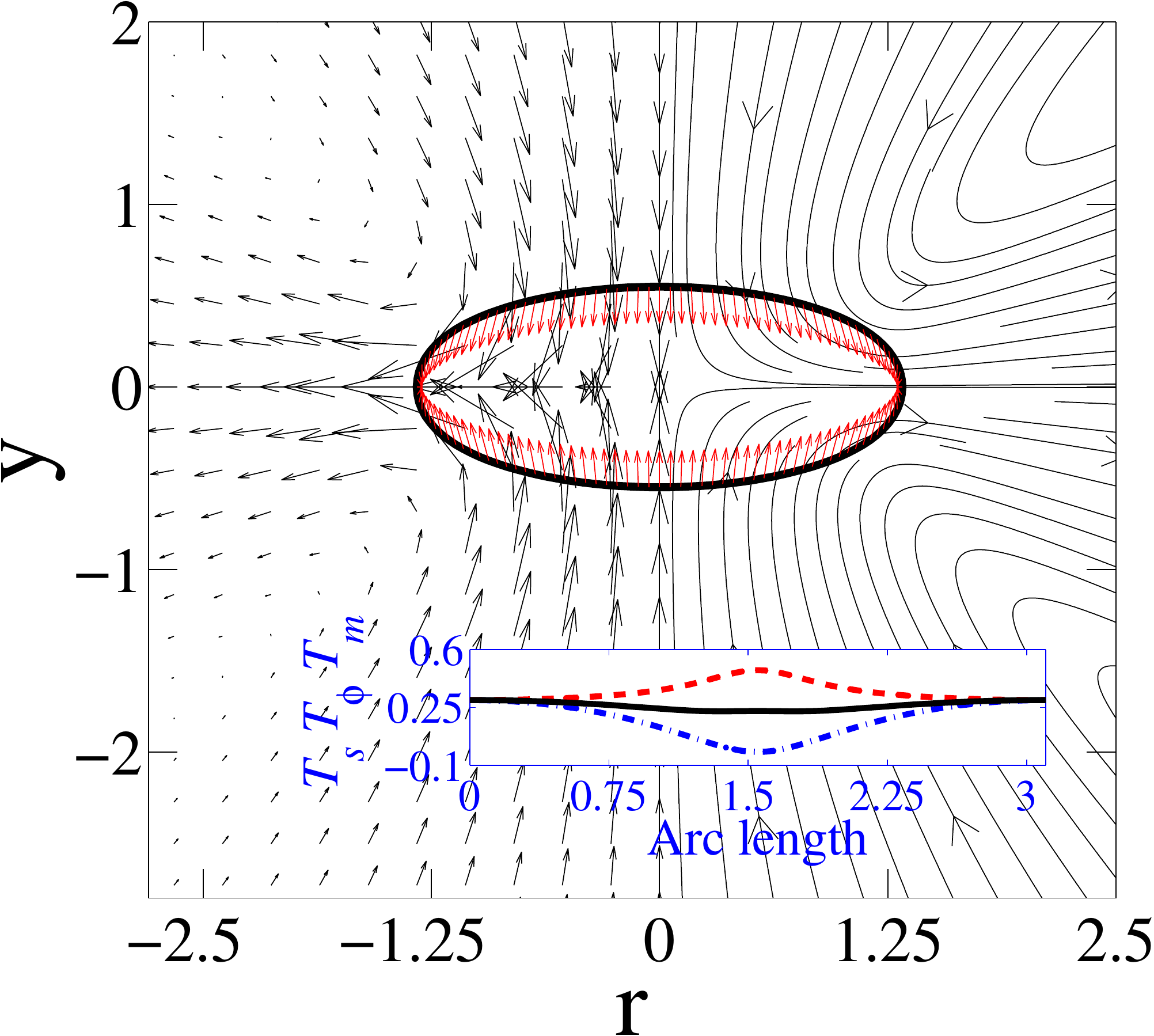}
  \caption{$t=10$}
  \label{fgr:streamb}
\end{subfigure}
\begin{subfigure}{.32\textwidth}
  \centering
  \includegraphics[width=1\textwidth]{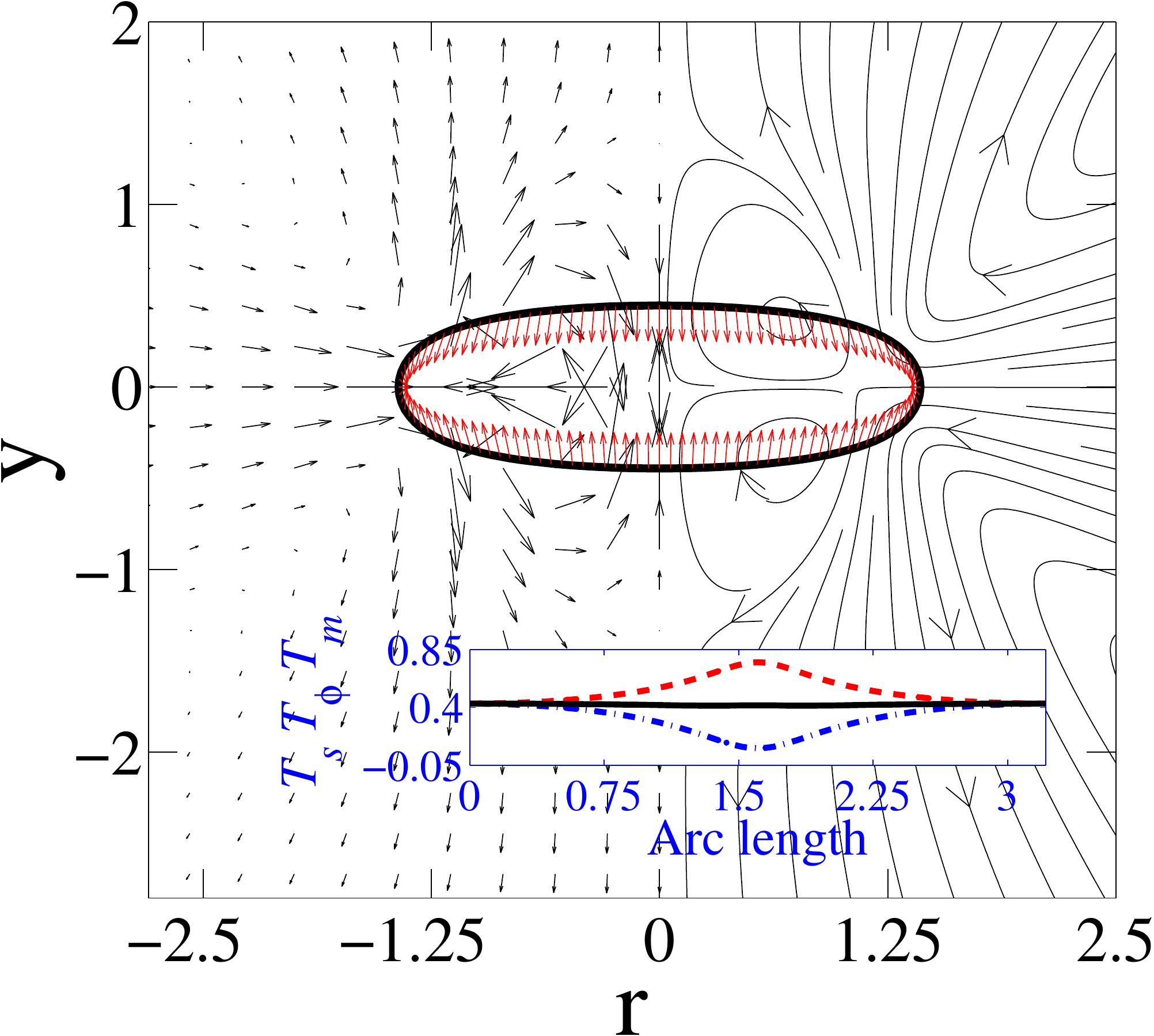}
  \caption{$t=23$}
  \label{fgr:streamc}
\end{subfigure}
\begin{subfigure}{.32\textwidth}
  \centering
  \includegraphics[width=1\textwidth]{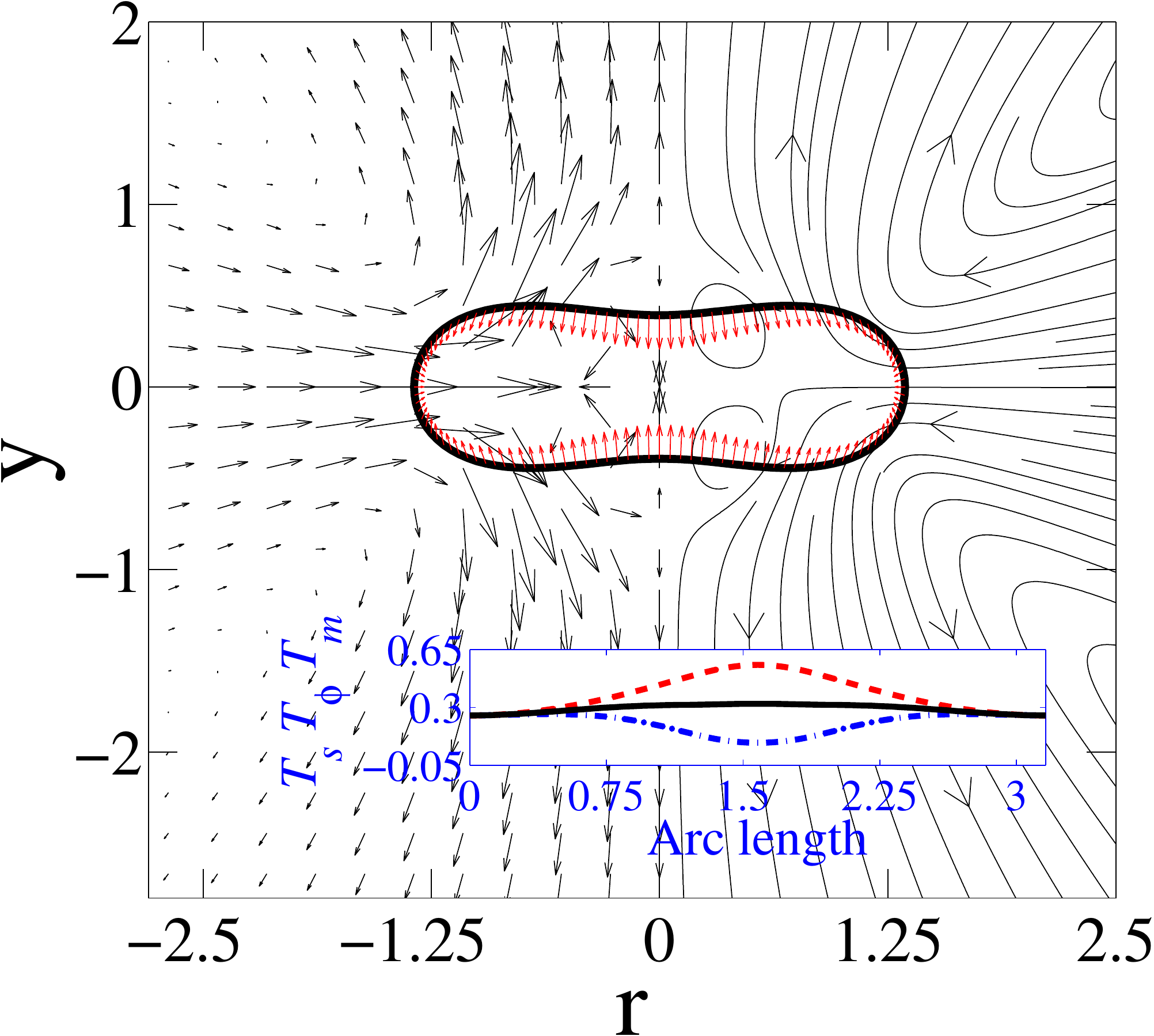}
  \caption{$t=50$}
  \label{fgr:streamd}
\end{subfigure}
\begin{subfigure}{.32\textwidth}
  \centering
  \includegraphics[width=1\textwidth]{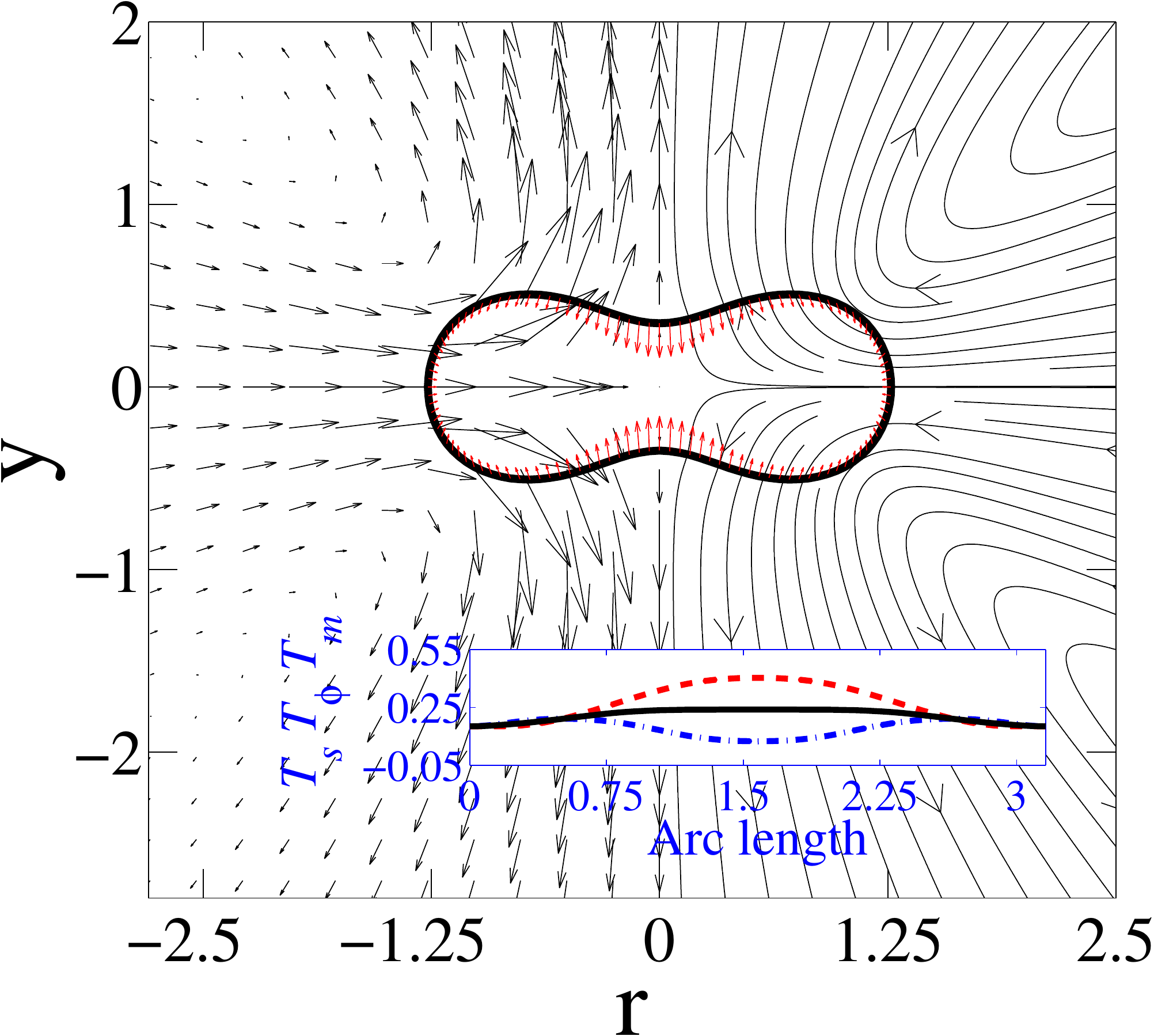}
  \caption{$t=75$}
  \label{fgr:streame}
\end{subfigure}
\begin{subfigure}{.32\textwidth}
  \centering
  \includegraphics[width=1\textwidth]{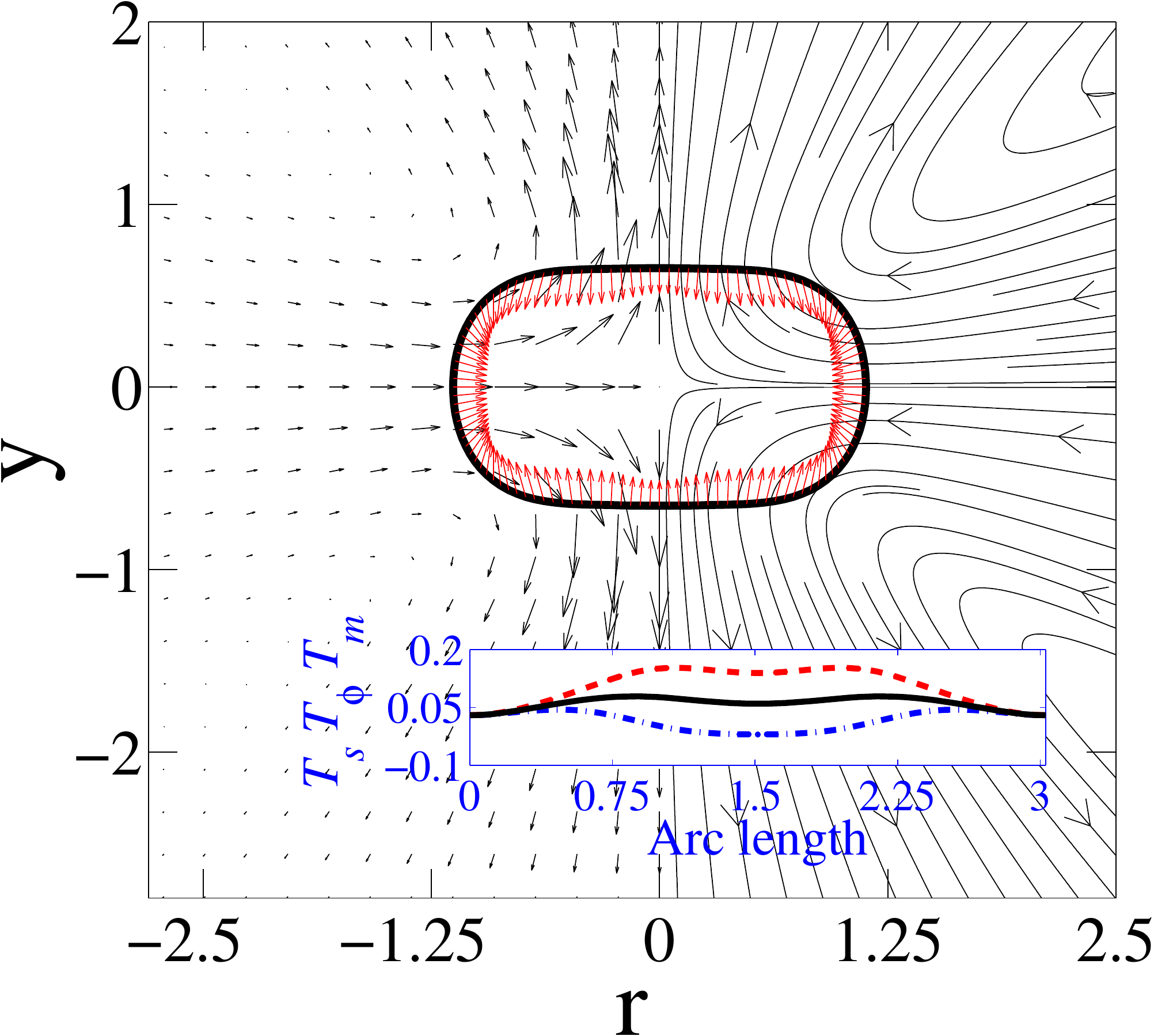}
  \caption{$t=120$}
  \label{fgr:streamf}
\end{subfigure}
\begin{subfigure}{.32\textwidth}
  \centering
  \includegraphics[width=1\textwidth]{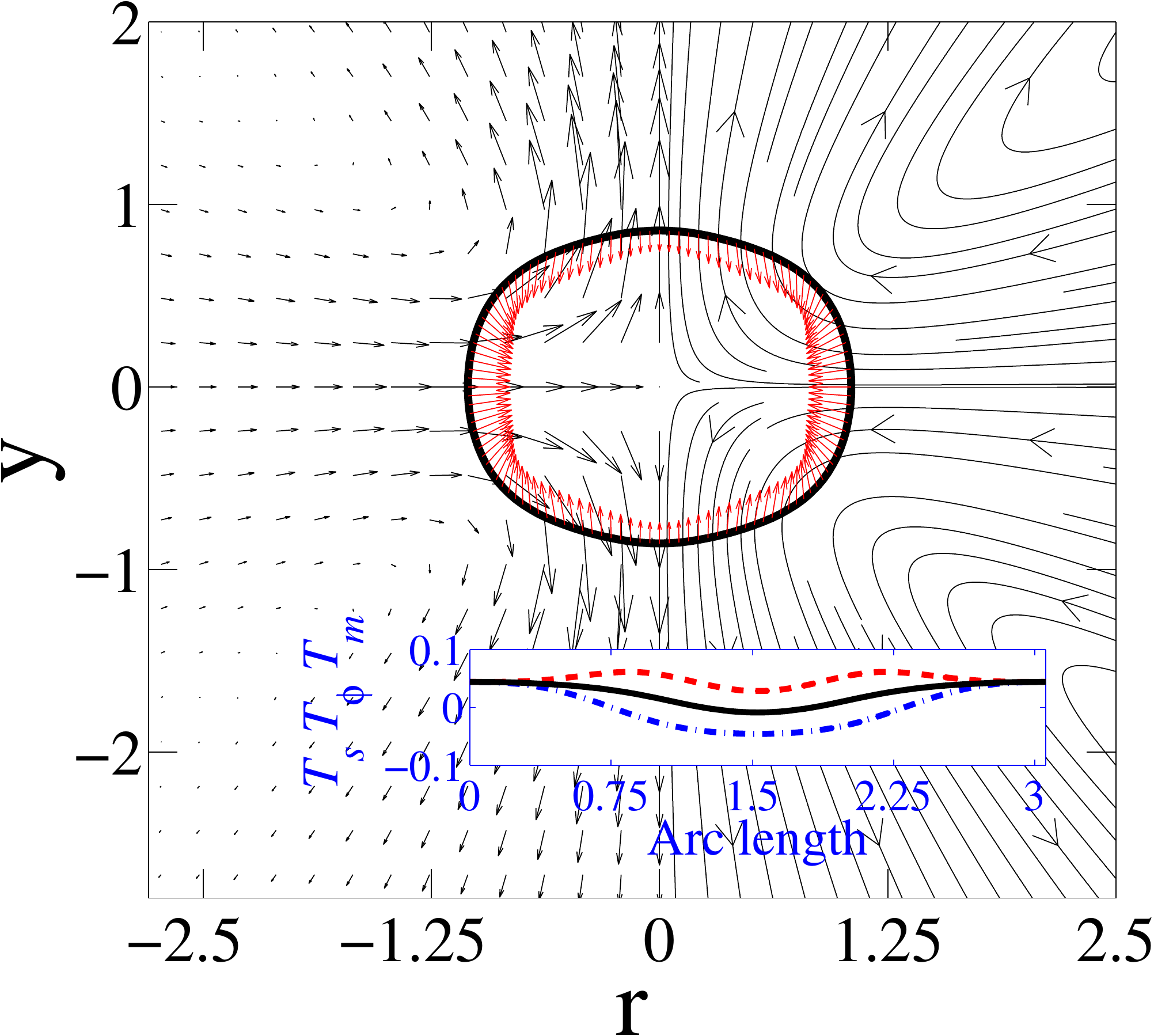}
  \caption{$t=150$}
  \label{fgr:streamg}
\end{subfigure}
\begin{subfigure}{.32\textwidth}
  \centering
  \includegraphics[width=1\textwidth]{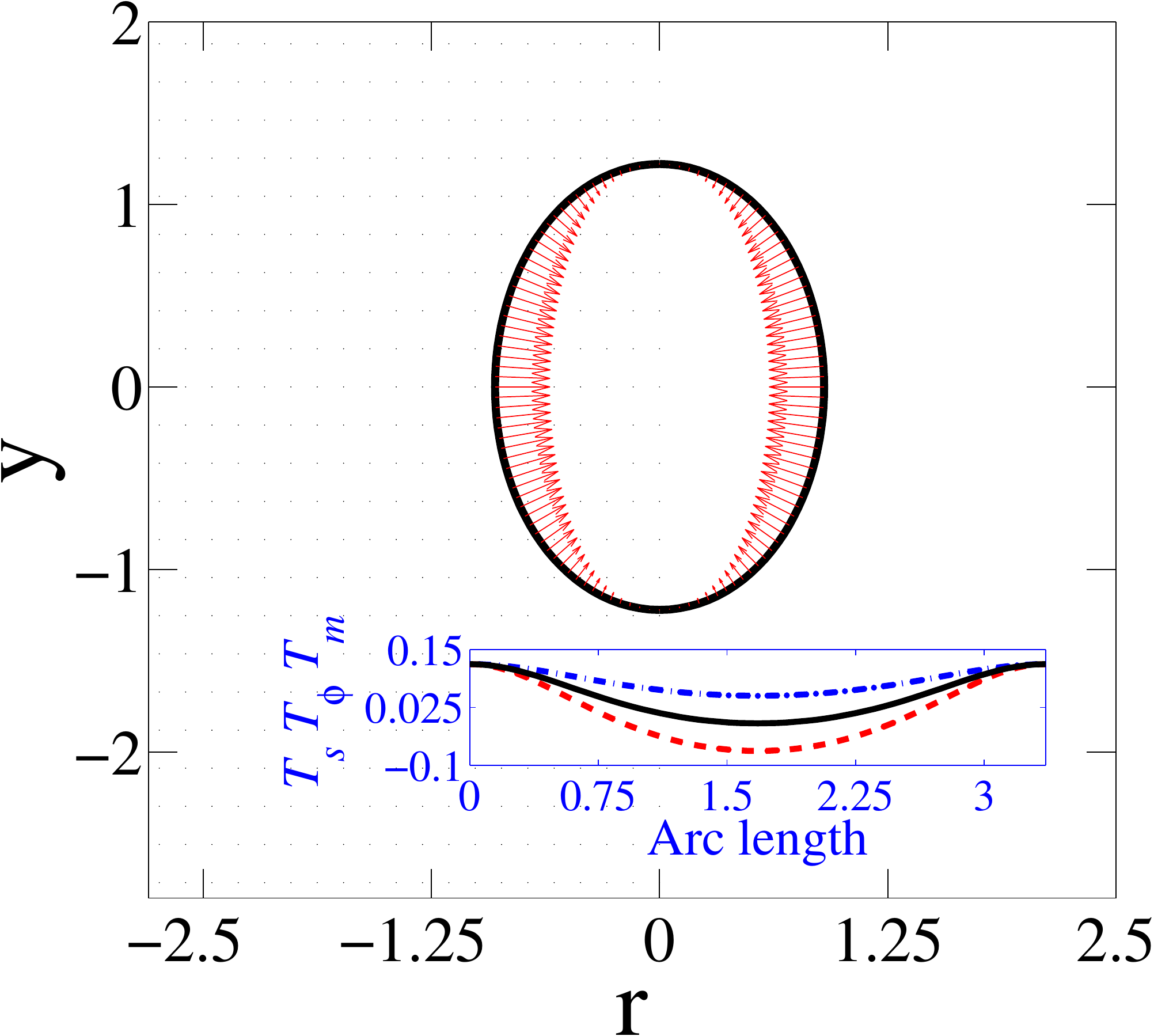}
  \caption{$t=\infty$}
  \label{fgr:streamh}
\end{subfigure}
\caption{Stream line plot (curves with arrow, shown only at right side of axis of symmetry), velocity profile (short arrows, magnitude represents extend of flow, shown only at left side of axis of symmetry) and electric stress (arrows from the interface) for $Ca=0.45$ and $\sigma_r=0.1$ at different times. Inset shows the variation of meridional, $T_s$ (\textcolor{blue}{$\pmb{-\cdot-}$}), azimuthal, $T_\phi$ (\textcolor{red}{$\pmb{--}$}) and mean, $T_m$ (\textcolor{black}{$\pmb{\mi}$}) membrane tensions along the arc length.}
\label{fig:streaml0p1}
\end{center}
\end{figure} 

\begin{figure}[H]
 \begin{center}
\begin{subfigure}{.32\textwidth}
  \centering
  \includegraphics[width=1\textwidth, trim=0.0in 0.0in 0.2in 0.0in, clip]{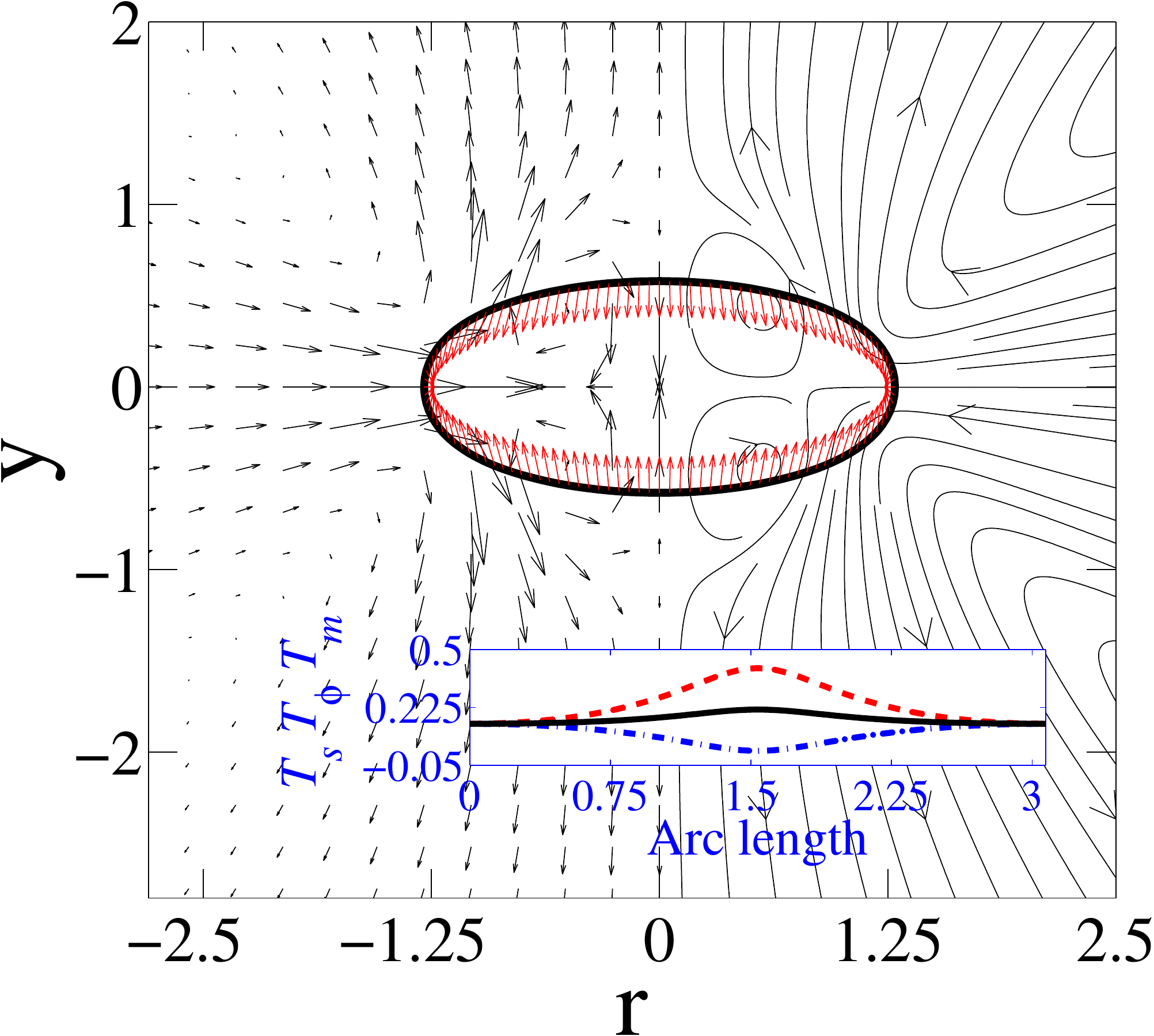}
  \caption{$t=28$}
  \label{fgr:streamca0p25a}
\end{subfigure}
\begin{subfigure}{.32\textwidth}
  \centering
  \includegraphics[width=1\textwidth, trim=0.0in 0.0in 0.2in 0in, clip]{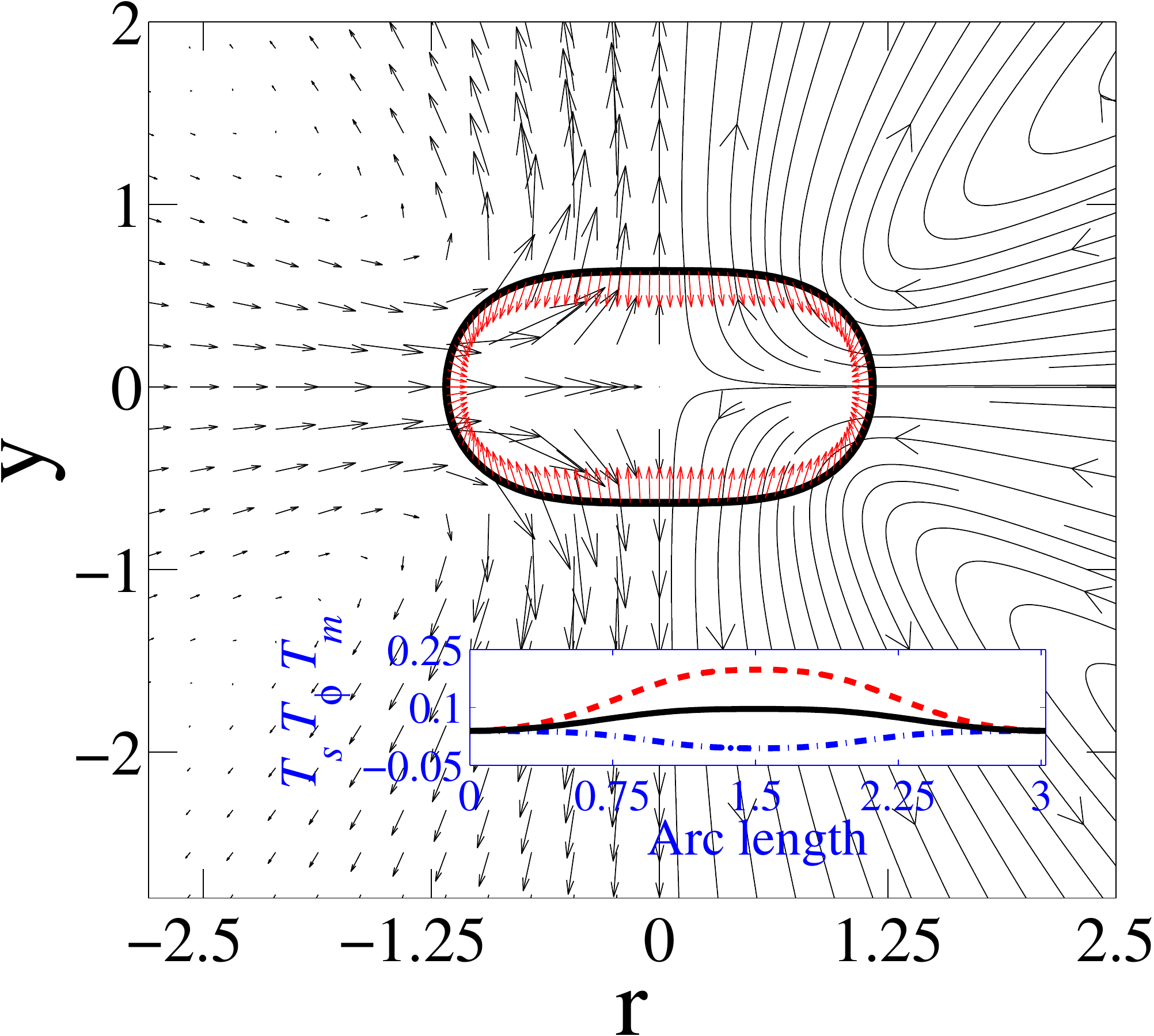}
  \caption{$t=85$}
  \label{fgr:streamca0p25b}
\end{subfigure}
\begin{subfigure}{.32\textwidth}
  \centering
  \includegraphics[width=1\textwidth, trim=0.0in 0.0in 0.2in 0in, clip]{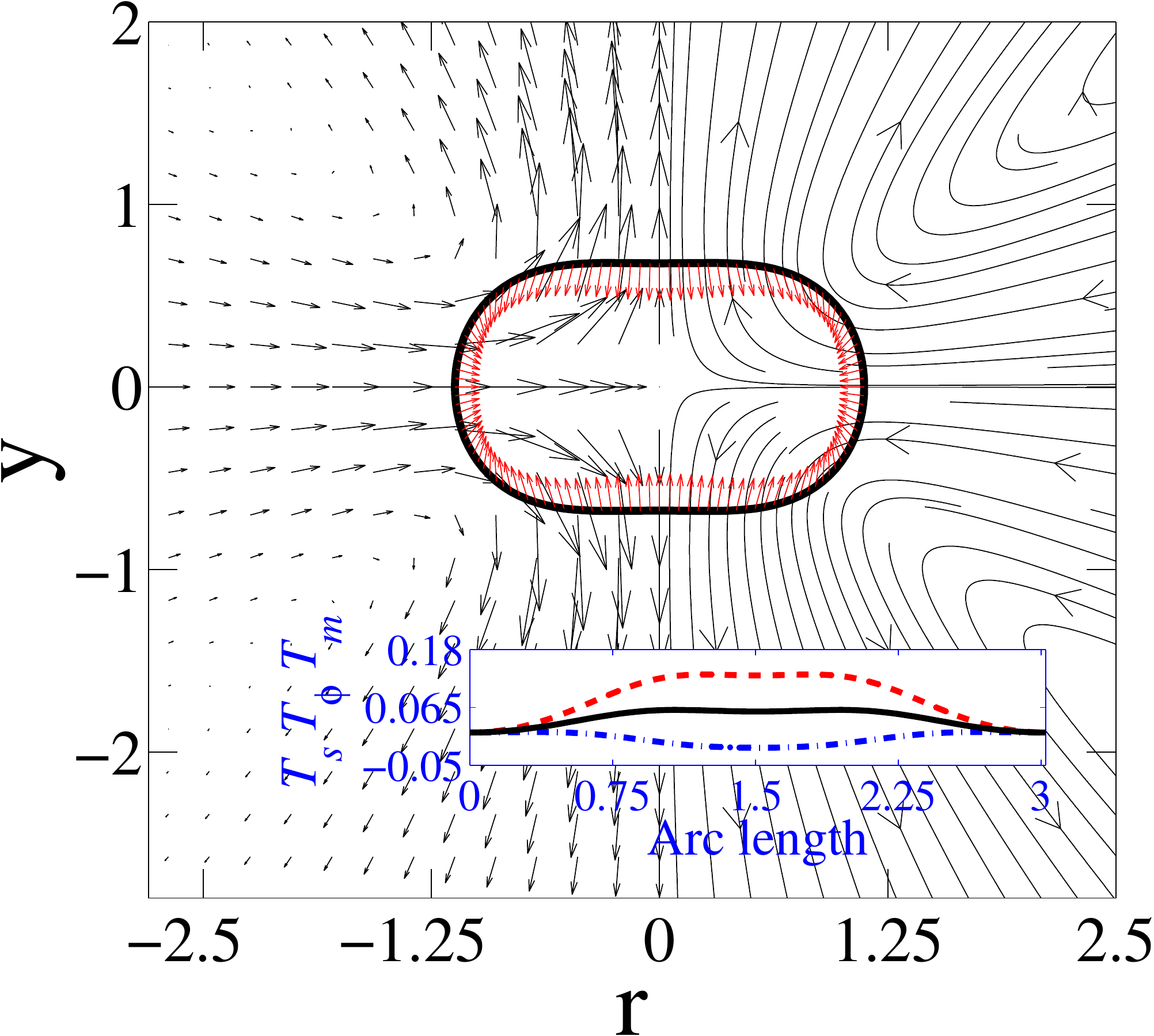}
  \caption{$t=110$}
  \label{fgr:streamca0p25c}
\end{subfigure}
\begin{subfigure}{.32\textwidth}
  \centering
  \includegraphics[width=1\textwidth, trim=0.0in 0.0in 0.2in 0in, clip]{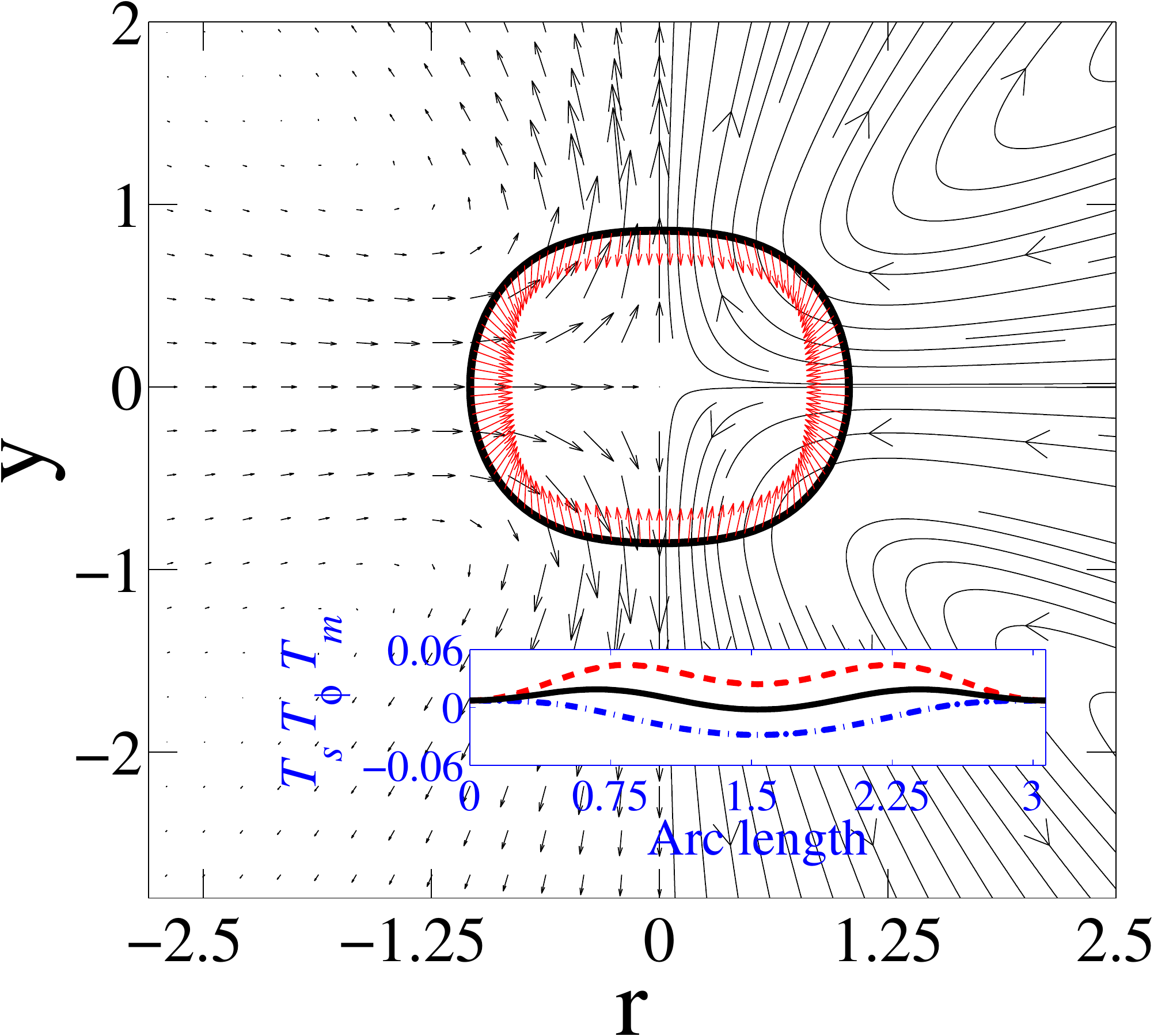}
  \caption{$t=170$}
  \label{fgr:streamca0p25d}
\end{subfigure}
\begin{subfigure}{.32\textwidth}
  \centering
  \includegraphics[width=1\textwidth, trim=0.0in 0.0in 0.2in 0in, clip]{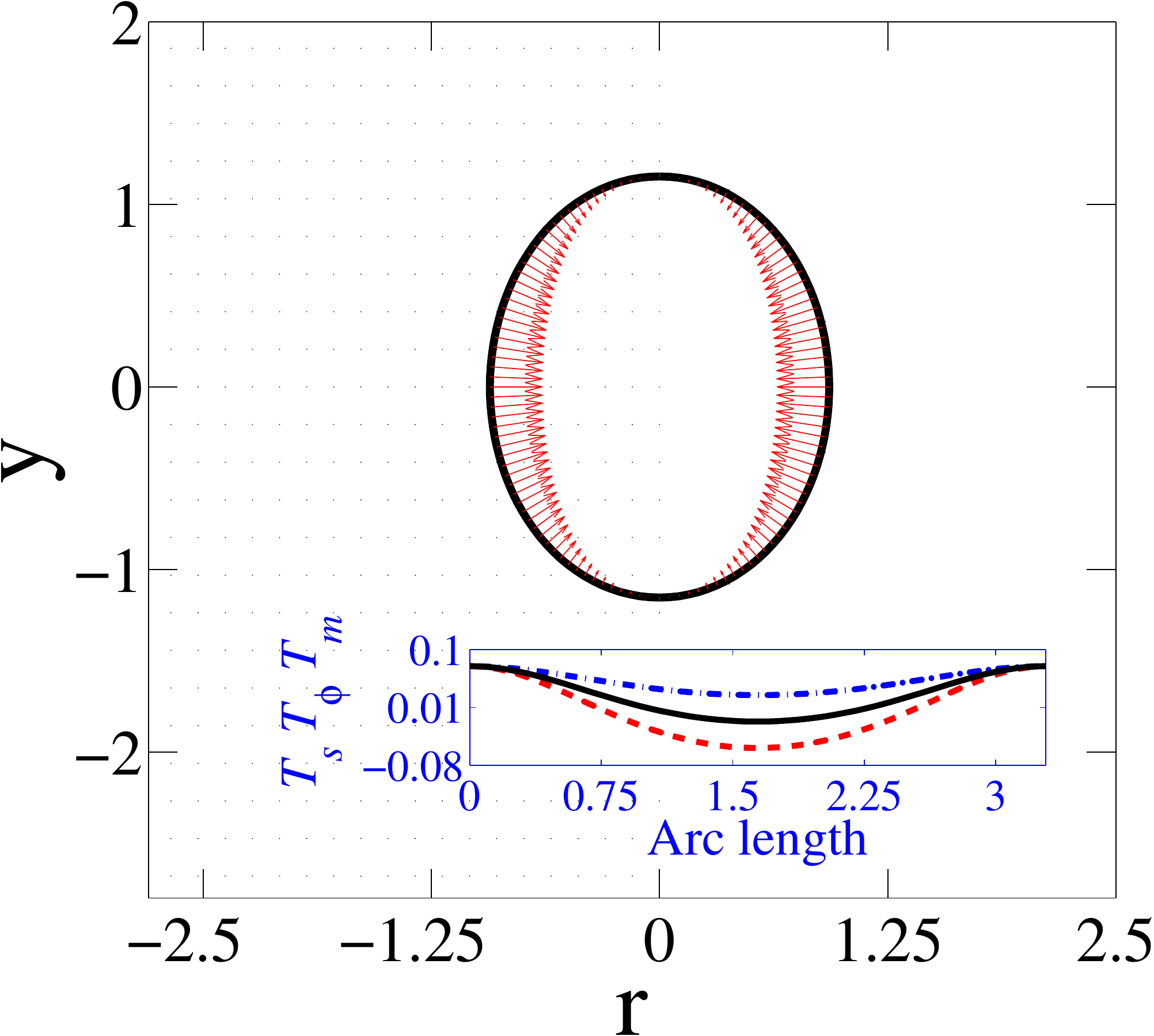}
  \caption{$t=\infty$}
  \label{fgr:streamca0p25e}
\end{subfigure}
\caption{Stream line plot (curves with arrow, shown only at right side of axis of symmetry), velocity profile (short arrows, magnitude represents extend of flow, shown only at left side of axis of symmetry) and electric stress (arrows from the interface) for $Ca=0.25$ and $\sigma_r=0.1$ at different times. Inset shows the variation of meridional, $T_s$ (\textcolor{blue}{$\pmb{-\cdot-}$}), azimuthal, $T_\phi$ (\textcolor{red}{$\pmb{--}$}) and mean, $T_m$ (\textcolor{black}{$\pmb{\mi}$}) membrane tensions along the arc length.}
\label{fig:streaml0p1ca0p25}
\end{center}
\end{figure}

The temporal evolution of a spherical capsule with $\sigma_r=0.1$ at a high capillary number, $(Ca=0.45)$, is shown in  \cref{fgr:shapellhca}. In \cref{fgr:streamb,fgr:streamc,fgr:streamd,fgr:streame,fgr:streamf,fgr:streamg,fgr:streamh} electric tractions, streamlines, velocity profiles are shown for \cref{fgr:shapellhcab,fgr:shapellhcac,fgr:shapellhcad,fgr:shapellhcae,fgr:shapellhcaf,fgr:shapellhcag,fgr:shapellhcah}, respectively, and meridional elastic tension ($T_s$), azimuthal elastic tension ($T_\phi$) and mean elastic tension ($T_m=(T_s+T_\phi)/2$) are shown in the insets of \cref{fgr:streamb,fgr:streamc,fgr:streamd,fgr:streame,fgr:streamf,fgr:streamg,fgr:streamh}, same data for $Ca=0.25$ and $\sigma_r=0.1$ corresponding to \cref{fgr:lcal0p1b,fgr:lcal0p1c,fgr:lcal0p1d,fgr:lcal0p1e,fgr:lcal0p1f} are shown in a sequence of  \cref{fgr:streamca0p25a,fgr:streamca0p25b,fgr:streamca0p25c,fgr:streamca0p25d,fgr:streamca0p25e}.

% To compare the  dynamics at Ca=0.45 with that at Ca=0.25,  a sequence of shape deformation with intermediate squaring of shapes (\cref{fgr:lcal0p1c,fgr:lcal0p1d,fgr:lcal0p1e}) before relaxation to prolate shapes (\cref{fgr:lcal0p1f}) is presented in a sequence of figures (\cref{fgr:streamca0p25a,fgr:streamca0p25b,fgr:streamca0p25c,fgr:streamca0p25d,fgr:streamca0p25e}). The streamlines, velocity profiles and electric tractions as well as elastic tensions, in the inset, are shown for $Ca=0.25$ and $\sigma_r=0.1$.
 
   \Cref{fgr:streamb} shows that the oblate shape at $\sigma_r=0.1$ is associated with compressive normal Maxwell's stress such that their distribution is of the $2^{nd}$ Legendre kind, the flow is from poles to equator due to tangential stresses, thereby assisting formation of oblate shapes.  (The elastic tensions (meridional and azimuthal) are tensile  for  oblate shapes at short times, whereas for biconcave, squaring and hexagonal shapes observed at intermediate times the  meridional tension is compressive at the equator while the azimuthal tension remains tensile. For the steady state prolate deformation, compressive azimuthal tension at the equator and tensile meridional tension over the entire arc length are observed.) 
   
   \Cref{fgr:streamc} shows further increase in the oblate deformation of the capsule because the normal Maxwell's stresses are compressive such that they are maximum at the poles and minimum at the equator. However, importantly, an onset of flow reversal (flow from equator to poles)  is observed.  A complex interplay of flow induced by both Maxwell stress as well as the elastic stress due to relaxation of the stretched capsule (indicated by the meridional tension also becoming tensile at the equator) leads to an initiation of biconcave shape formation of the capsule as shown in \cref{fgr:streamd}.

 At $t=75$, for Ca=0.45,  although the flow is from equator to poles, the large compressive normal Maxwell stress at the poles continues to bring together the poles, resulting in a biconcave shape (\cref{fgr:streame}). The biconcave shape is necessary to generate enough curvature so that the elastic stresses can balance the electric stress. On the other hand, for Ca=0.25, the compressive Maxwell stresses at poles are not high enough to cause a biconcave shape.

 For $Ca=0.45$, at $t=120$ the charging of the membrane leads to substantial compressive stress at the equator, with a corresponding lowering of stress at the poles, which results in further squaring of the capsule (\cref{fgr:streamf}). This is indicated by a greater region of high azimuthal tension around the equator. A peak in the azimuthal tension is now seen at the shoulders. The depression of the biconcave shape relaxes since the normal Maxwell stress at the poles reduce. The relaxation of the poles continues at t=150, leading to hexagonal shapes (\cref{fgr:streamg}). The shoulders seem to get  arrested, probably due to the absence of electric stress,  such that at $t=\infty$ the electric stresses are compressive, which as discussed earlier, can be decomposed into an isotropic part and a tensile $P_2(\cos{\theta})$ part which disappears at $54^{o}$, resulting in a prolate shape (\cref{fgr:streamh}). In the case of Ca=0.25 though, the insufficient compresseive  electrical stresses at the poles lead to an absence of biconcave and hexagonal shapes. This results in oblate shapes directly transforming into square shapes wherein high compressive electrical stresses  at the equator as compared to poles, leads to relaxation to a prolate  steady state shape. 

\subsubsection{{Case of $\sigma_r>1$}}
\begin{figure}[H]
\begin{center}
\begin{subfigure}{.22\textwidth}
  \centering
  \includegraphics[width=1\textwidth]{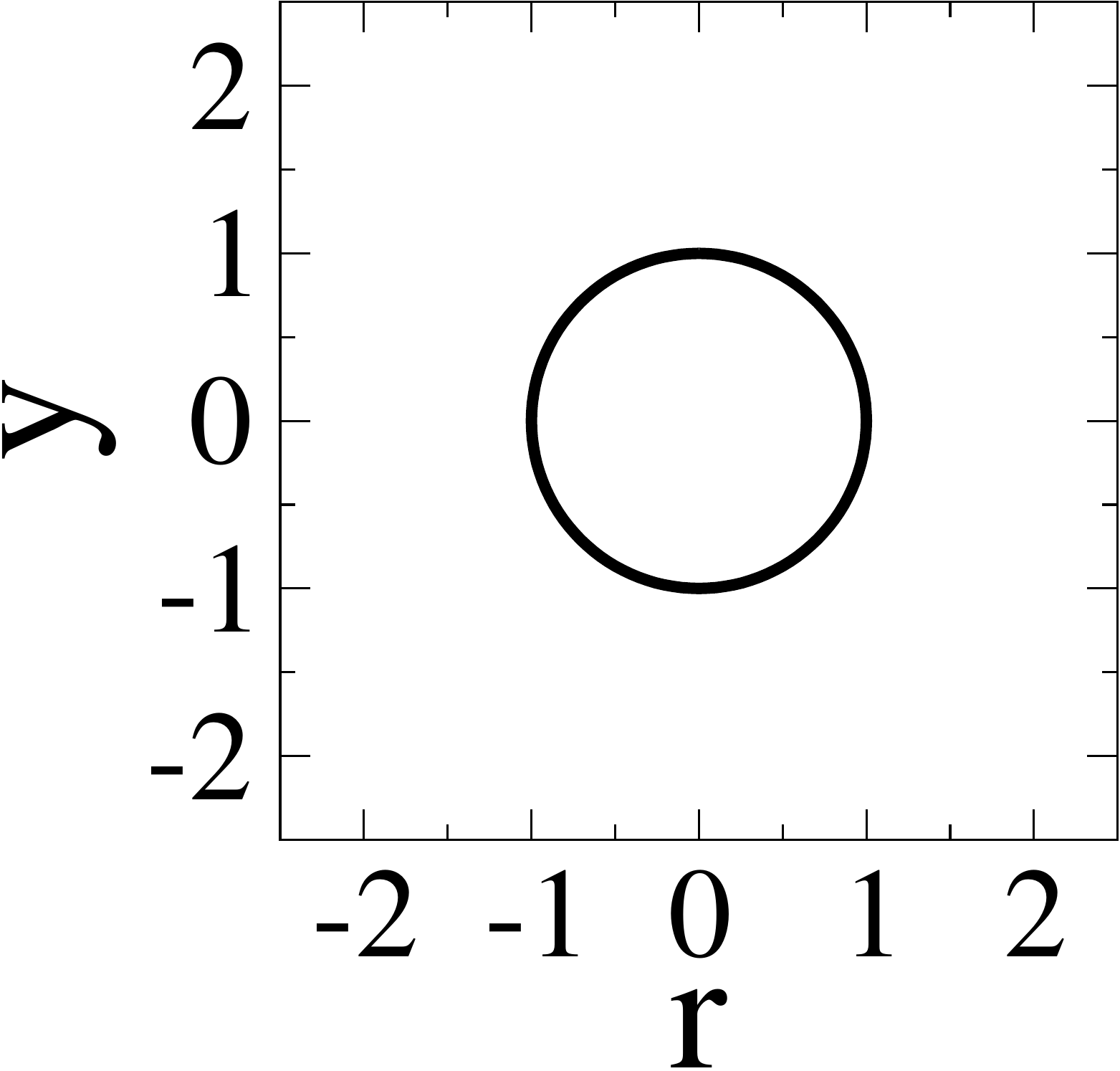}
  \caption{$t=0$}
  \label{fgr:shapel10a}
\end{subfigure}
\begin{subfigure}{.22\textwidth}
  \centering
  \includegraphics[width=1\textwidth]{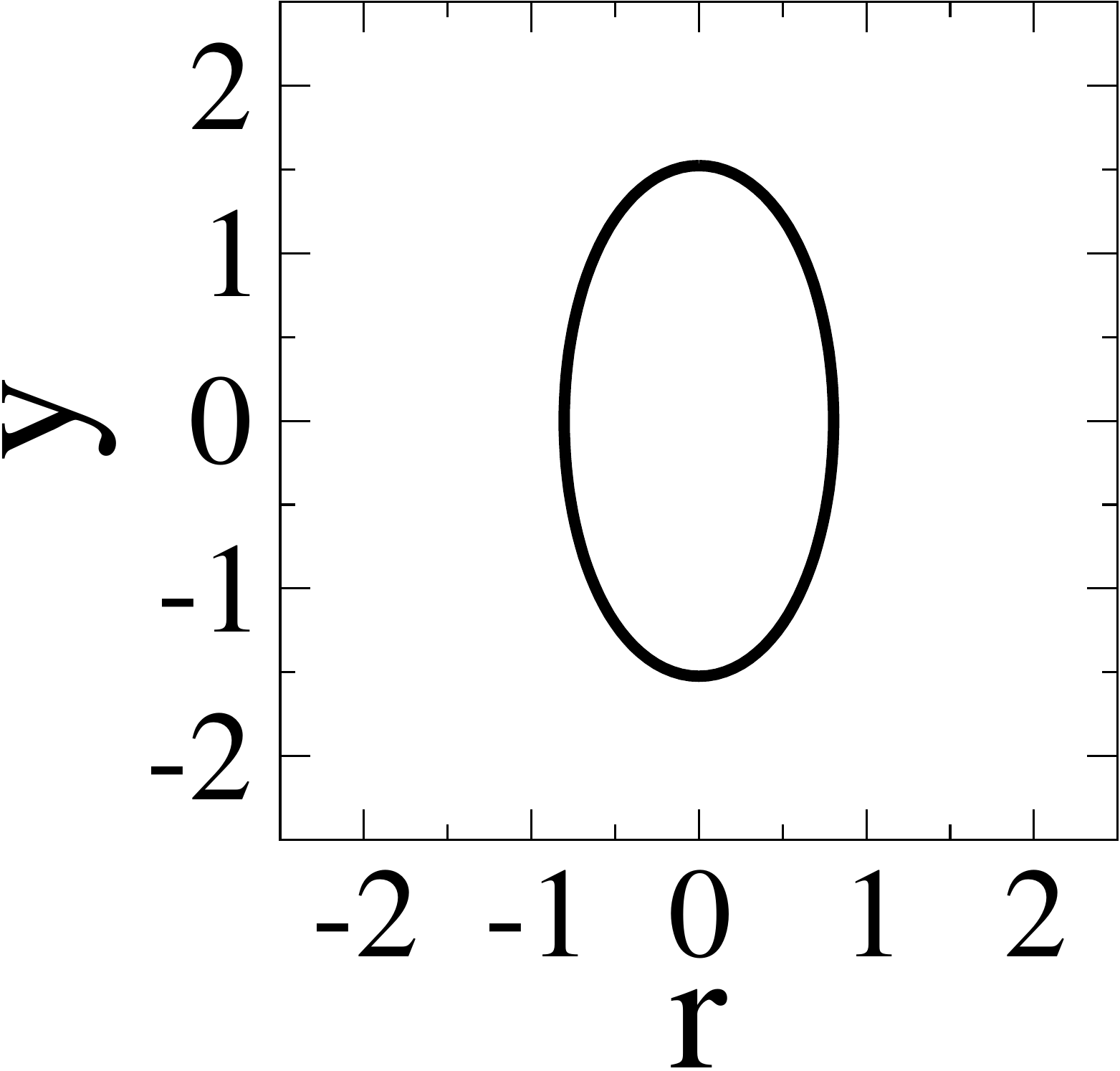}
  \caption{$t=1$}
  \label{fgr:shapel10b}
\end{subfigure}
\begin{subfigure}{.22\textwidth}
  \centering
  \includegraphics[width=1\textwidth]{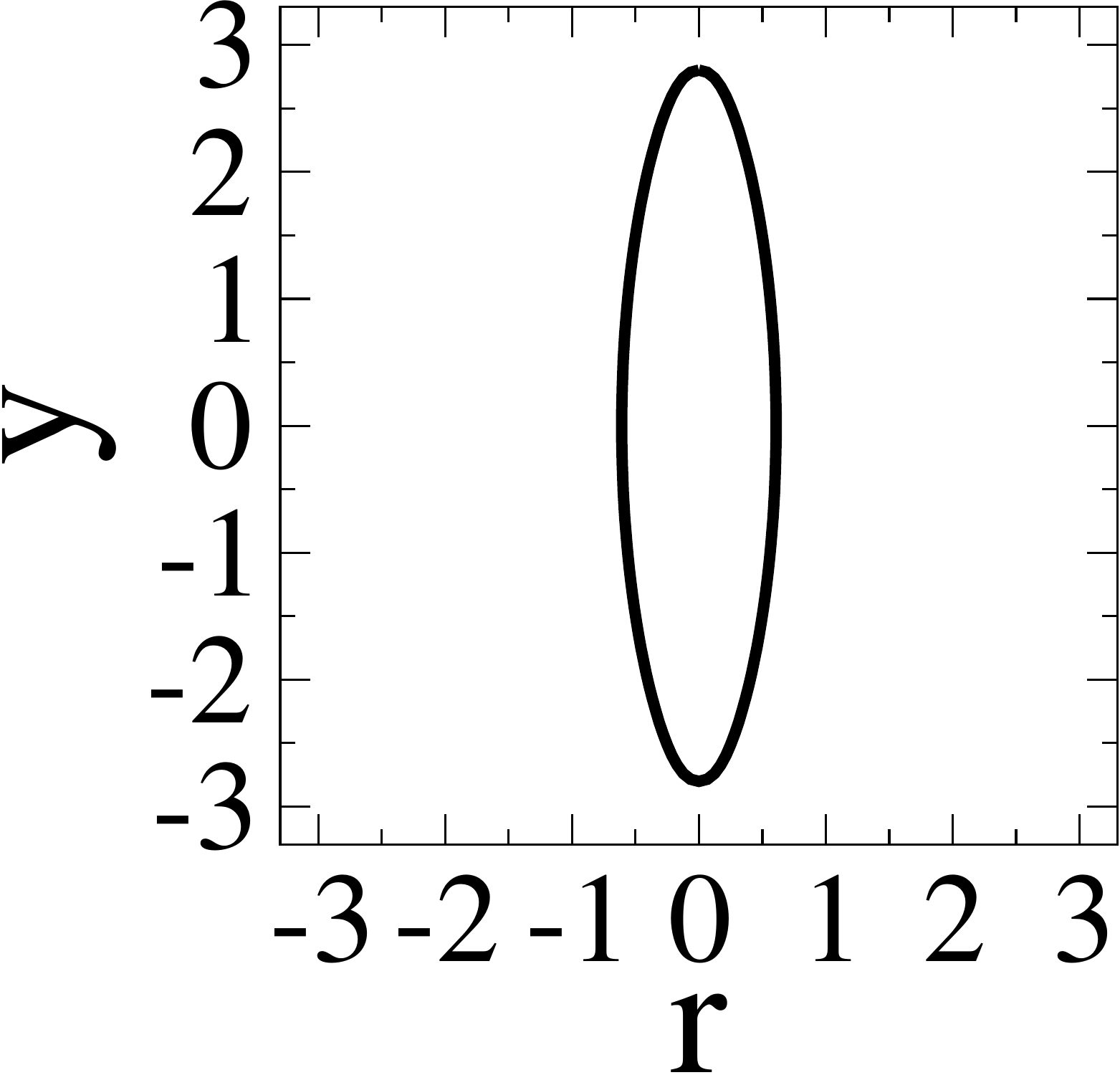}
  \caption{$t=4$}
  \label{fgr:shapel10c}
\end{subfigure}
\begin{subfigure}{.22\textwidth}
  \centering
  \includegraphics[width=1\textwidth]{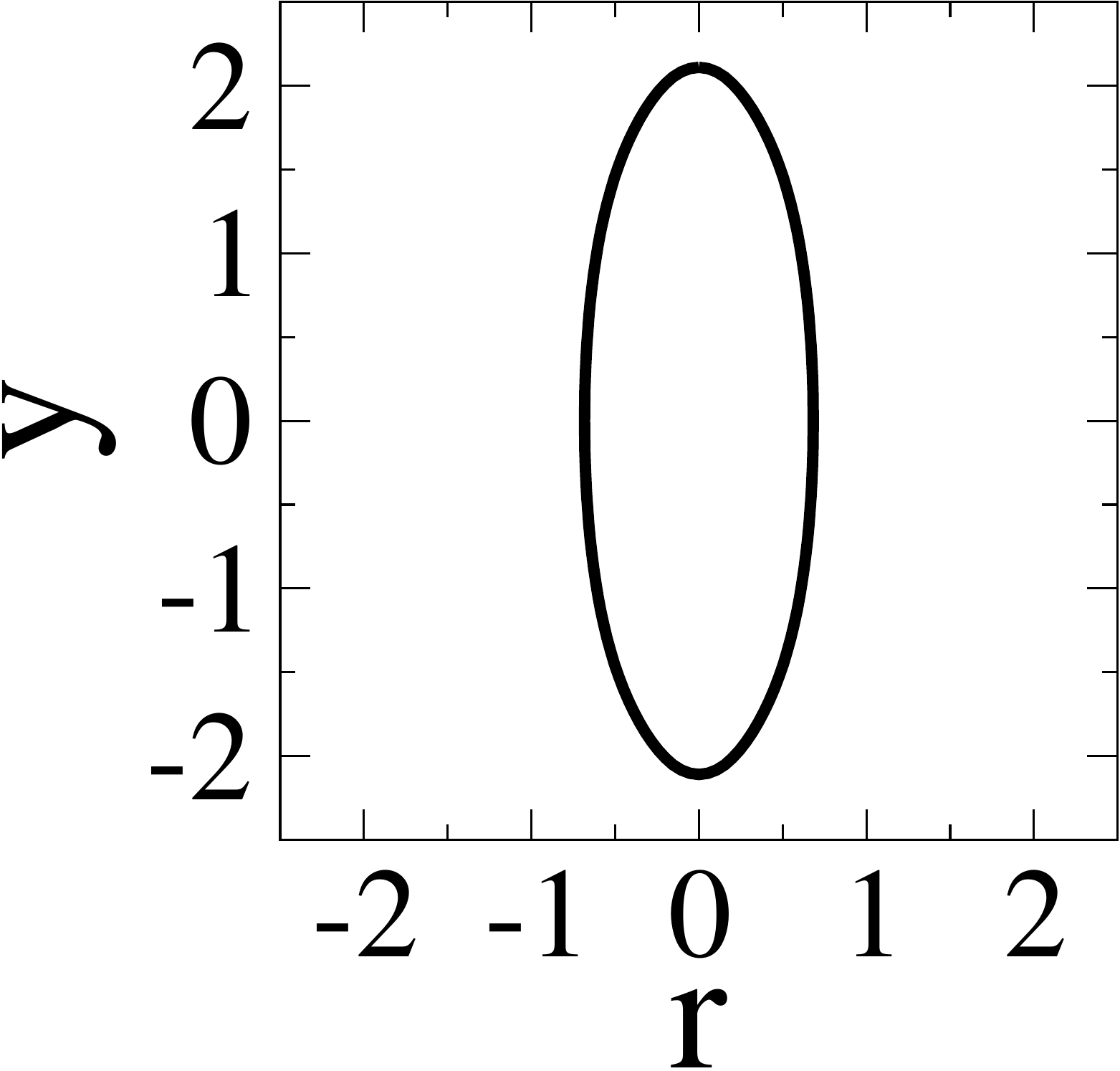}
  \caption{$t=40$}
  \label{fgr:shapel10d}
\end{subfigure}
\begin{subfigure}{.22\textwidth}
  \centering
  \includegraphics[width=1\textwidth]{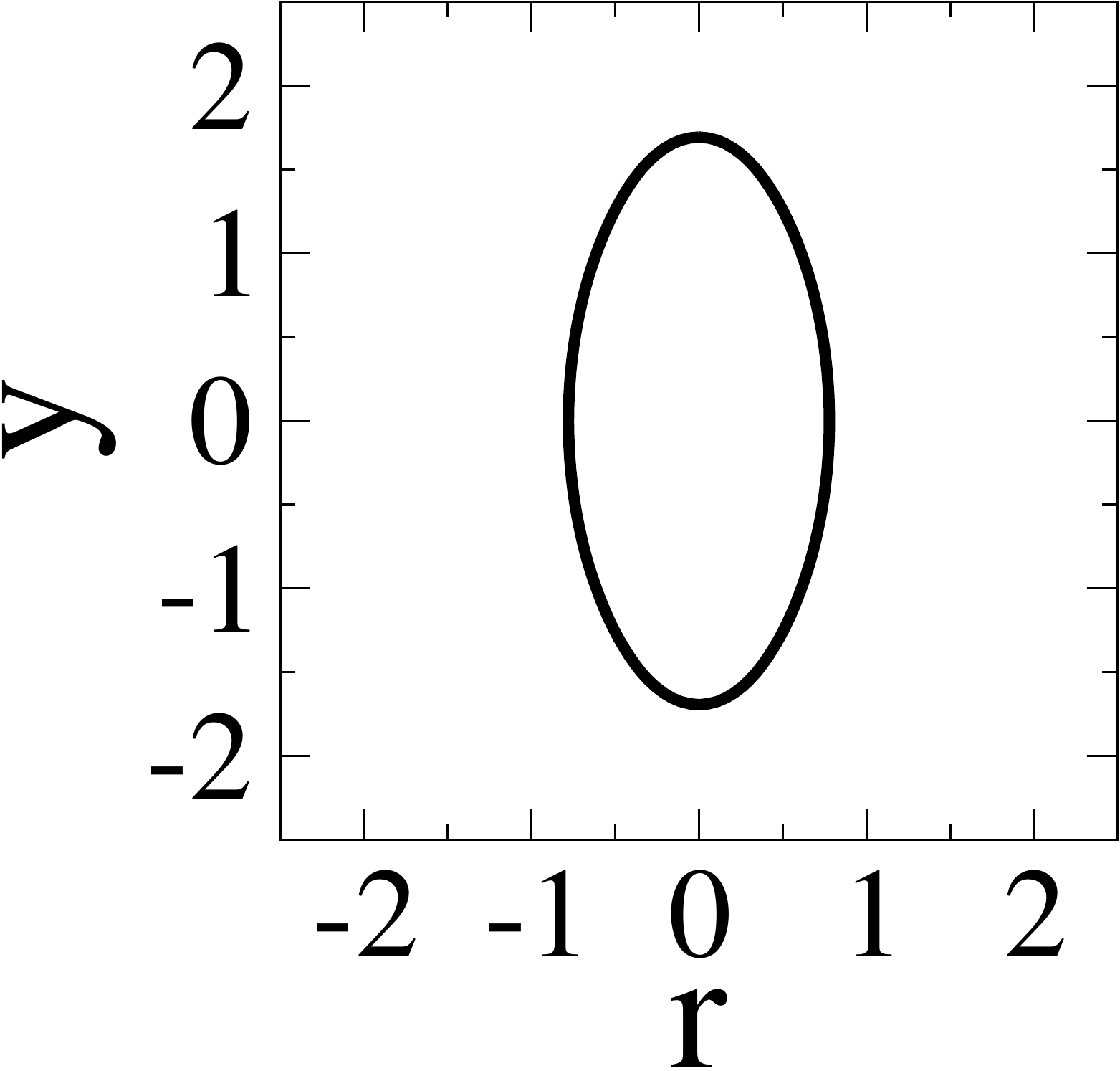}
  \caption{$t=70$}
  \label{fgr:shapel10e}
\end{subfigure}
\begin{subfigure}{.22\textwidth}
  \centering
  \includegraphics[width=1\textwidth]{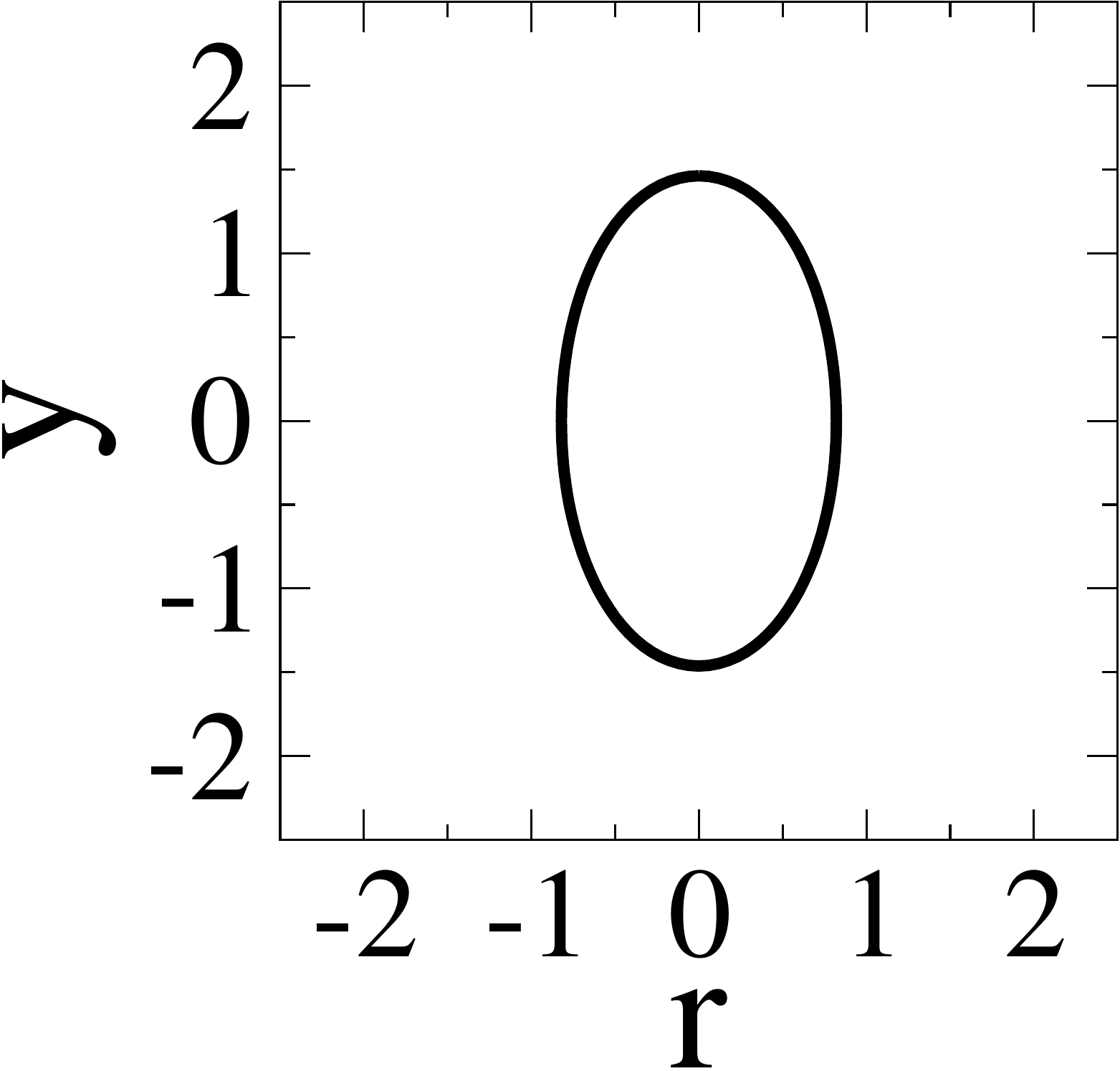}
  \caption{$t=\infty$}
  \label{fgr:shapel10f}
\end{subfigure}
\caption{Shape evolution of a capsule with Skalak membrane at $\sigma_r=10$ for $Ca=2$.}
\label{fgr:shapel10}
\end{center}
\end{figure}
\begin{figure}[H]
 \begin{center}
\begin{subfigure}{.32\textwidth}
  \centering
  \includegraphics[width=1\textwidth]{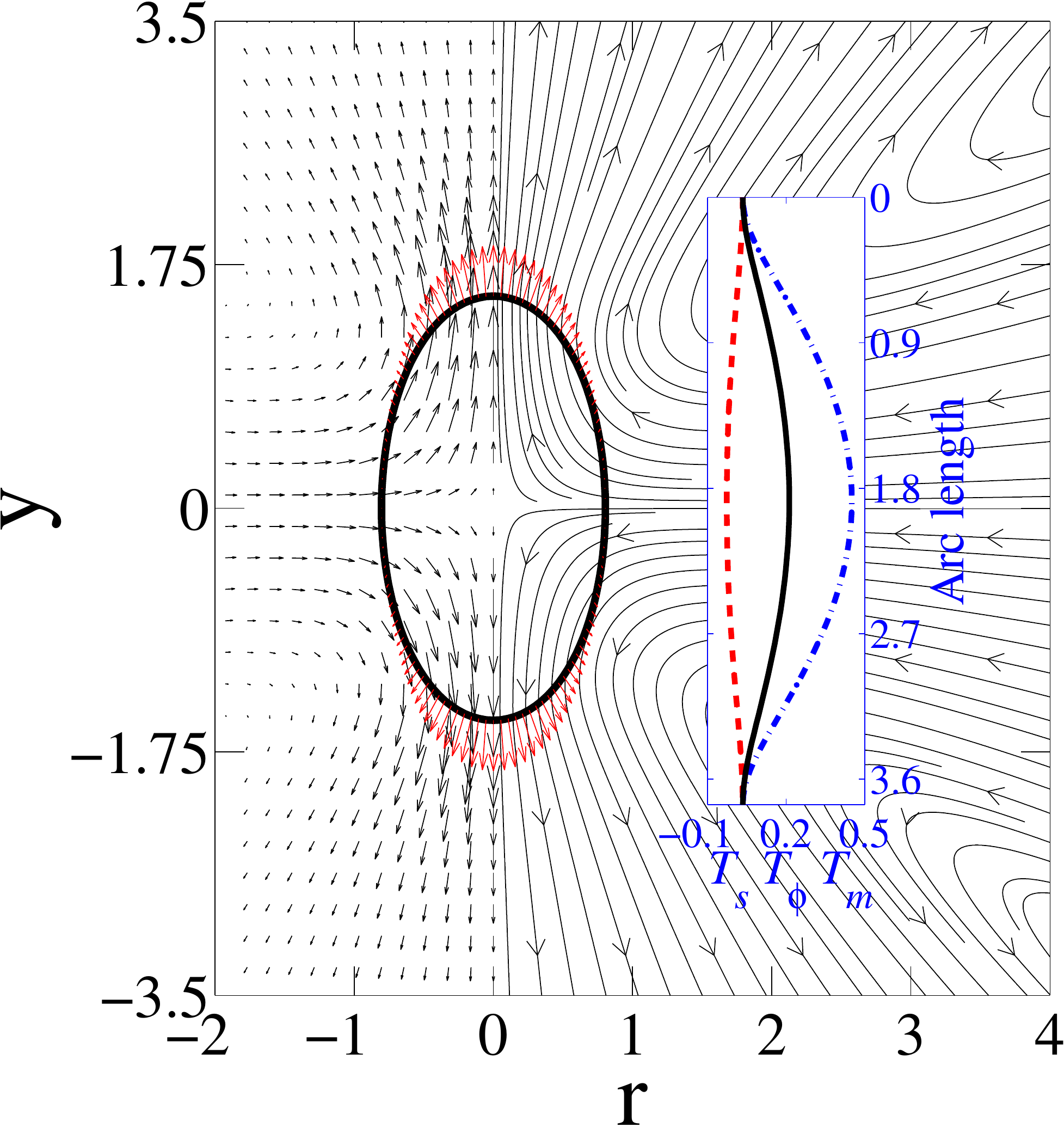}
  \caption{$t=1$}
  \label{fig:streaml10a}
\end{subfigure}
\begin{subfigure}{.32\textwidth}
  \centering
\includegraphics[width=1\textwidth]{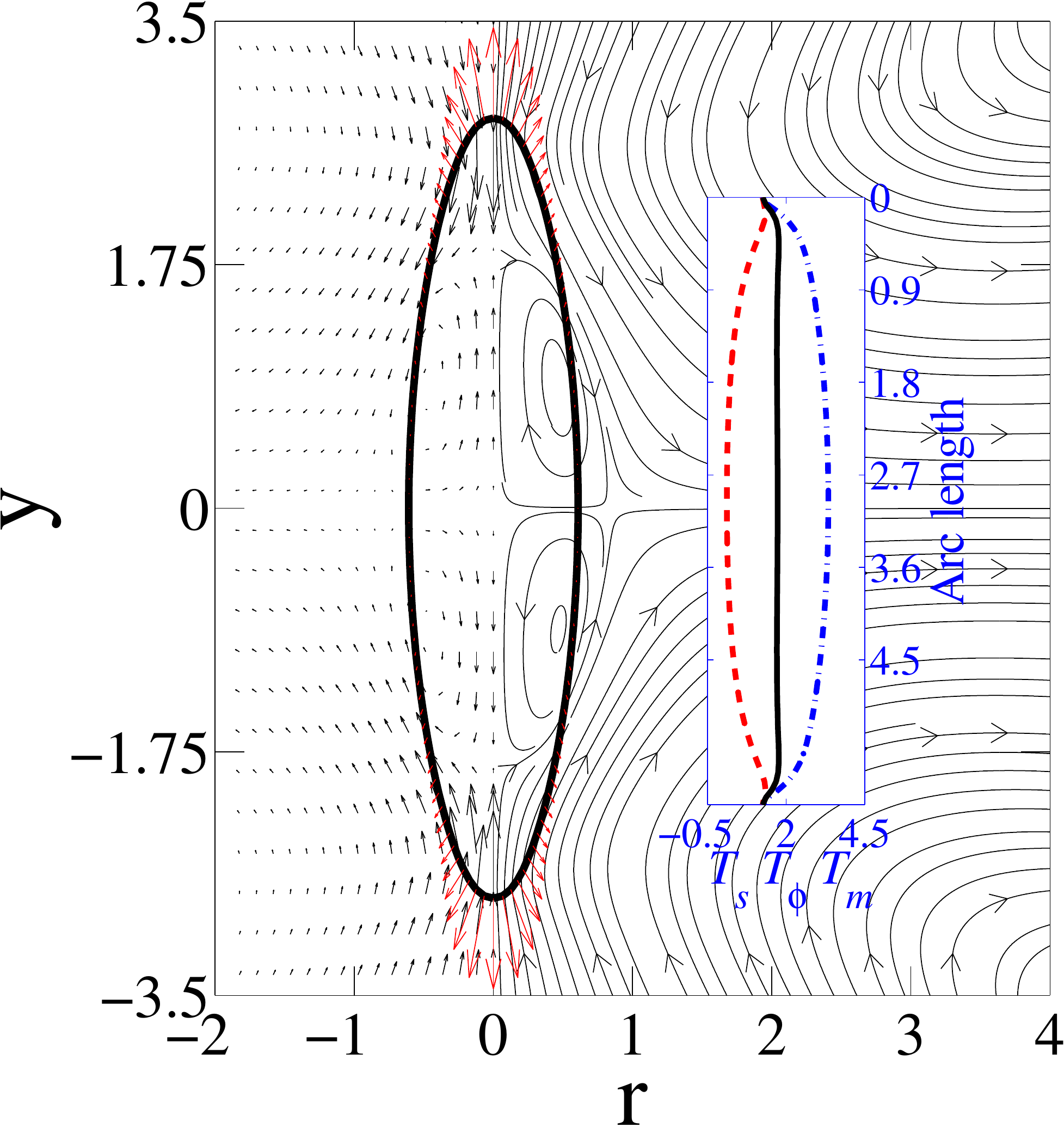}
  \caption{$t=4$}
  \label{fig:streaml10b}
\end{subfigure}
\begin{subfigure}{.32\textwidth}
  \centering
  \includegraphics[width=1\textwidth]{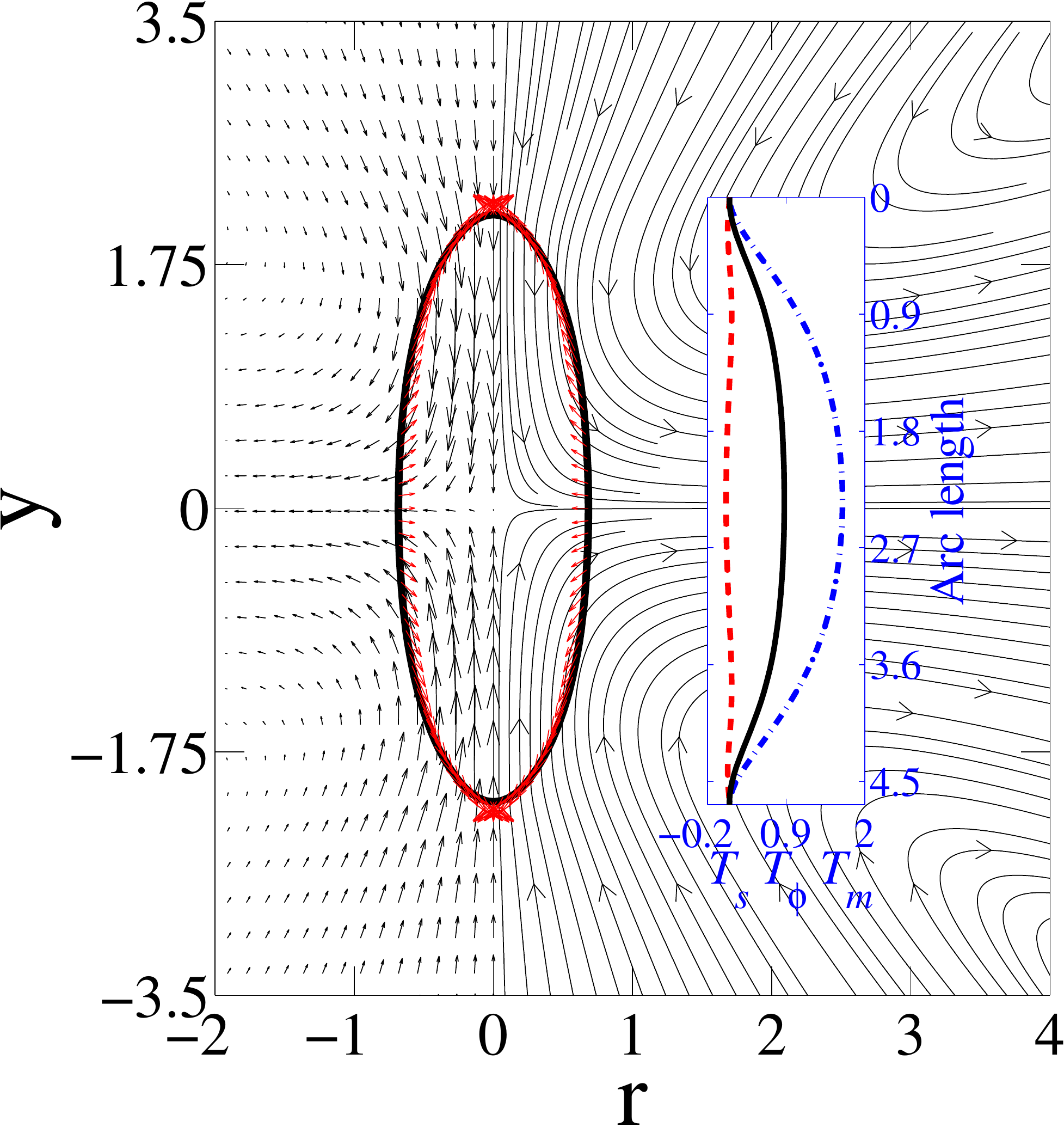}
  \caption{$t=40$}
  \label{fig:streaml10c}
\end{subfigure}
\begin{subfigure}{.32\textwidth}
  \centering
  \includegraphics[width=1\textwidth]{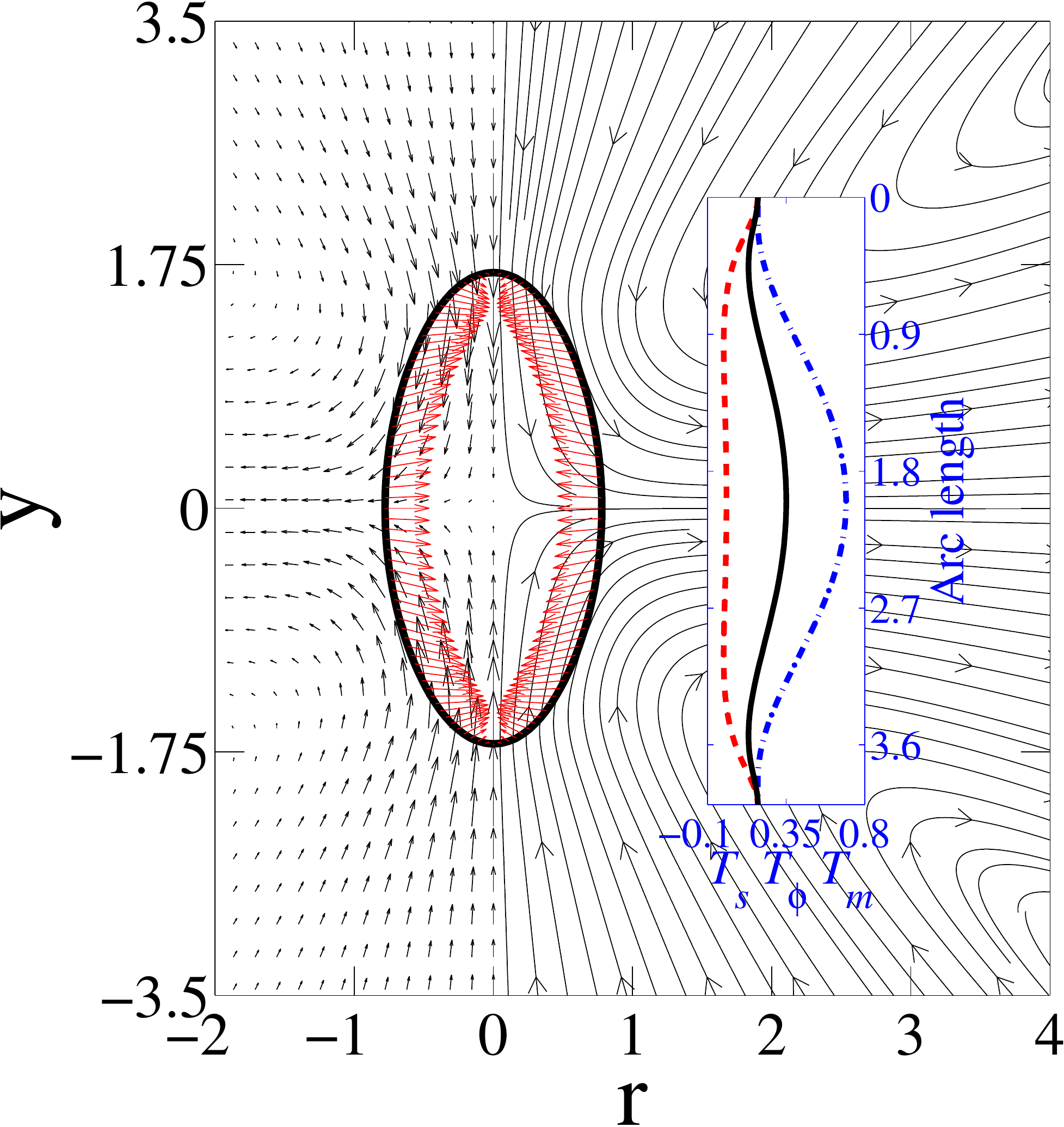}
  \caption{$t=70$}
  \label{fig:streaml10d}
\end{subfigure}
\begin{subfigure}{.32\textwidth}
  \centering
  \includegraphics[width=1\textwidth]{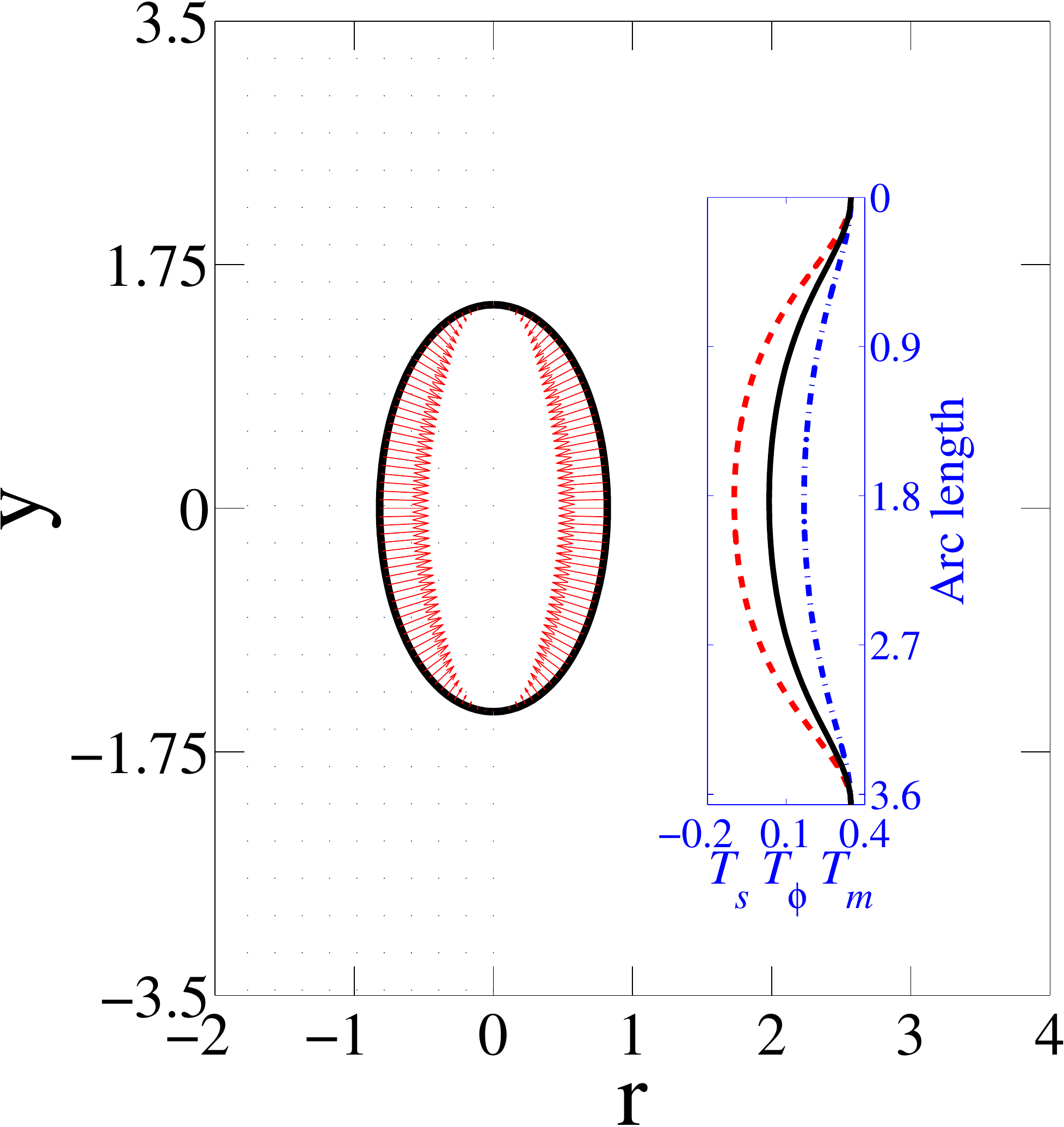}
  \caption{$t=\infty$}
  \label{fig:streaml10e}
\end{subfigure}
\caption{Stream line plot (curves with arrow, shown only at right side of axis of symmetry), velocity profile (short arrows, magnitude represents extend of flow, shown only at left side of axis of symmetry) and electric stress (arrows from the interface) for $Ca=2$ and $\sigma_r=10$ at different times. Inset shows the variation of meridional, $T_s$ (\textcolor{blue}{$\pmb{-\cdot-}$}), azimuthal, $T_\phi$ (\textcolor{red}{$\pmb{--}$}) and mean, $T_m$ (\textcolor{black}{$\pmb{\mi}$}) membrane tensions along the arc length.}
\label{fig:streaml10}
\end{center}
\end{figure}

 
In \cref{fgr:shapel10} the dynamics of a spherical capsule with a Skalak membrane at high capillary number ($Ca=2$) for $\sigma_r=10$ is shown. Unlike the case for $\sigma_r=0.1$, and as discussed earlier, a capsule does not attain oblate intermediate shapes for $\sigma_r=10$. Initially because of high tensile electric traction, a  capsule exhibits large prolate deformation (\cref{fig:streaml10a,fig:streaml10b}). Eventually because of the compressive electric tractions at long times, it relaxes back to a steady state prolate shape (\cref{fig:streaml10d,fig:streaml10e}).  At short times, the meridional and azimuthal tensions are tensile. While the meridional tension is maximum at the equator and smallest at the poles ( \cref{fig:streaml10a,fig:streaml10b}),  it is vice-versa for the azimuthal tension. At intermediate times, azimuthal tension is negligible around the equatorial region. At steady state, the elastic tensions are similar to that  at $\sigma_r=0.1$. Simultaneously, the flow changes from equator to poles at short time (\cref{fig:streaml10a}), to poles to equator at intermediate times (\cref{fig:streaml10b,fig:streaml10c,fig:streaml10d}), eventually disappearing at long times (\cref{fig:streaml10e}).

\subsection{Effect of membrane capacitance, conductance and hydrodynamics on capsule deformation}\label{sec:parametereffect}
\begin{figure}[H]
\centering
  \includegraphics[width=0.5\textwidth]{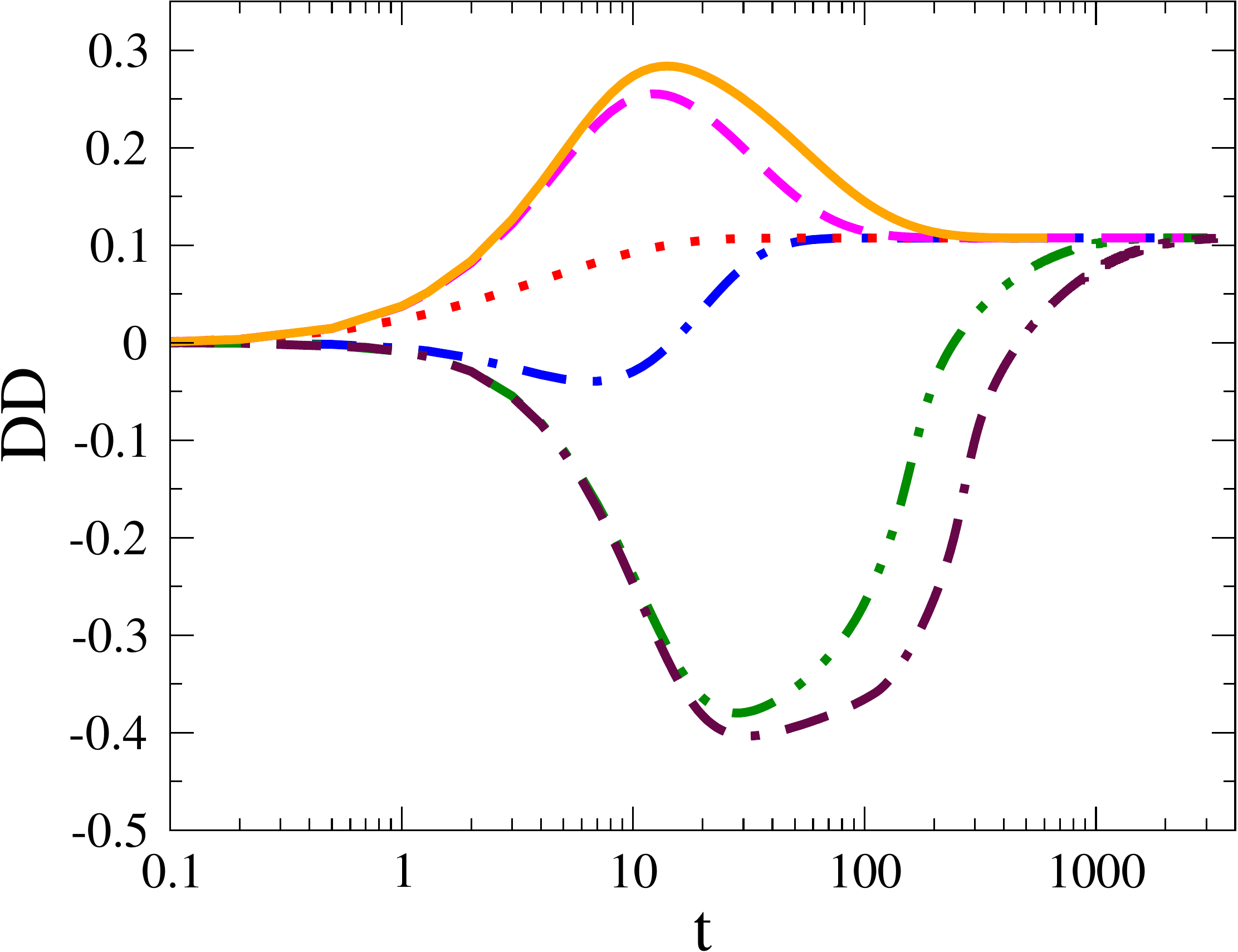}
  \caption{Effect of membrane capacitance on the dynamics of deformation of a capsule with Skalak membrane considering $\hat G_m=0$ at $Ca=0.25$. Curves with line-style (\textcolor{red}{$\pmb \cdots$}) for $\hat C_m=1$, (\textcolor{magenta}{$\pmb{--}$}) for $\hat C_m=50$ (\textcolor{orange}{$\pmb{\mi}$}) for $\hat C_m=100$ at $\sigma_r=10$ and curves with line-style (\textcolor{blue}{$\pmb{-\cdot-}$}) for $\hat C_m=1$, (\textcolor{forestgreen}{$\pmb{-\cdot\cdot}$}) for $\hat C_m=50$ and (\textcolor{brown}{$\pmb{--\cdot}$}) for $\hat C_m=100$ at $\sigma_r=0.1$. }
  \label{fgr:cmvarry}
\end{figure}

\begin{figure}[H]
\begin{center}
\begin{subfigure}{.22\textwidth}
  \centering
  \includegraphics[width=1\textwidth]{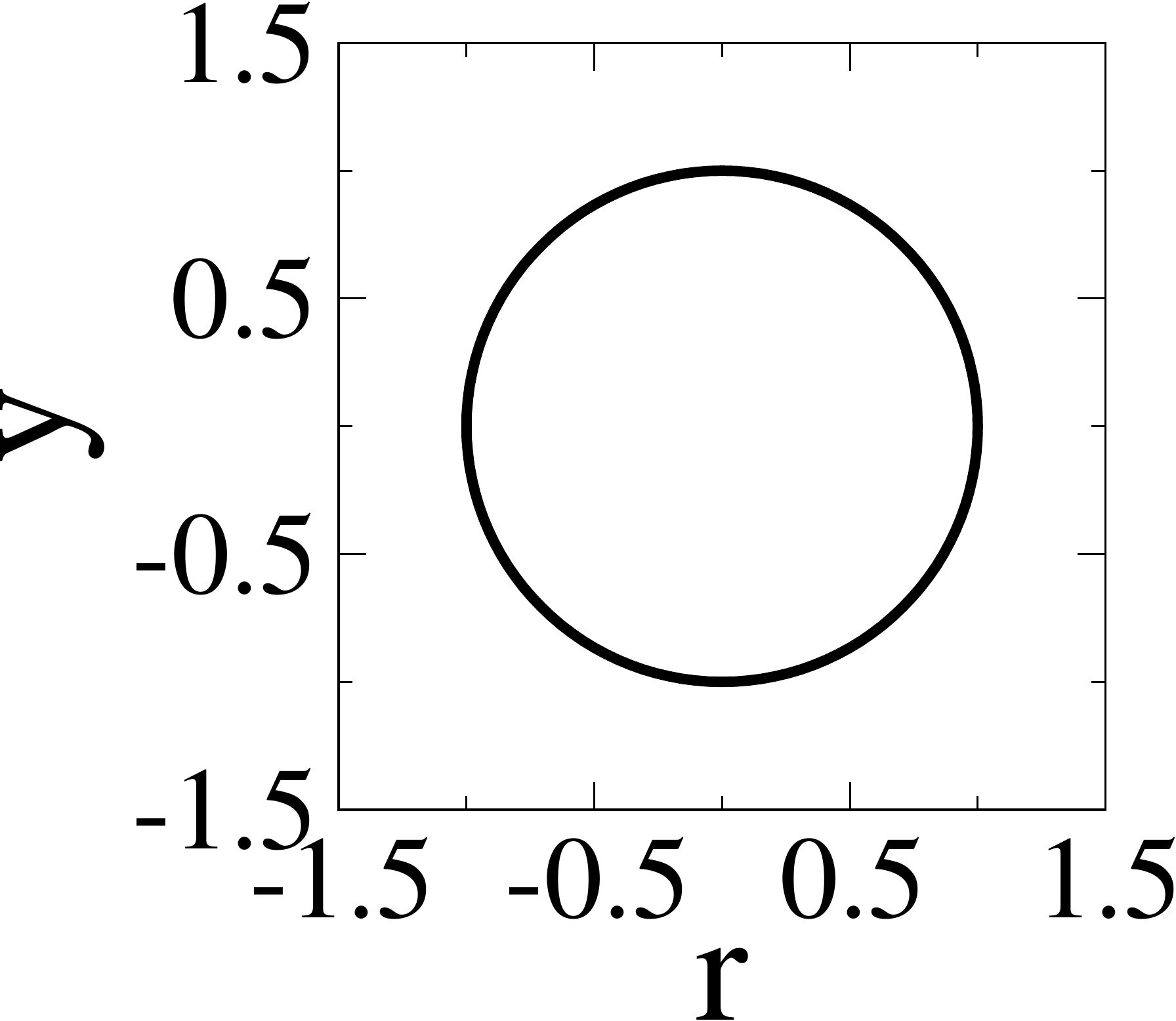}
  \caption{$t=0$}
  \label{fgr:shapel10cm1a}
\end{subfigure}
\begin{subfigure}{.22\textwidth}
  \centering
  \includegraphics[width=1\textwidth]{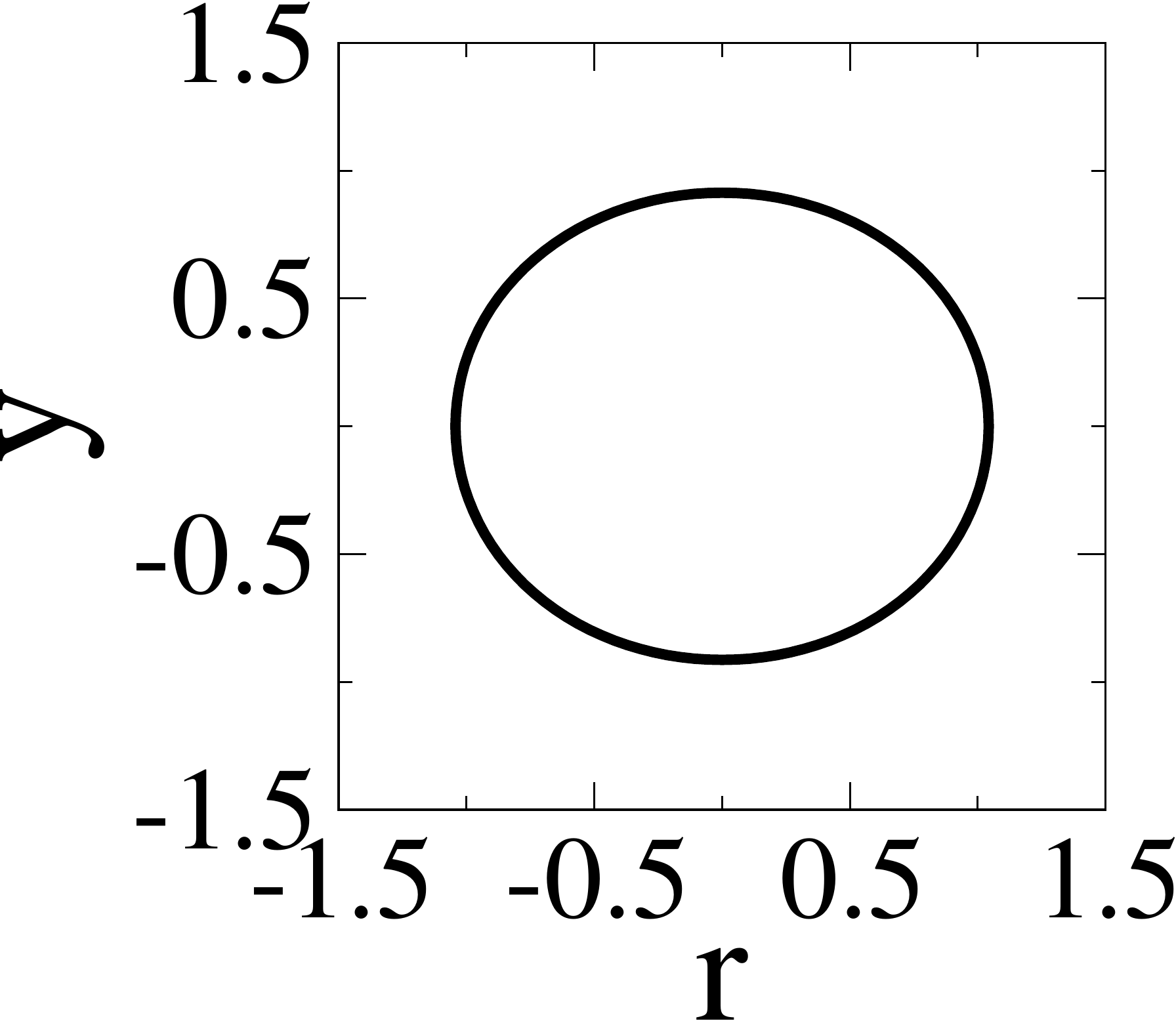}
  \caption{$t=12$}
  \label{fgr:shapel10cm1b}
\end{subfigure}
\begin{subfigure}{.22\textwidth}
  \centering
  \includegraphics[width=1\textwidth]{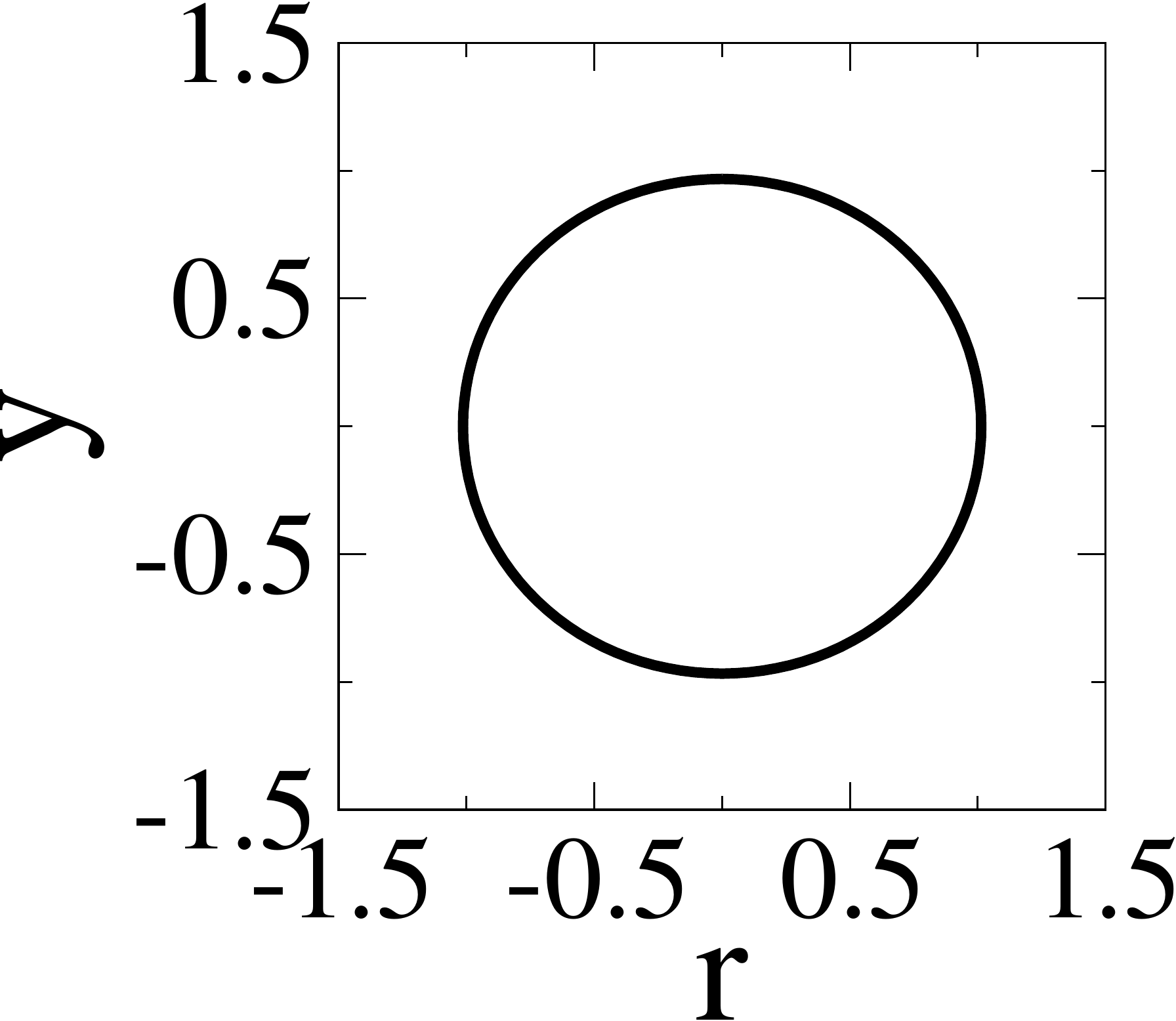}
  \caption{$t=40$}
  \label{fgr:shapel10cm1c}
\end{subfigure}
\begin{subfigure}{.22\textwidth}
  \centering
  \includegraphics[width=1\textwidth]{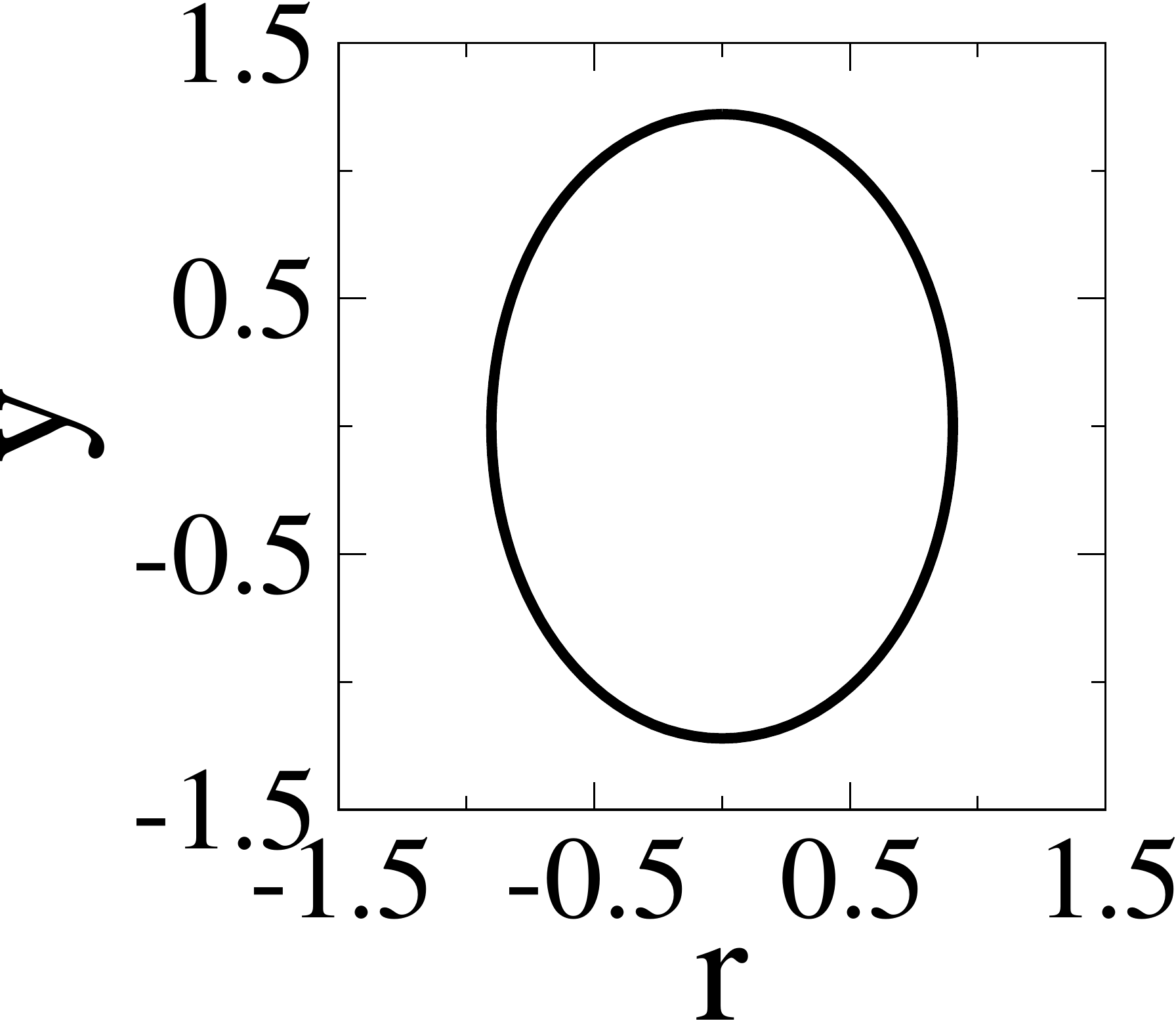}
  \caption{$t=\infty$}
  \label{fgr:shapel10cm1d}
\end{subfigure}
\caption{Shape evolution of a capsule with Skalak membrane at $\hat C_m=1$ for $\sigma_r=0.1$ and $Ca=0.45$.}
\label{fgr:shapel10cm1}
\end{center}
\end{figure} 

\Cref{fgr:cmvarry} shows the effect of membrane capacitance on the degree of deformation of a capsule. As the capacitance increases the response becomes slower, since the charging time increases. For $\sigma_r=10$, where prolate deformation is expected, the peak deformation for higher capacitances is higher because the continuity of Ohmic current in the fluid and the displacement current in the membrane, means that higher the capacitance higher is the  normal electric field in the two fluids. This leads to higher Maxwell stress. A similar explanation can be provided for the oblate deformation observed when $\sigma_r=0.1$. Moreover, as expected, the relaxation time is higher for higher capacitances. Lower capacitances, also mean faster charging of the membrane, higher membrane impedance at short times, which negate the drop-like behavior of a capsule at short times, since $t_{cap}\sim t_{MW}\sim t_H$. Very small oblate intermediate deformation are therefore observed at $\sigma_r=0.1$, when $\hat C_m=1$.

\Cref{fgr:shapel10cm1} shows that the intermediate shapes obtained during the deformation of capsules with small capacitances are quite different compared to those for high capacitances. A capsule   ($Ca=0.45$ and $\sigma_r=0.1$), deforms into an oblate spheroid (\cref{fgr:shapel10cm1b}) shape and through an intermediate near spherical shape (\cref{fgr:shapel10cm1c}) it reaches to a steady state prolate spheroid (\cref{fgr:shapel10cm1d}). The absence of squaring is really due to absence of high compressive stresses, observed in the high capacitance case, since the membrane already becomes charged enough, to negate the effect of build up of Maxwell stress due to conductivity contrast.  The steady state prolate shape is independent of the membrane capacitance. This suggests that for admitting intermediate squaring and biconcave shapes a sufficiently high capacitance of the membrane is essential. 

\begin{figure}[H]
\centering
  \includegraphics[width=0.5\textwidth]{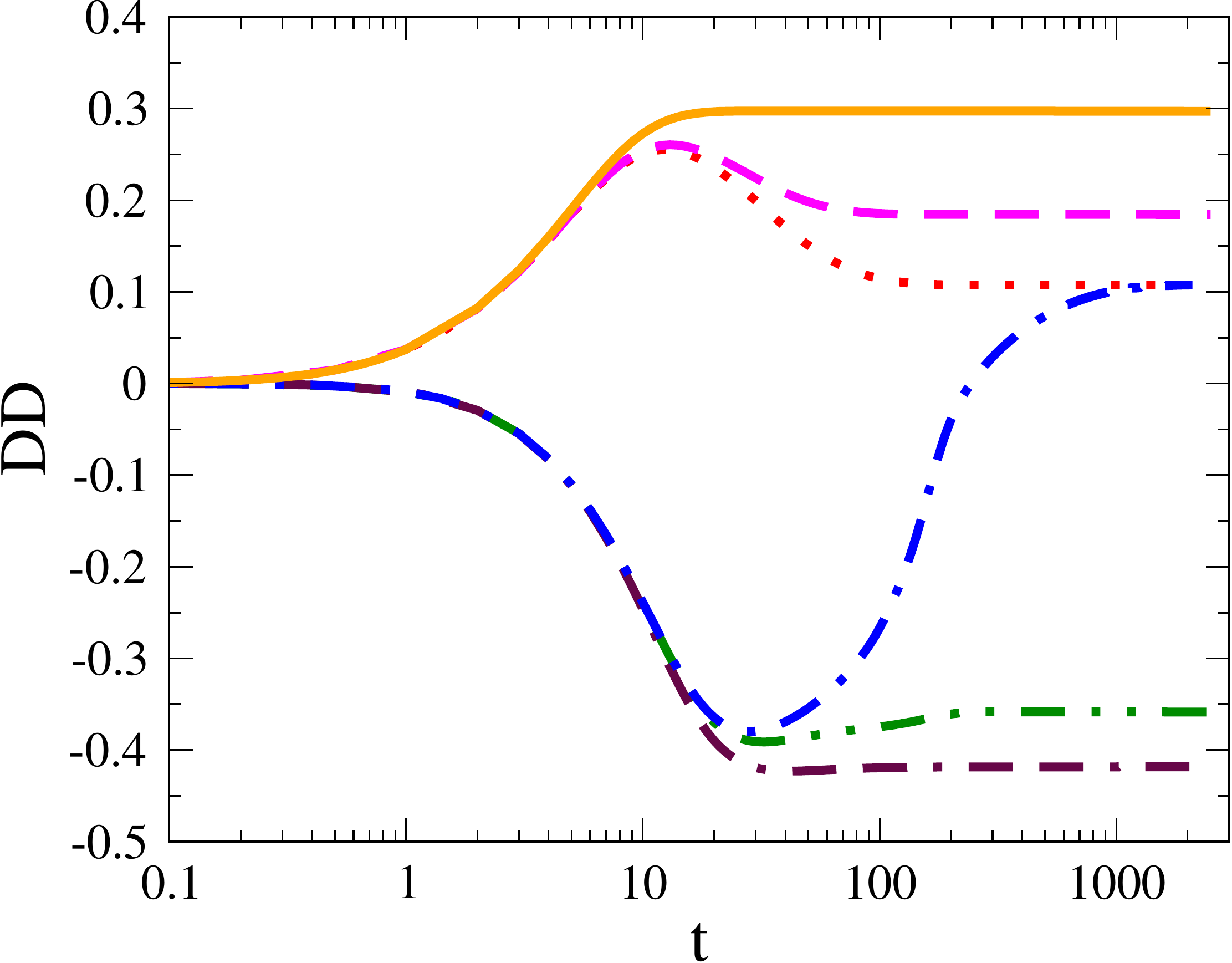}
  \caption{Effect of membrane conductance on the deformation of a capsule with Skalak membrane considering $\hat C_m=50$ at $Ca=0.25$. Curves with line-style (\textcolor{red}{$\pmb \cdots$}) for $\hat G_m=0$, (\textcolor{magenta}{$\pmb{--}$}) for $\hat G_m=1$ (\textcolor{orange}{$\pmb{\mi}$}) for $\hat G_m=10$ at $\sigma_r=10$ and curves with line-style (\textcolor{blue}{$\pmb{-\cdot-}$}) for $\hat G_m=0$, (\textcolor{forestgreen}{$\pmb{-\cdot\cdot}$}) for $\hat G_m=1$ and (\textcolor{brown}{$\pmb{--\cdot}$}) for $\hat G_m=10$ at $\sigma_r=0.1$.}
  \label{fgr:gmvarry}
\end{figure}

\begin{figure}[H]
\begin{center}
\begin{subfigure}{.22\textwidth}
  \centering
  \includegraphics[width=1\textwidth]{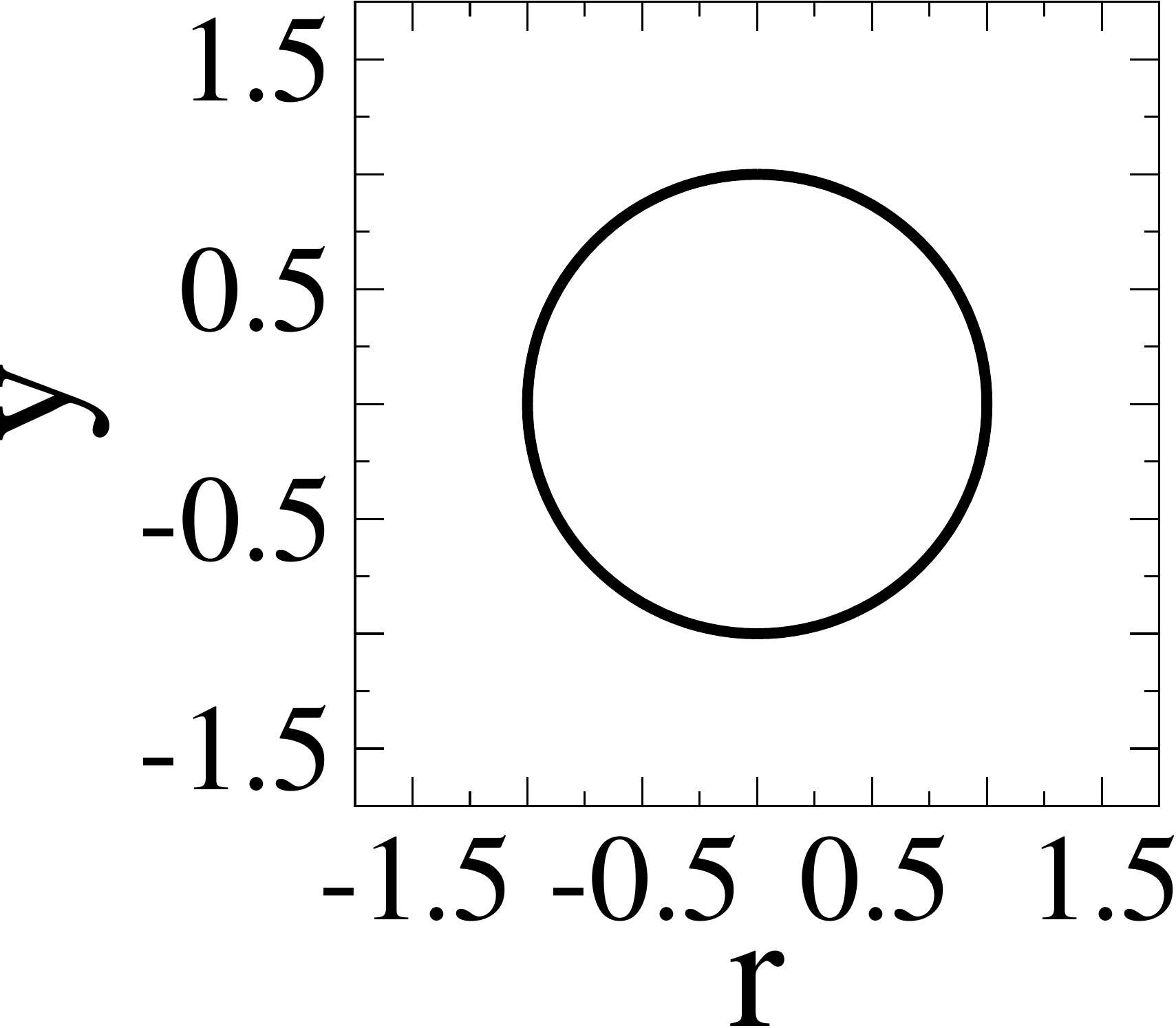}
  \caption{$t=0$}
  \label{fgr:gmeffecta}
\end{subfigure}
\begin{subfigure}{.22\textwidth}
  \centering
  \includegraphics[width=1\textwidth]{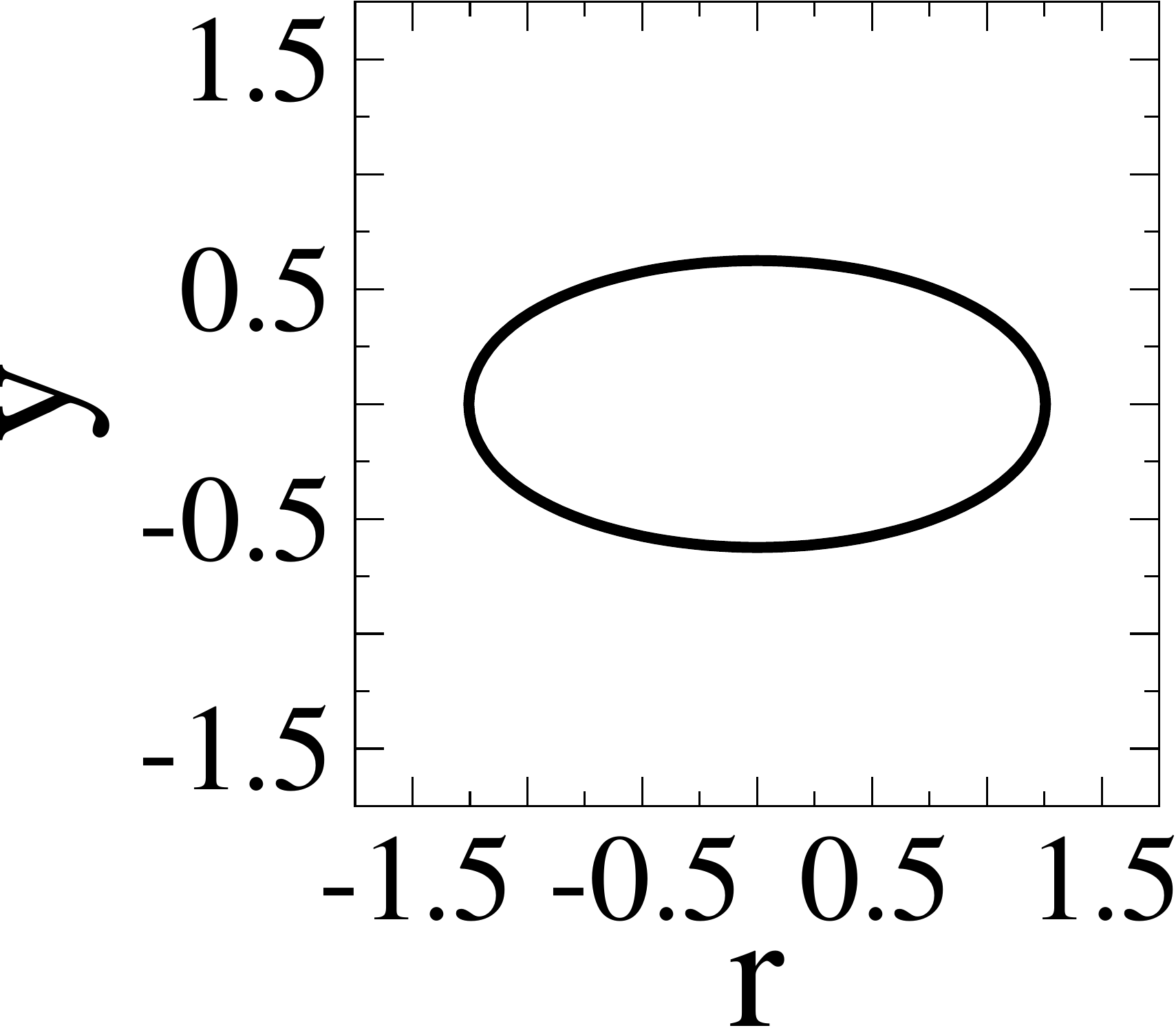}
  \caption{$t=10$}
  \label{fgr:gmeffectb}
\end{subfigure}
\begin{subfigure}{.22\textwidth}
  \centering
  \includegraphics[width=1\textwidth]{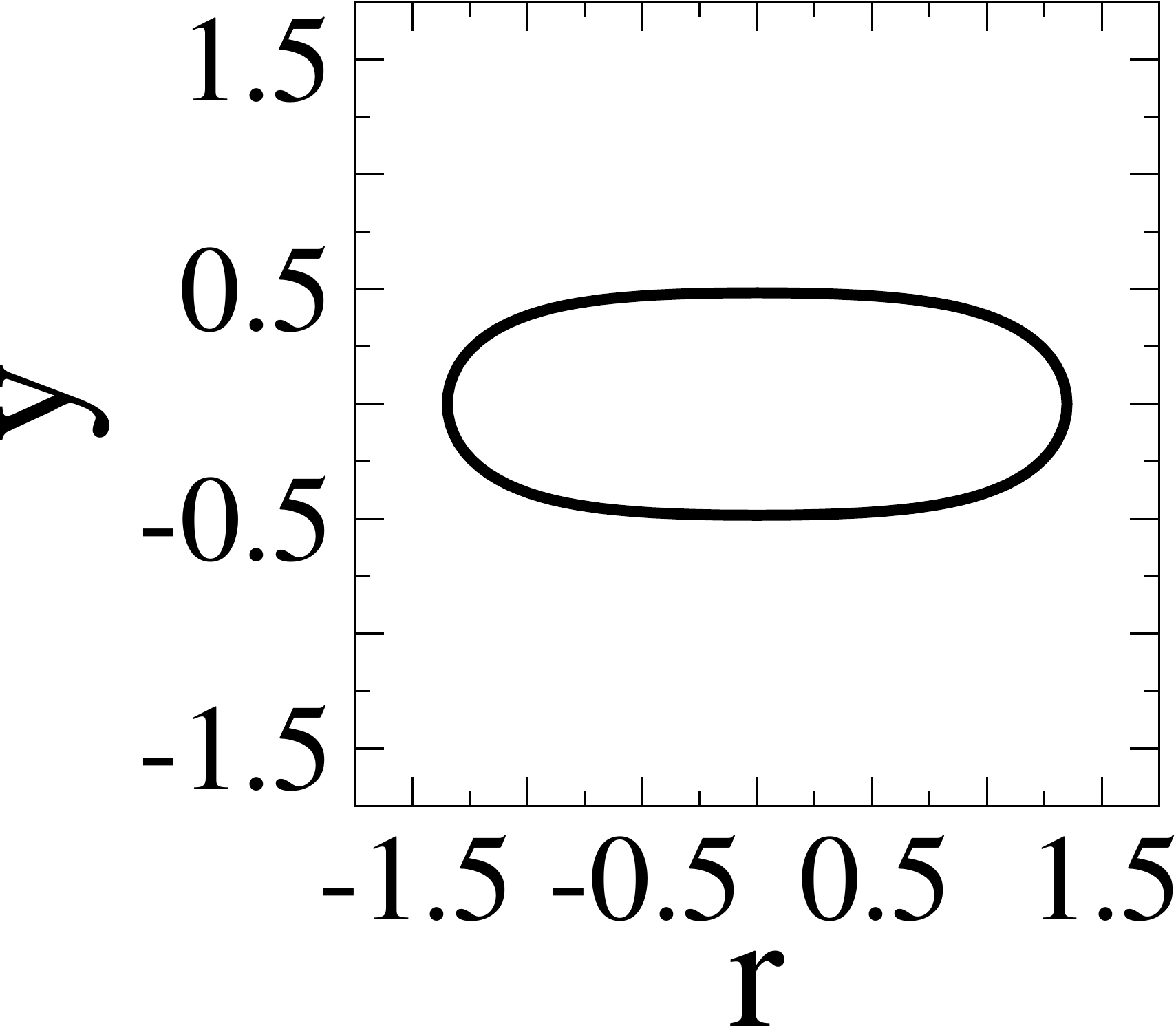}
  \caption{$t=50$}
  \label{fgr:gmeffectc}
\end{subfigure}
\begin{subfigure}{.22\textwidth}
  \centering
  \includegraphics[width=1\textwidth]{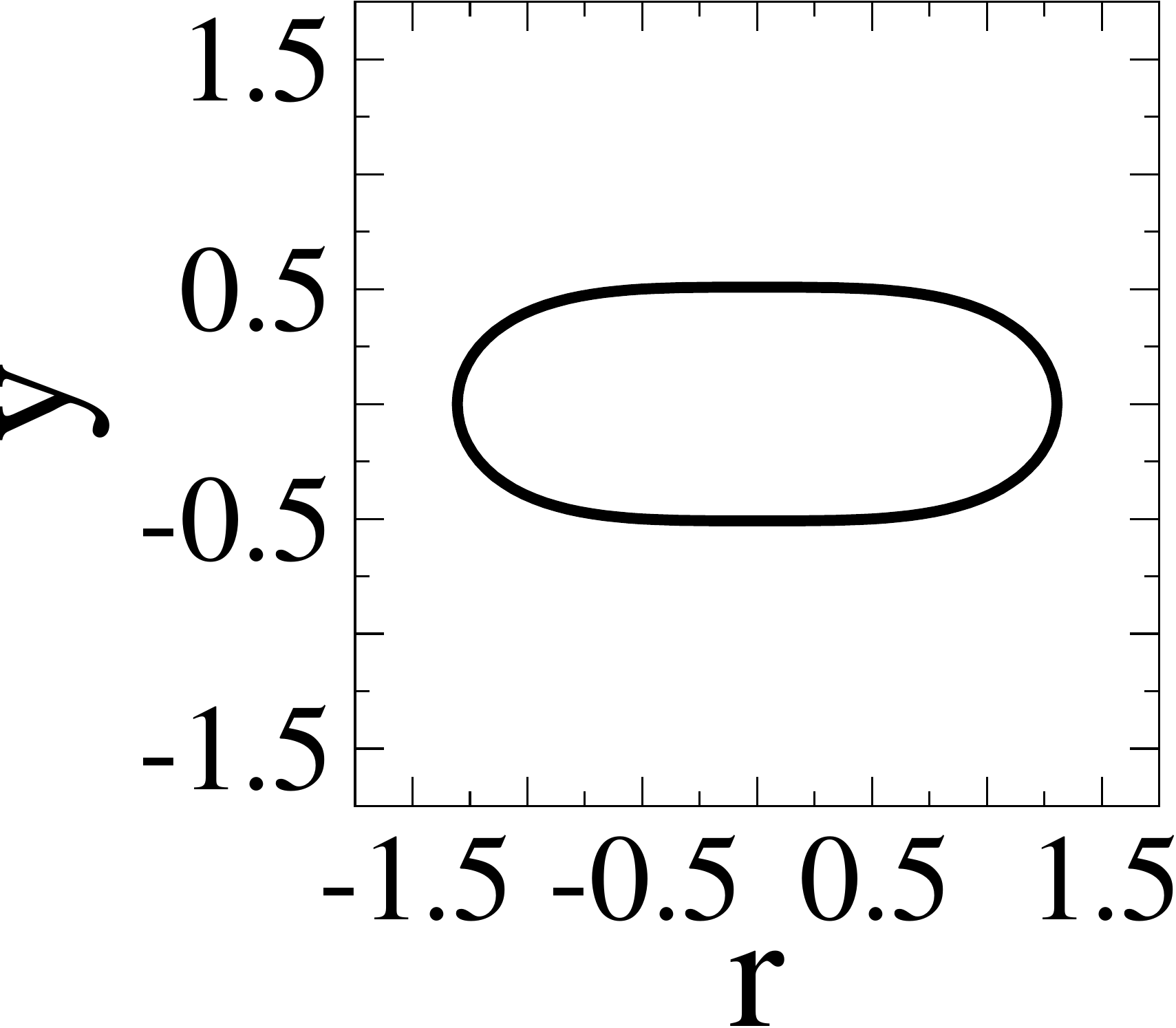}
  \caption{$t=\infty$}
  \label{fgr:gmeffectd}
\end{subfigure}
\begin{subfigure}{.22\textwidth}
  \centering
  \includegraphics[width=1\textwidth]{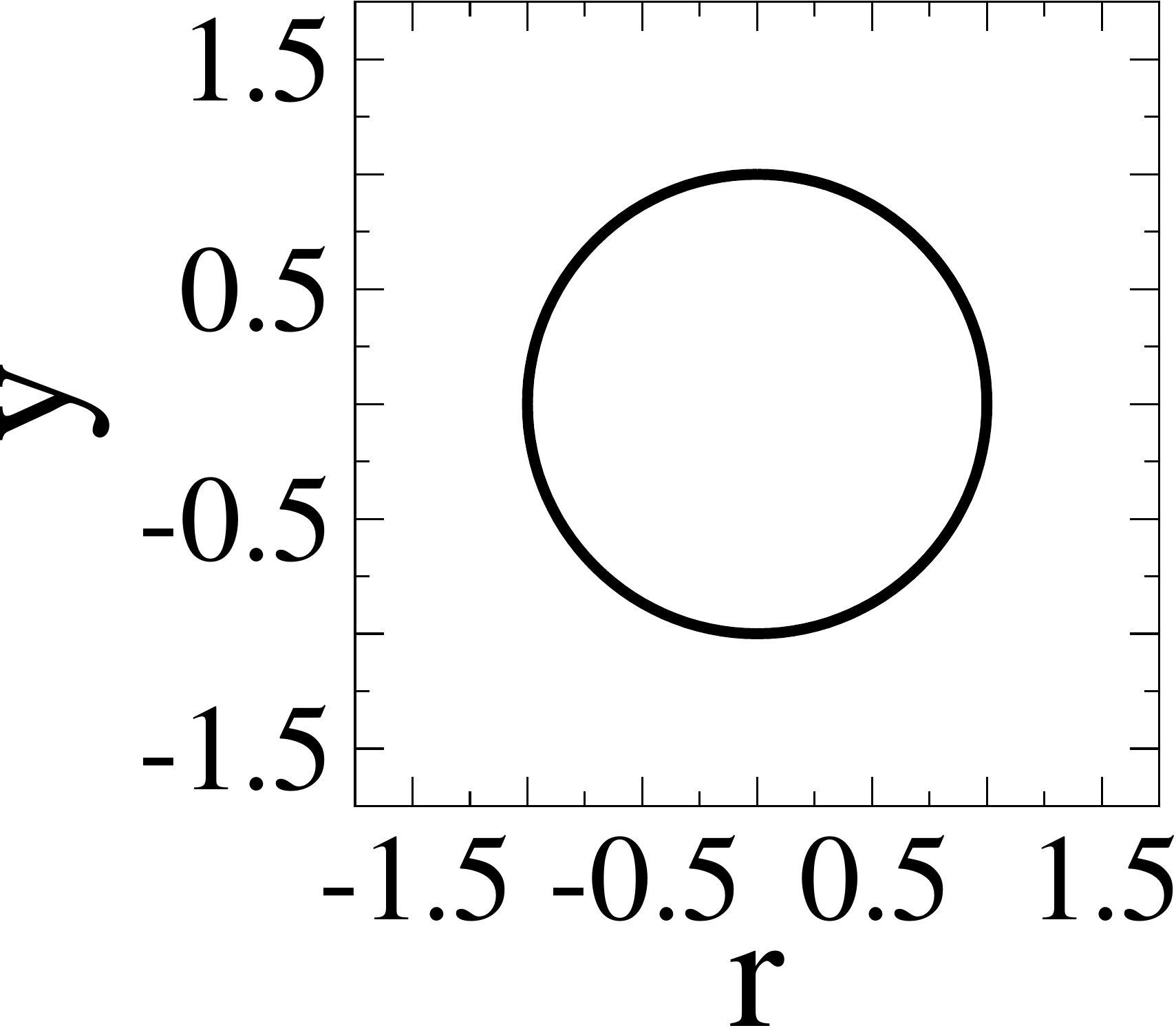}
  \caption{$t=0$}
  \label{fgr:gmeffecte}
\end{subfigure}
\begin{subfigure}{.22\textwidth}
  \centering
  \includegraphics[width=1\textwidth]{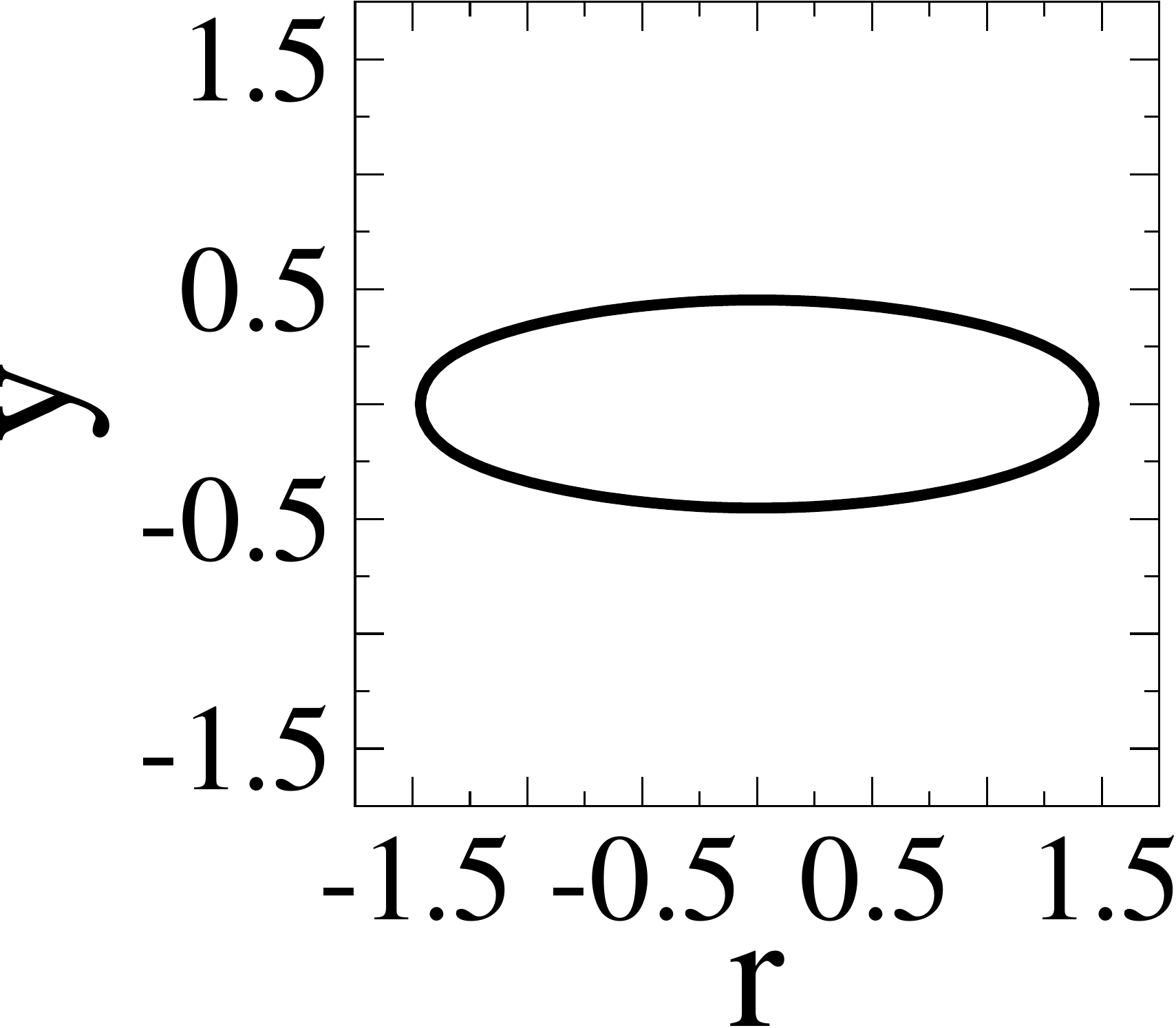}
  \caption{$t=30$}
  \label{fgr:gmeffectf}
\end{subfigure}
\begin{subfigure}{.22\textwidth}
  \centering
  \includegraphics[width=1\textwidth]{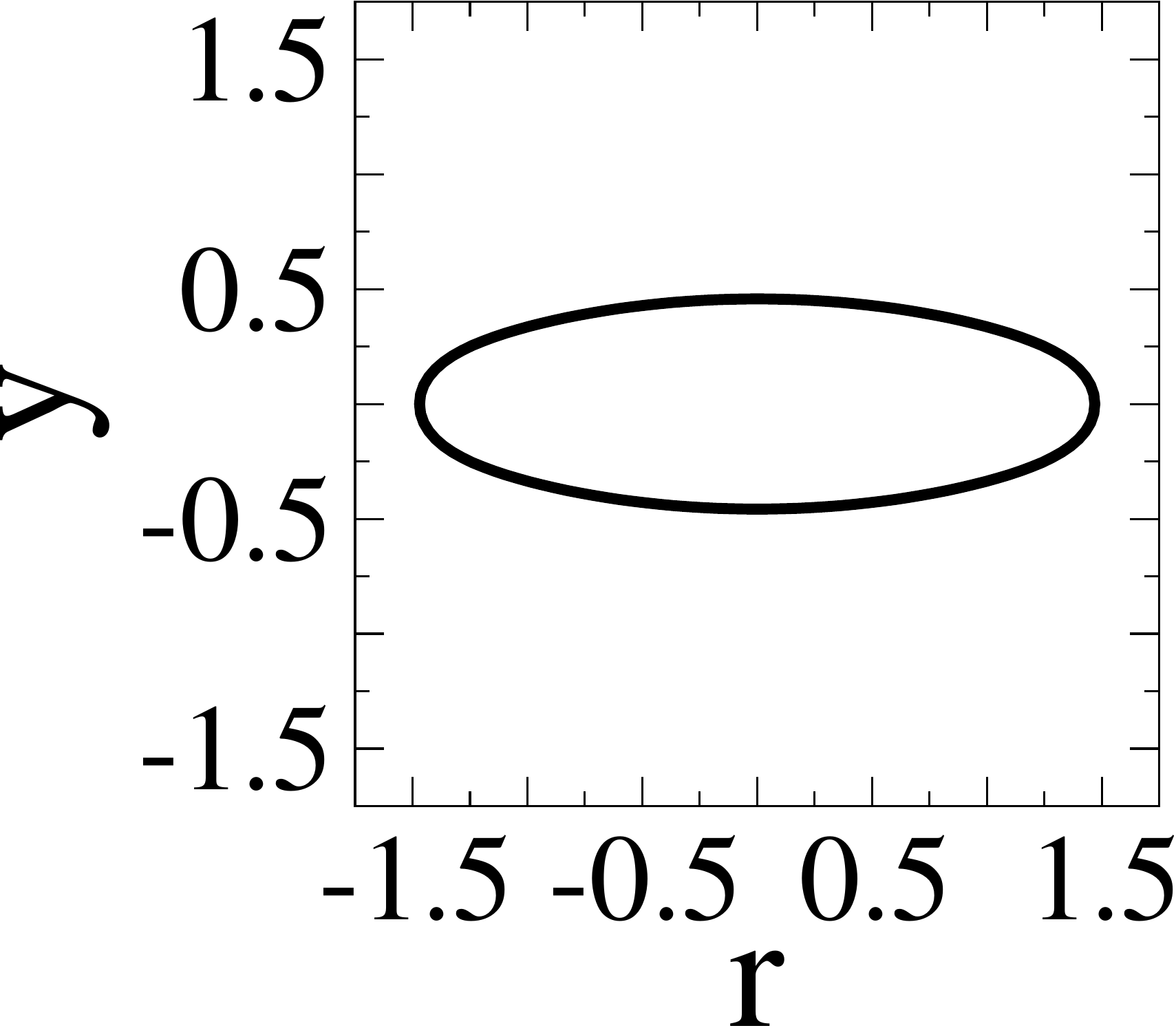}
  \caption{$t=110$}
  \label{fgr:gmeffectg}
\end{subfigure}
\begin{subfigure}{.22\textwidth}
  \centering
  \includegraphics[width=1\textwidth]{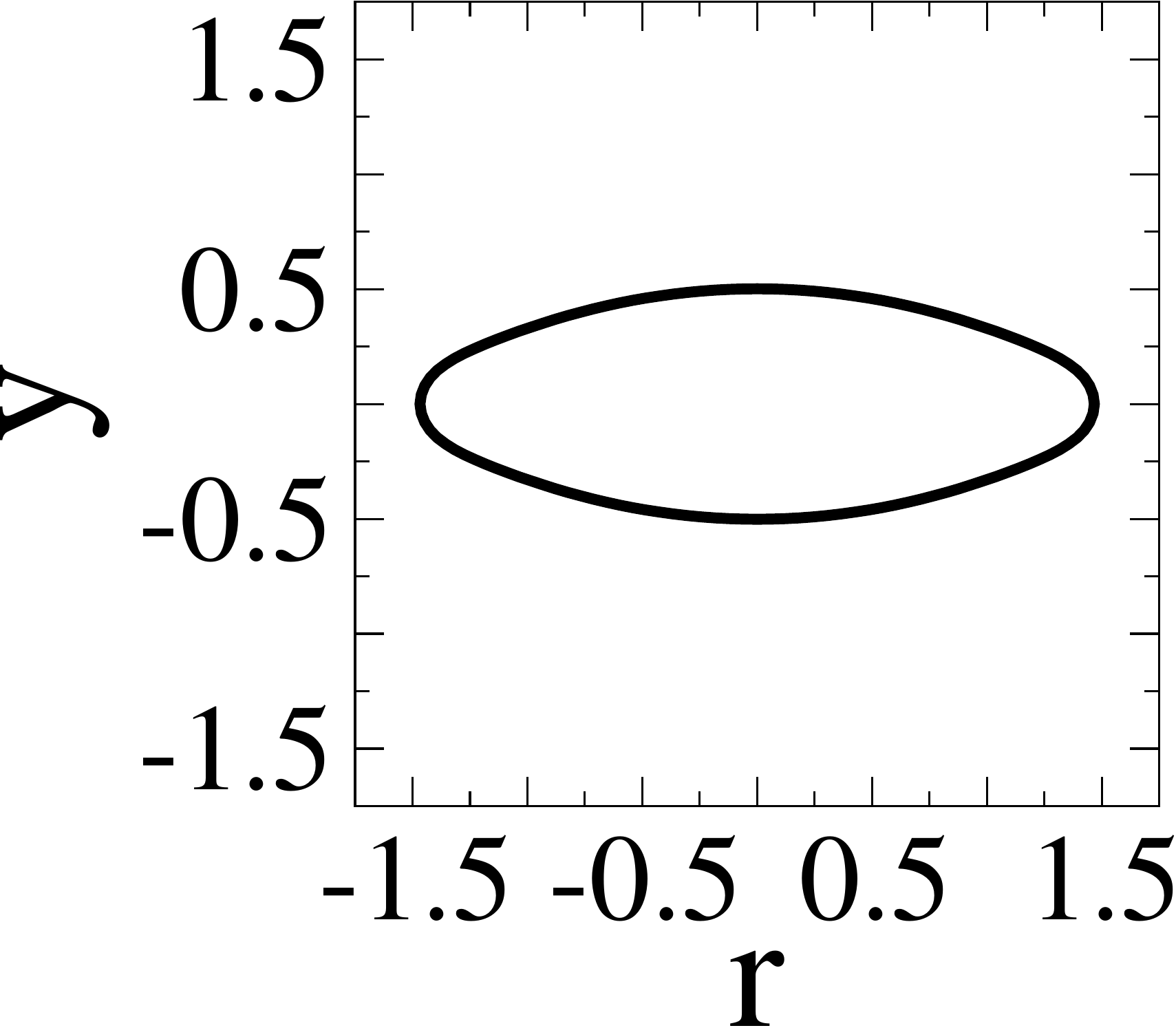}
  \caption{$t=\infty$}
  \label{fgr:gmeffecth}
\end{subfigure}
\caption{Shape evolution of a capsule with Skalak membrane at $\sigma_r=0.1$ for $Ca=0.45$ and $\hat C_m=50$ for conducting membrane. Figures a-d represent shapes during deformation at $\hat G_m=1$ and figs. e-h represent shapes during deformation at $\hat G_m=10$.}
\label{fgr:gmeffect}
\end{center}
\end{figure}

The effect of membrane conductance on the degree of deformation is presented in \cref{fgr:gmvarry}. For $\sigma_r=10$, showing prolate deformation, the steady state deformation depends upon the membrane conductance. In the limit of the membrane conductance going to infinity, the steady state deformation is same as that of a liquid drop. Thus when $\sigma_r=0.1$, the steady state deformation can be oblate, especially at high membrane conductances (note for $\hat G_m=0$, the steady state deformation is always prolate), which correspond to oblate deformation observed even in liquid drops. This is essentially due to a decrease in the membrane resistance and thereby the transmembrane potential going to zero.

For different values of membrane conductance, the dynamics (\cref{fgr:gmvarry}) as well as the final steady state deformation and shapes are different (\cref{fgr:gmeffect}). For both the small ($\hat G_m=1$ in \cref{fgr:gmeffecta,fgr:gmeffectb,fgr:gmeffectc,fgr:gmeffectd}) and large ($\hat G_m=10$ in \cref{fgr:gmeffecte,fgr:gmeffectf,fgr:gmeffectg,fgr:gmeffecth}) membrane conductance final steady state oblate shapes are observed ($\sigma_r=0.1$) unlike prolate steady state shapes for $\hat G_m=0$.

\begin{figure}[H]
\centering
  \includegraphics[width=0.5\textwidth]{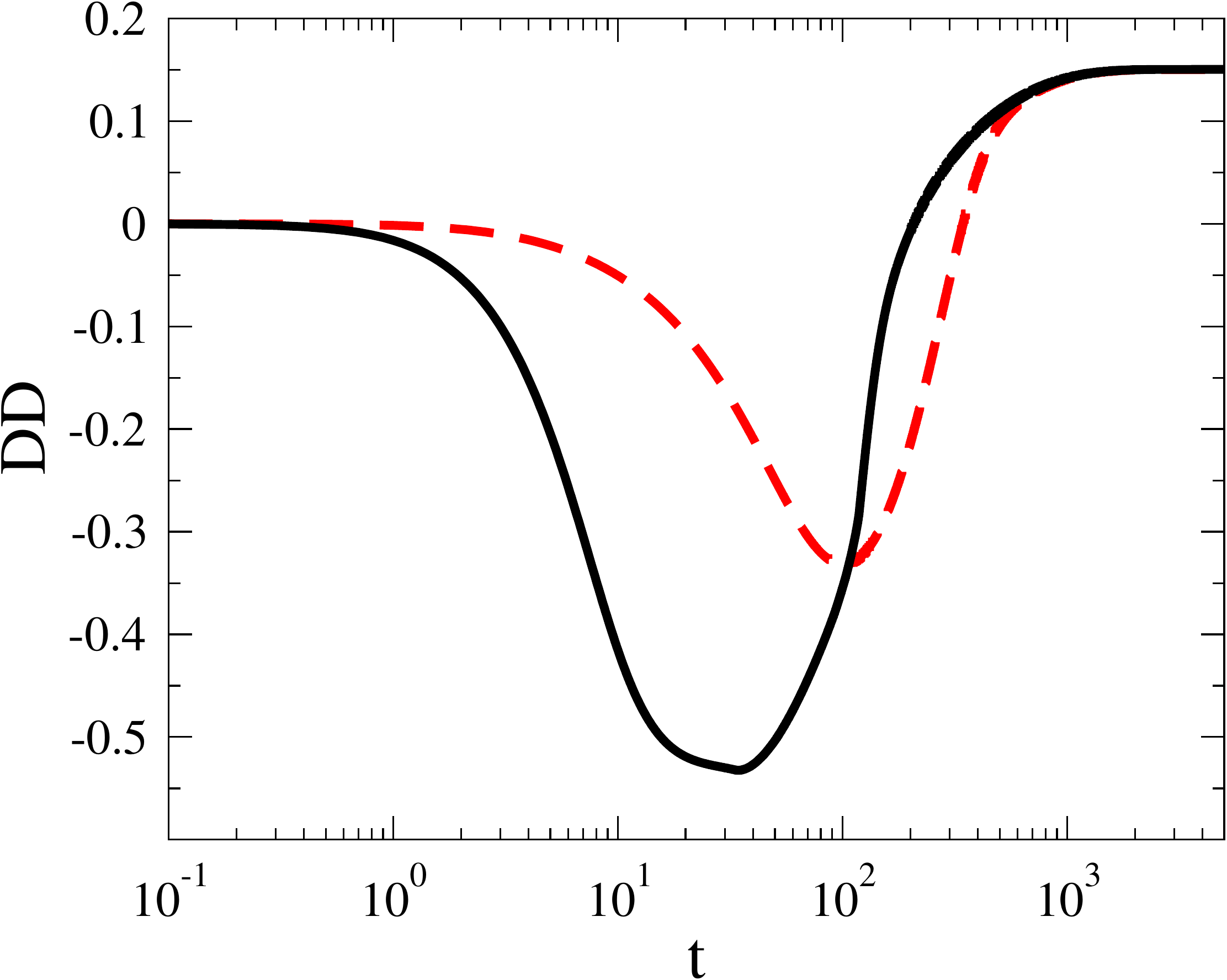}
  \caption{Comparison of the numerical simulation results for the dynamics of capsule deformation with $t_H=10$ (\textcolor{red}{$\pmb{--}$}) and $t_H=1$ ($\pmb{\mi}$) for $Ca=0.45$ and $\sigma_r=0.1$.}
  \label{fgr:kinematic10}
\end{figure} 

\begin{figure}[H]
\begin{center}
\begin{subfigure}{.22\textwidth}
  \centering
  \includegraphics[width=1\textwidth]{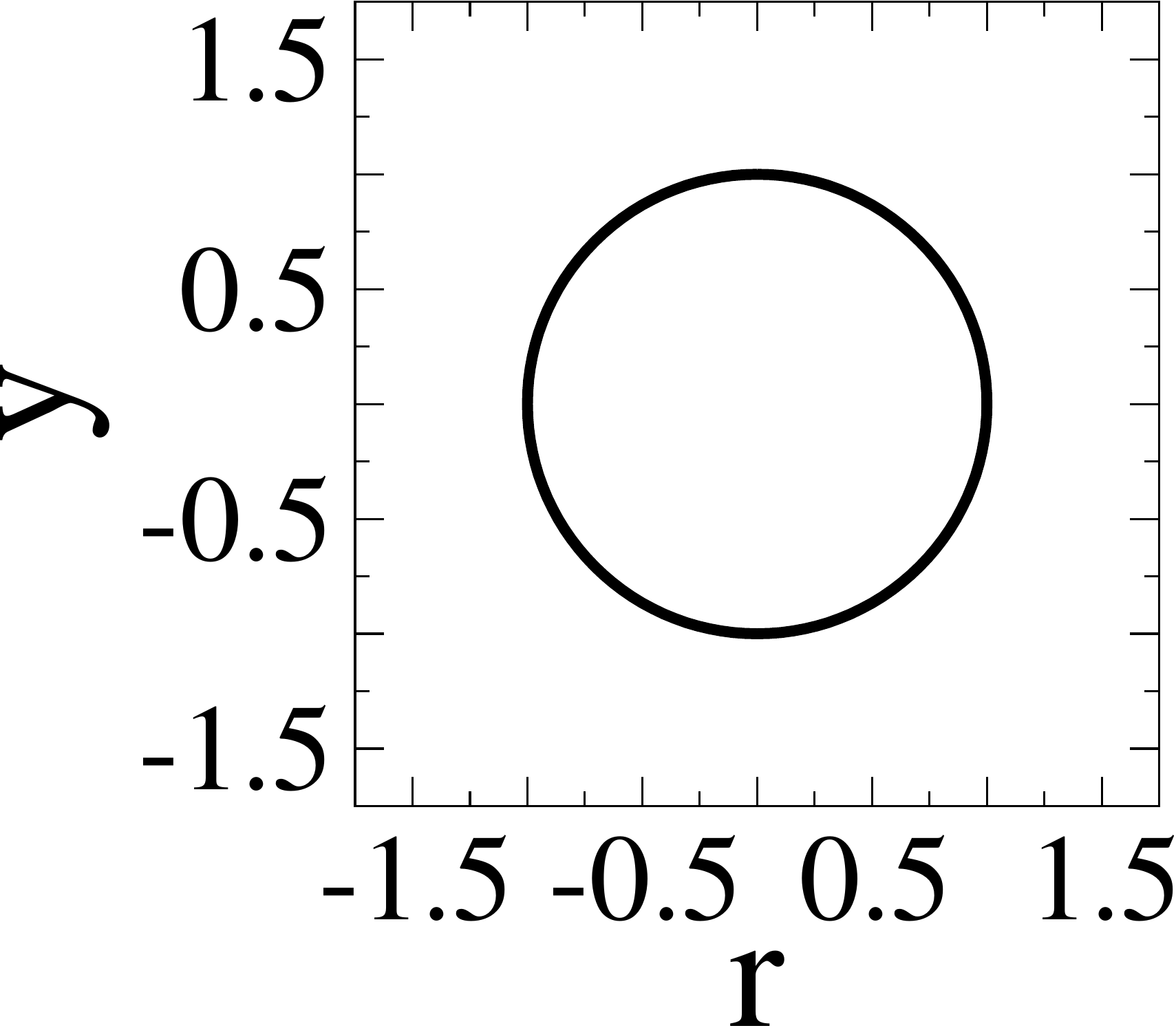}
  \caption{$t=0$}
  \label{fgr:tauha}
\end{subfigure}
\begin{subfigure}{.22\textwidth}
  \centering
  \includegraphics[width=1\textwidth]{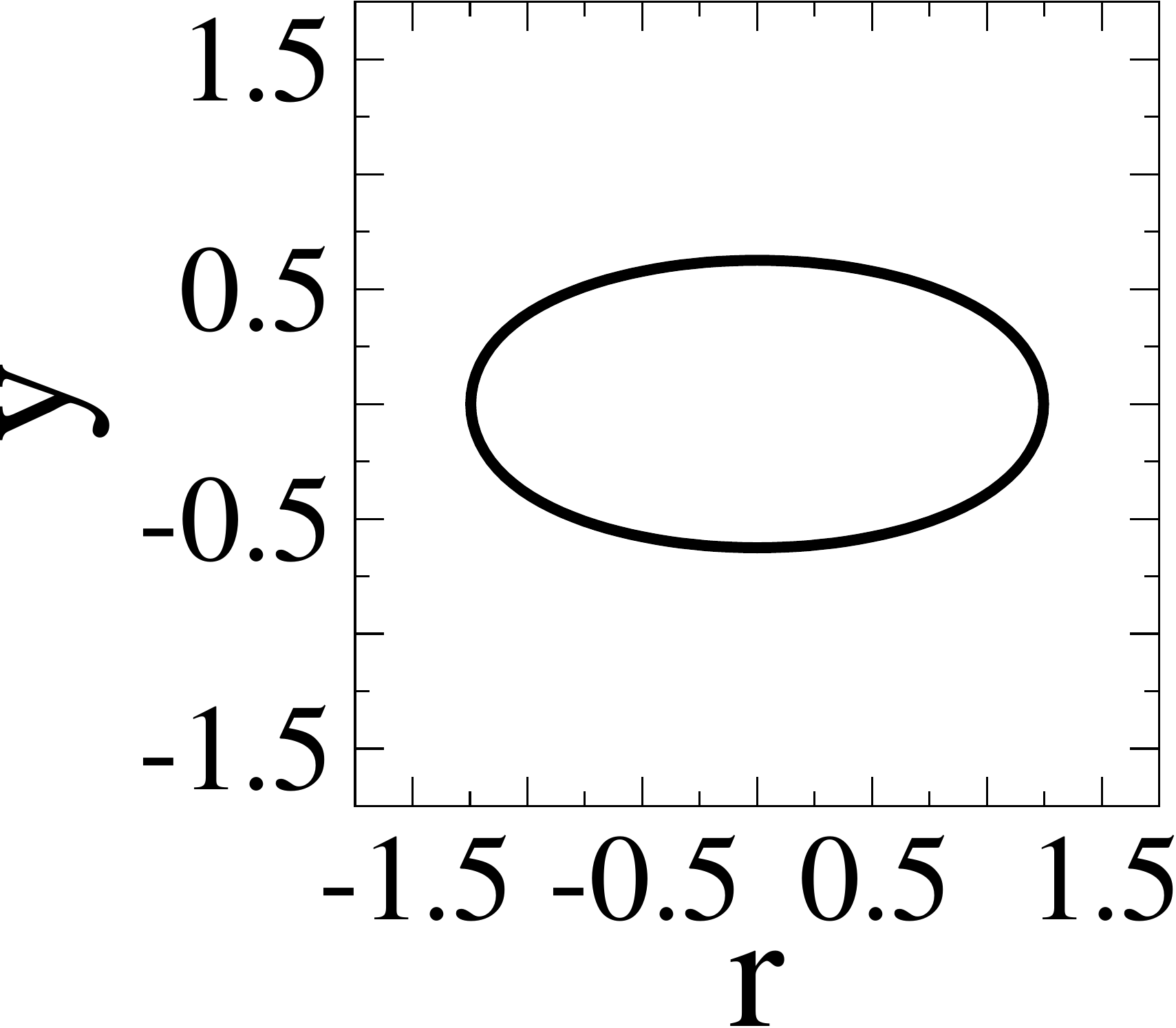}
  \caption{$t=105$}
  \label{fgr:tauhb}
\end{subfigure}
\begin{subfigure}{.22\textwidth}
  \centering
  \includegraphics[width=1\textwidth]{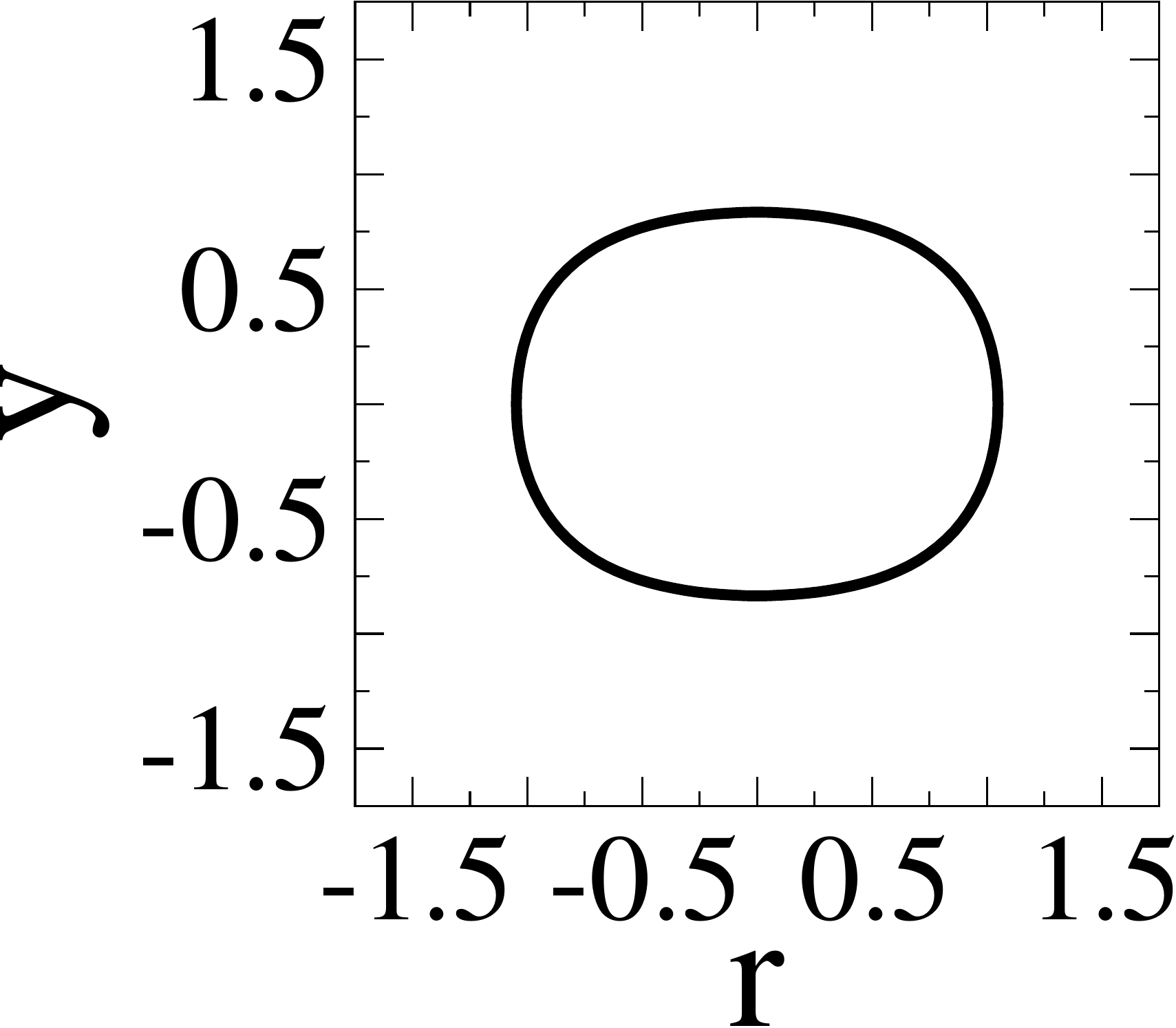}
  \caption{$t=260$}
  \label{fgr:tauhc}
\end{subfigure}
\begin{subfigure}{.22\textwidth}
  \centering
  \includegraphics[width=1\textwidth]{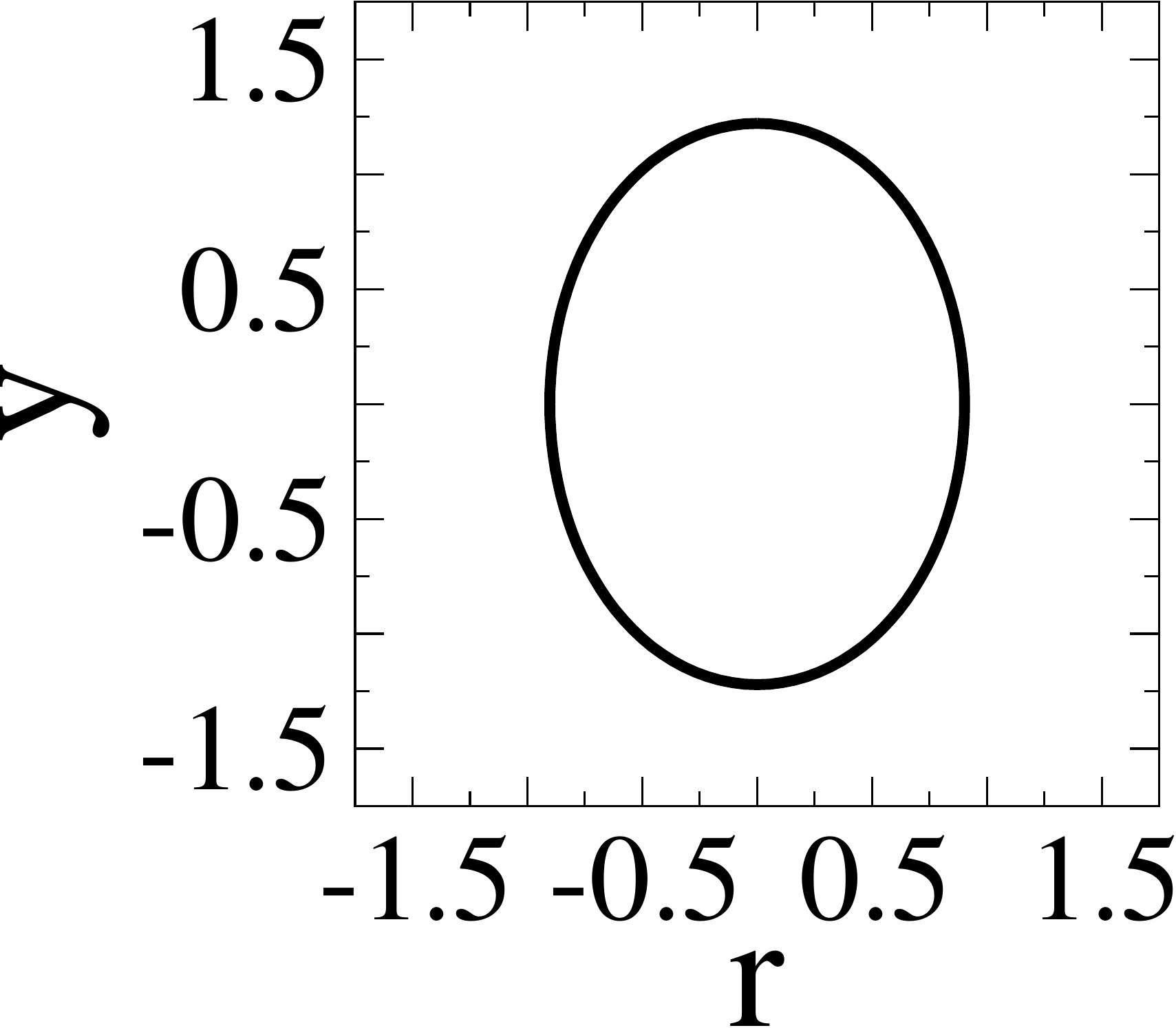}
  \caption{$t=\infty$}
  \label{fgr:tauhd}
\end{subfigure}
\begin{subfigure}{.22\textwidth}
  \centering
  \includegraphics[width=1\textwidth]{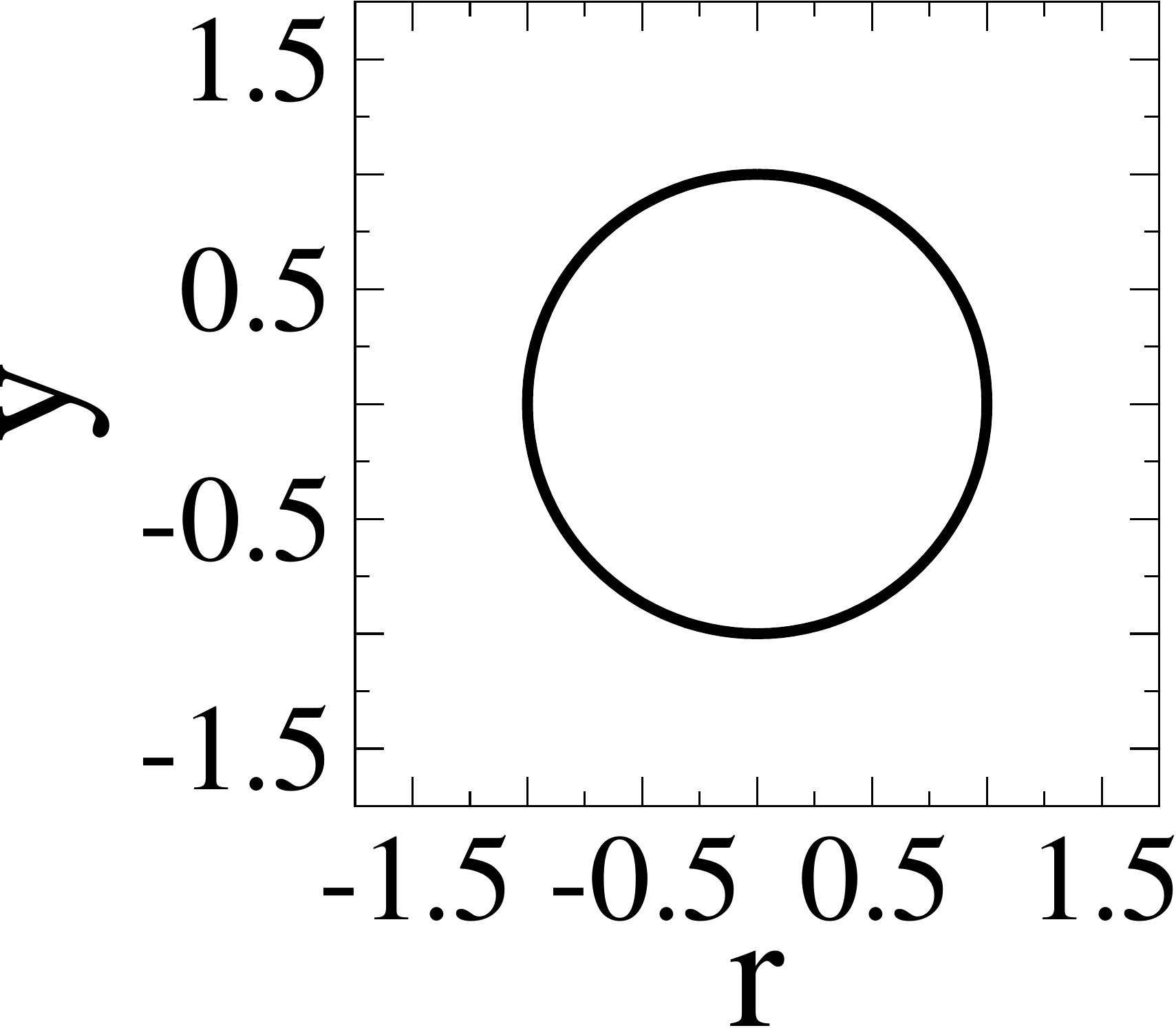}
  \caption{$t=0$}
  \label{fgr:tauhe}
\end{subfigure}
\begin{subfigure}{.22\textwidth}
  \centering
  \includegraphics[width=1\textwidth]{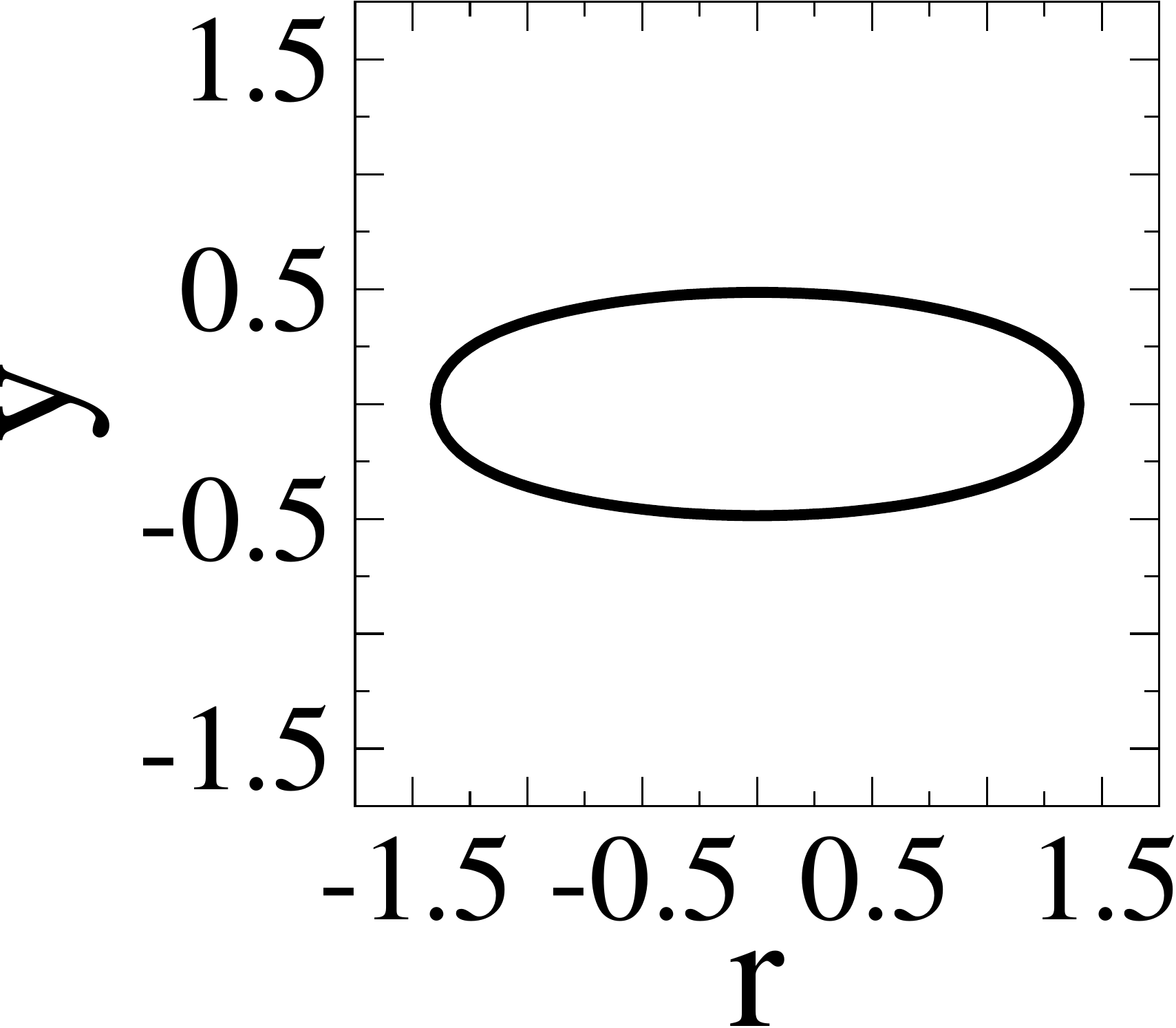}
  \caption{$t=78$}
  \label{fgr:tauhf}
\end{subfigure}
\begin{subfigure}{.22\textwidth}
  \centering
  \includegraphics[width=1\textwidth]{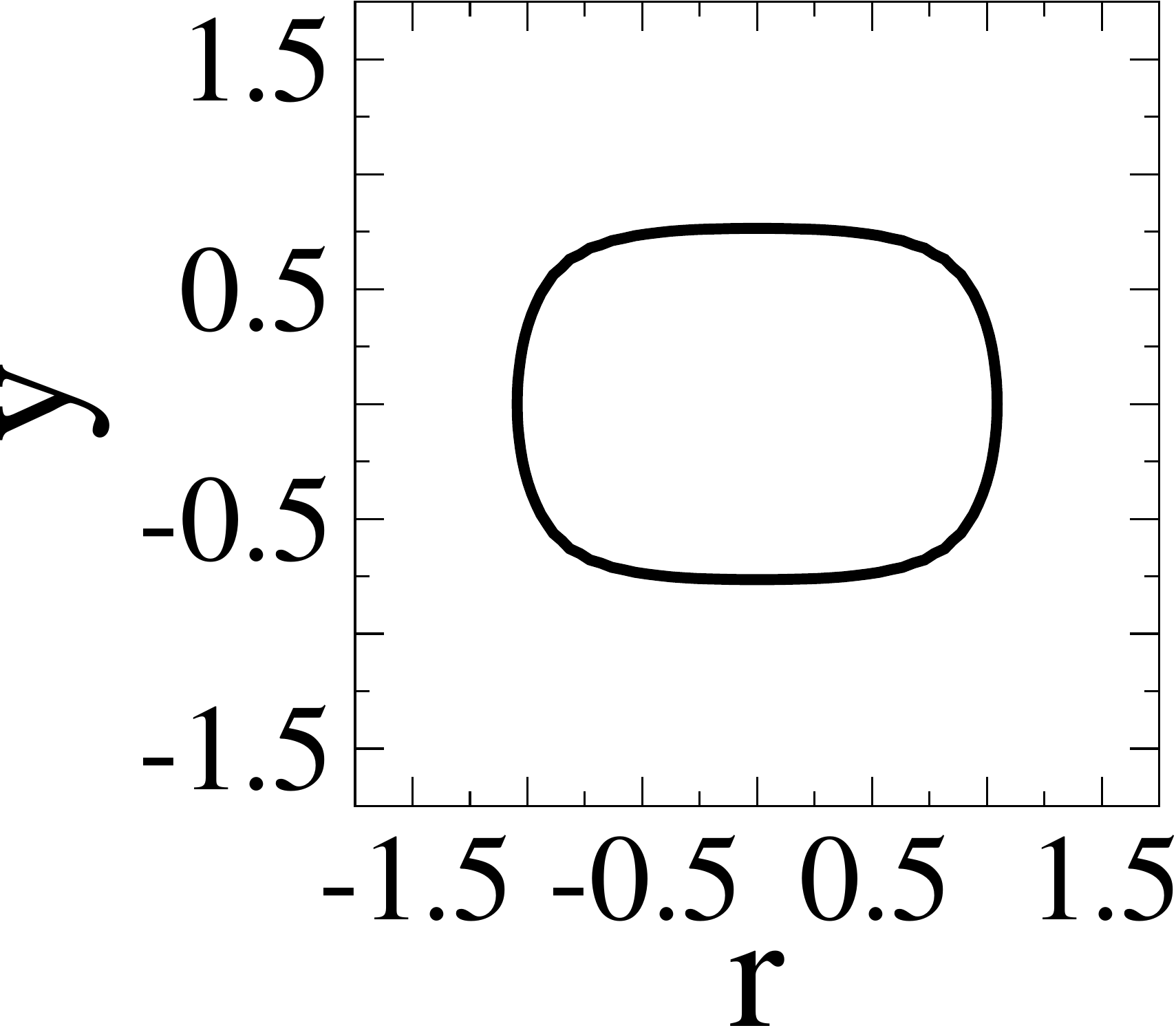}
  \caption{$t=230$}
  \label{fgr:tauhg}
\end{subfigure}
\begin{subfigure}{.22\textwidth}
  \centering
  \includegraphics[width=1\textwidth]{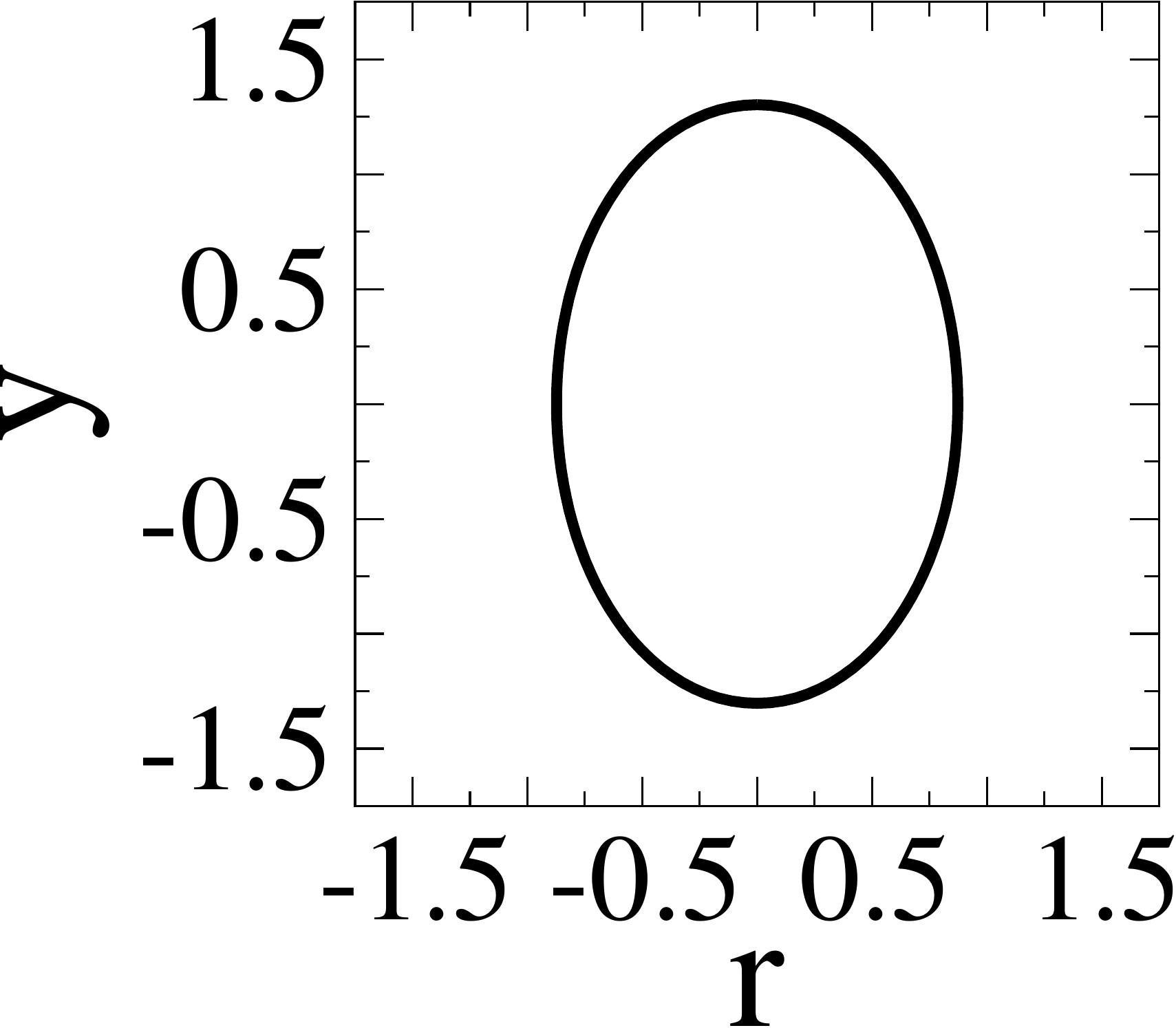}
  \caption{$t=\infty$}
  \label{fgr:tauhh}
\end{subfigure}
\caption{Shape evolution of a capsule with Skalak membrane at $\sigma_r=0.1$, $\epsilon_r=1$ with $t_H=10$ for nonconducting membrane with $\hat C_m=50$. Figures a-d represent shapes during deformation at $Ca=0.45$ and figures e-h represent deformed shapes at $Ca=0.8$.}
\label{fgr:tauh}
\end{center}
\end{figure}

When the hydrodynamic response time $(t_H)$ is greater than 1 (in this case $t_H=10$, $Ca=0.45$ and $\sigma_r=0.1$ are considered), a capsule undergoes much smaller intermediate deformation. \Cref{fgr:kinematic10} shows that although the membrane electrostatic stresses are high at $t \sim t_e$, a sphere does not deform due to slow hydrodynamic relaxation. By the time the capsule shape responds to the stresses, the electric stresses themselves have already relaxed over time scales of $t \sim t_{cap}$. Thus $t_H$ plays a critical role in the formation of biconcave intermediate shapes as well as squaring and hexagonal shapes discussed earlier. This is essentially due to the $t_H=t_e$ parameters chosen in those calculations. In that case,  the shape responds instantaneously to the applied electric stress. \Cref{fgr:tauha,fgr:tauhb,fgr:tauhc,fgr:tauhd}, show that a capsule does not attain intermediate biconcave and hexagonal shapes. Moreover, the squaring also become less prominent. At $Ca=0.8$ too, biconcave and hexagonal intermediate shapes are not admitted,  although squaring of the capsule is observed (\cref{fgr:tauhe,fgr:tauhf,fgr:tauhg,fgr:tauhh}). Further increase in capillary number results in failure of the boundary integral code  on account of numerical stiffness at the corners of the square shapes due to high curvature.  

\subsection{Deformation as a function of capillary number and capsule break up}
\begin{figure}[H]
\centering
  \includegraphics[width=0.5\textwidth]{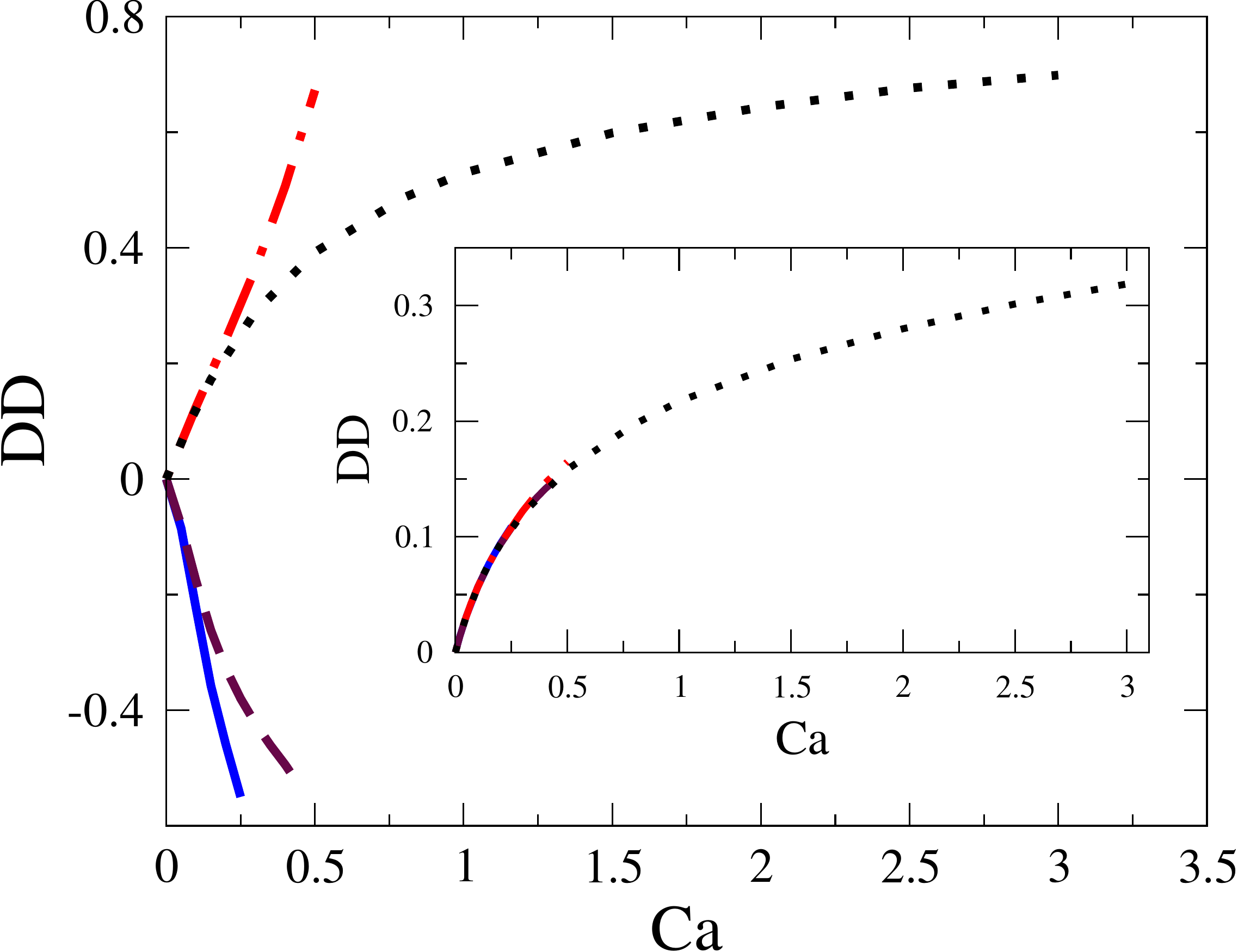}
  \caption{Intermediate maximum degree of deformation as a function of capillary number represented by curves (\textcolor{red}{$\pmb{-\cdot-}$}) and (\textcolor{blue}{$\pmb{\mi}$}) for neo-Hookean membrane  at $\sigma_r=10$ and $0.1$, respectively where as curves (\textcolor{black}{$\pmb{\cdots}$}) and (\textcolor{brown}{$\pmb{--}$}) for Skalak membrane  at at $\sigma_r=10$ and $0.1$, respectively. Inset shows the degree of deformation at steady state as a function of capillary number for the corresponding cases.}
  \label{fgr:ddvsca}
\end{figure}

\begin{figure}[H]
\begin{center}
\begin{subfigure}{.22\textwidth}
  \centering
  \includegraphics[width=1\textwidth]{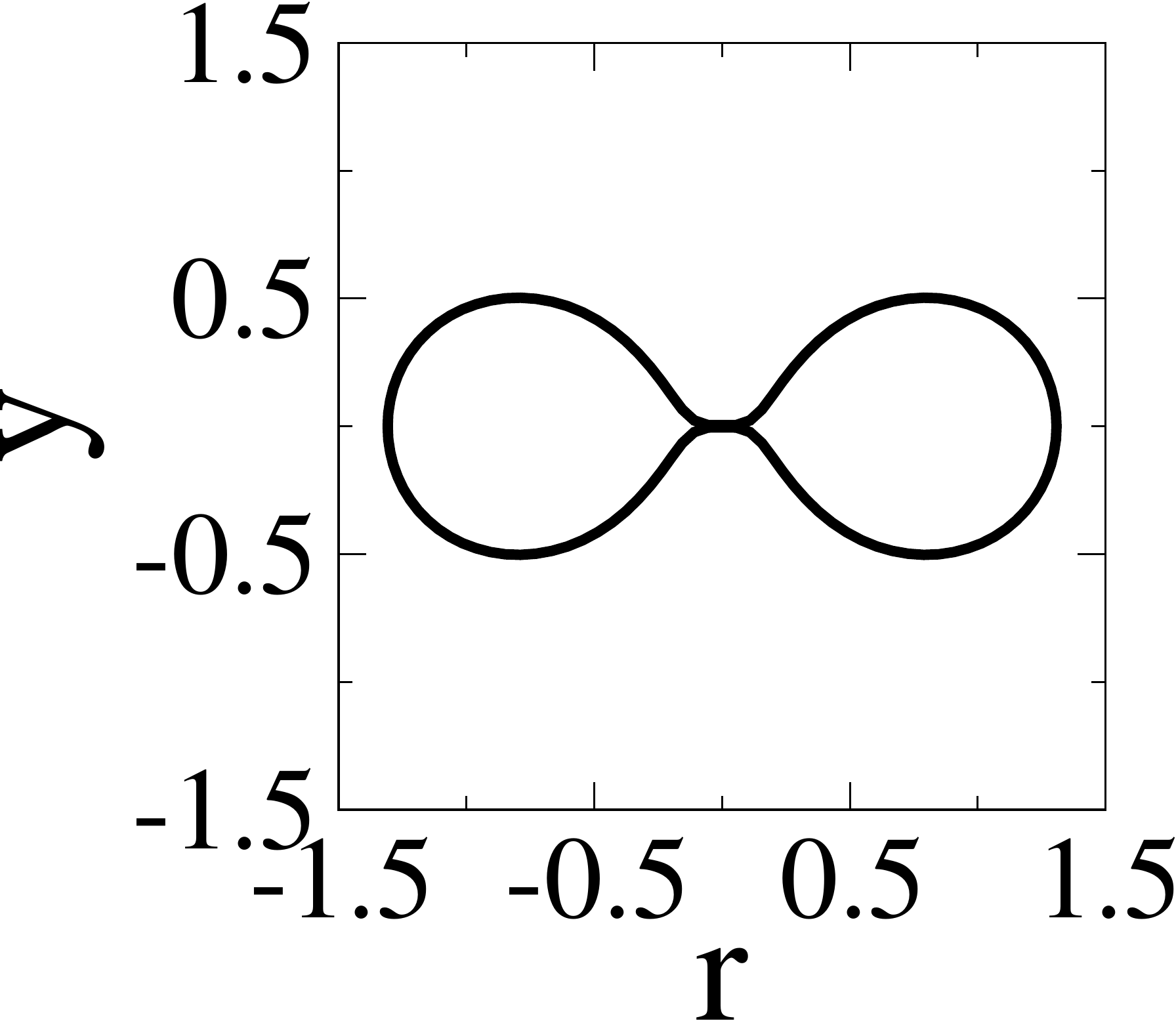}
  \caption{$\sigma_r=0.1$}
  \label{fgr:breakshapea}
\end{subfigure}
\begin{subfigure}{.22\textwidth}
  \centering
  \includegraphics[width=1\textwidth]{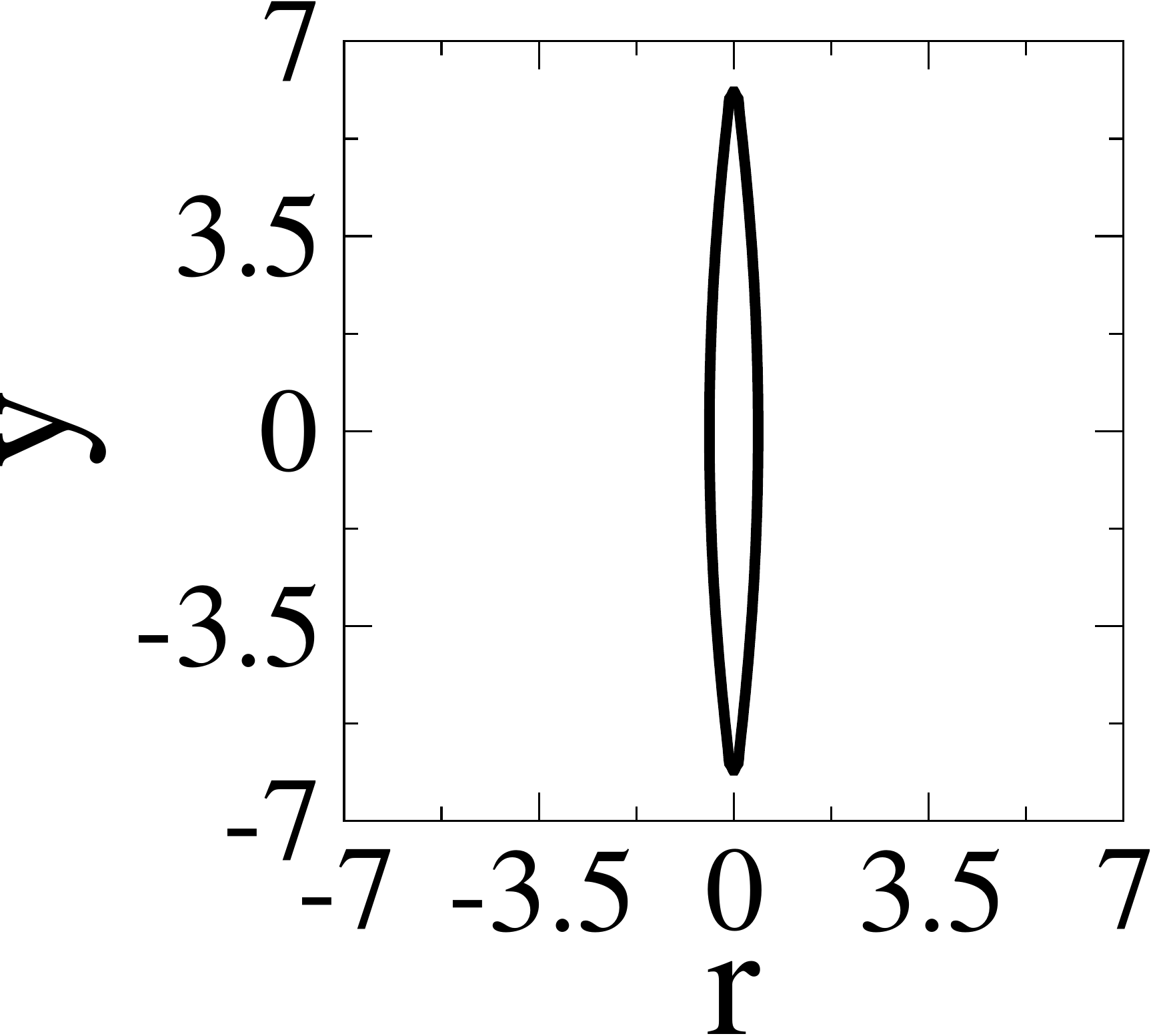}
  \caption{$\sigma_r=10$}
  \label{fgr:breakshapeb}
\end{subfigure}
\begin{subfigure}{.22\textwidth}
  \centering
  \includegraphics[width=1\textwidth]{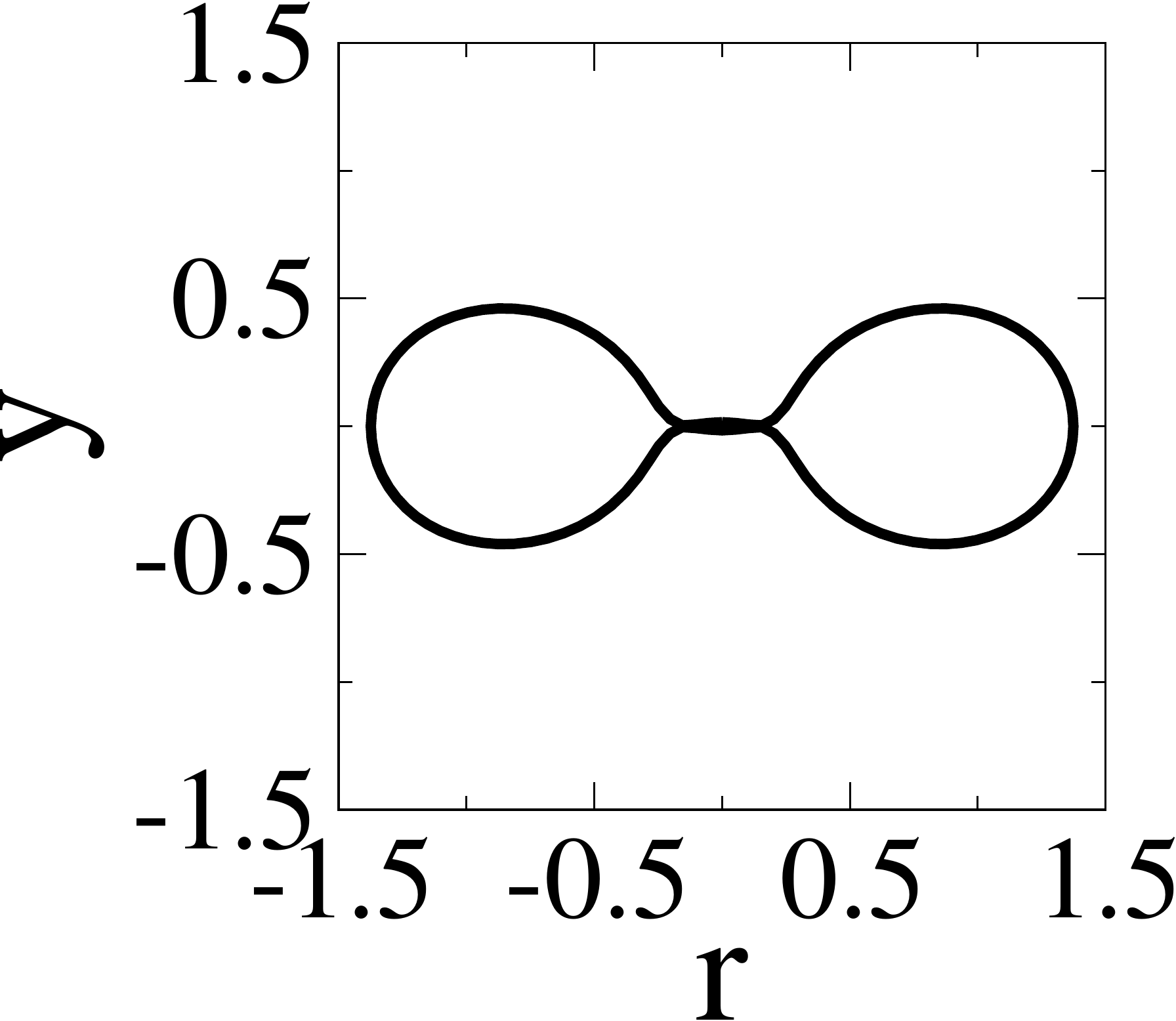}
  \caption{$\sigma_r=0.1$}
  \label{fgr:breakshapec}
\end{subfigure}
\begin{subfigure}{.22\textwidth}
  \centering
  \includegraphics[width=1\textwidth]{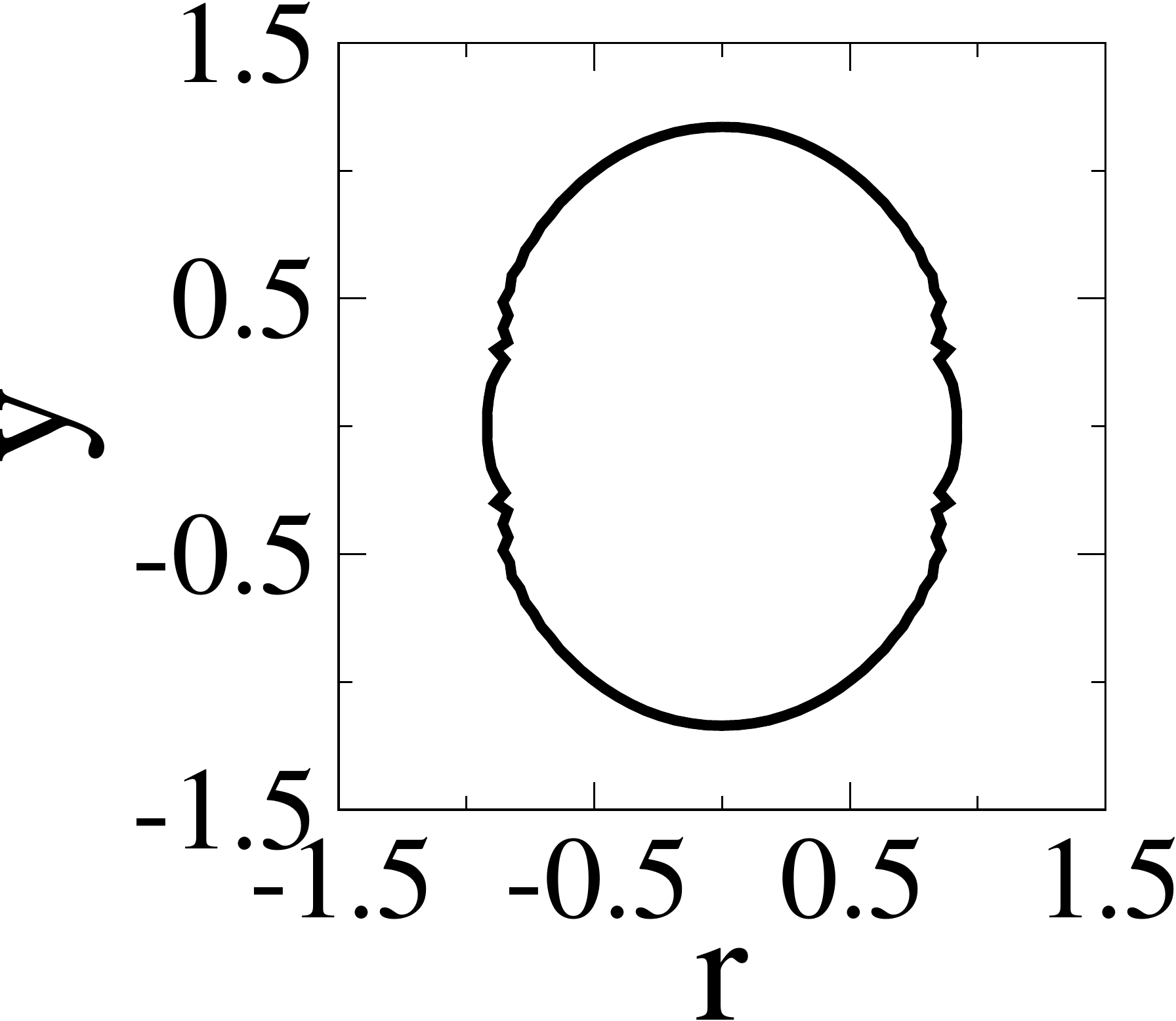}
  \caption{$\sigma_r=10$}
  \label{fgr:breakshaped}
\end{subfigure}
\caption{Break up shapes for a capsule with Skalak membrane at (a) $Ca=0.46$ and (b) $Ca=30$ and  neo-Hookean membrane at (c) $Ca=0.3$ and (d) $Ca=0.5$.}
\label{fgr:breakshape}
\end{center}
\end{figure}

It is interesting to understand the role of constitutive equations and resulting membrane stresses in deformation of a capsule. For $\sigma_r=0.1$, capsules with neo-Hookean membrane undergo a larger intermediate deformation compared to a capsule with Skalak membrane (\cref{fgr:ddvsca}), even though the final steady state deformations are same at a particular capillary number (inset of \cref{fgr:ddvsca}). Capsules undergo breakup at the intermediate maximum deformation at $Ca=0.46$ for Skalak membrane (\cref{fgr:breakshapea}) and at $Ca=0.3$ for neo-Hookean membrane (\cref{fgr:breakshapec}), therefore the calculations cannot be continued further.

For $\sigma_r=10$, at  large capillary numbers, a neo-Hookean capsule shows strain softening behavior and undergoes very large maximum deformation, whereas a capsule with Skalak membrane shows strain hardening behavior in the maximum deformation as well as in the steady state deformation (\cref{fgr:ddvsca}). Interestingly, both neo-Hookean membrane as well as Skalak membranes undergo possible breakup (which is manifested as a numerical singularity in our calculations) for $\sigma_r=0.1$. A capsule with strain hardening Skalak membrane does not show breakup for fairly high value of capillary number (numerical failure takes place at $Ca \sim 30$) for $\sigma_r=10$ (\cref{fgr:breakshapeb}), where as a capsule with neo-Hookean membrane shows a numerical singularity which is most likely a numerical artifact, at fairly low values of capillary number ($Ca = 0.5$ as shown in \cref{fgr:breakshaped}).

\section{Conclusions}
Our analysis of an elastic capsule under electric field indicates that several timescales are involved, $t_e$, $t_{cap}$, $t_{MW}$ and $t_H$ in the deformation of an elastic capsule undergoing deformation in an applied uniform electric field. The fluids undergo transition from perfect dielectrics to leaky dielectrics, over $t\sim t_{MW}$, the capsule relaxes over time scales $t_H$, whereas the capacitor is charged over time scales $t_{cap}$, wherein it builds a transmembrane potential $\phi_m=3/2$. An interplay of these time scales leads to a rich behavior in the response of elastic capsules to DC fields. Experiments on capsules in DC fields can thus enable estimation of electro-mechanical properties of these capsules.  

Moreover, experimental analysis of a capsule in millisecond to microsecond pulsed electric fields should be able to show all the intermediate shape transitions, since the typical time required for the shape transition, reported in this work, is of the order of non-dimensional time $t=200$ and the considered timescale $t_e$ varies in the range of $[10^{-5}s-10^{-7}s]$, depending upon the nature of the external fluid. 

The predictions of highly nonlinear shapes such as squaring of capsules, akin to those seen in vesicles (in experiments and simulations) confirms the appropriateness of this model to predict deformation of capsules in high fields, such as in electroporation, which has technological relevance.  This study also paves the way to understand simultaneous deformation and breakup of more complex biological elastic capsules such as blood cells, and will be a topic of a future study.

 \section*{Acknowledgment}
Authors would like to acknowledge Department of Science and Technology, India for financial support for this work.

% \clearpage
\normalsize
\bibliographystyle{unsrtnat}
\bibliography{dccapsule}

\begin{thebibliography}{68}
\providecommand{\natexlab}[1]{#1}
\providecommand{\url}[1]{\texttt{#1}}
\expandafter\ifx\csname urlstyle\endcsname\relax
  \providecommand{\doi}[1]{doi: #1}\else
  \providecommand{\doi}{doi: \begingroup \urlstyle{rm}\Url}\fi

\bibitem[Barth{\` e}s-Biesel and Rallison(1981)]{barthesbiesel81}
D.~Barth{\` e}s-Biesel and J.~M. Rallison.
\newblock The time-dependent deformation of a capsule freely suspended in a
  linear shear flow.
\newblock \emph{Journal of Fluid Mechanics}, 113:\penalty0 251--267, 12 1981.

\bibitem[Karyappa et~al.(2014{\natexlab{a}})Karyappa, Deshmukh, and
  Thaokar]{karyappa14}
R.~B. Karyappa, S.~D. Deshmukh, and R.~M. Thaokar.
\newblock Deformation of an elastic capsule in a uniform electric field.
\newblock \emph{Physics of Fluids}, 26\penalty0 (12):\penalty0 122108,
  2014{\natexlab{a}}.
\newblock \doi{10.1063/1.4903838}.

\bibitem[Kataoka et~al.(2001)Kataoka, Harada, and Nagasaki]{kataoka01}
K.~Kataoka, A.~Harada, and Y.~Nagasaki.
\newblock Block copolymer micelles for drug delivery: design, characterization
  and biological significance.
\newblock \emph{Advanced Drug Delivery Reviews}, 47\penalty0 (1):\penalty0 113
  -- 131, 2001.
\newblock ISSN 0169-409X.
\newblock Nanoparticulate Systems for Improved Drug Delivery.

\bibitem[Lighthill(1968)]{lighthill68}
M.~J. Lighthill.
\newblock Pressure-forcing of tightly fitting pellets along fluid-filled
  elastic tubes.
\newblock \emph{Journal of Fluid Mechanics}, 34\penalty0 (1):\penalty0
  113--143, 10 1968.

\bibitem[Hyman and Skalak(1972)]{hyman72}
W.~A. Hyman and R.~Skalak.
\newblock Non-newtonian behavior of a suspension of liquid drops in tube flow.
\newblock \emph{AIChE Journal}, 18\penalty0 (1):\penalty0 149--154, 1972.
\newblock ISSN 1547-5905.

\bibitem[Green and Adkins(1970)]{green60}
A.~E. Green and J.~E. Adkins.
\newblock \emph{large elastic deformation}.
\newblock The Oxford University Press, London, 2 edition, 1970.

\bibitem[Skalak et~al.(1973)Skalak, Tozeren, Zarda, and Chien]{skalak73}
R.~Skalak, A.~Tozeren, R.P. Zarda, and S.~Chien.
\newblock Strain energy function of red blood cell membranes.
\newblock \emph{Biophysical Journal}, 13\penalty0 (3):\penalty0 245 -- 264,
  1973.
\newblock ISSN 0006-3495.

\bibitem[Gao et~al.(2001)Gao, Leporatti, Moya, Donath, and M\"{o}hwald]{gao01}
C.~Gao, S.~Leporatti, S.~Moya, E.~Donath, and H.~M\"{o}hwald.
\newblock Stability and mechanical properties of polyelectrolyte capsules
  obtained by stepwise assembly of poly(styrenesulfonate sodium salt) and
  poly(diallyldimethyl ammonium) chloride onto melamine resin particles.
\newblock \emph{Langmuir}, 17\penalty0 (11):\penalty0 3491--3495, 2001.

\bibitem[Fery and Weinkamer(2007)]{fery07}
A.~Fery and R.~Weinkamer.
\newblock Mechanical properties of micro- and nanocapsules: Single-capsule
  measurements.
\newblock \emph{Polymer}, 48\penalty0 (25):\penalty0 7221 -- 7235, 2007.
\newblock ISSN 0032-3861.

\bibitem[Keller and Sottos(2006)]{keller2006}
M.~W. Keller and N.~R. Sottos.
\newblock Mechanical properties of microcapsules used in a self-healing
  polymer.
\newblock \emph{Experimental Mechanics}, 46\penalty0 (6):\penalty0 725--733,
  2006.
\newblock ISSN 1741-2765.

\bibitem[Dubreuil et~al.(2003)Dubreuil, Elsner, and Fery]{dubreuil2003}
F.~Dubreuil, N.~Elsner, and A.~Fery.
\newblock Elastic properties of polyelectrolyte capsules studied by
  atomic-force microscopy and ricm.
\newblock \emph{The European Physical Journal E}, 12\penalty0 (2):\penalty0
  215--221, 2003.
\newblock ISSN 1292-895X.

\bibitem[Chang and Olbricht(1993)]{chang93}
K.~S. Chang and W.~L. Olbricht.
\newblock Experimental studies of the deformation of a synthetic capsule in
  extensional flow.
\newblock \emph{Journal of Fluid Mechanics}, 250:\penalty0 587--608, 005 1993.

\bibitem[Barth{\` e}s-Biesel(1980)]{barthesbiesel80}
D.~Barth{\` e}s-Biesel.
\newblock Motion of a spherical microcapsule freely suspended in a linear shear
  flow.
\newblock \emph{Journal of Fluid Mechanics}, 100\penalty0 (4):\penalty0
  831--853, 10 1980.

\bibitem[Barth{\` e}s-Biesel and Chhim(1981)]{BARTHESBIESEL81a}
D.~Barth{\` e}s-Biesel and V.~Chhim.
\newblock The constitutive equation of a dilute suspension of spherical
  microcapsules.
\newblock \emph{International Journal of Multiphase Flow}, 7\penalty0
  (5):\penalty0 493 -- 505, 1981.
\newblock ISSN 0301-9322.

\bibitem[Barth{\`e}s-Biesel(1991)]{barthesbiesel91}
D.~Barth{\`e}s-Biesel.
\newblock Role of interfacial properties on the motion and deformation of
  capsules in shear flow.
\newblock \emph{Physica A: Statistical Mechanics and its Applications},
  172\penalty0 (1):\penalty0 103 -- 124, 1991.
\newblock ISSN 0378-4371.

\bibitem[Ha and Yang(2000)]{jong00}
J.-W. Ha and S.-M. Yang.
\newblock Electrohydrodynamic effects on the deformation and orientation of a
  liquid capsule in a linear flow.
\newblock \emph{Physics of Fluids}, 12\penalty0 (7):\penalty0 1671--1684, 2000.

\bibitem[Thaokar(2016)]{rt16}
R.~M. Thaokar.
\newblock Time-dependent electrohydrodynamics of a compressible viscoelastic
  capsule in the small-deformation limit.
\newblock \emph{Phys. Rev. E}, 94:\penalty0 042607, Oct 2016.

\bibitem[Kessler et~al.(2009)Kessler, Finken, and Seifert]{kessler09}
S.~Kessler, R.~Finken, and U.~Seifert.
\newblock Elastic capsules in shear flow: Analytical solutions for constant and
  time-dependent shear rates.
\newblock \emph{The European Physical Journal E}, 29\penalty0 (4):\penalty0
  399--413, Aug 2009.
\newblock ISSN 1292-895X.

\bibitem[Finken et~al.(2011)Finken, Kessler, and Seifert]{seifert11}
R.~Finken, S.~Kessler, and U.~Seifert.
\newblock Micro-capsules in shear flow.
\newblock \emph{Journal of Physics: Condensed Matter}, 23\penalty0
  (18):\penalty0 184113, 2011.

\bibitem[Zhou and Pozrikidis(1995)]{zhou95}
H.~Zhou and C.~Pozrikidis.
\newblock Deformation of liquid capsules with incompressible interfaces in
  simple shear flow.
\newblock \emph{Journal of Fluid Mechanics}, 283:\penalty0 175--200, 001 1995.
\newblock \doi{10.1017/S0022112095002278}.

\bibitem[Rao et~al.(1994)Rao, Zahalak, and Sutera]{rao94}
P.~R. Rao, G.~I. Zahalak, and S.~P. Sutera.
\newblock Large deformations of elastic cylindrical capsules in shear flows.
\newblock \emph{Journal of Fluid Mechanics}, 270:\penalty0 73--90, 007 1994.

\bibitem[Pozrikidis(1995)]{pozrikidis95}
C.~Pozrikidis.
\newblock Finite deformation of liquid capsules enclosed by elastic membranes
  in simple shear flow.
\newblock \emph{Journal of Fluid Mechanics}, 297:\penalty0 123--152, 008 1995.

\bibitem[Ramanujan and Pozrikidis(1998)]{ramanujan98}
S.~Ramanujan and C.~Pozrikidis.
\newblock Deformation of liquid capsules enclosed by elastic membranes in
  simple shear flow: large deformations and the effect of fluid viscosities.
\newblock \emph{Journal of Fluid Mechanics}, 361:\penalty0 117--143, 04 1998.

\bibitem[Pozrikidis(2001)]{pozrikidis01}
C.~Pozrikidis.
\newblock Effect of membrane bending stiffness on the deformation of capsules
  in simple shear flow.
\newblock \emph{Journal of Fluid Mechanics}, 440:\penalty0 269--291, 08 2001.

\bibitem[Le(2010)]{duc10}
D.~V. Le.
\newblock Effect of bending stiffness on the deformation of liquid capsules
  enclosed by thin shells in shear flow.
\newblock \emph{Phys. Rev. E}, 82:\penalty0 016318, Jul 2010.

\bibitem[Kessler et~al.(2008)Kessler, Finken, and Seifert]{kessler08}
S.~Kessler, R.~Finken, and U.~Seifert.
\newblock Swinging and tumbling of elastic capsules in shear flow.
\newblock \emph{Journal of Fluid Mechanics}, 605:\penalty0 207--226, 006 2008.
\newblock \doi{10.1017/S0022112008001493}.

\bibitem[Sui et~al.(2008{\natexlab{a}})Sui, Low, Chew, and Roy]{sui08}
Y.~Sui, H.~T. Low, Y.~T. Chew, and P.~Roy.
\newblock Tank-treading, swinging, and tumbling of liquid-filled elastic
  capsules in shear flow.
\newblock \emph{Phys. Rev. E}, 77:\penalty0 016310, Jan 2008{\natexlab{a}}.

\bibitem[Sui et~al.(2007)Sui, Chew, Roy, and Low]{sui07}
Y.~Sui, Y.~T. Chew, P.~Roy, and H.~T. Low.
\newblock Effect of membrane bending stiffness on the deformation of elastic
  capsules in extensional flow: A lattice boltzmann study.
\newblock \emph{International Journal of Modern Physics C}, 18\penalty0
  (08):\penalty0 1277--1291, 2007.

\bibitem[Sui et~al.(2010)Sui, Chen, Chew, Roy, and Low]{sui10}
Y.~Sui, X.B. Chen, Y.T. Chew, P.~Roy, and H.T. Low.
\newblock Numerical simulation of capsule deformation in simple shear flow.
\newblock \emph{Computers \& Fluids}, 39\penalty0 (2):\penalty0 242 -- 250,
  2010.
\newblock ISSN 0045-7930.

\bibitem[Skotheim and Secomb(2007)]{skotheim07}
J.~M. Skotheim and T.~W. Secomb.
\newblock Red blood cells and other nonspherical capsules in shear flow:
  Oscillatory dynamics and the tank-treading-to-tumbling transition.
\newblock \emph{Phys. Rev. Lett.}, 98:\penalty0 078301, Feb 2007.

\bibitem[Navot(1998)]{navot98}
Y.~Navot.
\newblock Elastic membranes in viscous shear flow.
\newblock \emph{Physics of Fluids}, 10\penalty0 (8):\penalty0 1819--1833, 1998.
\newblock \doi{10.1063/1.869702}.

\bibitem[Lac et~al.(2004)Lac, Barth{\` e}s-Biesel, Pelekasis, and
  Tsamopoulos]{lac04}
E.~Lac, D.~Barth{\` e}s-Biesel, N.~A. Pelekasis, and J.~Tsamopoulos.
\newblock Spherical capsules in three-dimensional unbounded stokes flows:
  effect of the membrane constitutive law and onset of buckling.
\newblock \emph{Journal of Fluid Mechanics}, 516:\penalty0 303--334, 10 2004.

\bibitem[Lac and Barth{\` e}s-Biesel(2005)]{lac05}
E.~Lac and D.~Barth{\` e}s-Biesel.
\newblock Deformation of a capsule in simple shear flow: Effect of membrane
  prestress.
\newblock \emph{Physics of Fluids}, 17\penalty0 (7):\penalty0 072105, 2005.
\newblock \doi{10.1063/1.1955127}.

\bibitem[Sui et~al.(2008{\natexlab{b}})Sui, Chew, Roy, and Low]{sui08a}
Y.~Sui, Y.T. Chew, P.~Roy, and H.T. Low.
\newblock A hybrid method to study flow-induced deformation of
  three-dimensional capsules.
\newblock \emph{Journal of Computational Physics}, 227\penalty0 (12):\penalty0
  6351 -- 6371, 2008{\natexlab{b}}.
\newblock ISSN 0021-9991.

\bibitem[Finken and Seifert(2006)]{seifert06}
R.~Finken and U.~Seifert.
\newblock Wrinkling of microcapsules in shear flow.
\newblock \emph{Journal of Physics: Condensed Matter}, 18\penalty0
  (15):\penalty0 L185--L191, 2006.

\bibitem[Dodson and Dimitrakopoulos(2009)]{dodson09}
W.~R. Dodson and P.~Dimitrakopoulos.
\newblock Dynamics of strain-hardening and strain-softening capsules in strong
  planar extensional flows via an interfacial spectral boundary element
  algorithm for elastic membranes.
\newblock \emph{Journal of Fluid Mechanics}, 641:\penalty0 263--296, 2009.
\newblock \doi{10.1017/S0022112009991662}.

\bibitem[Kwak and Pozrikidis(2001)]{kwak01}
S.~Kwak and C.~Pozrikidis.
\newblock Effect of membrane bending stiffness on the axisymmetric deformation
  of capsules in uniaxial extensional flow.
\newblock \emph{Physics of Fluids}, 13\penalty0 (5):\penalty0 1234--1242, 2001.
\newblock \doi{10.1063/1.1352629}.

\bibitem[Dupont et~al.(2015)Dupont, Salsac, Barth{\` e}s-Biesel, Vidrascu, and
  Tallec]{dupont15}
C.~Dupont, A.-V. Salsac, D.~Barth{\` e}s-Biesel, M.~Vidrascu, and P.~L. Tallec.
\newblock Influence of bending resistance on the dynamics of a spherical
  capsule in shear flow.
\newblock \emph{Physics of Fluids}, 27\penalty0 (5):\penalty0 051902, 2015.

\bibitem[Crowley(1973)]{crowley73}
J.~M. Crowley.
\newblock Electrical breakdown of bimolecular lipid membranes as an
  electromechanical instability.
\newblock \emph{Biophysical Journal}, 13\penalty0 (7):\penalty0 711 -- 724,
  1973.
\newblock ISSN 0006-3495.

\bibitem[Tieleman(2004)]{tieleman04}
D.~P. Tieleman.
\newblock The molecular basis of electroporation.
\newblock \emph{BMC Biochemistry}, 5\penalty0 (1):\penalty0 10, Jul 2004.
\newblock ISSN 1471-2091.

\bibitem[Neumann et~al.(1989)Neumann, Sowers, and Jordan]{neumann89}
E.~Neumann, A.E. Sowers, and C.A. Jordan.
\newblock \emph{Electroporation and Electrofusion in Cell Biology}.
\newblock Plenum press, New York, 1 edition, 1989.

\bibitem[Pethig(1996)]{ronald96}
R.~Pethig.
\newblock Dielectrophoresis: Using inhomogeneous ac electrical fields to
  separate and manipulate cells.
\newblock \emph{Critical Reviews in Biotechnology}, 16\penalty0 (4):\penalty0
  331--348, 1996.

\bibitem[Kang et~al.(2008)Kang, Li, Kalams, and Eid]{kang08}
Y.~Kang, D.~Li, S.~A. Kalams, and J.~E. Eid.
\newblock Dc-dielectrophoretic separation of biological cells by size.
\newblock \emph{Biomedical Microdevices}, 10\penalty0 (2):\penalty0 243--249,
  Apr 2008.
\newblock ISSN 1572-8781.

\bibitem[Yamamoto et~al.(2010)Yamamoto, Aranda-Espinoza, Dimova, and
  Lipowsky]{dimova10}
T.~Yamamoto, S.~Aranda-Espinoza, R.~Dimova, and R.~Lipowsky.
\newblock Stability of spherical vesicles in electric fields.
\newblock \emph{Langmuir}, 26\penalty0 (14):\penalty0 12390--12407, 2010.

\bibitem[Thaokar and Deshmukh(2010)]{rtsd10}
R.~M. Thaokar and S.~D. Deshmukh.
\newblock Rayleigh instability of charged drops and vesicles in the presence of
  counterions.
\newblock \emph{Physics of Fluids}, 22\penalty0 (3):\penalty0 034107, 2010.

\bibitem[Deshmukh and Thaokar(2012{\natexlab{a}})]{shivrajrt12}
S.~D. Deshmukh and R.~M. Thaokar.
\newblock Deformation, breakup and motion of a perfect dielectric drop in a
  quadrupole electric field.
\newblock \emph{Physics of Fluids}, 24\penalty0 (3):\penalty0 032105,
  2012{\natexlab{a}}.

\bibitem[Deshmukh and Thaokar(2013{\natexlab{a}})]{deshmukh_thaokar_2013}
S.~D. Deshmukh and R.~M. Thaokar.
\newblock Deformation and breakup of a leaky dielectric drop in a quadrupole
  electric field.
\newblock \emph{Journal of Fluid Mechanics}, 731:\penalty0 713--733,
  2013{\natexlab{a}}.

\bibitem[Karyappa et~al.(2014{\natexlab{b}})Karyappa, Deshmukh, and
  Thaokar]{karyappa_deshmukh14}
R.~B. Karyappa, S.~D. Deshmukh, and R.~M. Thaokar.
\newblock Breakup of a conducting drop in a uniform electric field.
\newblock \emph{Journal of Fluid Mechanics}, 754:\penalty0 550--589,
  2014{\natexlab{b}}.

\bibitem[Das et~al.(2015)Das, Mayya, and Thaokar]{skd15}
S.~Das, Y.~S. Mayya, and R.~Thaokar.
\newblock Dynamics of a charged drop in a quadrupole electric field.
\newblock \emph{EPL (Europhysics Letters)}, 111\penalty0 (2):\penalty0 24006,
  2015.

\bibitem[McConnell et~al.(2015{\natexlab{a}})McConnell, Vlahovska, and
  Miksis]{mcconnell15sm}
L.~C. McConnell, P.~M. Vlahovska, and M.~J. Miksis.
\newblock Vesicle dynamics in uniform electric fields: squaring and breathing.
\newblock \emph{Soft Matter}, 11:\penalty0 4840--4846, 2015{\natexlab{a}}.
\newblock \doi{10.1039/C5SM00585J}.

\bibitem[Ouriemi and Vlahovska(2015)]{vlahovska15}
M.~Ouriemi and P.~M. Vlahovska.
\newblock Electrohydrodynamic deformation and rotation of a particle-coated
  drop.
\newblock \emph{Langmuir}, 31\penalty0 (23):\penalty0 6298--6305, 2015.
\newblock \doi{10.1021/acs.langmuir.5b00774}.
\newblock PMID: 26011225.

\bibitem[Riske and Dimova(2006)]{karin06}
K.~A. Riske and R.~Dimova.
\newblock Electric pulses induce cylindrical deformations on giant vesicles in
  salt solutions.
\newblock \emph{Biophysical Journal}, 91\penalty0 (5):\penalty0 1778 -- 1786,
  2006.
\newblock ISSN 0006-3495.

\bibitem[Dimova et~al.(2007)Dimova, Riske, Aranda, Bezlyepkina, Knorr, and
  Lipowsky]{dimova07}
R.~Dimova, K.~A. Riske, S.~Aranda, N.~Bezlyepkina, R.~L. Knorr, and
  R.~Lipowsky.
\newblock Giant vesicles in electric fields.
\newblock \emph{Soft Matter}, 3:\penalty0 817--827, 2007.

\bibitem[Dimova et~al.(2009)Dimova, Bezlyepkina, Jordo, Knorr, Riske, Staykova,
  Vlahovska, Yamamoto, Yang, and Lipowsky]{dimova09}
R.~Dimova, N.~Bezlyepkina, M.~D. Jordo, R.~L. Knorr, K.~A. Riske, M.~Staykova,
  P.~M. Vlahovska, T.~Yamamoto, P.~Yang, and R.~Lipowsky.
\newblock Vesicles in electric fields: Some novel aspects of membrane behavior.
\newblock \emph{Soft Matter}, 5:\penalty0 3201--3212, 2009.

\bibitem[McConnell et~al.(2015{\natexlab{b}})McConnell, Miksis, and
  Vlahovska]{mcconnell15}
L.~C. McConnell, M.~J. Miksis, and P.~M. Vlahovska.
\newblock Continuum modeling of the electric-field-induced tension in deforming
  lipid vesicles.
\newblock \emph{The Journal of Chemical Physics}, 143\penalty0 (24):\penalty0
  243132, 2015{\natexlab{b}}.

\bibitem[McConnell et~al.(2013)McConnell, Miksis, and Vlahovska]{mcconnell13}
L.~C. McConnell, M.~J. Miksis, and P.~M. Vlahovska.
\newblock Vesicle electrohydrodynamics in dc electric fields†.
\newblock \emph{IMA Journal of Applied Mathematics}, 78\penalty0 (4):\penalty0
  797, 2013.

\bibitem[Veerapaneni(2016)]{Veerapaneni16}
S.~Veerapaneni.
\newblock Integral equation methods for vesicle electrohydrodynamics in three
  dimensions.
\newblock \emph{Journal of Computational Physics}, 326:\penalty0 278 -- 289,
  2016.
\newblock ISSN 0021-9991.

\bibitem[Kolahdouz and Salac(2015{\natexlab{a}})]{ebrahim15}
E.~M. Kolahdouz and D.~Salac.
\newblock Electrohydrodynamics of three-dimensional vesicles: A numerical
  approach.
\newblock \emph{SIAM Journal on Scientific Computing}, 37\penalty0
  (3):\penalty0 B473--B494, 2015{\natexlab{a}}.
\newblock \doi{10.1137/140988966}.

\bibitem[Kolahdouz and Salac(2015{\natexlab{b}})]{ebrahim15a}
E.~M. Kolahdouz and D.~Salac.
\newblock Dynamics of three-dimensional vesicles in dc electric fields.
\newblock \emph{Phys. Rev. E}, 92:\penalty0 012302, Jul 2015{\natexlab{b}}.

\bibitem[Grosse and Schwan(1992)]{grosse92}
C.~Grosse and H.~P. Schwan.
\newblock Cellular membrane potentials induced by alternating fields.
\newblock \emph{Biophysical Journal}, 63\penalty0 (6):\penalty0 1632 -- 1642,
  1992.
\newblock ISSN 0006-3495.

\bibitem[DeBruin and Krassowska(1999)]{debruin99}
K.~A. DeBruin and W.~Krassowska.
\newblock Modeling electroporation in a single cell. i. effects of field
  strength and rest potential.
\newblock \emph{Biophysical Journal}, 77\penalty0 (3):\penalty0 1213 -- 1224,
  1999.
\newblock ISSN 0006-3495.

\bibitem[Barth{\`e}s-Biesel et~al.(2002)Barth{\`e}s-Biesel, Diaz, and
  Dhenin]{barthesbiesel02}
D.~Barth{\`e}s-Biesel, A.~Diaz, and E.~Dhenin.
\newblock Effect of constitutive laws for two-dimensional membranes on
  flow-induced capsule deformation.
\newblock \emph{Journal of Fluid Mechanics}, 460:\penalty0 211--222, 2002.

\bibitem[Hu et~al.(2014)Hu, Kim, and Lai]{hu14}
W.~F. Hu, Y.~Kim, and M.-C. Lai.
\newblock An immersed boundary method for simulating the dynamics of
  three-dimensional axisymmetric vesicles in navier-stokes flows.
\newblock \emph{Journal of Computational Physics}, 257, Part A:\penalty0 670 --
  686, 2014.
\newblock ISSN 0021-9991.

\bibitem[Rallison and Acrivos(1978)]{rallison78}
J.~M. Rallison and A.~Acrivos.
\newblock A numerical study of the deformation and burst of a viscous drop in
  an extensional flow.
\newblock \emph{Journal of Fluid Mechanics}, 89:\penalty0 191--200, 11 1978.
\newblock ISSN 1469-7645.

\bibitem[Deshmukh and Thaokar(2012{\natexlab{b}})]{shivraj12}
S.~D. Deshmukh and R.~M. Thaokar.
\newblock Deformation, breakup and motion of a perfect dielectric drop in a
  quadrupole electric field.
\newblock \emph{Physics of Fluids}, 24\penalty0 (3):\penalty0 032105,
  2012{\natexlab{b}}.

\bibitem[Deshmukh and Thaokar(2013{\natexlab{b}})]{shivraj13}
S.~D. Deshmukh and R.~M. Thaokar.
\newblock Deformation and breakup of a leaky dielectric drop in a quadrupole
  electric field.
\newblock \emph{Journal of Fluid Mechanics}, 731:\penalty0 713--733, 9
  2013{\natexlab{b}}.

\bibitem[Pozrikidis(1990)]{poz90}
C.~Pozrikidis.
\newblock The axisymmetric deformation of a red blood cell in uniaxial
  straining stokes flow.
\newblock \emph{Journal of Fluid Mechanics}, 216:\penalty0 231--254, 1990.
\newblock ISSN 1469-7645.

\bibitem[Schwalbe et~al.(2011)Schwalbe, Vlahovska, and Miksis]{Schwalbe11}
J.~T. Schwalbe, P.~M. Vlahovska, and M.~J. Miksis.
\newblock Vesicle electrohydrodynamics.
\newblock \emph{Phys. Rev. E}, 83:\penalty0 046309, Apr 2011.

\end{thebibliography}

\begin{appendices}
 \section{Analytical solution}
Electrostatics and hydrodynamics of capsule deformation are solved analytically in spherical coordinate system $(r,\theta,\phi)$, where $r$, $\theta$ and $\phi$ are the measurements of radial distance, meridional and azimuthal angles. 
\subsection{Solution for the electrostatics}\label{appA}
\subsubsection{Analytical electrostatic theory (AET)}\label{aapAcm}
The external and internal potentials because of the external uniform electric field of intensity $E_0$ of a spherical interface can be represented using spherical harmonics.
\begin{eqnarray}\label{eq:poteqns}
   \phi_i &= A_1 r \cos(\theta)\\
  \phi_e &= B_1 \cos(\theta)/r^2-r \cos(\theta), 
 \end{eqnarray}
where $A_1$ and $B_1$ are time dependent coefficients of spherical harmonics. The time dependent transmembrane potential is expressed as
\begin{equation}\label{eq:tmpot}
 \phi_m=\phi_{amp}\cos(\theta),
\end{equation}
where $\phi_{amp}=\phi_i-\phi_e$ is the amplitude of the transmembrane potential. Current continuity condition across the membrane is expressed by
\begin{equation}\label{eq:current}
  \sigma_r E_{n,i}+\epsilon_r\frac{dE_{n,i}}{dt}=E_{n,e}+\frac{dE_{n,e}}{dt}=\hat C_m \frac{d \phi_m}{dt}+\hat G_m \phi_m
\end{equation}
In these non-dimensional equations, one can recall that $\hat C_m=C_m a/(\epsilon_e \epsilon_0)$ and $\hat G_m=aG_m/\sigma_e$.
Solving \cref{eq:poteqns} along with \cref{eq:tmpot} and \cref{eq:current} coefficients $A_1$ and $B_1$ are obtained, and the analytical expressions are very large and cumbersome to evaluate. Internal and external normal electric fields are obtained as
\begin{eqnarray}
  {\bf E}_{n,i} &=& -{\bf n}\cdot{\bf \nabla}\phi_i\\
 {\bf E}_{n,e} &=& -{\bf n}\cdot{\bf \nabla}\phi_e
\end{eqnarray}
and tangential fields as
\begin{eqnarray}
  {\bf E}_{t,i} &=& -{\bf t}\cdot{\bf \nabla}\phi_i\\
 {\bf E}_{t,e} &=& -{\bf t}\cdot{\bf \nabla}\phi_e.
\end{eqnarray}
Normal and tangential electric stresses can be obtained from 
\begin{eqnarray}
 {\bf \tau}_n^E &=& \frac{1}{2}Ca\left[({\bf E}_{n,e}^2-{\bf E}_{t,e}^2)-\epsilon_r({\bf E}_{n,i}^2-{\bf E}_{t,i}^2)\right],\\
 {\bf \tau}_t^E &=& Ca\left[{\bf E}_{n,e}{\bf E}_{t,e}-\epsilon_r{\bf E}_{n,i}{\bf E}_{t,i}\right].
\end{eqnarray}
Electric tractions acting on the interface are given by
\begin{eqnarray}
 f_x^E &=& {\bf \tau}_n^E n_x+\tau_t^E t_x,\\
 f_y^E &=& {\bf \tau}_n^E n_y+\tau_t^E t_y.
\end{eqnarray}
Transmembrane potential at $\hat G_m=0$, $\hat C_m=50$, $\epsilon_r=1$ and $\sigma_r=1$ is obtained as 
\begin{equation}
 \phi_m=\frac{3}{2}(1-e^{-t/76})\cos\theta,
\end{equation}
and normal and tangential stresses are obtained as
\begin{eqnarray}
 \tau_n &=&\frac{1}{8}Ca\left(3e^{-t/38}\sin^2\theta+6e^{-t/76}\sin^2\theta-9\sin^2\theta\right)\\
 \tau_t &=&-\frac{3}{4}Ca \left(e^{-t/76}-1\right)e^{-t/38}\sin(2\theta). 
\end{eqnarray}
For $\sigma_r=0.1$ and $10$ analytical solutions for normal and tangential electrical stresses are very complicated and obtained as large expressions. Figures comparing results for transmembrane potential, tangential electric stress and normal electric stress obtained from analytical theory and boundary integral simulation can be obtained in \cref{sec:fieldvalidation} (\cref{fgr:vmcheck,fgr:taunl1,fgr:tautl1} for $\sigma_r=1$, \cref{fgr:vmcheckl10,fgr:taunl10,fgr:tautl10} for $\sigma_r=10$ and \cref{fgr:vmcheckl0p1,fgr:taunl0p1,fgr:tautl0p1} $\sigma_r=0.1$). 

\subsubsection{Simplified electrostatic model (SEM)}\label{aapAsm}
It is often a good approximation to neglect the displacement current in the two fluids, since it is expected to be much smaller than the capacitive current of the membrane, ($\epsilon_e/\sigma_e\ll 1$ and $\epsilon_i/\sigma_i\ll 1$). In this case, the electrostatics equations reduce to,
\begin{equation}\label{eq:redcurrent}
  \sigma_r E_{n,i}=E_{n,e}=\hat C_m \frac{d \phi_m}{dt}+\hat G_m \phi_m
\end{equation}
and for a non conducing membrane the transmembrane potential is obtained as 
\begin{equation}
 \phi_m=\frac{3}{2} (1-e^{-t/t_{cap}}),
\end{equation}
known as Schwan equation~\cite{grosse92} and normal and tangential stresses are obtained as
\begin{eqnarray}
 \tau_t &=& \frac{9}{4}\left[\frac{(2+\sigma_r^2)e^{\frac{-2t}{t_{cap}}}-\sigma_r(2+\sigma_r)e^{\frac{-t}{t_{cap}}}}{(2+\sigma_r)^2}\right]\sin 2\theta\\
 \tau_n &=& \frac{9}{8}\frac{4(\sigma_r^2-1)e^{\frac{-2t}{t_{cap}}}+\left[(5\sigma_r^2-8)e^{\frac{-2t}{t_{cap}}}+2\sigma_r(2+\sigma_r)e^{\frac{-t}{t_{cap}}}-(2+\sigma_r)^2\right]\sin^2\theta}{(2+\sigma_r)^2}.
\end{eqnarray}
Obtained form of the expressions for tangential and normal stresses are different but produce same result obtained with the reported expressions of stresses by~\citet{Schwalbe11}.
To analytically solve the Electrohydrodynamics of capsule, simplified electrostatic model is used to calculate interfacial electric stresses. Formulation of the elastic traction (considering liner elasticity model with Poisson's ratio $\nu_s=1$), hydrodynamics and the solution methodology for the deformation of an elastic capsule are followed as reported by~\citet{rt16}. The general expression for the degree of deformation is very large and complex. For $\hat G_m=0$, $\hat C_m=50$, $\epsilon_r=1$ degrees of deformation of different conductivity ratios are obtained as
\begin{eqnarray}
 DD_{\sigma_r=0.1} &=& Ca \Big[-0.000193+0.0882e^{\frac{1}{5}\sqrt{\frac{31}{7}}t}+0.703e^{\frac{1}{70}(21+\sqrt{217})t}-\nonumber\\
 & & 2.125e^{\frac{311+15\sqrt{217}}{1050}t}+0.014e^{\frac{313+15\sqrt{217}}{1050}t}\Big]e^{-\frac{1}{70}(21+\sqrt{217})t}\\
 DD_{\sigma_r=1} &=& Ca \Big[-0.00504+0.319e^{\frac{1}{5}\sqrt{\frac{31}{7}}t}+0.703e^{\frac{1}{70}(21+\sqrt{217})t}-\nonumber\\
 & & 1.122e^{\frac{287+15\sqrt{217}}{1050}t}+0.106e^{\frac{301+15\sqrt{217}}{1050}t}\Big]e^{-\frac{1}{70}(21+\sqrt{217})t}\\
 DD_{\sigma_r=10} &=& Ca \Big[-0.00344-3.775e^{\frac{1}{5}\sqrt{\frac{31}{7}}t}+\nonumber\\
 & & e^{\frac{1}{210}(49+3\sqrt{217})t}(4.887+0.333e^{\frac{1}{30}t}+0.703e^{\frac{1}{15}t})\Big]e^{-\frac{1}{70}(21+\sqrt{217})t}. 
\end{eqnarray}
Analytical and boundary integral simulation results for degree of deformation as a function of time at small capillary number are compared (\cref{fgr:thnumddfl1}, \cref{fgr:thnumddfl0p1} and \cref{fgr:thnumddfl10} for $\sigma_r=1$, $0.1$ and $10$, respectively) considering $\epsilon_r=1$, $\hat C_m=50$ and $\hat G_m=0$ for the validation of the boundary integral code, a very good match is obtained between numerical and analytical values.  
\subsection{Validation of boundary integral code solving electric field}\label{sec:fieldvalidation}
\begin{figure}[H]
\begin{center}
\begin{subfigure}{.32\textwidth}
  \centering
  \includegraphics[width=1\textwidth]{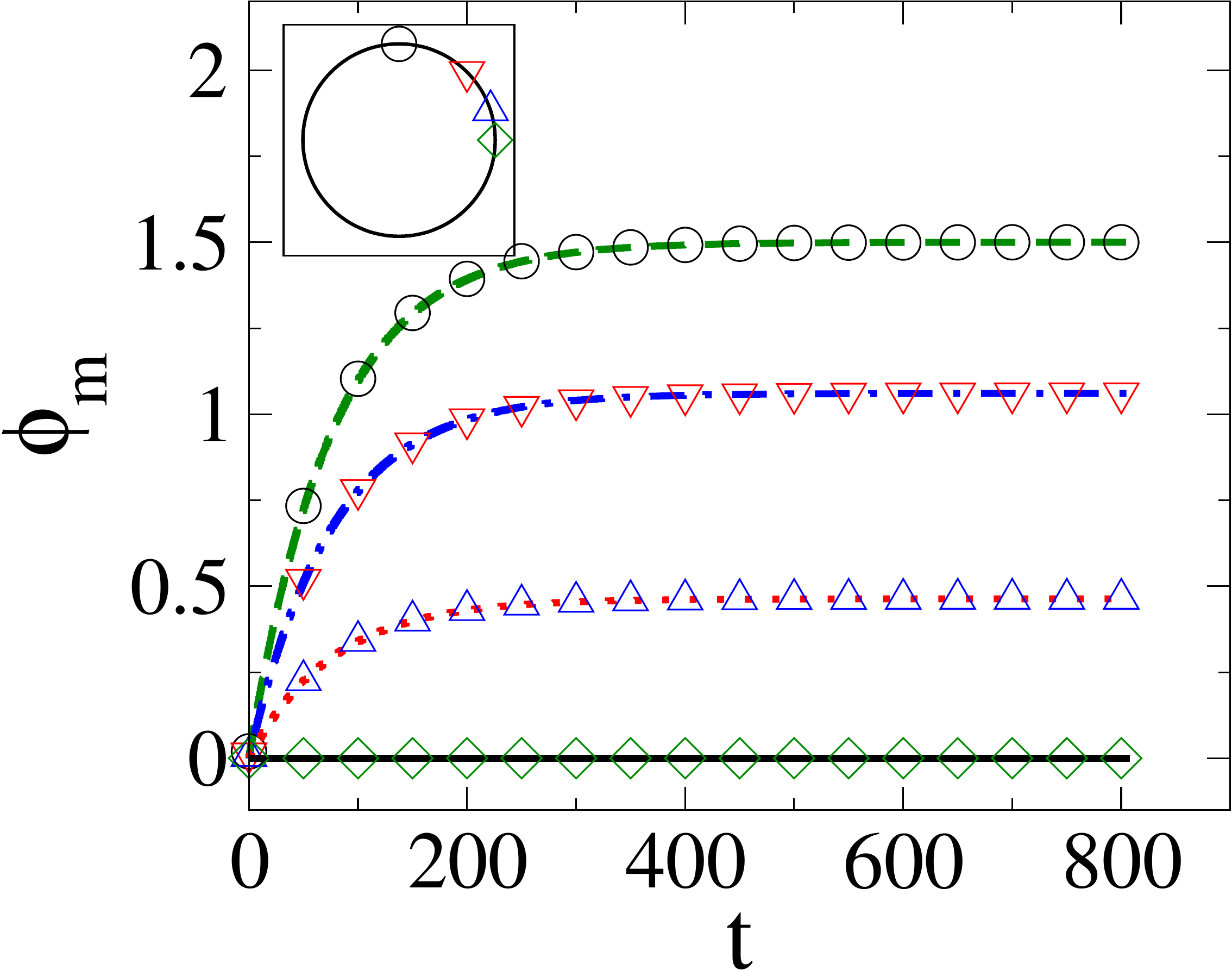}
  \caption{$\phi_m$ vs. $t$}
  \label{fgr:vmcheck}
\end{subfigure}
\begin{subfigure}{.32\textwidth}
  \centering
  \includegraphics[width=1\textwidth]{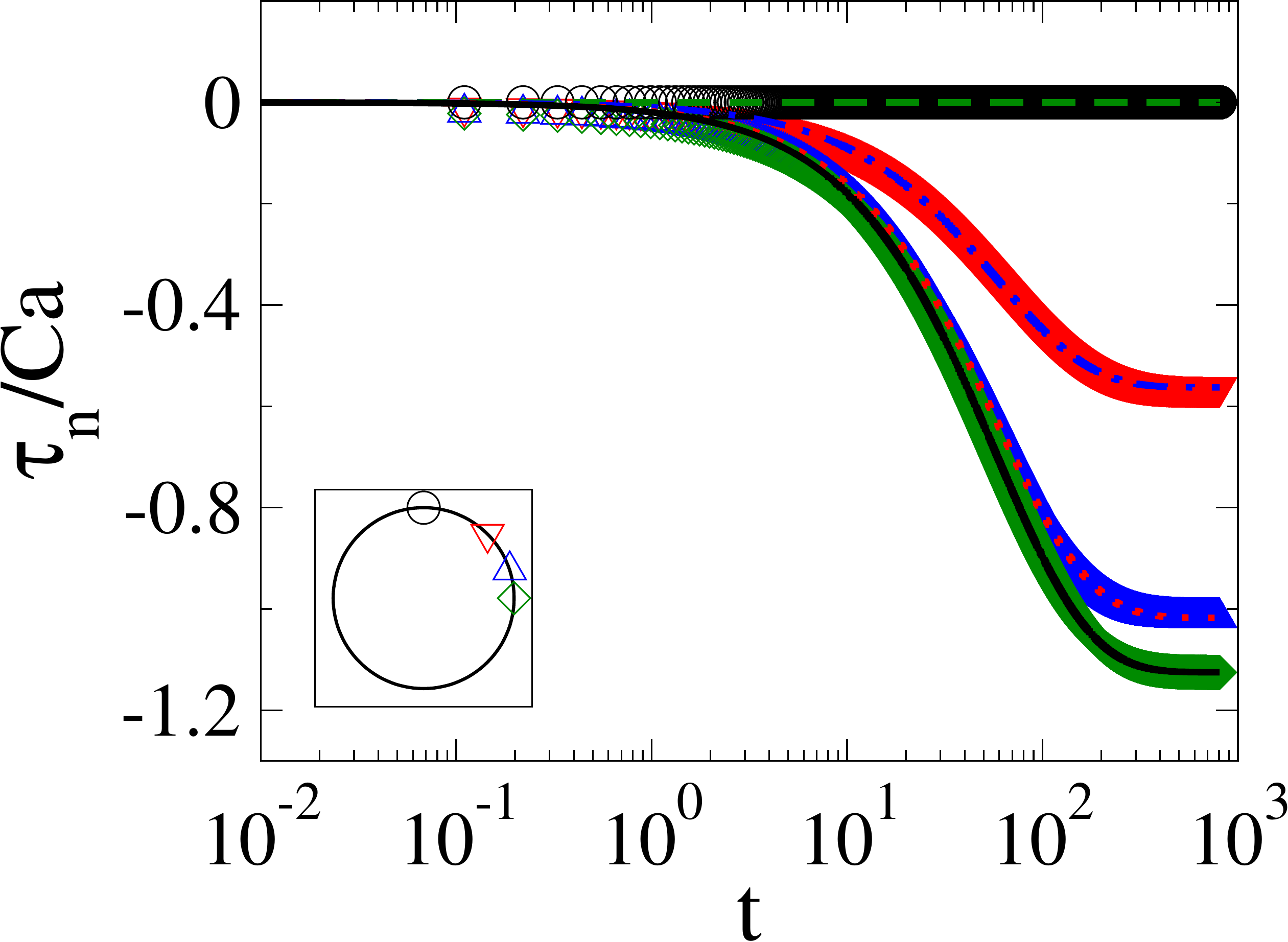}
  \caption{$\tau_n/Ca$ vs. $t$}
  \label{fgr:taunl1}
\end{subfigure}
\begin{subfigure}{.32\textwidth}
  \centering
  \includegraphics[width=1\textwidth]{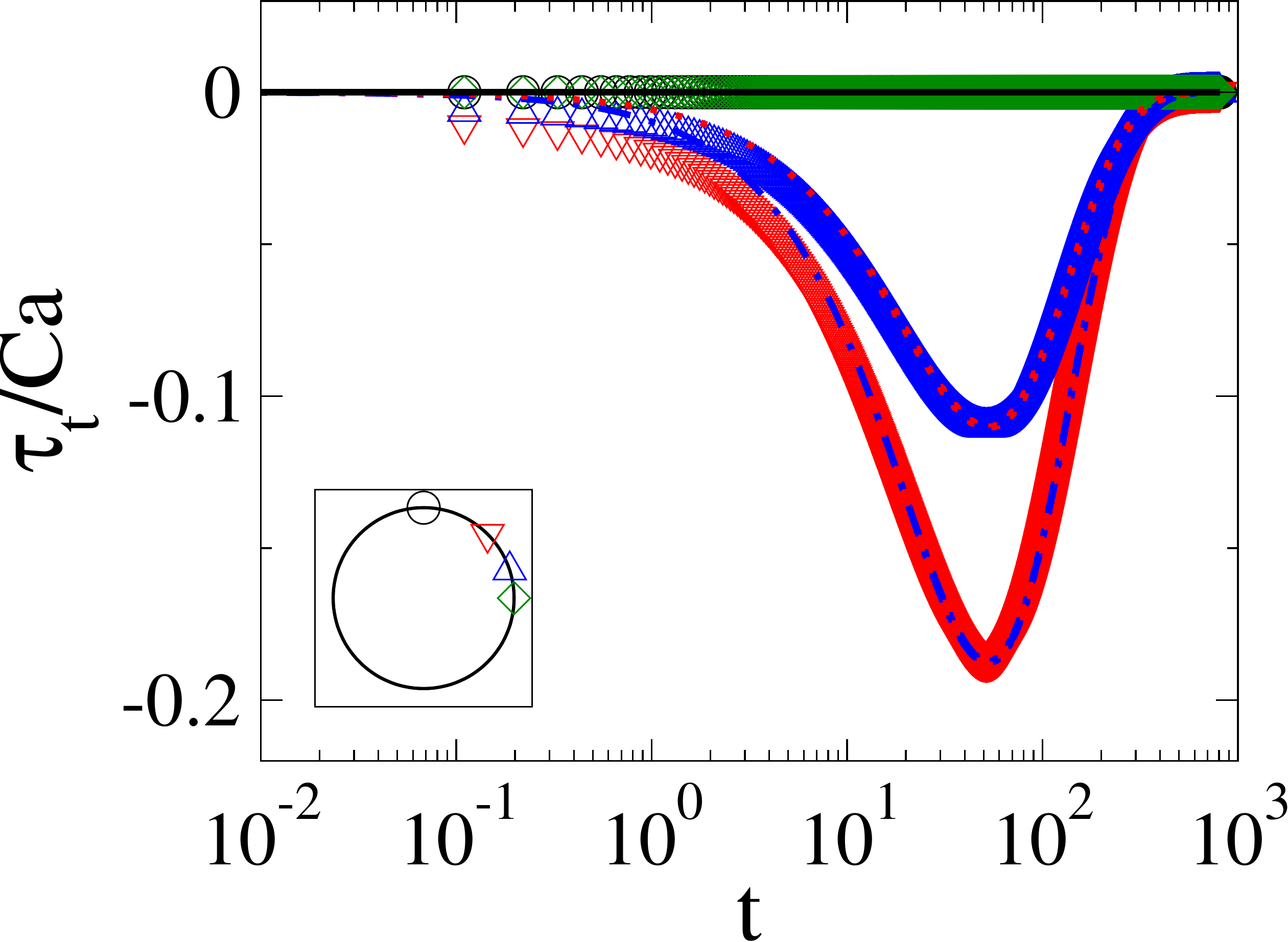}
  \caption{$\tau_t/Ca$ vs. $t$}
  \label{fgr:tautl1}
\end{subfigure}
\caption{Comparison of numerically (\textcolor{black}{$\bigcirc$}, \textcolor{red}{$\triangledown$}, \textcolor{blue}{$\vartriangle$} and \textcolor{forestgreen}{$\diamondsuit$} at $\theta=0$, $\pi/4$, $\pi/2.5$ and $\pi/2$, respectively) and analytically (\textcolor{forestgreen}{$\pmb{--}$}, \textcolor{blue}{$\pmb{-\cdot-}$}, \textcolor{red}{$\pmb{\cdots}$} and \textcolor{black}{$\pmb{\mi}$} at $\theta=0$, $\pi/4$, $\pi/2.5$ and $\pi/2$, respectively) obtained (a) transmembrane potential, (b) normal electric stress  and (c) tangential electric stress as a function of time for $\sigma_r=1$.}
\label{fgr:fieldl1}
\end{center}
\end{figure}

\begin{figure}[H]
\begin{center}
\begin{subfigure}{.32\textwidth}
  \centering
  \includegraphics[width=1\textwidth]{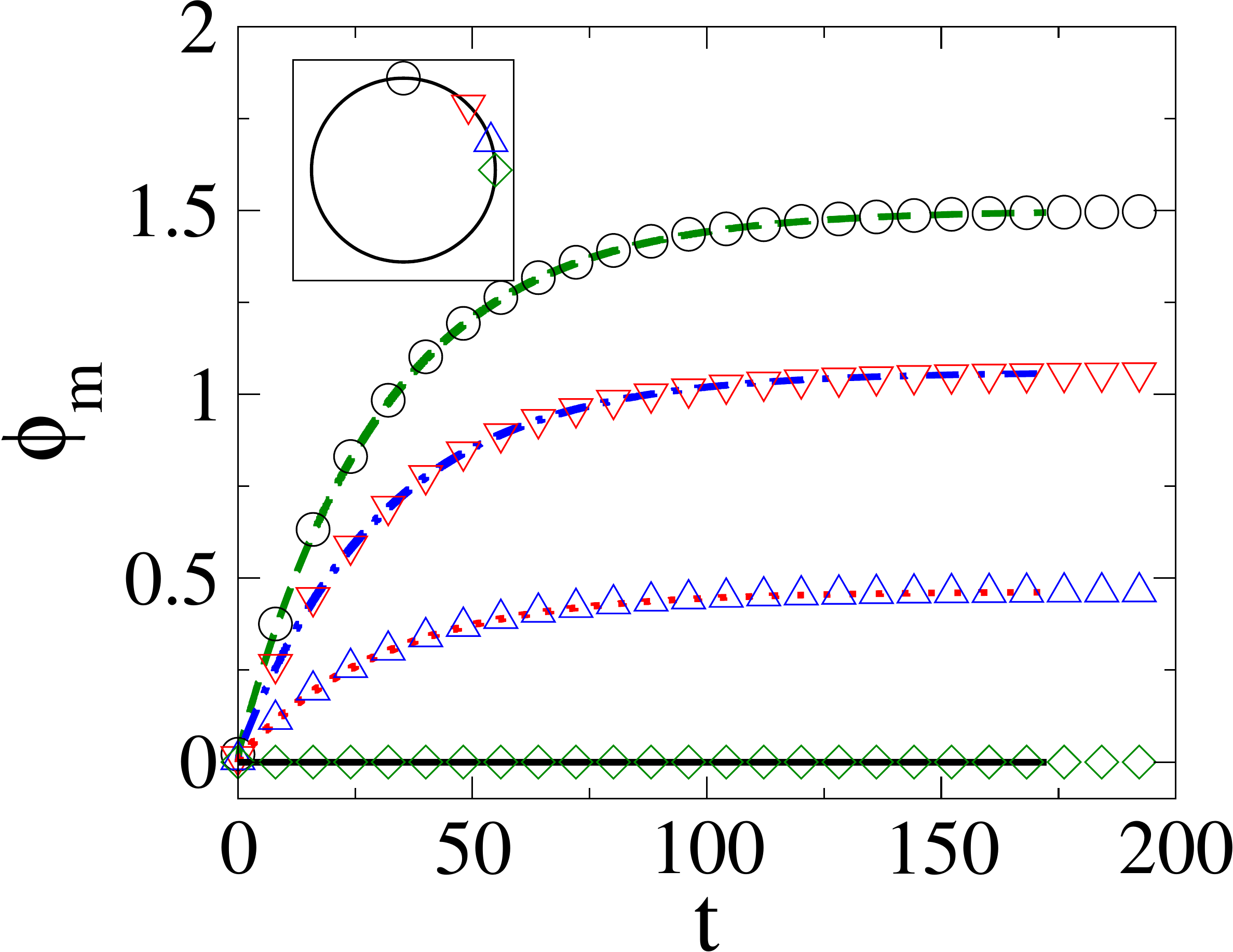}
  \caption{$\phi_m$ vs. $t$}
  \label{fgr:vmcheckl10}
\end{subfigure}
\begin{subfigure}{.32\textwidth}
  \centering
  \includegraphics[width=1\textwidth]{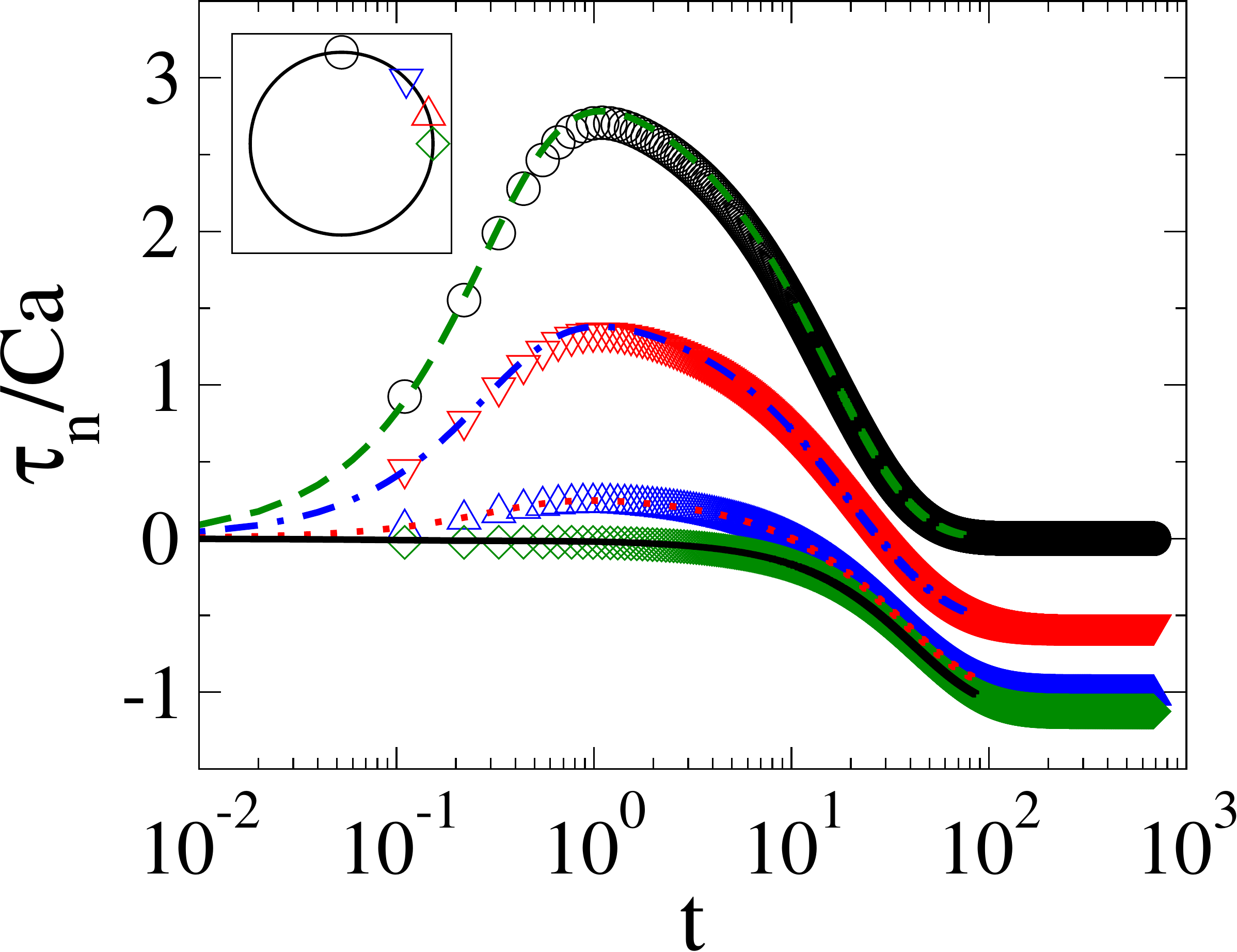}
  \caption{$\tau_n/Ca$ vs. $t$}
  \label{fgr:taunl10}
\end{subfigure}
\begin{subfigure}{.32\textwidth}
  \centering
  \includegraphics[width=1\textwidth]{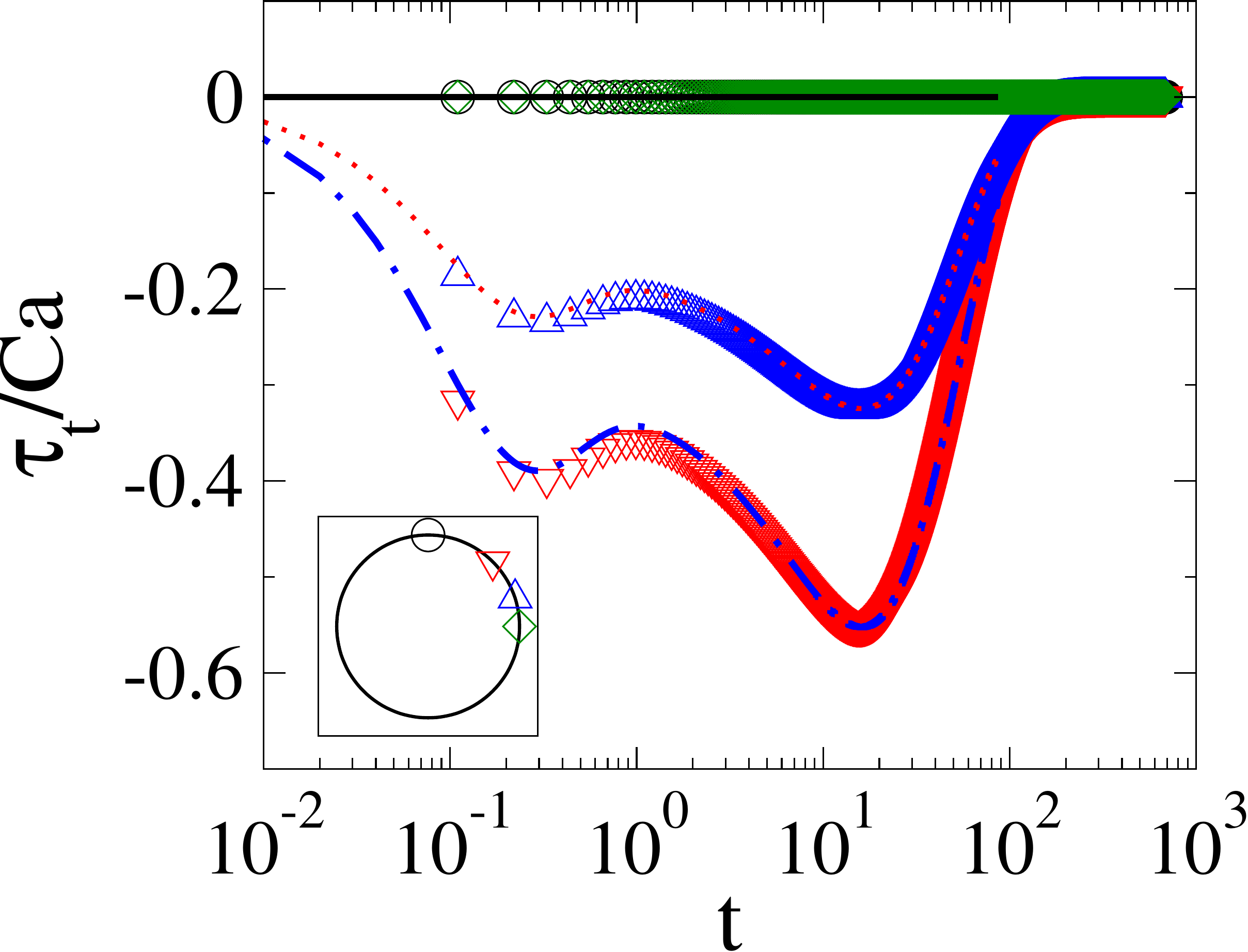}
  \caption{$\tau_t/Ca$ vs. $t$}
  \label{fgr:tautl10}
\end{subfigure}
\caption{Comparison of numerically (\textcolor{black}{$\bigcirc$}, \textcolor{red}{$\triangledown$}, \textcolor{blue}{$\vartriangle$} and \textcolor{forestgreen}{$\diamondsuit$} at $\theta=0$, $\pi/4$, $\pi/2.5$ and $\pi/2$, respectively) and analytically (\textcolor{forestgreen}{$\pmb{--}$}, \textcolor{blue}{$\pmb{-\cdot-}$}, \textcolor{red}{$\pmb{\cdots}$} and \textcolor{black}{$\pmb{\mi}$} at $\theta=0$, $\pi/4$, $\pi/2.5$ and $\pi/2$, respectively) obtained (a) transmembrane potential, (b) normal electric stress  and (c) tangential electric stress as a function of time for $\sigma_r=10$.}
\label{fgr:fieldl10}
\end{center}
\end{figure}

\begin{figure}[H]
\begin{center}
\begin{subfigure}{.32\textwidth}
  \centering
  \includegraphics[width=1\textwidth]{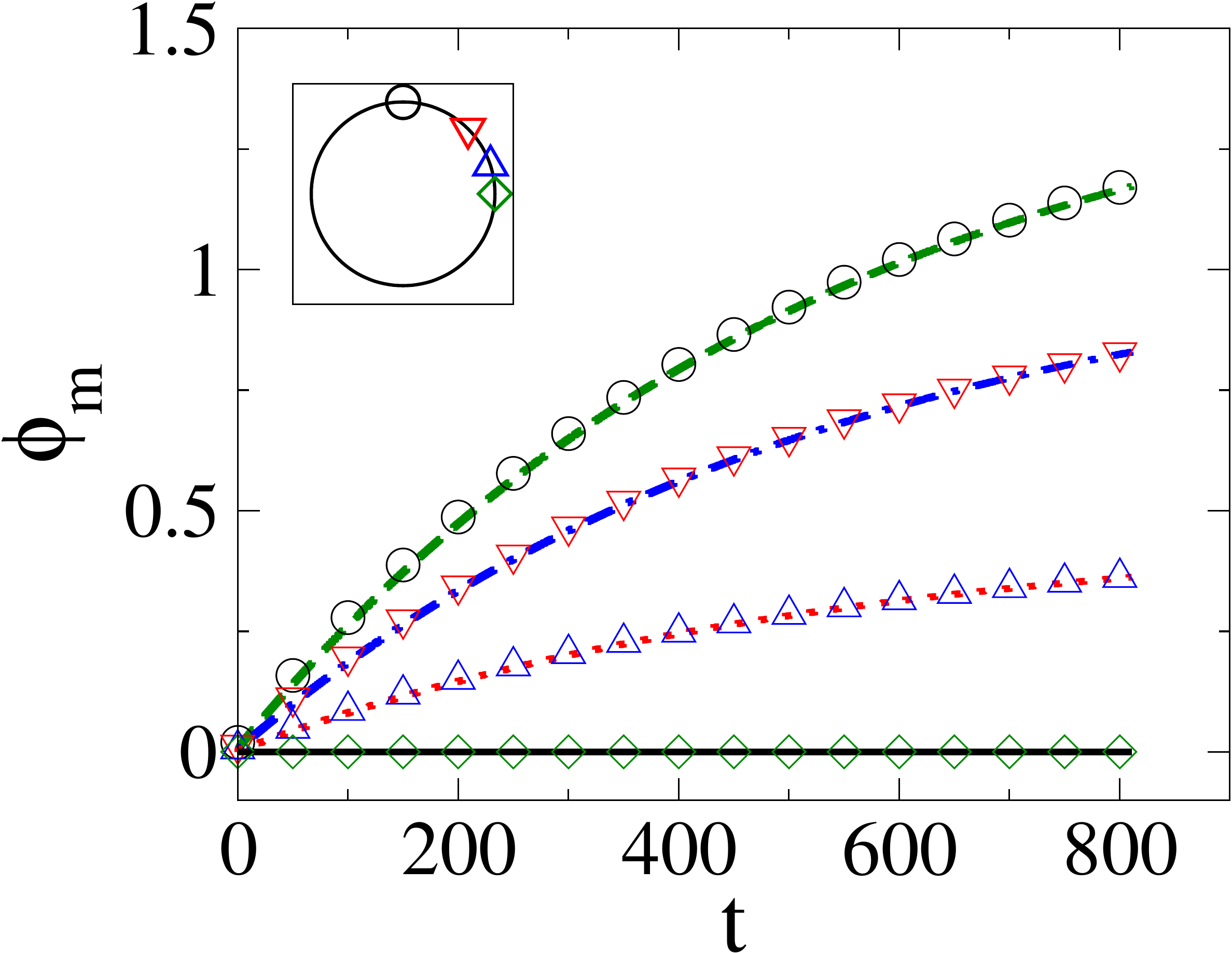}
  \caption{$\phi_m$ vs. $t$}
  \label{fgr:vmcheckl0p1}
\end{subfigure}
\begin{subfigure}{.32\textwidth}
  \centering
  \includegraphics[width=1\textwidth]{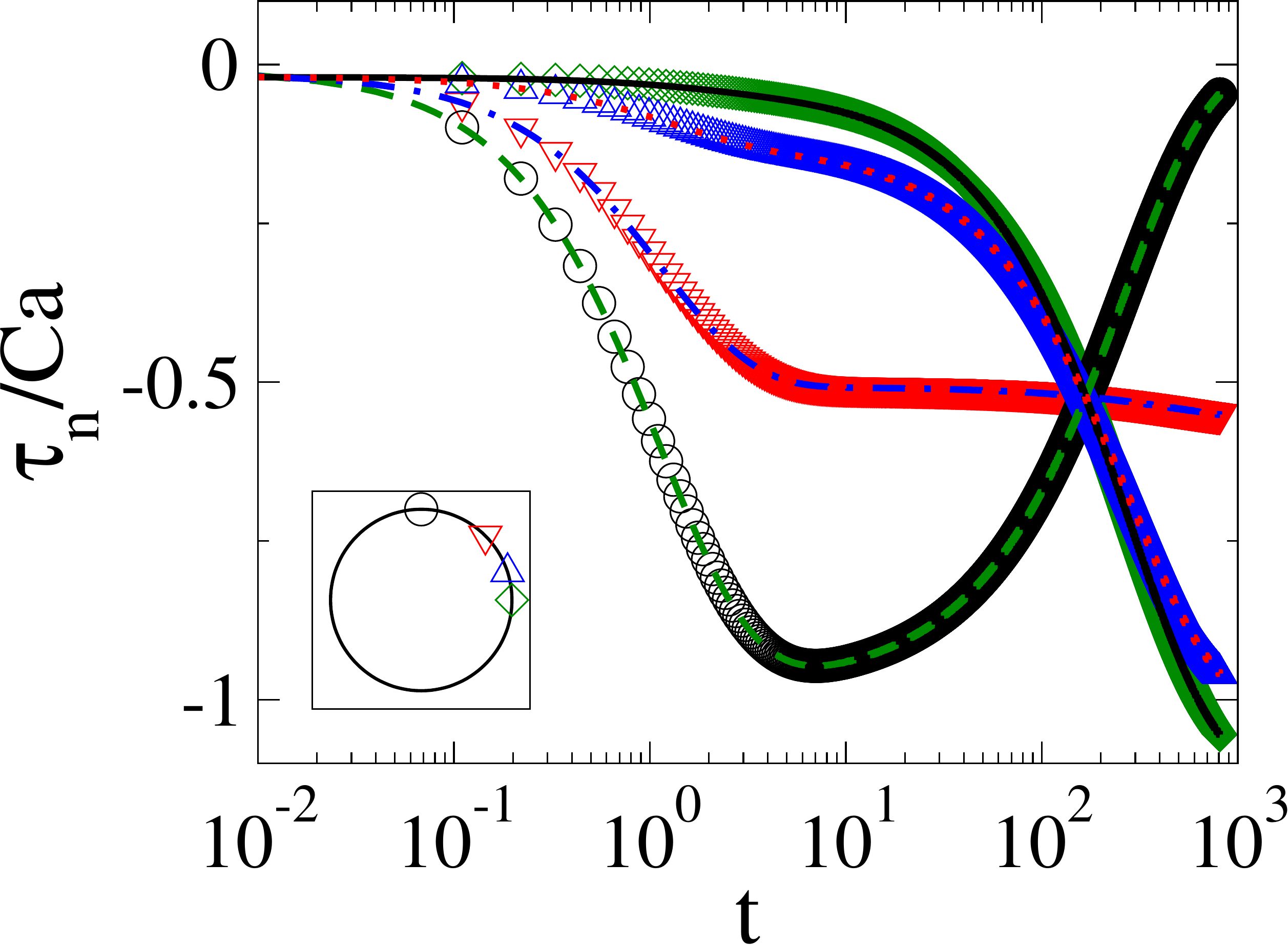}
  \caption{$\tau_n/Ca$ vs. $t$}
  \label{fgr:taunl0p1}
\end{subfigure}
\begin{subfigure}{.32\textwidth}
  \centering
  \includegraphics[width=1\textwidth]{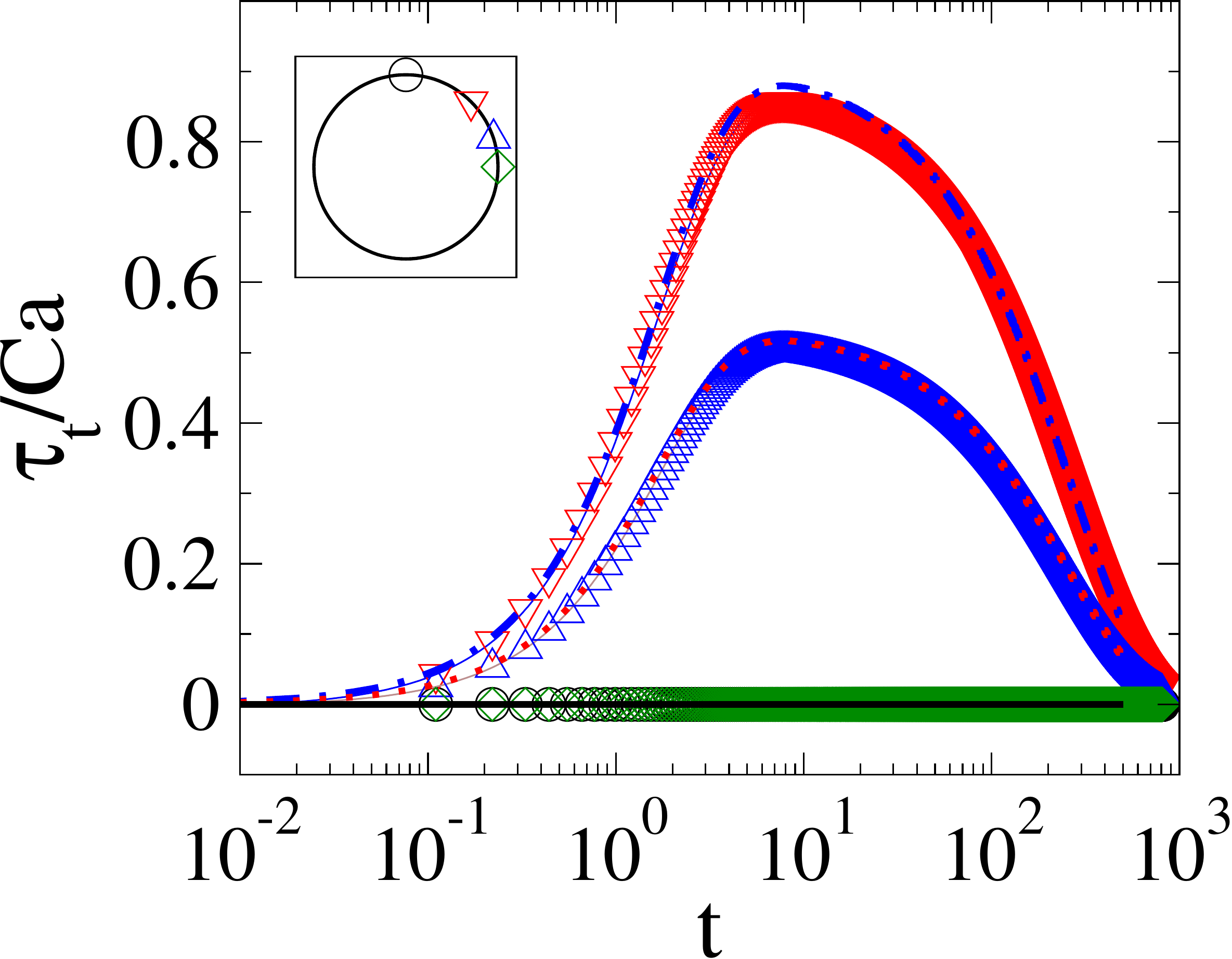}
  \caption{$\tau_t/Ca$ vs. $t$}
  \label{fgr:tautl0p1}
\end{subfigure}
\caption{Comparison of numerically (\textcolor{black}{$\bigcirc$}, \textcolor{red}{$\triangledown$}, \textcolor{blue}{$\vartriangle$} and \textcolor{forestgreen}{$\diamondsuit$} at $\theta=0$, $\pi/4$, $\pi/2.5$ and $\pi/2$, respectively) and analytically (\textcolor{forestgreen}{$\pmb{--}$}, \textcolor{blue}{$\pmb{-\cdot-}$}, \textcolor{red}{$\pmb{\cdots}$} and \textcolor{black}{$\pmb{\mi}$} at $\theta=0$, $\pi/4$, $\pi/2.5$ and $\pi/2$, respectively) obtained (a) transmembrane potential, (b) normal electric stress  and (c) tangential electric stress as a function of time for $\sigma_r=0.1$.}
\label{fgr:fieldl0p1}
\end{center}
\end{figure}

 Boundary integral code with capacitor model (BEM-CM) to solve the electrostatics is validated comparing results with the analytical electrostatic theory (AET) considering $\hat{G}_m=0$, $\hat{C}_m=50$ and $\epsilon_r=1$. The evolution of transmembrane potential, normal and tangential stresses are shown in \cref{fgr:vmcheck,fgr:taunl1,fgr:tautl1}, respectively for $\sigma_r=1$, \cref{fgr:vmcheckl10,fgr:taunl10,fgr:tautl10}, respectively for $\sigma_r=10$ and \cref{fgr:vmcheckl0p1,fgr:taunl0p1,fgr:tautl0p1}, respectively for $\sigma_r=0.1$. 
 
% \pagebreak
% \bibliographystyle{jfm}
% % Note the spaces between the initials
% \bibliography{dccapsule}
% 
% \newpage

% \section{Supplementary information}
\subsection{Validation of boundary integral code for the deformation of capsule}
\begin{figure}[H]
\begin{center}
\begin{subfigure}{.32\textwidth}
  \centering
  \includegraphics[width=1\textwidth]{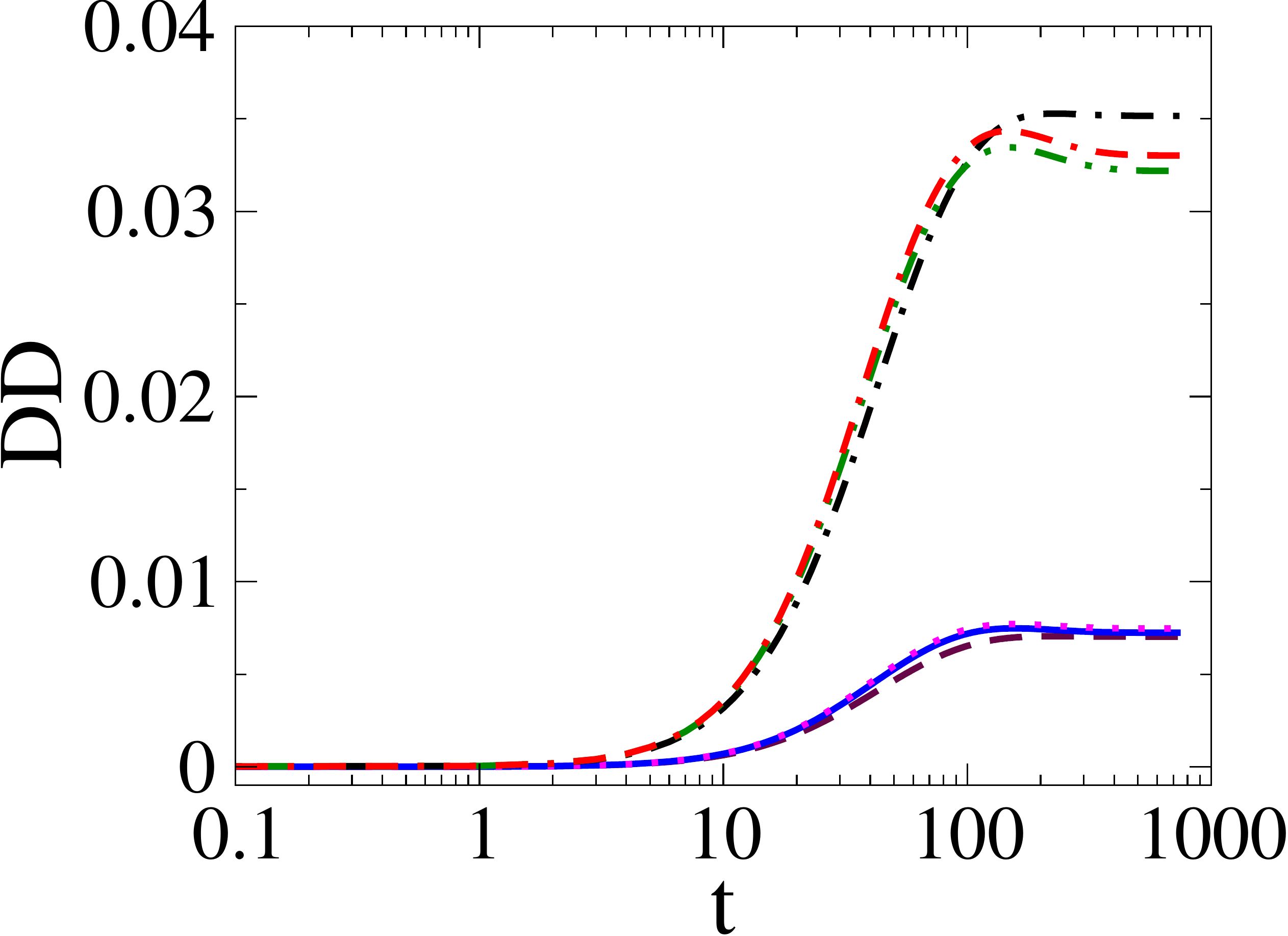}
  \caption{$\sigma_r=1$}
  \label{fgr:thnumddfl1}
\end{subfigure}
\begin{subfigure}{.32\textwidth}
  \centering
  \includegraphics[width=1\textwidth]{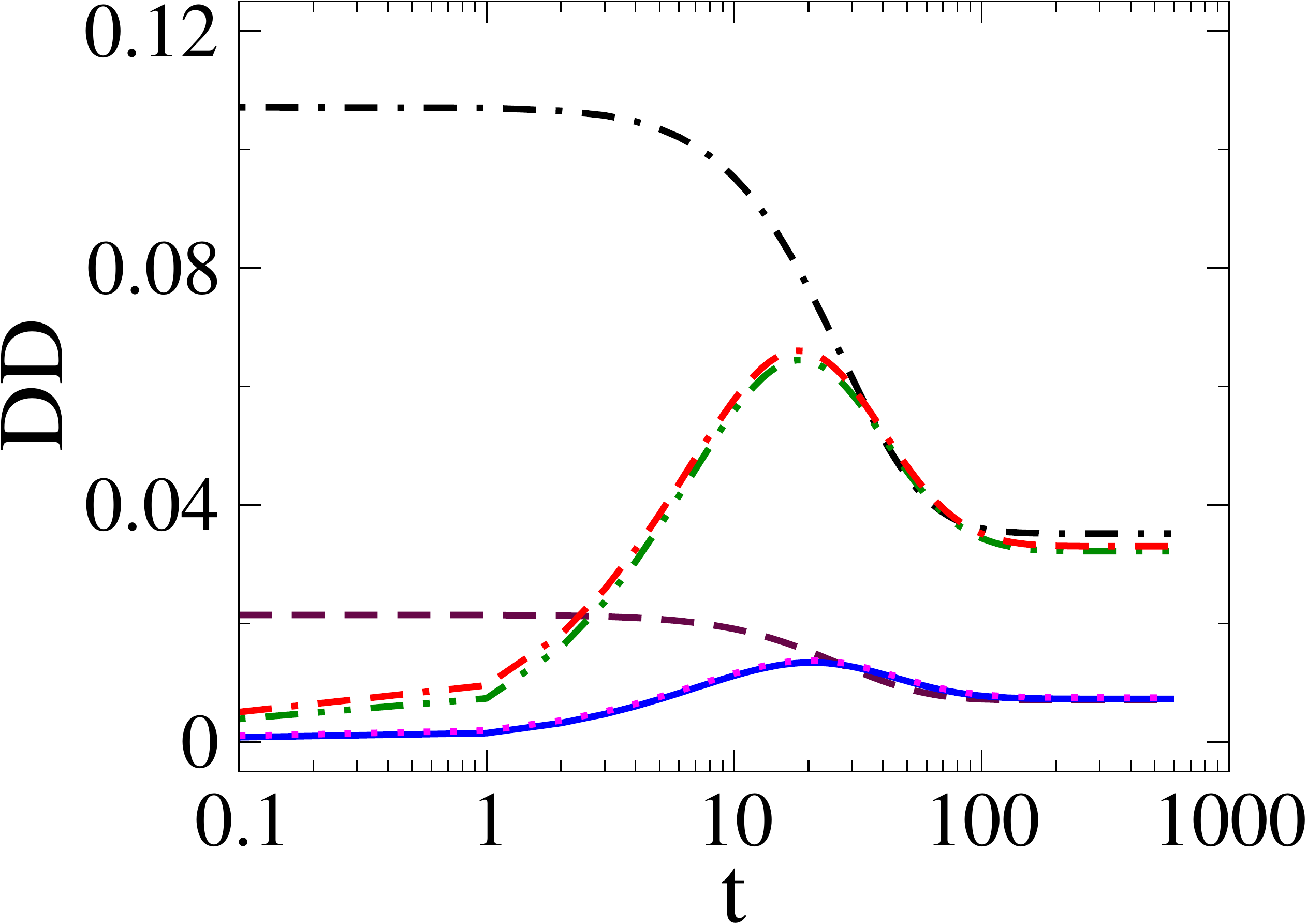}
  \caption{$\sigma_r=10$}
  \label{fgr:thnumddfl10}
\end{subfigure}
\begin{subfigure}{.32\textwidth}
  \centering
  \includegraphics[width=1\textwidth]{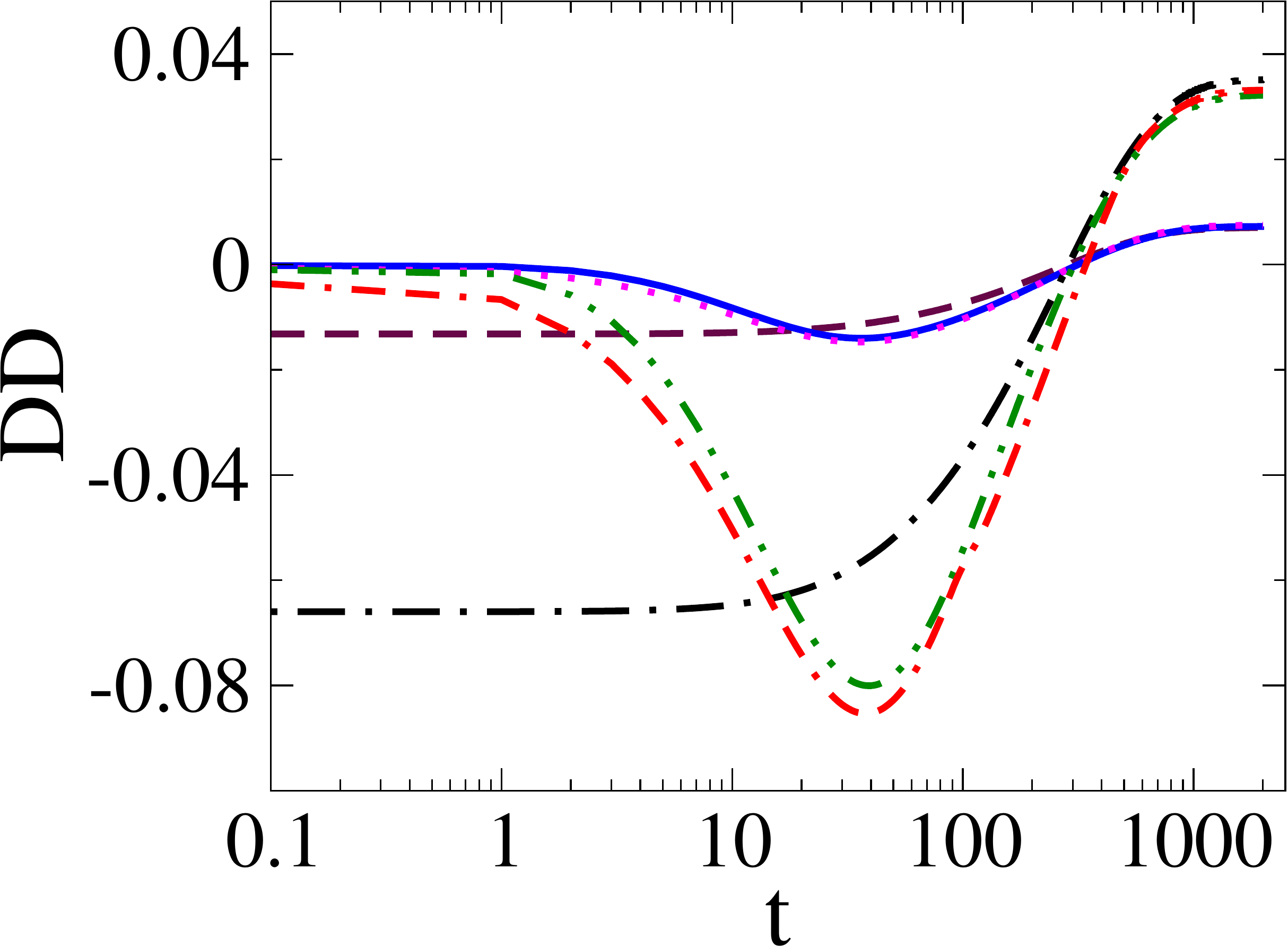}
  \caption{$\sigma_r=0.1$}
  \label{fgr:thnumddfl0p1}
\end{subfigure}
\caption{Comparison of dynamics obtained from SEM, BEM-SM and BEM-CM. At $Ca=0.01$ dynamics are represented by line-style (\textcolor{brown}{$\pmb{--}$}) for SEM, (\textcolor{magenta}{$\pmb{\cdots}$})  for BEM-SM and (\textcolor{blue}{$\pmb{\mi}$}) for BEM-CM. At $Ca=0.05$ dynamics are represented by line-style (\textcolor{black}{$\pmb{-\cdot-}$}) for SEM, (\textcolor{red}{$\pmb{--\cdot}$})  for BEM-SM and (\textcolor{forestgreen}{$\pmb{-\cdot\cdot}$}) for BEM-CM.}
\label{fgr:thnumddf}
\end{center}
\end{figure}

Dynamics of deformation of a capsule with Skalak membrane obtained from analytical theory considering simplified electrostatic model (SEM), boundary integral method considering simplified capacitor model (BEM-SM) and boundary integral method considering capacitor model (BEM-CM) are compared at small capillary numbers ($Ca=0.01$ and $Ca=0.05$) for the validation of the BEM-CM which is used to study the dynamics of capsule deformation in this work. Considering $\epsilon_r=1$, $\hat C_m=50$ and $\hat G_m=0$ dynamics are compared for $\sigma_r=1$, $\sigma_r=10$ and $\sigma_r=0.1$ in \cref{fgr:thnumddfl1}, \cref{fgr:thnumddfl10} and \cref{fgr:thnumddfl0p1}, respectively. 

\subsection{Importance of incorporating membrane bending}\label{sec:bending}
\begin{figure}[H]
\centering
  \includegraphics[width=0.4\textwidth]{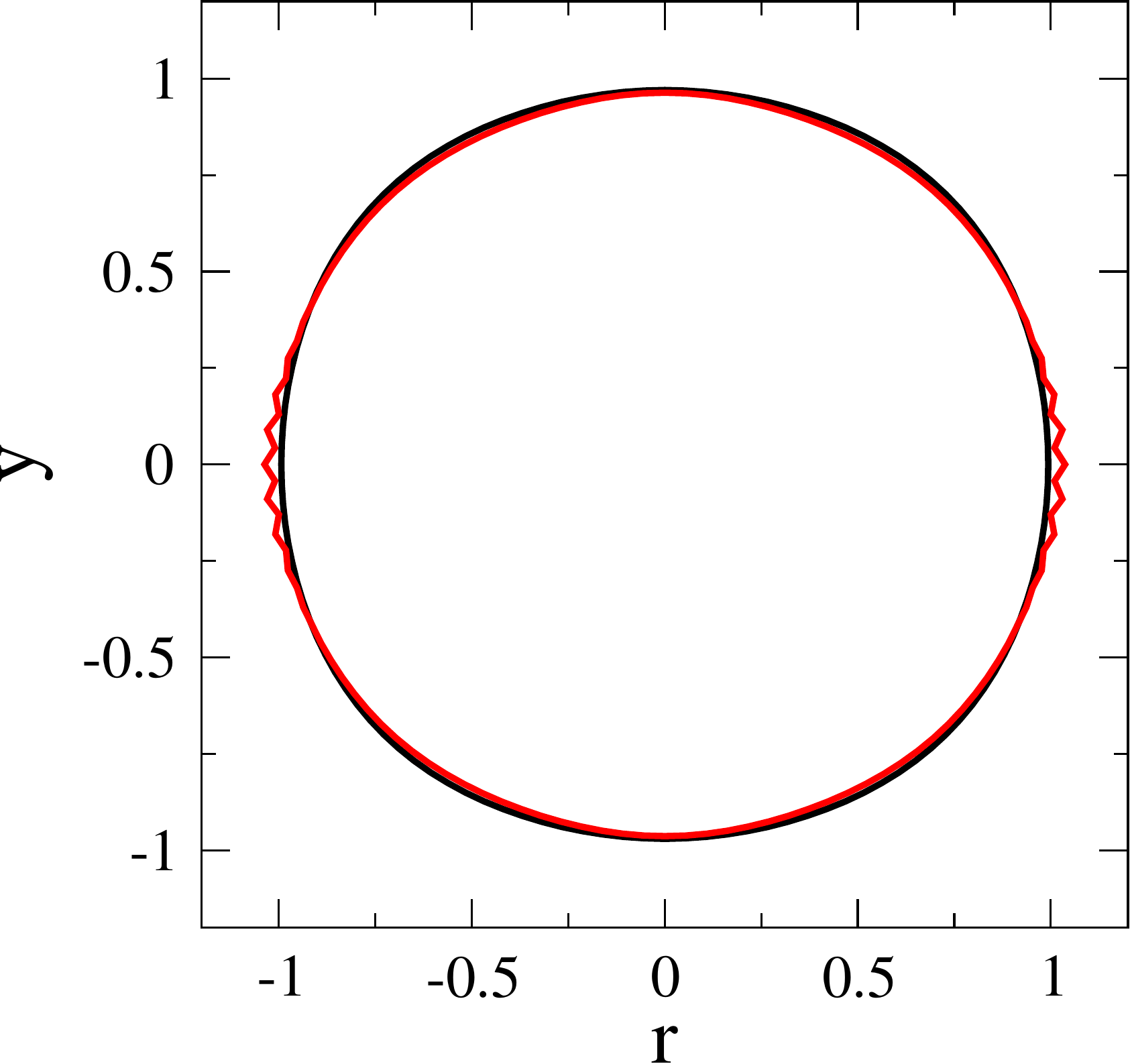}
  \caption{Effect of membrane bending rigidity on the deformation of a capsule with Skalak membrane considering $\sigma_r=0.1$ and $Ca=0.45$. Continuous curve represents shape obtained from the boundary integral simulation without considering effect of bending rigidity and the dash-dotted curve represents the same with $\kappa_b=0.001$ at $t=196$.}
  \label{fgr:bendingeffect}
\end{figure}

In the present study, the effect of bending rigidity is clearly demonstrated in \cref{fgr:bendingeffect}. Incorporating bending forces eliminate the formation of wiggles. However, it should be mentioned that for deformations with nominal curvatures, it was observed that the results of simulations with and without bending did not have significant differences.

\subsection{Dynamics of capsule with neo-Hookean membrane}
\begin{figure}[H]
\begin{center}
\begin{subfigure}{.22\textwidth}
  \centering
  \includegraphics[width=1\textwidth]{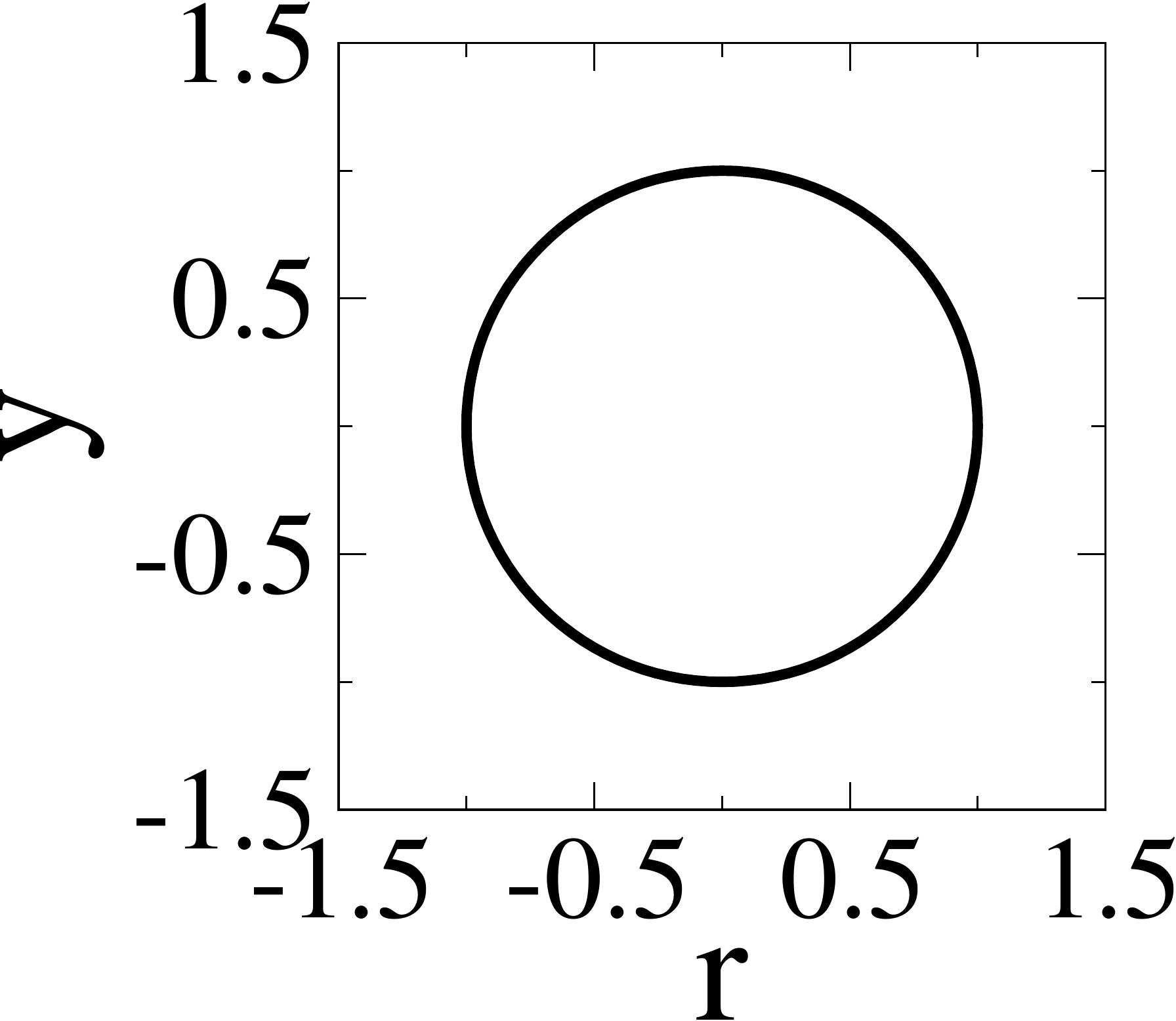}
  \caption{$t=0$}
  \label{fgr:shapel10nha}
\end{subfigure}
\begin{subfigure}{.22\textwidth}
  \centering
  \includegraphics[width=1\textwidth]{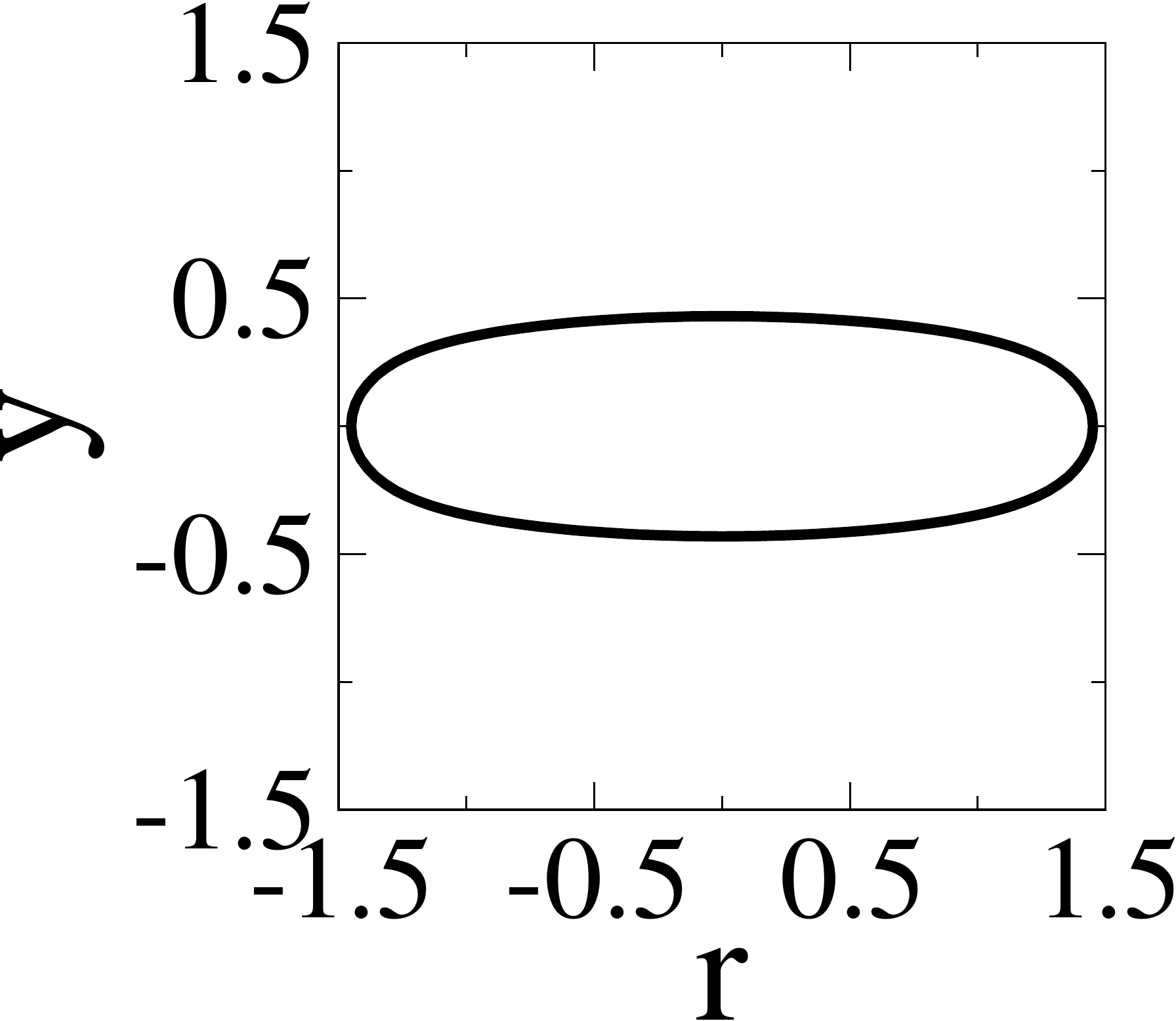}
  \caption{$t=32$}
  \label{fgr:shapel10nhb}
\end{subfigure}
\begin{subfigure}{.22\textwidth}
  \centering
  \includegraphics[width=1\textwidth]{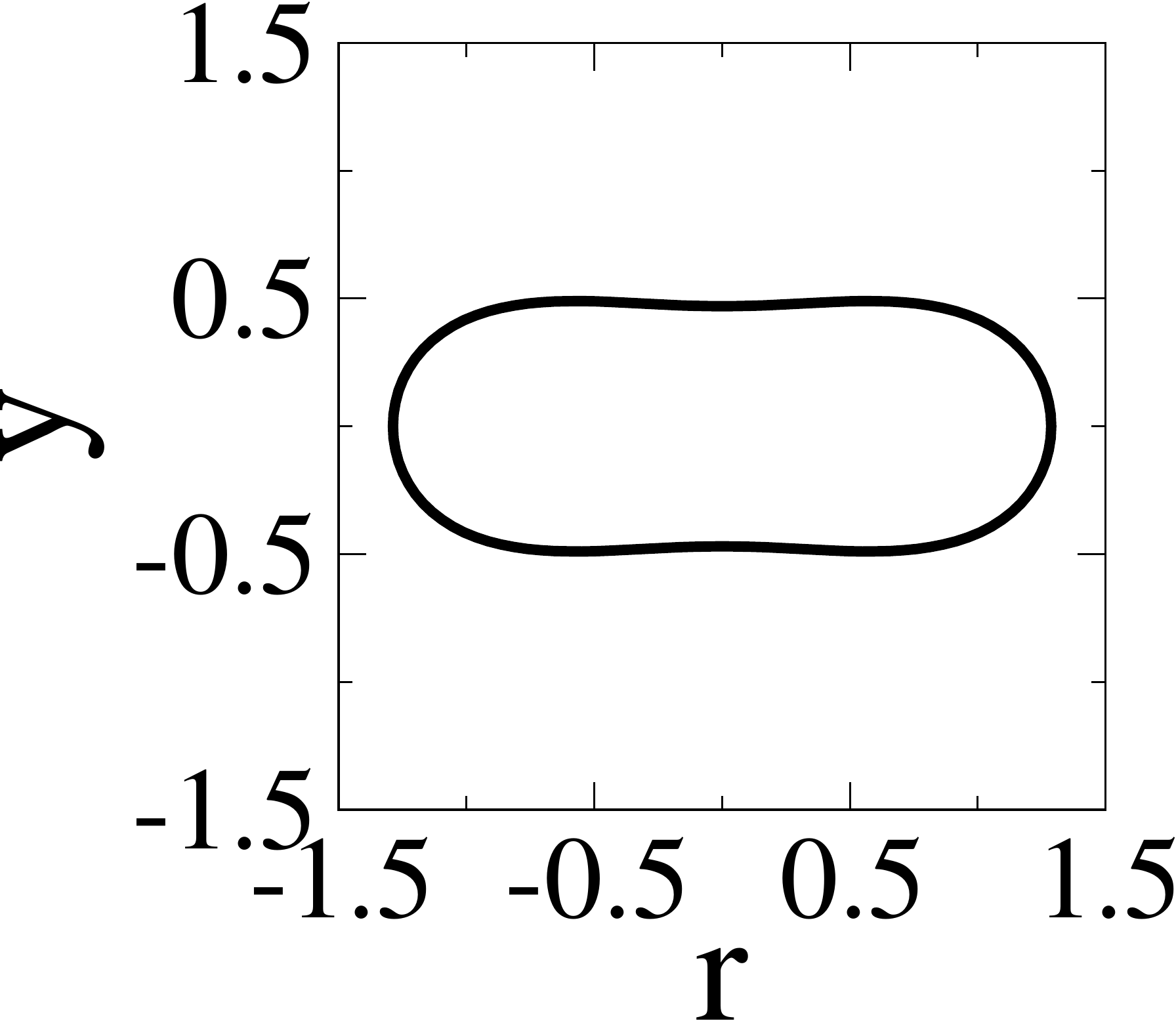}
  \caption{$t=75$}
  \label{fgr:shapel10nhc}
\end{subfigure}
\begin{subfigure}{.22\textwidth}
  \centering
  \includegraphics[width=1\textwidth]{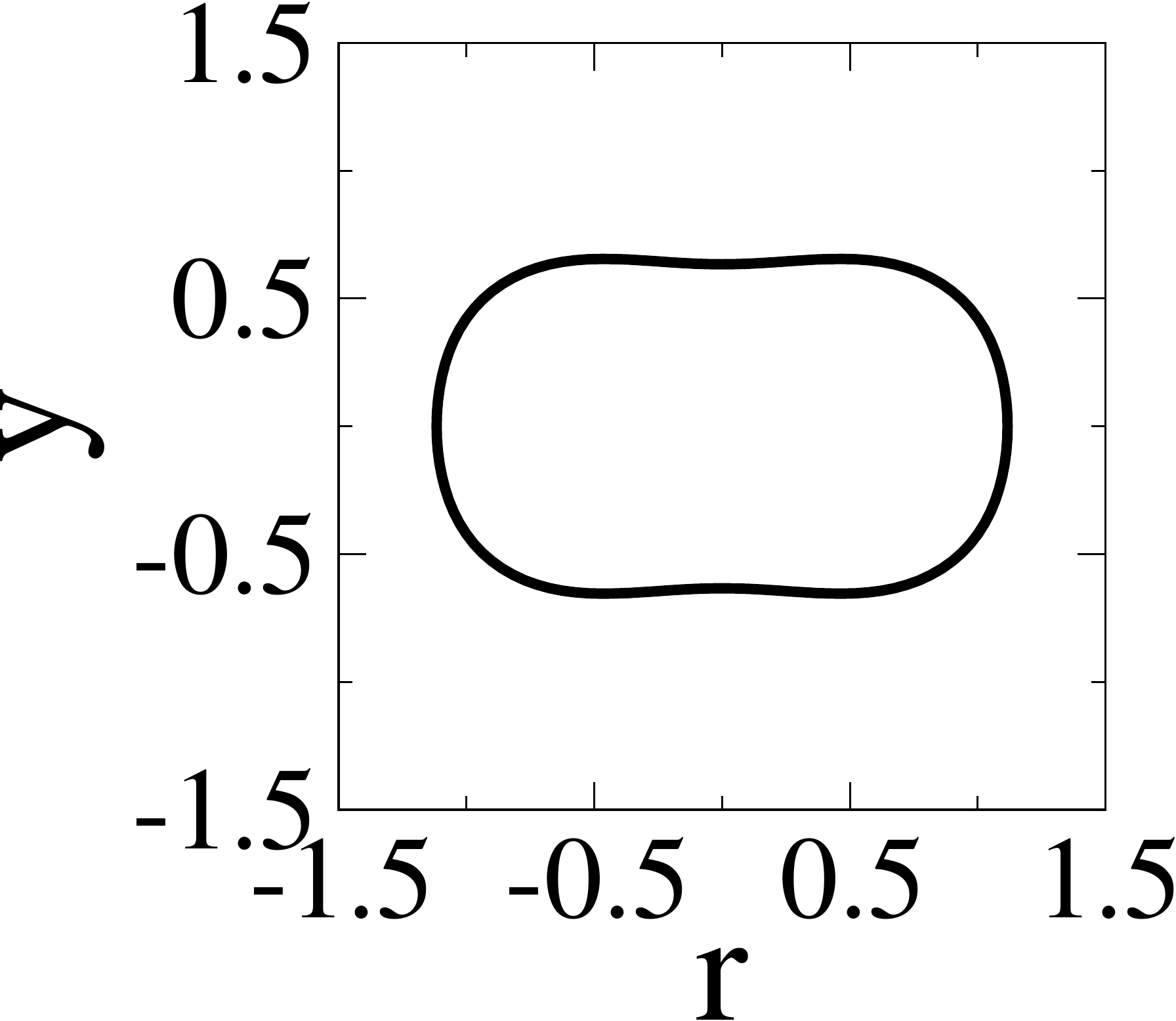}
  \caption{$t=120$}
  \label{fgr:shapel10nhd}
\end{subfigure}
\begin{subfigure}{.22\textwidth}
  \centering
  \includegraphics[width=1\textwidth]{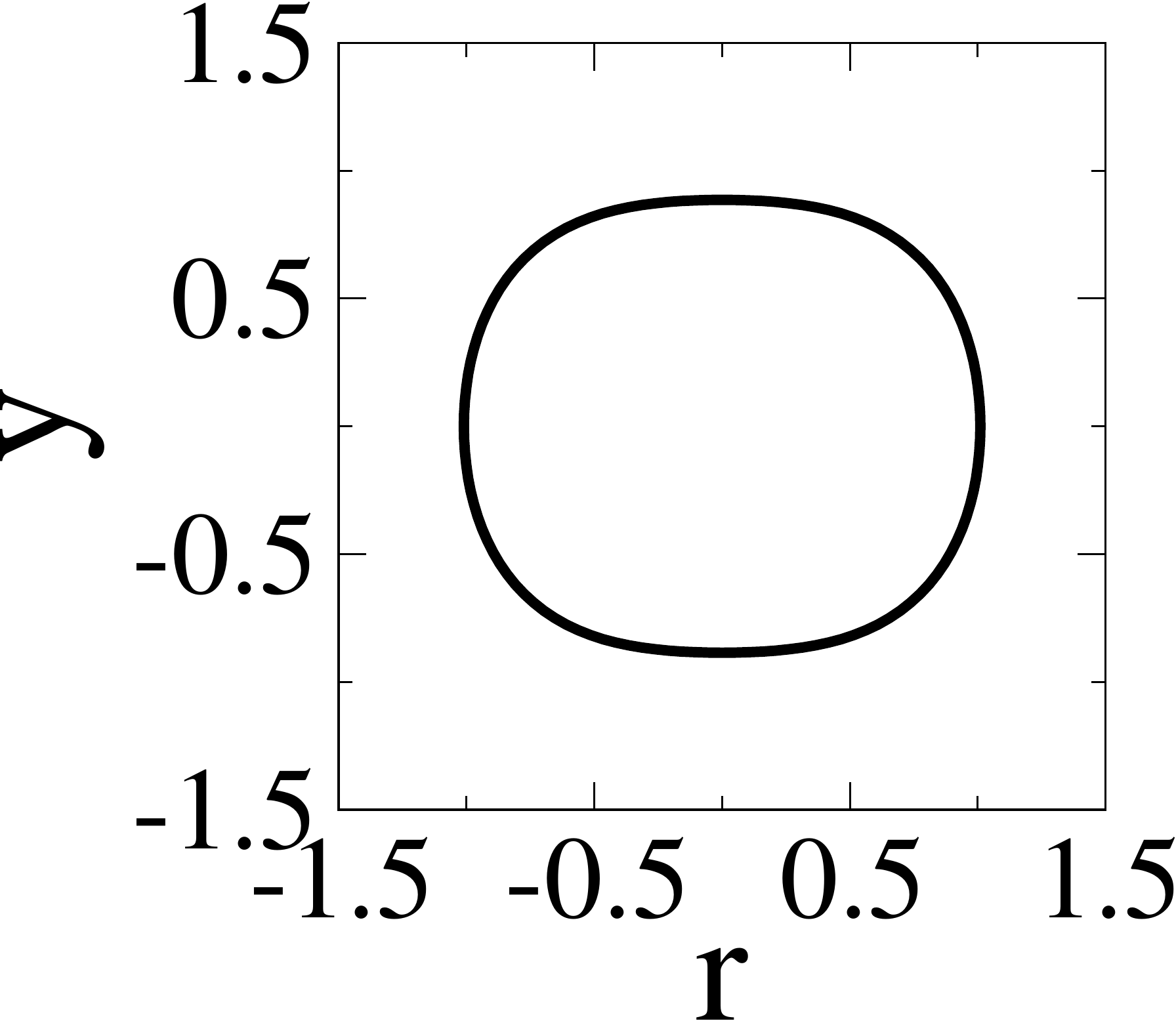}
  \caption{$t=175$}
  \label{fgr:shapel10nhe}
\end{subfigure}
\begin{subfigure}{.22\textwidth}
  \centering
  \includegraphics[width=1\textwidth]{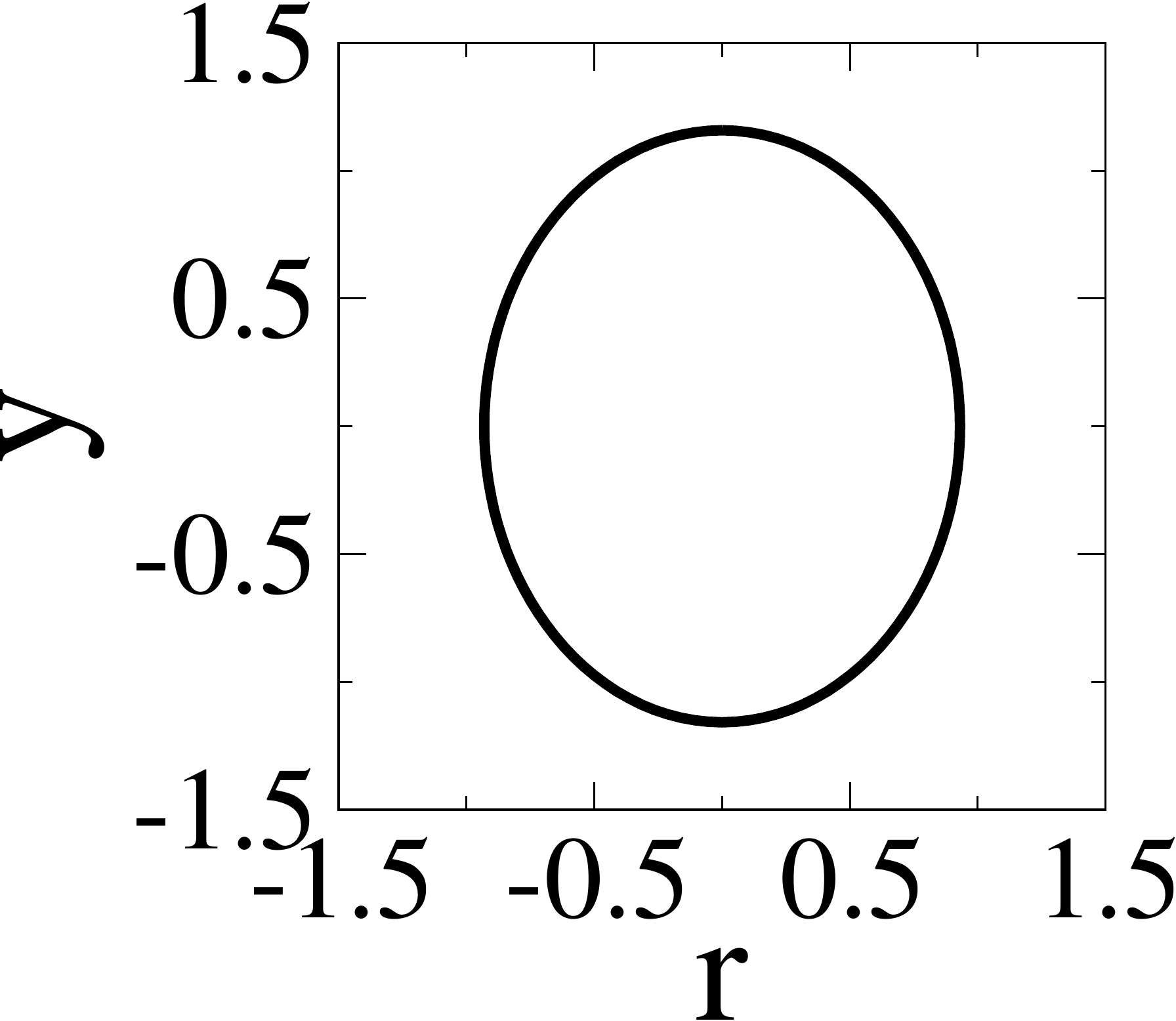}
  \caption{$t=\infty$}
  \label{fgr:shapel10nhf}
\end{subfigure}
\caption{Shape evolution of a capsule with neo-Hookean membrane at $\sigma_r=0.1$ for $Ca=0.25$.}
\label{fgr:shapel10nh}
\end{center}
\end{figure}

Capsules with a neo-Hookean membrane show very similar intermediate shapes (\cref{fgr:shapel10nha,fgr:shapel10nhb,fgr:shapel10nhc,fgr:shapel10nhd,fgr:shapel10nhe,fgr:shapel10nhf}) as observed for capsules with a Skalak membrane (\cref{fgr:lcal0p1a,fgr:lcal0p1b,fgr:lcal0p1c,fgr:lcal0p1d,fgr:lcal0p1e,fgr:lcal0p1f}) at $Ca=0.25$ and $\sigma_r=0.1$. From the simulated shapes it can be observed that a neo-Hookean capsule passes through the squaring (\cref{fgr:shapel10nhd,fgr:shapel10nhe}) to a steady state prolate shape (\cref{fgr:shapel10nhf}). 

\end{appendices}

%\correctpage
% \printnomenclature[4cm]

\end{document}